\pdfoutput=1
\pdfmapfile{+stix2.map}
\documentclass[12pt, twoside]{report}

\usepackage{color,slashed,booktabs,braket,bm,authblk,graphicx,bbold, adjustbox}
\usepackage{amsmath}
\usepackage{feynmp-auto}
\usepackage{hyperref}
\usepackage[a4paper,width=150mm,top=30mm,bottom=30mm,bindingoffset=6mm]{geometry}

\usepackage{fancyhdr}
\usepackage{stix2}
\usepackage[T1]{fontenc}
\usepackage[utf8]{inputenc} 
\DeclareUnicodeCharacter{2212}{-}

\usepackage{CJKutf8}

\usepackage[sort&compress,numbers, merge]{natbib}
\usepackage[nottoc]{tocbibind}

\definecolor{azure}{rgb}{0.0, 0.5, 1.0}
\definecolor{green}{rgb}{0.0, 0.5, 0.0}
\hypersetup{
  linktoc=page,
  colorlinks = true,
  linkcolor = azure,
  citecolor = azure,
}

\usepackage{afterpage}

\newcommand\blankpage{
  \null
  \thispagestyle{empty}
  \addtocounter{page}{-1}
  \newpage
}

\newcommand{\vev}[1]{\left\langle #1\right\rangle}

\newcommand\Order{\mathop{\mathcal{O}}}

\newcommand\unit[1]{\,\mathrm{#1}}

\newcommand\keV{\unit{keV}}
\newcommand\MeV{\unit{MeV}}
\newcommand\GeV{\unit{GeV}}
\newcommand\TeV{\unit{TeV}}

\newcommand\Msun{M_\odot}

\newcommand\bcm[3]{\left(\frac{#1}{#2}\right)^{#3}}

\begin{document}


\begin{titlepage}
  \begin{center}
    \vspace*{1cm}
    
    \Large
    Doctoral Dissertation\\
    \vspace{0.2cm}
    \begin{CJK}{UTF8}{ipxg}
      博士論文
    \end{CJK}

    \vspace{1.5cm}
    \LARGE
    \textbf{Thermal Evolution of Neutron Stars as a Probe of Physics beyond the Standard Model}
    
    \vspace{0.5cm}
    \large
    \begin{CJK}{UTF8}{ipxg}
      （中性子星の熱的進化を用いた標準模型を超える物理の探索）
    \end{CJK}
    
    \large
    \vspace{1.5cm}
    A Dissertation Submitted for the Degree of Doctor of Philosophy\\
    \vspace{0.2cm}
    December 2019

    \large
    \vspace{0.2cm}
    \begin{CJK}{UTF8}{ipxg}
      令和元年12月　博士（理学）申請
    \end{CJK}

    \large
    \vspace{1.5cm}
    Department of Physics, Graduate School of Science,\\
    The University of Tokyo

    \large
    \vspace{0.3cm}
    \begin{CJK}{UTF8}{ipxg}
    東京大学 大学院 理学系研究科\\
    物理学専攻
    \end{CJK}

    \vspace{2cm}
    \LARGE
    \textbf{Keisuke Yanagi}
    \vspace{0.2cm}

    \begin{CJK}{UTF8}{ipxg}柳　圭祐\end{CJK}

    \vfill

  \end{center}
\end{titlepage}

\afterpage{\blankpage}
\clearpage

\begin{abstract}
  The physics beyond the standard model (BSM) has not appeared in extensive experimental and observational searches.
  Although the problems in the standard model strongly suggest the existence of a more fundamental theory, its detail is still unclear.
  In such a situation, it becomes more and more important to further extend the area of new physics search beyond the conventional ones.
  In this dissertation, we study the thermal evolution of neutron stars (NSs) as a probe of new physics.

  The NS cooling is well studied and the current standard theory is successful in explaining many of the observed surface temperatures of NSs.
  If a new particle is light enough to be created in NSs, and has small interactions with ordinary matter, its emission can be an extra cooling source.
  If this enhancement of cooling spoils the success of the standard cooling, the model should be excluded.
  An important candidate of such a new particle is axion, which dynamically solves the strong CP problem of the standard model.
  The axion-nucleon coupling induces the axion emission from the NS core, and it affects the cooling of young NSs. 
  The stringent constraint is obtained by comparing the predicted cooling curve to the observed surface temperature of the NS in the supernova remnant of Cassiopeia A (Cas A).
  We constrain the axion models by taking account of the temperature evolution in the whole life of the Cas A NS.
  The resultant limit on the axion decay constant is $f_a \gtrsim 10^8\GeV$, which is comparable to the existing limit from SN1987A.

  The NS heating provides yet another example of probing new physics in NSs.
  In particular, it is known that the dark matter (DM) accretion leads to the heating of old NSs.
  It occurs through the DM accumulation in NSs by the scattering with nucleons, and their subsequent annihilation inside the star.
  The surface temperatures of old NSs ($t\gtrsim 10^7\unit{yr}$) are predicted to be $T_s = (2-3)\times 10^3\unit{K}$, which is stark contrast to the standard cooling theory.
  Thus the measurement of the surface temperatures of old NSs can provide a hint/constraint for DM models.
  This conclusion is, however, drawn with the assumption that there is no other heating source.
  In fact, it is known that the non-equilibrium beta process induces the late-time heating through the imbalance of chemical potentials among nucleons and leptons: this is called \textit{rotochemical heating} and is inevitable for pulsars.
  The rotochemical heating typically predicts $T_s \sim 10^{5-6}\unit{K}$ for old NSs, and hence can be much stronger than the DM heating.
  This possibility has been overlooked in the studies of DM heating.
  In this dissertation, we address the condition in which DM heating dominates the rotochemical heating.
  For that purpose, we first perform detailed analysis of the rotochemical heating; extending the previous studies, we perform numerical calculation with both the proton and neutron pairing gaps, and compare the predictions to the observations.
  Then we include the DM heating, and investigate whether the DM heating is visible or not, varying the NS parameters.
  We find that the DM heating can still be detected in old ordinary pulsars.

  These results for axion and DM demonstrate that the thermal evolution of NSs is useful to probe the physics beyond the standard model.
  Further development in nuclear theory, experiments and observations will help reduce the uncertainties of these predictions.
\end{abstract}


\clearpage
\setcounter{tocdepth}{2}

\chapter*{Acknowledgment}
I would like to express my greatest appreciation to my supervisor, Koichi Hamaguchi, for helping me in all aspects during my graduate course.
He has always encouraged me to start new things and collaborate with others.
I am truly happy to have spent years of my graduate course in his group.

I would also like to express my sincere gratitude to Natsumi Nagata for a lot of helpful advice and suggestions in our collaborative studies of neutron stars and particle physics.
Without his great insight in physics research, I could not have completed those works.
I have learned a lot from him about how to tackle research problems.

I would like to show my great gratitude to my collaborator, Jiaming Zheng, for discussing with me and answering many questions during the collaboration.

I am also grateful to all my collaborators, who do not have direct relation to the works which this dissertation is based on, but have helped me on many occasions in my graduate course.
I would also like to thank all the members in the particle physics theory group at University of Tokyo. 

I would like to thank Teruaki Enoto, Kenji Fukushima, Kazuhiro Nakazawa, Hideyuki Umeda, and Satoshi Yamamoto for valuable discussions and suggestions when we tried to begin our neutron star project.

The works on which this dissertation bases were supported by JSPS KAKENHI Grant Number JP18J10202.

\clearpage
\tableofcontents
\clearpage

\chapter{Introduction}
\label{chap:introduction}


The physics beyond the standard model (BSM) of particle physics has been pursued for several decades in all over the world.
Theorists have proposed many attractive models to resolve the puzzles in the standard model (SM), and experimentalists have searched for predicted new particles in various ways including huge collider experiments.
Despite such tremendous efforts in both theoretical and experimental sides, we have not discovered any BSM particle. 
Many models once thought to be promising are now excluded, and the scale of new physics is pushed higher and higher.
We of course need to continue the current experimental searches, but even after the future upgraded experiments, we might end up without any discovery.
Therefore, it is also necessary to look for new directions to completely study the attractive models and their parameter spaces.

Thermal evolution of neutron stars (NSs) provides such an opportunity.
As is well known, a NS is a compact astrophysical object that consists mainly of degenerate neutrons.
It is as heavy as the sun, and the total mass is enclosed in a small sphere of radius $R\sim10\unit{km}$.
The mass density reaches $\sim 10^{12}\unit{kg/cm^3}$, and hence NSs offer ultra-dense environments which cannot be achieved on the earth.
The study of the NS cooling began in the mid 60's before the first discovery of the pulsar~\cite{1966CaJPh..44.1863T}, and has been updated until today along with the development of the theory of microphysics in the ultra-dense environment (see, e.g., Refs.~\cite{Yakovlev:2004iq, Yakovlev:2007vs}).
There are also a few tens of measurements of surface temperatures, and thus this NS cooling theory can now be tested against the observations.
The current standard theory of the NS cooling, which is dubbed as \textit{minimal cooling}, properly takes account of the effects of nucleon superfluidity, and it can explain many observed surface temperatures~\cite{Page:2004fy, Gusakov:2004se, Page:2009fu}.
If a BSM particle couples to the NS matter, it can alter the prediction of this standard cooling theory.
Comparing the prediction to the observed temperatures, we can constrain models of new physics.
This is the approach that we discuss in this dissertation.

There are studies that probe the new physics models using NS thermal evolution.
In the standard cooling, a young NS cools down by the various neutrino emission processes (see Ref.~\cite{Yakovlev:2000jp} for a comprehensive review).
If a new light particle couples to nucleons, its emission can also contribute to the cooling.
An important such candidate is axion. It is predicted associated with the spontaneous breaking of PQ symmetry, which resolves the strong CP problem in the SM~\cite{Weinberg:1977ma, Wilczek:1977pj, Peccei:1977hh, Peccei:1977ur}.
Axions are pseudo-scalar particles and weakly couple to nucleons. They are emitted from NSs by the nucleon-axion bremsstrahlung~\cite{Iwamoto:1984ir, Nakagawa:1987pga, Nakagawa:1988rhp, Iwamoto:1992jp, Umeda:1997da, Paul:2018msp} and the nucleon Cooper pair reformation process~\cite{Leinson:2014ioa, Sedrakian:2015krq}.
These axion emission processes indeed enhance the cooling rate if the axion decay constant $f_a$, which characterize the axion-nucleon interaction strength, is sufficiently small.
Comparing the predicted temperature to the observed ones of several young and middle-aged NSs, Ref.~\cite{Sedrakian:2015krq} obtained the bound $f_a > \Order(10^7)\GeV$.

The thermal evolution of NSs can also be used to search for the dark matter (DM).
Although the existence of DM is established in the cosmology and astrophysics, its non-gravitational nature is poorly known so far.
A popular DM candidate is weakly interacting massive particles (WIMPs), which have weak-scale mass and weak interaction with ordinary matter (see Refs~\cite{Roszkowski:2017nbc, Arcadi:2017kky} for recent reviews).
WIMPs are naturally realized in the well-motivated models such as the supersymmetric SM, and the WIMP abundance is determined by the famous freeze-out mechanism independently of the initial condition of the universe.
For many years, DM direct search experiments have been conducted to detect WIMP scattering off to ordinary matter.
Nevertheless we do not have any signal of such events (see, e.g., the latest XENON1T results~\cite{Aprile:2018dbl}), and the WIMP parameter space is more and more constrained.
This WIMP-ordinary matter scattering is also probed in NSs~\cite{Kouvaris:2007ay, Bertone:2007ae, Kouvaris:2010vv, deLavallaz:2010wp}. 
WIMPs accrete onto a NS by its strong gravity, and they can scatter nucleons by the weak interaction; the conventional WIMPs, whose mass is at the weak scale, lose its initial kinetic energy just by a single scattering, and are eventually trapped by the NS gravitational potential.
The energy of WIMPs is transferred to the star through the scattering and/or annihilation~\cite{Kouvaris:2007ay, Baryakhtar:2017dbj}.
This affects the thermal evolution of old NSs ($t\gtrsim 10^7\unit{yr}$), whose dominant cooling source is surface photon emission, and without DM heating the surface temperature drops much below $10^3\unit{K}$.
With the DM heating, however, their surface temperatures are predicted to be $T_s = \Order(10^3)\unit{K}$ if the DM-nucleon scattering cross section is $\sigma \gtrsim 10^{-45}\unit{cm^2}$~\cite{Kouvaris:2007ay, Baryakhtar:2017dbj}, and thus we can probe the WIMP DMs by the future infrared telescopes~\cite{Baryakhtar:2017dbj}.
See Refs.~\cite{Bramante:2017xlb, Raj:2017wrv, Chen:2018ohx, Bell:2018pkk, Camargo:2019wou, Bell:2019pyc, Acevedo:2019agu} for recent discussion of the application of the DM heating.

In this dissertation, we will further develop the search of axions and WIMP DMs through NS thermal evolution.
We first consider the constraint on axion from the young NS in the supernova remnant of Cassiopeia A (Cas A)~\cite{1999IAUC.7246....1T}.
It provides a stringent constraint because its about 10-year observations provide not only the temperature itself, but also its cooling rate~\cite{Heinke:2010cr}.
By properly incorporating all the relevant processes of axion emission and taking the whole life of the Cas A NS into account, we obtain the bound of $f_a\gtrsim 10^8\GeV$, which is comparable to that from SN1987A.
Regarding the DM search, the previous studies of DM heating do not take account of potential heating sources.
Heating sources other than DM have been discussed in the astrophysics community (see, e.g., Ref.~\cite{Gonzalez:2010ta}), and also recent observations suggest the presence of old warm NSs which cannot be explained by the standard cooling theory~\cite{Kargaltsev:2003eb, Durant:2011je, Rangelov:2016syg, Mignani:2008jr, Pavlov:2017eeu}.
Among the proposed heating mechanisms, the \textit{rotochemical heating}, heating caused by the rotationally induced imbalance among chemical potentials, should be considered in all the pulsars since it does not assume any exotic physics~\cite{Reisenegger:1994be}.
If this heating is stronger than the DM heating, we have no hope to use NSs as DM detectors.
To address this issue, we first show the importance of simultaneous inclusion of both proton and neutron superfluidity in the rotochemical heating.
Then we will clarify the condition in which the DM heating surpasses the rotochemical heating.

The organization of this dissertation is as follows:
\begin{itemize}
\item
  Chapter~\ref{chap:neutron-star-cooling} is a review of the standard cooling theory of NSs. From Sec.~\ref{sec:structure-neutron-star} to \ref{sec:neutrino-emis}, we explain the basics necessary for the calculation of the cooling, with particular emphasis on the roles of nucleon superfluidity.
  In Sec.~\ref{sec:standard-cooling}, we integrate them and solve the time evolution equations of the NS temperature, showing that the so-called minimal cooling paradigm explains many observed surface temperatures consistently.

\item
  In Chapter~\ref{chap:limit-axion-decay}, we use the minimal cooling theory to constrain the axion model.
  We will show the bound from the Cas A NS is as strong as that from SN1987A.
  This chapter is based on the author's work~\cite{Hamaguchi:2018oqw}.
\item
  In Chapter~\ref{chap:neutron-star-heating}, we study the rotochemical heating.
  We incorporate the superfluidity of both protons and neutrons, and compare the theoretical predictions to the observed surface temperatures.
  It turns out that the rotochemical heating explains old warm NSs which are much hotter than the prediction of the standard cooling.
  This chapter is based on the author's work~\cite{Yanagi:2019vrr}.
\item
  In Chapter~\ref{chap:dm-heating-vs-roto}, we reevaluate the DM heating in the presence of the rotochemical heating.
  We will show that the signature of DM heating can still be detected in old ordinary pulsars, while it is concealed by the rotochemical heating for old millisecond pulsars.
  This chapter is based on the author's work~\cite{Hamaguchi:2019oev}. 
\item
  Chapter~\ref{chap:conclusion} is devoted to the conclusion of the dissertation.
\end{itemize}

\chapter{Neutron star cooling}
\label{chap:neutron-star-cooling}
In this chapter, we review the standard cooling theory of NSs, which is necessary to understand the results in the following chapters.
In Sec.~\ref{sec:structure-neutron-star}, we briefly review the equation of state (EOS), and its consequence.
In Sec.~\ref{sec:microphysics}, we explain the Fermi liquid theory and nucleon superfluidity, which are very important for the thermal evolution.
Then in Sec.~\ref{sec:thermodynamics} and \ref{sec:neutrino-emis}, we provide thermodynamic quantities such as specific heat or luminosities.
Finally in Sec.~\ref{sec:standard-cooling}, we use all these results to solve the thermal evolution equations, and compare the resultant surface temperatures to the observed ones.

\section{Structure of neutron stars}
\label{sec:structure-neutron-star}

\subsection{Overview}
\label{sec:overview}
NSs are the last stage of massive stars. It is widely accepted
that when the mass of the star is $(3-6)\Msun \lesssim M \lesssim (5-8)\Msun$, it collapses to a NS~\cite{Shapiro:1983du}.
A NS is a compact object whose typical mass is $1.4\Msun$ and typical radius is $10\unit{km}$.
The dominant component of NSs is degenerate neutrons, and their degenerate pressure largely support the NS against the gravity.

The static structure of NSs is calculated by solving the Einstein equations. We usually assume the spherical symmetry, with which it is called Tolman-Oppenheimer-Volkof (TOV) equation.
The property of matter components is provided by the form of equation of state (EOS).
Although there are uncertainties in EOS due to uncertain nuclear interaction, we have qualitative understanding of the large part of the inner structure.
In this subsection, we provide the overview of the structure of NSs.
We leave the detailed discussion of TOV equation and EOS in Sec.~\ref{sec:tov-equation} and \ref{sec:eos} respectively. 

\paragraph{Atmosphere}

The outer most layer of a NS is called \textit{atmosphere}, consisting of gas elements.
They emit thermal photons to the outer space, which are finally observed on the earth.
We infer the surface luminosity and temperature by fitting this photon flux.
Since the spectral shape depends on the composition of the atmosphere, building the atmosphere models is important to interpret the observation.
We discuss it in Sec.~\ref{sec:surface-temperature}.
In addition, the surface photon emission is a major source of cooling for middle-aged and old NSs, and will be discussed in Sec.~\ref{sec:envelopew}.

\paragraph{Envelope}

Below the atmosphere, there is a thin region called \textit{envelope}, where matters are not fully ionized.
The envelope works as a thermal insulator which shields the surface from the hot interior; there is a large temperature gradient between the top and bottom of the envelope.
This gradient is determined by the amount of light elements such as H or He.
The photon cooling is thus dependent on the composition of the envelope.
See Sec.~\ref{sec:envelopew}.

\paragraph{Outer crust}

The \textit{outer crust} is about $ 0.3 \unit{km}$ thick, and consists of localized nuclei and degenerate electrons.
In terms of the mass density, it extends from $\rho \sim 10^{7} \unit{g/cm^3}$ to $4\times 10^{11}\unit{g/cm^3}$.
At $\rho \sim 10^7\unit{g/cm^3}$, ${}^{56}\mathrm{Fe}$ is the dominant species, sitting in the sea of degenerate electrons.
As the density becomes larger, the electron capture proceeds, and more neutron-rich nuclei, such as ${}^{62}\mathrm{Ni}$, appear.
This continues to $\rho \sim 4\times 10^{11}\unit{g/cm^3}$, and constituent nuclei become more and more neutron-rich.

\paragraph{Inner crust}

When the density becomes as large as $\rho \sim 4\times 10^{11}\unit{g/cm^3}$, part of neutrons created by electron capture cannot be bounded in a nucleus. 
This is called \textit{neutron drip},%
\footnote{The neutron dripline is studied in the experiments up to $Z=10$~\cite{Ahn:2019xgh}.}
and the system consists not only of electrons and nuclei but also of dripped neutrons.
This region is called \textit{inner crust}. It extends to $\sim 1\unit{km}$ below the outer crust.
At sufficiently low temperature (typically $T \lesssim 10^{10} \unit{K}$), dripped neutrons become ${}^1S_0$ \textit{superfluid} state.

\paragraph{Outer core}

For the density $\rho \gtrsim 2.8\times 10^{14}\unit{g/cm^3}$, the nuclei are broken into fluid consisting of neutrons, protons and electrons.
This threshold density is called nuclear saturation density, and as a number density, it is $n_0\sim 0.16\unit{fm^{-3}}$, which is a typical nucleon density inside a nucleus.
This region is called (outer) core.
Due to the attractive nuclear force, protons form Cooper pair and the core becomes the superconductor.
Neutrons also form Copper pairs, but the pairing type is different from that in the crust; in the crust $s$-wave interaction is responsible for the pairing, while in the core, $p$-wave is the dominant channel of attractive force.
These pairings affect the heat capacity and neutrino emissivity.
Muons are also produced in the high density region in the core where the electron chemical potential exceeds the muon mass.

\paragraph{Inner core}
In addition to nucleons and charged leptons (electrons and muons), pion, hyperon or quark matter may appear in the extremely dense region close to the NS center.
The condition for these exotic particles to appear is highly uncertain because the EOS for such high density is not well understood.
In Sec.~\ref{sec:eos}, we will briefly review the current status of the constraint on the EOS and its consequence in the NS interior.


\subsection{TOV equation}
\label{sec:tov-equation}

The static structure of a neutron star is determined by the Einstein equations.
The actual neutron star is rotating, but the centrifugal force is small compared to the gravitational force for a typical NS.%
\footnote{The ratio of the centrifugal force to the gravity is estimated as
  \begin{align}
    \label{eq:ratio-cent-grav}
    \frac{R^3\Omega^2}{GM} \sim 10^{-7}
    \bcm{1\unit{s}}{P}{2}
    \bcm{R}{10\unit{km}}{3}
    \bcm{1.4\Msun}{M}{}\,,
  \end{align}
  where $G$ is the gravitational constant and $\Omega$ is the angular velocity.
  This is much smaller than unity for a typical NS.
  }
Thus we consider the spherically symmetric solution of the Einstein equations. For matter components, we assume the energy-momentum tensor of the perfect fluid form:
\begin{align}
  \label{eq:energy-mom-tensor}
  T_{\mu\nu} = \rho u_\mu u_\nu +P(g_{\mu\nu} +u_\mu u_\nu)\,,
\end{align}
where $u_\mu$ is the fluid four-velocity, $\rho$ the energy density and $P$ the pressure.
Then the spherically symmetric metric is written as (e.g. Ref.~\cite{Wald:1984rg})
\begin{align}
  ds^2
  &=
    -e^{2\Phi(r)/c^2}c^2dt^2 + e^{2\Lambda(r)}dr^2 + r^2(d\theta^2+\sin^2\theta d\varphi^2)\,,
    \label{eq:int-metric}
\end{align}
It has two unknown functions $\Phi$ and $\Lambda$.
Thus in total, we have four functions to be determined: $\rho$, $P$, $\Phi$ and $\Lambda$.
The Einstein equation provides three independent equations:
\begin{align}
  e^{2\Lambda(r)}
   &=
     \left( 1-\frac{2Gm(r)}{c^2r} \right)^{-1}\,,
     \text{ where }
  m(r)
  =
    4\pi\int_0^r dr^\prime r^{\prime 2}\rho(r^\prime)\,,
     \label{eq:lambda-def}\\
  \frac{d\Phi(r)}{dr}
    &=
    G\frac{m(r) + 4\pi r^3 P(r)/c^2}{r^2}e^{2\Lambda(r)}\,,
    \label{eq:phi-eq}\\
  \frac{dP(r)}{dr}
    &=
    -G\left(\rho(r) + \frac{P(r)}{c^2}\right)\frac{m(r) + 4\pi r^3P(r)/c^2}{r^2}e^{2\Lambda(r)}\,.
    \label{eq:press-eq}
\end{align}
In particular, Eq.~\eqref{eq:press-eq} is called TOV equation.
In the non-relativistic limit, Eq.~\eqref{eq:phi-eq} becomes $d\Phi/dr \simeq Gm/r^2$ and Eq.~\eqref{eq:press-eq} $dP/dr \simeq -G\rho m/r^2$; the former is the ordinary Poisson equation, and the latter is familiar expression of the hydrostatic equilibrium condition in the Newtonian gravity.
We need another equation to solve these equations, and it is usually provided by EOS of the form $P = P(\rho)$.
We will discuss the EOS of NSs in the next subsection.

The metric obtained above is connected to the solution outside the star
\begin{align}
  \label{eq:schwarzschild}
  ds^2
  =
  -\left( 1-\frac{2GM}{c^2R} \right)c^2dt^2
  +\left( 1-\frac{2GM}{c^2R} \right)^{-1}dr^2
  +r^2(d\theta^2+\sin^2\theta d\varphi^2)\,,
\end{align}
where $M$ is the total mass of the NS.
Note that the Schwarzschild radius is $r_c = 2GM/c^2\sim
 3(M/\Msun)\unit{km}$, so the typical NS radius, $R\sim10\unit{km}$, is only
 a few times larger than $r_c$.

\subsection{Equation of state}
\label{sec:eos}
An EOS is necessary to solve the TOV equation.
The property of nearly isospin-symmetric nuclei close to the nuclear saturation density ($n_0\sim 0.16\unit{fm^{-3}}$) is well studied in many experiments.
On the other hand, the NS interior is highly neutron-rich environment; the typical proton fraction is only $Y_p \equiv n_p/n_B \sim 0.1$, where $n_B$ and $n_p$ are baryon and proton number density, respectively.
In addition, $n_B$ can be several times higher than $n_0$ in the core.
Thus, it is necessary to rely on a theoretical model to study NS bulk property.
The difficulty of building such a model from microphysics is due to the non-perturbative nature of the nuclear interaction.
There are a number of approaches to address this problem: see Refs.~\cite{Ozel:2016oaf, Lattimer:2015nhk, Oertel:2016bki} for recent reviews.

Given an EOS, one can integrate the TOV equation~\eqref{eq:press-eq} from the center to the surface.
Changing the central density $\rho_c$, one can obtain NSs with different mass and radius.
Thus different EOSs provide different $M$-$R$ relation, which is compared to the measurements. 
Since the gravity has to be supported by the pressure in a static NS, the \textit{stiff} EOSs, where the pressure tends to be higher than the \textit{soft} ones for a given density, realizes heavier NSs.
Thus every EOS predicts its own maximum NS mass depending on its stiffness; if a NS heavier than this maximum is observed, such an EOS is excluded.
The most stringent constraints are obtained from PSR J1614-2230 ($M=1.97\pm0.04\Msun$)~\cite{Demorest:2010bx}, PSR J0348+0432 ($2.01\pm0.04\Msun$)~\cite{Antoniadis:2013pzd} and PSR J0740+6620 ($2.14^{+0.20}_{-0.18}\Msun$)~\cite{Cromartie:2019kug}.
After these measurements, many EOSs are excluded, but we still have a number of candidates of EOSs: see Ref.~\cite{Ozel:2016oaf}.

Following Ref.~\cite{Page:2004fy}, we use the Akmal-Pandharipande-Ravenhall (APR) equation of state~\cite{Akmal:1998cf} in this dissertation.
This EOS uses the Argonne $v_{18}$ potential~\cite{Wiringa:1994wb} for the nucleon two-body interaction, which well fits the nucleon scattering data below $350\MeV$, and the Urbana model IX (UIX)~\cite{Pudliner:1995wk} for three nucleon interaction.
The special relativistic effect is incorporated up to the quadratic order.
The resultant EOS is called A18$+\delta v+$UIX${}^*$, which we will refer to as the APR EOS.
The APR EOS predicts the maximum NS mass of $2.2\Msun$.
Ref.~\cite{Akmal:1998cf} also discusses transitions to exotic phases such as the pion condensation, or the emergence of hyperons and quarks. 
In the APR EOS, pion condensation may occur near the center of a NS with the radial size of $\sim10\unit{m}$.
Hyperons or quarks may also be produced since the chemical potentials of nucleons and leptons can exceed the threshold.
The emergence of such exotic phases will, however, soften the EOS, and in general such an EOS is not favored by the recent discoveries of the NSs heavier than $2.0\Msun$.%
\footnote{
  In particular, recent discovery of $2\Msun$ NSs raised the so-called \textit{hyperon puzzle}.
  In such a heavy NS, we expect the emergence of hyperons since the chemical potential of neutrons can exceed, e.g., the mass of $\Lambda$ hyperon.
  Converting nucleons to hyperons softens the EOS dramatically due to the reduction of baryonic Fermi pressure.
  It is pointed out that the three-body repulsive interaction among nucleons and hyperons is important to reconcile the EOSs with the observed heavy NSs; Ref.~\cite{Lonardoni:2014bwa} shows that incorporating such repulsion increases the maximum NS mass.  However, there is a large theoretical uncertainty in this three-body interaction, and the actual effect of hyperonic degrees of freedom is still unclear.
}
In addition, it is also disfavored by the observations of NS surface temperatures because the exotic particles tend to enhance the neutrino emission and thus predict the NSs colder than many observations (see Sec.~\ref{sec:enhanced-cooling}).
To study the significance of such exotic particles, we need to understand the behavior of the chemical potential and effective mass of nuclear matter in an extremely dense environment~\cite{Oertel:2016bki}. 
This is currently an open issue, and we do not discuss it any more; in this dissertation we use the APR EOS with $npe\mu$ matter.%
\footnote{
  For instance, Ref.~\cite{Lonardoni:2014bwa} shows that in a certain model of the three-body interaction, the appearance of hyperons is not energetically favorable up to highly dense region.
  Thus one can obtain $npe\mu$ matter even with an hyperonic EOS.
}

Regarding the NS thermal evolution, EOSs determine whether the fast neutrino emission process, called direct Urca process (See Sec.~\ref{sec:direct-urca}), occurs or not for a given NS mass.
The direct Urca process occurs when the proton fraction exceeds about $10\%$, and this fraction in a NS depends on EOSs.
Given a nuclear interaction, we usually have the energy per nucleon in the form of
\footnote{
  $n_B \mathcal{E}$ is the baryonic part of the energy density $\rho$ in TOV equation.
}
\begin{align}
  \label{eq:nucl-energy}
  \mathcal{E}(n_B, Y_p) = \mathcal E(n_B, 1/2) + S(n_B, Y_p),
\end{align}
where $\mathcal E(n_B, 1/2)$ is the energy of the isospin symmetric matter and $S(n_B, Y_p)$ is the energy loss due to the isospin asymmetry, called symmetry energy.
If this symmetry energy is large, converting neutrons to proton is energetically favorable, and the threshold of the direct Urca process is overcome more easily.
To see this more explicitly, let us expand the symmetry energy as
\begin{align}
  \label{eq:sym-energy}
  S(n_B, Y_p)
  &\simeq
    S_2(n_B)(1-2Y_p)^2 + \cdots\,.
\end{align}
The thermodynamics gives $\mu_n - \mu_p = -\partial \mathcal E/\partial Y_p$.
Once we assume the chemical equilibrium between nucleons and electrons by the weak interaction,\footnote{This is called beta equilibrium. See Chap.~\ref{chap:neutron-star-heating} for the validity of this assumption.} we can determine $\mu_e$ by $\mu_e = \mu_n - \mu_p$. On the other hand the electrons chemical potential is expressed by the free relativistic gas approximation as $\mu_e \simeq (3\pi^2n_BY_p)^{1/3}$, where we assume charge neutrality.\footnote{For simplicity, we ignore muons.}
Thus we obtain the equilibrium condition
\begin{align}
  \label{eq:eq-yp}
  4S_2(n_B)(1-2Y_p)
  \simeq
  (3\pi^2n_B Y_p)^{1/3}\,,
\end{align}
where we neglect the higher order terms in $S$.
This equation tells the proton fraction as the increase of $n_B$; if $S_2(n_B)$ grows very rapidly, then $Y_p$ increases, and if not, $Y_p$ remains small.
Since laboratory data are available only near the saturation density, the constraints on $S_2$ are given to the coefficients of the following expansion: 
\begin{align}
  \label{eq:sym-en-exp}
  S_2(n_B)
  =
  S_v + \frac{1}{3}\frac{n_B-n_0}{n_0}L
  +\frac{1}{18}\left( \frac{n_B-n_0}{n_0} \right)^2K_{\mathrm{sym}}+\cdots\,.
\end{align}
Combining the constraints from several experimental results, the allowed ranges are $S_v\simeq29-33\MeV$ and $L\simeq40-60\MeV$ (for the detail, see Ref.~\cite{Lattimer:2012xj} and references therein).
Higher order terms such as $K_{\mathrm{sym}}$ is difficult to constrain~\cite{Lattimer:2012xj}.



\section{Microphysics}
\label{sec:microphysics}

The calculation of thermodynamic quantities requires the knowledge of microphysics.
Since the nuclear force among neutrons and protons are strong, we cannot treat it perturbatively.
In such a strongly interacting many body system, we need to use Landau's \textit{Fermi liquid theory}~\cite{Landau:1956zuh}.
There the fundamental constituent is not elementary particles but \textit{quasiparticles}, which look like elementary particles but are actually their collective excitation.
Moreover, nucleons can condensate due to the attractive nature of nuclear force, which further complicates the thermal evolution.

In this section, we briefly review the condensed matter physics necessary for the calculation of NS cooling.
The Fermi liquid theory and nucleon pairings are frequently used in the later sections.

\subsection{Fermi liquid theory}
\label{sec:fermi-liquid-theory}

Due to the strong nuclear force, nucleons cannot be treated as free Fermi gas.
Such an interacting many body system is called Fermi liquid, and 
calculating physical quantities is generically very difficult.
In a very cold system, however, some quantities require only the information in the vicinity of the
Fermi surface. In such a case, we can use Landau's Fermi liquid
theory~\cite{Landau:1956zuh}. In this subsection, we review the Fermi
liquid theory. See a classic textbook~\cite{lifshitz2013statistical} for the detail. 

To introduce the Fermi liquid theory, let us first consider a free Fermi gas.
It consists of free fermions of spin 1/2, and the energy eigenstate is specified
by momentum $\bm p$ and spin $\alpha=\uparrow\downarrow$. This state corresponds to
individual particles. At absolute zero temperature, the energy levels are occupied from
below, which forms Fermi sphere in momentum space. The momentum of the highest occupied level is
called Fermi momentum, $p_F$, and is related to the number density of gas as
\begin{align}
  \label{eq:fermi-gas-n-kf}
  n = \frac{N}{V} = \frac{p_F^3}{3\pi^2\hbar^3}\,,
\end{align}
where $N$ is the number of particles and $V$ the volume.

In the Fermi liquid, due to interaction terms, the eigenstate of $(\bm p, \alpha)$ is no longer the eigenstate of the interaction Hamiltonian.
However, we can use this label as a \textit{classification} of the energy levels of the Fermi liquid as follows.
Suppose that we have a free Hamiltonian at $t=0$, and introduce interactions gradually, which eventually evolves to the interaction Hamiltonian of the Fermi liquid at $t\to \infty$.
An engenstate of $(\bm p, \alpha)$ is an engenstate of the Hamiltonian at $t=0$, and according to the adiabatic theorem, this state evolves to an eigenstate of the interaction Hamiltonian at $t\to\infty$.
This new state is not the eigenstate of momentum (and generically spin), but it has
one-to-one correspondence to the original free state $(\bm p, \alpha)$. It is
collective excitation of many particles, but looks like one-particle state with
definite momentum and spin if their fluctuations are small.
Thus it is called a \textit{quasiparticle}.

Because quasiparticles are classified by the momentum and spin, we can consider
the distribution function of the form $f(\bm p)$. Here we consider isotropic case for
simplicity, but the spin degrees of freedom is easily introduced to the
distribution function~\cite{lifshitz2013statistical}. From the construction
above, the number of quasiparticles is the same as the number of particles:
\begin{align}
  \label{eq:nubber-qs}
  \frac{N}{V}
  &=
    2\int \frac{d^3p }{(2\pi\hbar)^3}f(\bm p)\,.
\end{align}
From the Pauli principle, the ground state is determined by
occupying the energy levels from the lowest one, which again forms Fermi sphere.
Therefore, we can define the Fermi momentum for Fermi liquid by the same formula
~\eqref{eq:fermi-gas-n-kf}.

Since quasiparticles have definite momenta, their energy spectrum,
$\varepsilon(\bm p)$, is defined as follows.
In the free theory, the energy spectrum is $\varepsilon(\bm p) = p^2/2m$. Thus
when a single particle is added on the Fermi surface, the total energy of the
system increases by $\sim p_F^2/2m$.
In the strongly interacting system, the potential energy also contributes.
Furthermore, the added particle interacts the nearby particles and alter their configuration.
Considering this, we define the energy of the quasiparticles in the functional manner:
\begin{align}
  \label{eq:energy}
  \frac{\delta E}{V}
  =
  2\int \frac{d^3p}{(2\pi\hbar)^3} \varepsilon(\bm p) \delta f(\bm p),
\end{align}
where $\delta f(\bm p)$ is the infinitesimal variation of the distribution
function. The quasiparticle energy is defined such that the total energy changes
by the amount of $\varepsilon(\bm p)$ when a single quasiparticle of momentum $\bm p$ is added on the Fermi surface.

We can determine the form of the distribution function as follows. In an
isolated system, thermodynamic equilibrium is determined by maximizing the entropy
\begin{align}
  \label{eq:entropy}
  \frac{S}{V}
  =
  - 2\int \frac{d^3p}{(2\pi\hbar)^3} \left[ f\ln f + (1-f)\ln(1-f) \right]
\end{align}
with the constraints
\begin{align}
  \label{eq:constraint}
  \frac{\delta N}{V} = 0\,,
  \frac{\delta E}{V} = 0\,.
\end{align}
Using the Lagrange multipliers, we obtain
\begin{align}
  \label{eq:dist}
  f(\bm p) = \frac{1}{e^{(\varepsilon(\bm p) - \mu)/T}+1},
\end{align}
where $T$ and $\mu$ are identified by the thermodynamic relation:$dS =
(1/T)dE + (\mu/T)dN$.
Note that this is the same expression as the Fermi distribution, but the energy is
functional of $f(\bm p)$ determined by Eq.~\eqref{eq:energy}, so
Eq.~\eqref{eq:dist} is in fact the functional equation of $f(\bm p)$.

So far, we have assumed that quasiparticles have definite momenta. 
In general, however, the momentum fluctuates due to the finite mean free path of
quasiparticles. Thus the notion of quasiparticles is
valid only when the fluctuation of the momentum, $\delta p$, is smaller than the momentum
itself, and the thermal width $\Delta p$ around Fermi surface. Otherwise we
cannot distinguish respective quasiparticles excited on the Fermi surface.  Landau argues
that the momentum fluctuation is proportional to $1/\tau$, where $\tau$ is the
mean free time, and it is suppressed as $1/\tau \propto T^2$ due to the
Pauli blocking~\cite{Landau:1956zuh, lifshitz2013statistical}. Thus at very low temperature, momentum fluctuation is suppressed
by the temperature squared, $\delta p \propto T^2$, while the thermal width is
$\Delta p \propto T$, which realizes $\delta p \ll \Delta p$.

\subsection{Effective masses}
\label{sec:effective-mass}

The advantage to introduce quasiparticles is that we can use the particle
picture in a strongly interacting system. Since particles inside a NS are degenerate
($T \ll \mu$), the thermodynamic quantities such as specific heat or neutrino
emissivities are determined by the excitations near the Fermi surface.
In the Fermi liquid theory, the effect of interaction is incorporated, e.g., by renormalizing the mass.
This renormalized mass is called effective mass.

The energy near the Fermi surface is expanded as
\begin{align}
  \label{eq:en-near-fermi-surface}
  \varepsilon(\bm p) \simeq \mu + \left. \frac{\partial \varepsilon}{\partial p}\right |_{p=p_F}
  \cdot (p-p_F)\,.
\end{align}
We define the Fermi velocity
by
\begin{align}
  \label{eq:fermi-velocity}
  \bm v_F \equiv \left. \frac{\partial \varepsilon}{\partial \bm p}\right |_{p=p_F}\,,
\end{align}
and the effective mass by
\begin{align}
  \label{eq:effective-mass}
  m^* \equiv \frac{p_F}{v_F}\,.
\end{align}
Thus in the quasiparticle picture, quasiparticles near the Fermi surface move with the velocity $v_F$ and mass $m^*$

The lepton effective mass in a NS is easily determined.
Unlike the nuclear force, the weak interaction can be treated as a perturbation,
so leptons
are almost free Fermi gas. Their Fermi energy is $\varepsilon_{F,\ell} =
\sqrt{m_\ell^2 + p_{F,\ell}^2}$%
\footnote{
  For leptons, we take relativistic energy including mass, which changes the origin of the chemical potential but does not change the effective mass.
  For nucleons, we use non-relativistic energy and chemical potential.
}. 
Hence the effective mass is
\begin{align}
  \label{eq:mstar-lepton}
  m_\ell^* = \varepsilon_{F,\ell} = \mu_\ell\,.
\end{align}

\begin{figure}
  \centering
  \includegraphics[width=0.5\linewidth]{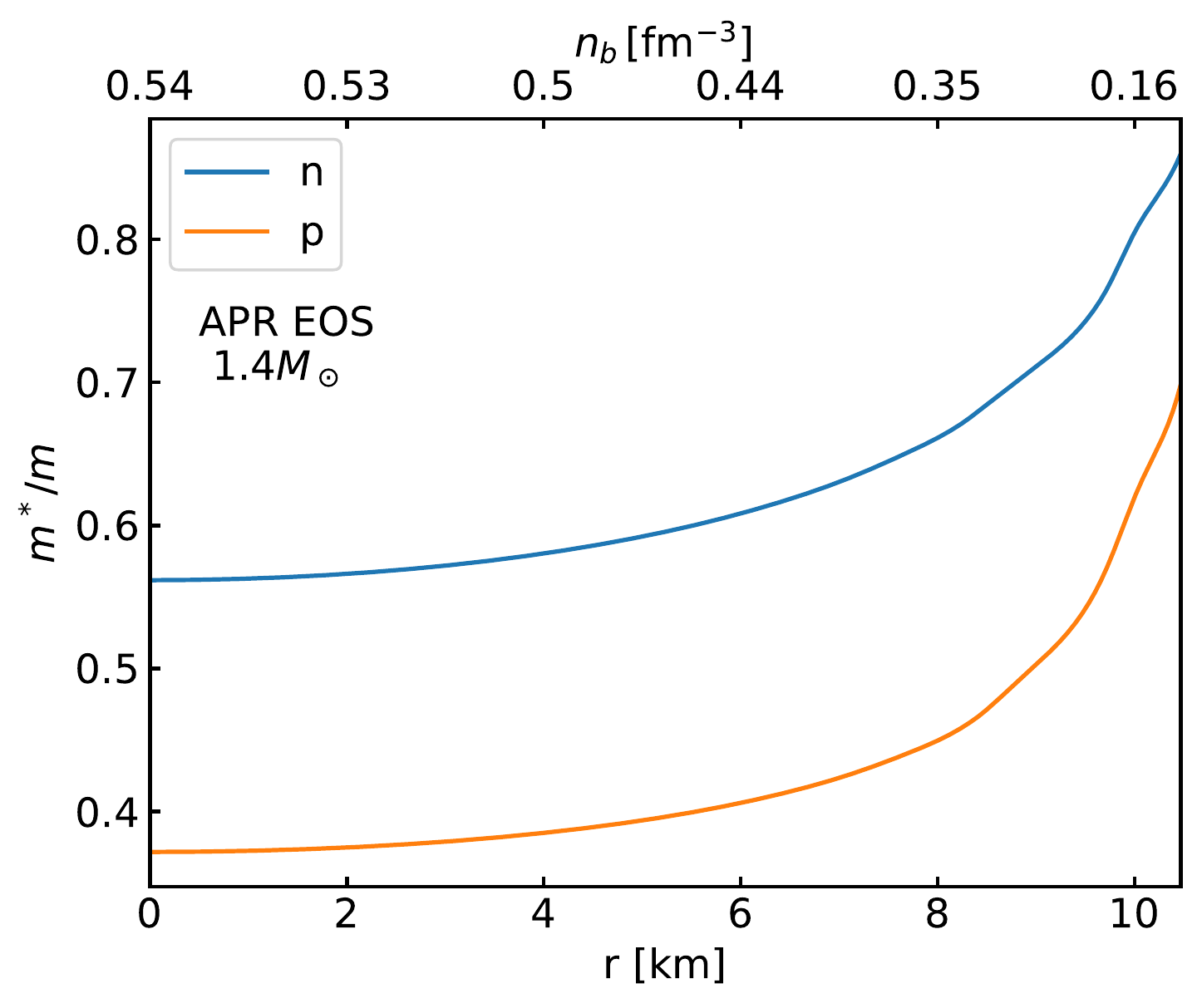}
  \caption{Effective masses of nucleons as functions of radius and baryon density for $M=1.4\Msun$. We use the table of APR EOS and the solutions of TOV equation in \texttt{NSCool}~\cite{NSCool}.}
  \label{fig:mst}
\end{figure}
Nucleon effective masses are determined by the nuclear interaction; its value depends on
the choice of EOSs. For instance, in the case of APR EOS~\cite{Akmal:1998cf}, the effective
Hamiltonian is given, so that one can explicitly calculate the effective mass of
nucleons for given baryon density.
We plot $m^*/m$ of nucleons in Fig.~\ref{fig:mst} for $M=1.4\Msun$ using APR EOS.
We see that the effective masses are smaller than the bare mass in the whole NS, and proton effective mass is smaller than the neutrons everywhere.

\subsection{Nucleon pairing}
\label{sec:nucleon-pairing}
In Sec.~\ref{sec:fermi-liquid-theory}, we introduce  quasiparticles through the adiabatic evolution from a Fermi gas.
This discussion holds if the non-interacting system goes to interacting
system smoothly through the adiabatic evolution. This is, however, not always true for
the fermionic system with attractive interaction. At very low temperature,
attractive force on the Fermi surface destabilizes the normal ground state;
below the critical temperature, fermions form the Cooper pairs.
This is known as \textit{Cooper theorem}~\cite{Cooper:1956zz}.
The BCS theory~\cite{Bardeen:1957mv} shows that this pairing generates the energy gap between the ground state and excited state, and the system becomes superconductor.
Later the Bogoliubov transformation~\cite{Bogolyubov:1958km} is developed for the Fermionic system.
It provides the transformation from the creation/annihilation operators of individual particles to those of collective excitations on the superconductive ground state, which explicitly shows that the ground state is reorganized by the pairing.

Historically, these theories are developed for the electron superconductivity in metal, where the attractive interaction is provided by the lattice vibration, or phonon exchange.
In neutron stars, nucleons interact by nuclear force.
If this is attractive, nucleons can also form the Cooper pair.
Since the pairing occurs through the reorganization of states around the Fermi surface, and the Fermi momentum of proton and neutron is very different, the pairing occurs only for neutron-neutron or proton-proton pairs in a NS. 

For protons in the core, $s$-wave interaction is the dominant channel, which is attractive for their corresponding Fermi energy.
Thus protons form ${}^1S_0$ pairing ($\ell = 0$ and spin-singlet pairing) and shows the superconductivity.
Since neutrons have larger Fermi momenta than protons, neutrons in the core form the Cooper pair by the $p$-wave interaction.
This is called ${}^3P_2$ pairing ($\ell = 1$ and spin-triplet pairing).
Unbounded neutrons also exist in the crust, and due to their small Fermi momentum, they form ${}^1S_0$ pairing there.
These paired neutrons become superfluid state since they are electrically neutral.

There are several observational supports for the pairing in a NS.
First, neutron pairing in the inner crust explains the sudden period change of pulsars, which is called a \textit{glitch}.
Usually the pulsar period increases very slowly in a constant rate, but sometimes it rapidly drops and then relaxes to the original value (see, e.g., Ref.~\cite{2011MNRAS.414.1679E}).
It is believed that the neutron \textit{superfluid vortex} in the crust caused by the pairing is responsible for this phenomenon~\cite{Anderson:1975zze}. 
In addition, the cooling rate of the NS in the supernova remnant Cas A is explained by the Cooper pair breaking and formation process.
We will discuss it in Chap.~\ref{chap:limit-axion-decay}.

In this subsection, we review the nucleon pairing inside NSs (see also Refs.~\cite{Page:2013hxa, Sedrakian:2018ydt} for recent reviews).
We need the theory not only for $s$-wave but also $p$-wave pairing.
In App.~\ref{chap:gap-equation}, we derive the gap equation by using generalized Bogoliubov transformation following Ref.~\cite{1970PThPh..44..905T}.

\subsubsection{Gap equation}
\label{sec:gap-eq}

For the thermal evolution of a NS, the effect of nucleon pairing enters through
the distribution function. In the presence of the energy gap, energy spectrum near the
Fermi surface is expressed as
\begin{align}
  \label{eq:energy-spectrum-gap}
  \varepsilon(\bm p) \simeq \mu + \mathrm{sign}(p-p_F)\sqrt{v_F^2(p-p_F)^2 + \delta^2}\,,
\end{align}
where the gap amplitude $\delta$ depends on the Fermi momentum, $p_F$, and its
direction around the quantization axis, $\hat p_F$, which we collectively denote
$\bm p_F$.

\begin{table}
  \centering
  \begin{tabular}{lcccc}\toprule
    Name & Superfluid state & $\lambda$ & $F(\theta)$ & $\Delta(0)/k_BT_C$ \\\midrule
    A    & $^1S_0$          & $1$      & $1$ & $1.76388$ \\
    B    & $^3P_2(m_J=0)$   & $1/2$     & $1+3\cos^2\theta$ & $1.18867$\\
    C    & $^3P_2(|m_J|=2)$ & $3/2$     & $\sin^2\theta$ & $2.02932$\\
    \bottomrule
  \end{tabular}
  \caption{The parameters of energy gap. $\theta$ is the polar angle around the
    quantization axis. We follow the convention of Ref.~\cite{Yakovlev:2000jp}.}
  \label{tab:gap-angle}
\end{table}
As we discuss in App.~\ref{chap:gap-equation}, the ${}^1S_0$ pairing is isotropic, which means $\delta$ does not depend on $\hat p_F$.

On the other hand, the ${}^3P_2$ pairing is
anisotropic.
In fact, the state of total angular momentum $j=2$ consists also of ${}^3F_2$ ($\ell = 3$ and spin triplet).
Hence it is in fact ${}^3P_2 - {}^3 F_2$ mixed state.
The contribution from ${}^3F_2$ is, however, so small that it is often neglected~\cite{1970PThPh..44..905T}.
Even with this simplification, the triplet pairing is still complicated because the gap is sum of the contributions from different angular momentum $m_j$, and several equations are coupled for the determination of gap amplitudes.
Thus it is assumed that only one $m_j$ dominates the pairing gap.
Furthermore, $m_j = 0$ and $\pm2$ are often used in the NS study because of its simple angular dependence (see App.~\ref{chap:gap-equation}, in particular Eqs.~\eqref{eq:gap-mj-0} and \eqref{eq:gap-mj-2}).
Then the gap amplitude, including ${}^1S_0$ case, is collectively denoted by
\begin{align}
  \label{eq:gap-angle-decomp}
  \delta^2 = \Delta(T)^2F(\theta)\,,
\end{align}
where $\theta$ is the polar angle of Fermi momentum with respect to the quantization axis.
The function $F$ is listed in Tab.~\ref{tab:gap-angle}.
We note that $\Delta(T)$ depends on $p_F$ but not on $\hat{p}_F$.

The gap $\delta$ is determined by solving the gap equation. 
The equation for $s$-wave pairing is studied in the context of the electron superconductivity in metals~\cite{Bardeen:1957mv, Bogolyubov:1958km}.
It is generalized in Ref.~\cite{1970PThPh..44..905T} to the $p$-wave pairing.
This generalized gap equation is written as
\begin{align}
  \label{eq:gap-eq}
  \frac{g}{2}\int\frac{d^3p}{(2\pi\hbar)^3}
  \frac{1}{\sqrt{\eta_p^2 + \delta^2}}
  \tanh\left( \frac{\sqrt{\eta_p^2 + \delta^2}}{2T} \right)\cdot\lambda F(\theta)
  = 1\,,
\end{align}
where $\eta_p = p^2/2m^* - \mu$ and $g$ denotes the interaction strength.
The interaction is attractive (repulsive) for $g > 0$ ($g < 0$).
The coefficient $\lambda$ is different for different superfluid types and also shown in Tab.~\ref{tab:gap-angle}.
We provide the derivation of Eq.~\eqref{eq:gap-eq} in App.~\ref{chap:gap-equation}.

At $T=0$, the gap equation reads
\begin{align}
  \label{eq:gap-eq-0}
  \frac{g}{2}\int\frac{d^3p}{(2\pi\hbar)^3}
  \frac{1}{\sqrt{\eta_p^2 + \delta^2}}
 \cdot\lambda F(\theta)
  = 1\,,
\end{align}
By solving this, one can obtain $\Delta(T=0)$ as the function of $p_F$.
We see that it has $\Delta \neq 0$ solution only for an attractive ($g > 0$) interaction.

The zero temperature gap provided by Eq.~\eqref{eq:gap-eq-0} is related to the critical temperature above which the pairing vanishes.
To see this, it is convenient to reduce the gap equation into the following form:
\begin{align}
  \label{eq:gap-eq-2}
  \log\left( \frac{\Delta(0)}{\Delta(T)} \right)
  =
  2\lambda\int\frac{d\Omega}{4\pi}F(\theta)\cdot I\left( \frac{\delta}{T} \right)\,,
\end{align}
where
\begin{align}
  \label{eq:gap-integ}
  I(y)
  &=
    \int_0^\infty dx\frac{1}{\sqrt{x^2+y^2}\left( \exp(\sqrt{x^2+y^2})+1 \right)}\,.
\end{align}
Note that we extend the range of energy integration to infinity because the integrand rapidly converges.
The expansion close to the critical temperature gives the ratio $\Delta(0)/T_c$.
For $y \ll 1$~\cite{lifshitz2013statistical},
\begin{align}
  \label{eq:gap-integ-expand}
  I(y) \simeq
  \frac{1}{2}\left[ \ln\left( \frac{\pi}{y} \right) -\gamma \right]
  +\frac{7\zeta(3)}{16\pi^2}y^2\,.
\end{align}
Using this expression in Eq.~\eqref{eq:gap-eq-2}, and taking $\delta \to 0$, we obtain
\begin{align}
  \label{eq:tc-gap-rel}
  T_c = \frac{e^\gamma}{\pi\Gamma}\Delta(0)\,,
  \text{ where }
  \ln\Gamma = -\frac{1}{2}\lambda\int\frac{d\Omega}{4\pi}F(\theta)\ln F(\theta)
\end{align}
We show the numerical values of $\Delta(0)/T_c$ in Tab.~\ref{tab:gap-angle} (see also Ref.~\cite{Amundsen:1984qc}).
As we will see in the next subsection, $\Delta(0)$ for nucleons in a NS is about $0.1 - 1\MeV$, which corresponds to the critical temperature of $10^9 - 10^{10}\unit{K}$.

The gap amplitude at intermediate temperature is obtained by solving Eq.~\eqref{eq:gap-eq-2} numerically.
For later convenience, we define the dimensionless gap and temperature as
\begin{align}
  \label{eq:dimless-gap}
  v \equiv \frac{\Delta(T)}{k_BT}\,,\quad & \tau \equiv \frac{T}{T_c}\,.
\end{align}
The numerical results for respective paring types are fitted by the following formulae~\cite{Yakovlev:2000jp}
\begin{align}
  v_A &= \sqrt{1-\tau}\left(1.456 - \frac{0.157}{\sqrt \tau} + \frac{1.764}{\tau}\right)\,,
  \label{eq:v-tau-a}
  \\
  v_B &= \sqrt{1-\tau}\left(0.7893 + \frac{1.188}{\tau}\right)\,,
        \label{eq:v-tau-b}
  \\
  v_C &= \frac{\sqrt{1-\tau^4}}{\tau}\left(2.030 - 0.4903\tau^4 + 0.1727\tau^8\right)\,.
        \label{eq:v-tau-c}
\end{align}
Note that these expressions recover the relation between the critical temperature $T_c$ and the zero temperature gap amplitude $\Delta(0)$ by taking the limit of $\tau \to 0$ in $\tau\cdot v$.

\subsubsection{Gap models}
\label{sec:gap-models}

\begin{figure}[t]
  \centering
  \begin{minipage}{0.45\linewidth}
    \includegraphics[width = 1.0\linewidth]{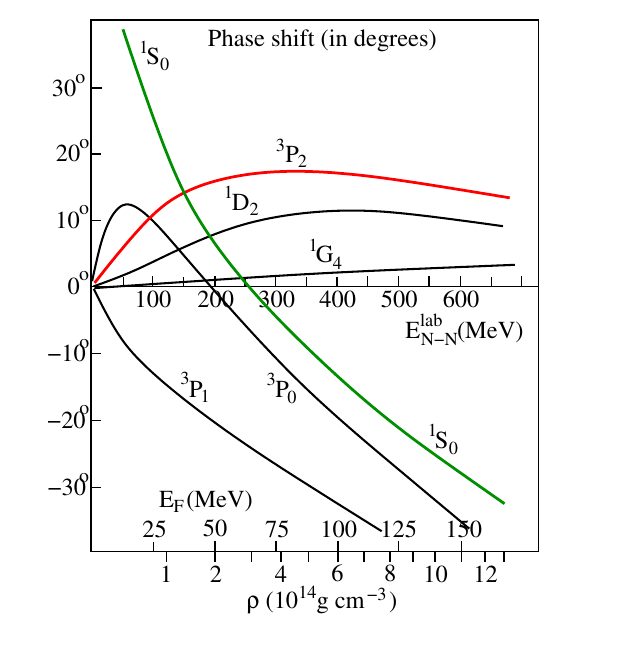}
  \end{minipage}%
  \begin{minipage}{0.55\linewidth}
    \includegraphics[width=1.0\linewidth]{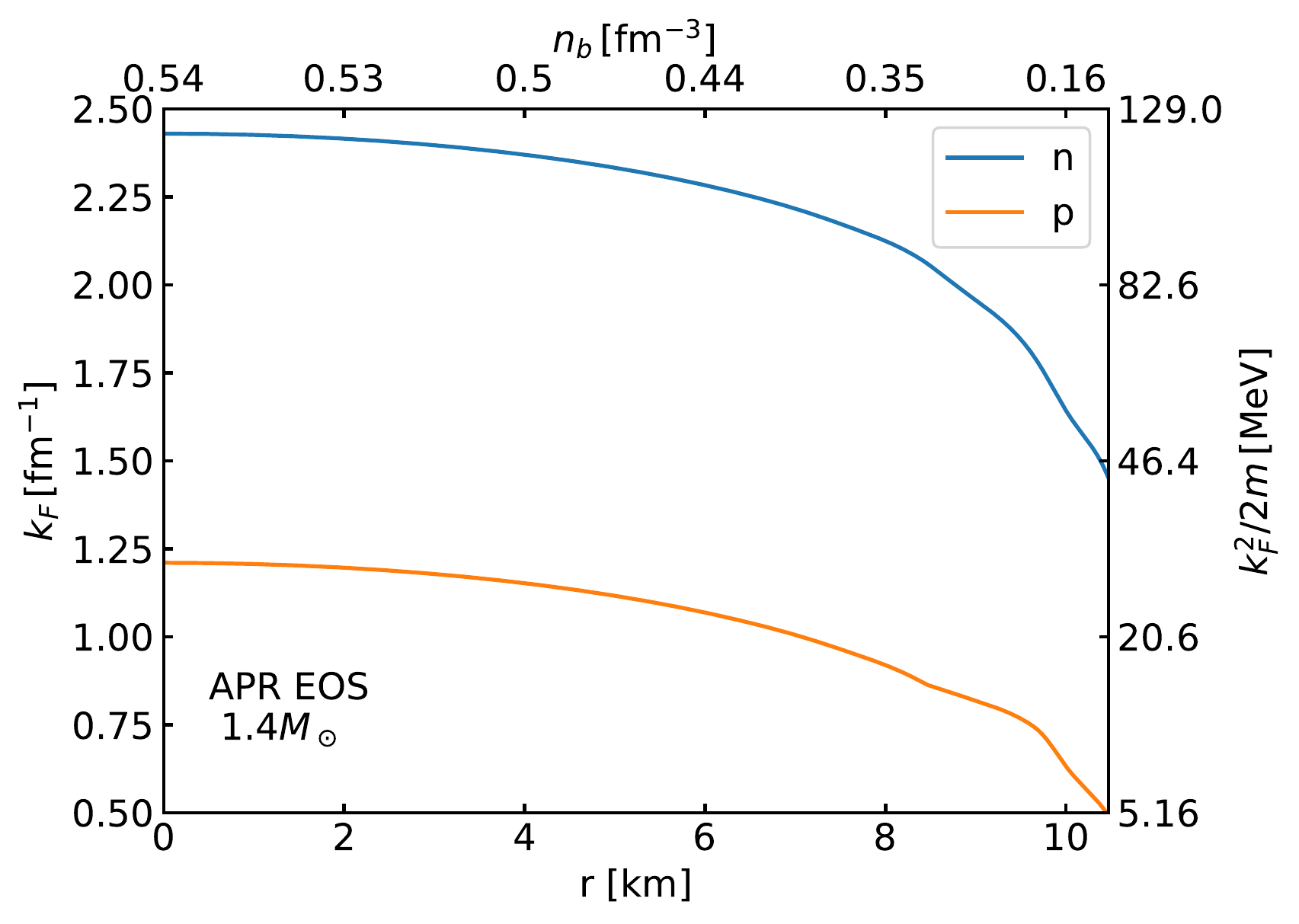}
  \end{minipage}
  \caption{Left: Phase shift taken from Ref.~\cite{Page:2013hxa}. The lines are calculated in Ref.~\cite{1970PThPh..44..905T} using the potential model in Ref.~\cite{Tamagaki:1968zz}.
    Right: The nucleon Fermi momentum in the core for $1.4\Msun$ NS using APR EOS.
    The corresponding Fermi energy is also shown in the right vertical axis using the bare mass $m_n \simeq m_p$.
  Note that the effective mass is smaller than the bare mass (see Fig.~\ref{fig:mst}), so the actual Fermi energy is larger.}
  \label{fig:phase-shift}
\end{figure}
Given Eqs.~\eqref{eq:v-tau-a} - \eqref{eq:v-tau-c}, all we have to do is to solve $T=0$ gap equation \eqref{eq:gap-eq-0} for a given nuclear interaction.
The pair-interaction between nucleons is often provided by the non-relativistic potential.
It is caused by the exchange of mesons such as pion, but calculating the potential from the first principle (i.e., QCD) is difficult.
Thus we usually use the results of nucleon scattering experiments.
There are a lot of experiments that measured the $pp$ and/or $np$ elastic scattering cross section (see, e.g., Ref.~\cite{Arndt:1997if} and references therein), and to construct the potential, the data below $350\MeV$ is usually used.
Note that these are the fixed target experiments, so $E_{\mathrm{CM}}=E_{\mathrm{lab}}/2$ in the non-relativistic limit, where $E_{\mathrm{CM}}$ is the center-of-mass energy.
Since our interest is the scattering of nucleons near the Fermi surface, we can translate the experimental results by $\varepsilon_F \simeq E_{\mathrm{lab}}/4$.

In the inner crust, the matter density is $\rho \sim 4\times 10^{11}\unit{g/cm^3} - 10^{14}\unit{g/cm^3}$, corresponding to the neutron Fermi momentum $p_F \sim 4\MeV - 200\MeV$, or Fermi energy $\varepsilon_F \sim 0.7 - 20 \MeV$.
In this energy range, the interaction between neutrons is dominated by attractive $s$-wave channel (see Fig.~\ref{fig:phase-shift}), and the potential is well constrained by the scattering experiments.
However, there are still several sources of uncertainties.
The BCS gap equation, Eq.~\eqref{eq:gap-eq-0}, does not incorporate the momentum dependence of the potential.
For some potential models, the potential slightly off the Fermi surface may provide an important contribution~\cite{Khodel:1996mob}.
Also, the medium effect changes the resultant gap amplitude (see, e.g., Ref.~\cite{Sedrakian:2018ydt}).
These, and other development of the many body calculation have been incorporated to the numerical calculations, and for neutron ${}^1S_0$ pairing, the gap is suppressed compared to the BCS theory with the bare potential~\cite{Sedrakian:2018ydt}. 
The typical gap amplitude is $\Delta = \Order(1)\MeV$~\cite{Sedrakian:2018ydt}. 

\begin{figure}[t]
  \centering
  \begin{minipage}{0.5\linewidth}
    \includegraphics[width=1.0\linewidth]{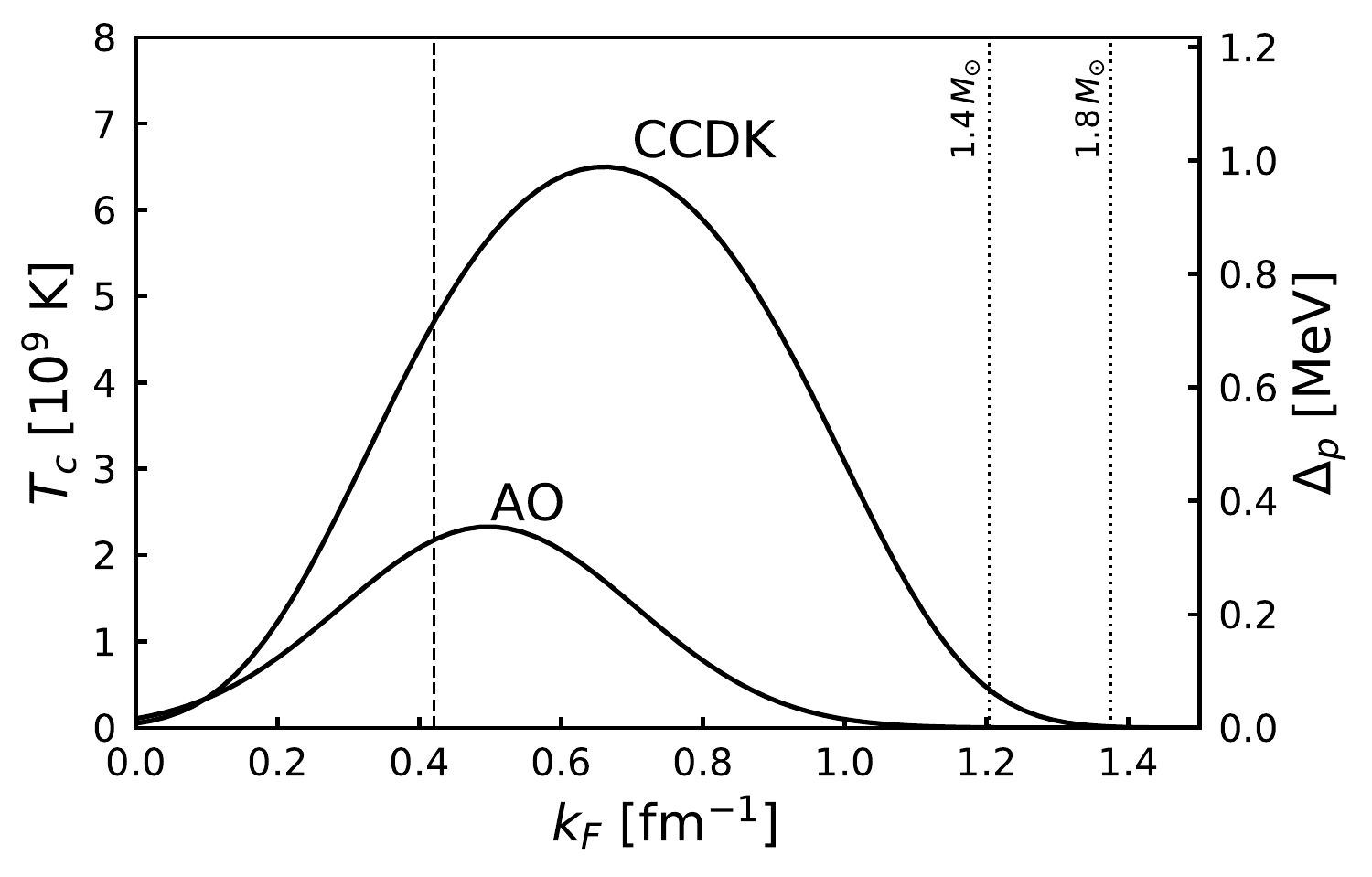}
  \end{minipage}%
  \begin{minipage}{0.5\linewidth}
    \includegraphics[width=1.0\linewidth]{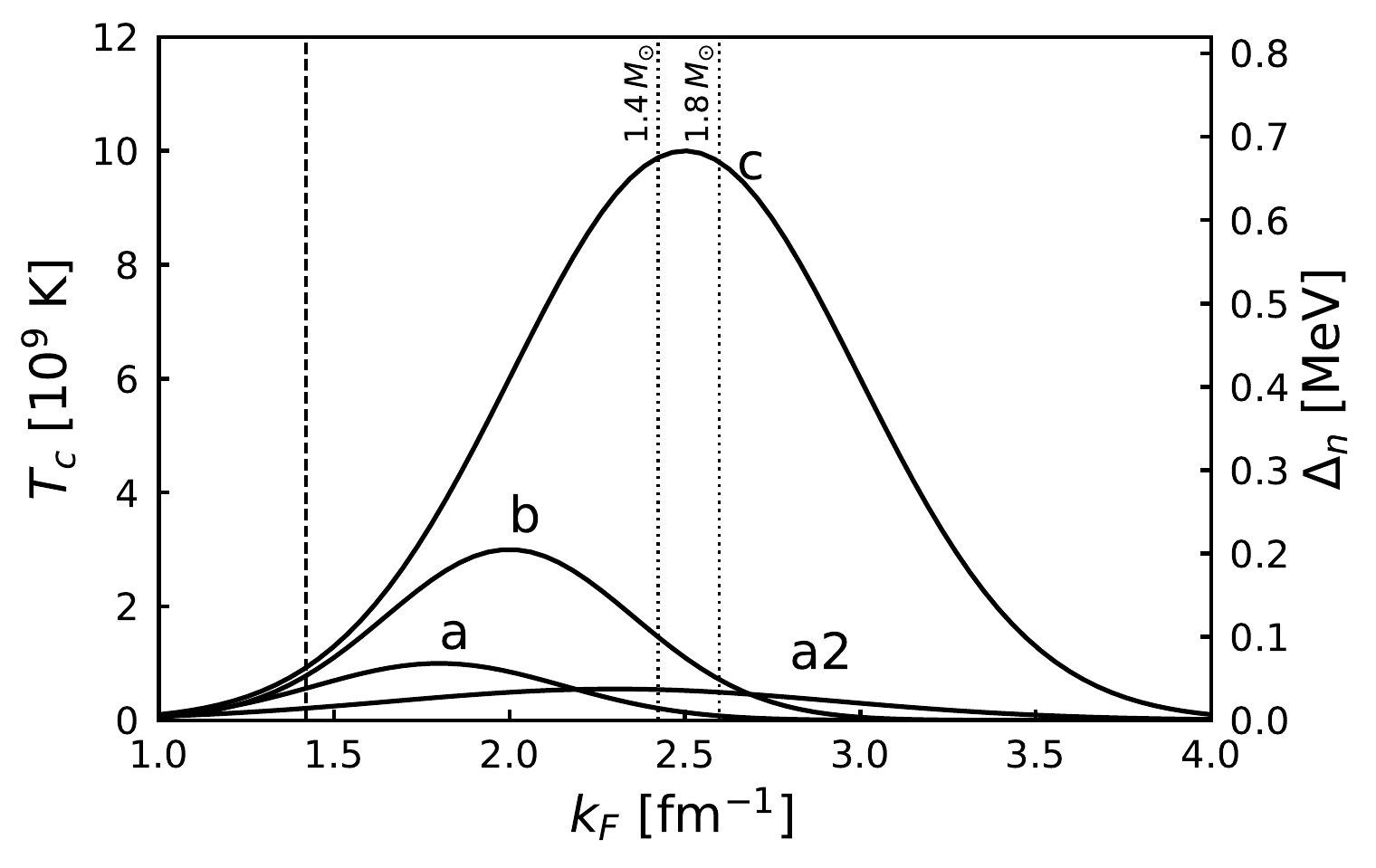}
  \end{minipage}
  \caption{Left: proton ${}^1\mathrm{S}_0$ gap models. Right: neutron ${}^3\mathrm{P}_2(|m_J|=0)$ gap models.
    The critical temperature and the zero temperature gap are shown.
    The right vertical axis for neutron triplet pairing corresponds to the gap amplitude for $\cos\theta = 0$.
    The dashed lines show the boundary between the crust and core.
  The dotted lines correspond to the NS center density for $M=1.4\,M_\odot$ and $1.8\,M_\odot$.}
  \label{fig:gap-3p2}
\end{figure}
In the core,  protons typically have Fermi energy of $\Order(10)\MeV$ (see the right panel of Fig.~\ref{fig:phase-shift}). Their main interaction channel is ${}^1S_0$. Although the pairing is similar to the neutron singlet pairing, the gap amplitude of protons is smaller. 
In addition, protons are surrounded by abundant neutrons, and interact via the electromagnetic force.
These provide the medium correction different from neutron ${}^1S_0$ gap.
In the left panel of Fig.~\ref{fig:gap-3p2}, we show two gap calculations: the AO~\cite{Amundsen:1984qq} gap model is one of the smallest gap reported in the literature, while the CCDK~\cite{Chen:1993bam} model is one of the largest.
We see that the critical temperature is $\Order(10^9)\unit{K}$. Near the core-crust boundary (dashed line), protons are already in the superconductor, while near the NS center the proton ${}^1S_0$ pairing can vanish depending on the gap models and NS masses.

As we see in the right panel of Fig.~\ref{fig:phase-shift}, neutrons in the core have much larger chemical potential than protons due to the large number density.
Their interaction is dominated by the $p$-wave, and $j=2$ interaction becomes attractive due to the spin-orbit potential~\cite{1970PThPh..44..905T} (see also App.~\ref{sec:potential}).
Although ${}^3P_2$ interaction is certainly attractive, the precise form of the potential is still unclear.
The main limitation comes from the available energy range of experiments, $E_{\mathrm{lab}} < 350\MeV$, corresponding to $k \lesssim 2\unit{fm^{-1}}$.
Reference~\cite{Baldo:1998ca} presents the study of the uncertainty in ${}^3P_2$ gap amplitude. The authors compare the gap amplitudes derived by the different potentials which are phase shift-equivalent for $E_{\mathrm{lab}} < 350\MeV$. They show that the resultant gap is almost the same for the range of validity of the potentials, $k_F \lesssim 1.8\unit{fm^{-1}}$, while for $k \gtrsim 1.8\unit{fm^{-1}}$, different potentials provide the very different gap amplitude.
Thus the uncertainty of neutron ${}^3P_2$ pairing is much larger than neutron/proton ${}^1S_0$ pairing.
In the right panel of Fig.~\ref{fig:gap-3p2}, we show several phenomenological gap models used in the literature~\cite{Page:2004fy, Page:2013hxa}; they cover the uncertainty range found in Ref.~\cite{Baldo:1998ca}.
Unlike the proton singlet pairing, the critical temperature of neutron ${}^3P_2$ pairing can be lower than $10^9\unit{K}$ if the actual gap is as small as ``a2'', which provide an important consequence for the study of Cas A NS in Chap.~\ref{chap:limit-axion-decay}.
To reduce the uncertainty of the neutron triplet gap, we need to construct the potential model which takes account of the inelasticity for $E_{\mathrm{lab}} > 350\MeV$~\cite{Baldo:1998ca}.

\begin{table}
  \centering
  \begin{tabular}{lccccc}\toprule
    & $T_{c,\mathrm{max}}\unit{[K]}$ & $k_{F,0}\unit{[fm^{-1}]}$ & $\delta k_F\unit{[fm^{-1}]}$ & $r$\\\midrule
    AO & $2.35\times10^9$ & $0.49$ & $0.31$ &$0$ \\
    CCDK & $6.6\times10^9$ & $0.66$ & $0.46$ & $0.69$ \\
    a & $1\times10^9$  & $1.8$ & $0.5$ & $0$\\
    b & $3\times10^9$ & $2.0$ & $0.5$ & $0$\\
    c & $1\times10^{10}$ & $2.5$ & $0.7$ & $0$\\
    a2 & $5.5\times10^{9}$ & $2.3$ & $0.9$ & $0$\\
    \bottomrule
  \end{tabular}
  \caption{Parameters of gap models in Fig.~\ref{fig:gap-3p2} for Eq.~\eqref{eq:p-gap-mod-gauss}.}
  \label{tab:fit-para-p-2}
\end{table}
Considering the uncertainty of nucleon gaps, in particular of neutron ${}^3P_2$ gap, it is sometimes convenient to treat the gaps as free parameters using simple analytic functions, which is also useful for the implementation in the code.
One way to parametrize the nucleon gap is to use the following modified Gaussian function
\begin{align}
  \label{eq:p-gap-mod-gauss}
  T_c(k_F) = T_{c,\mathrm{max}}\exp\left(-\left(\frac{k_F - k_{F,0}}{\delta k_F}\right)^2
  -r \left(\frac{k_F - k_{F,0}}{\delta k_F}\right)^4\right)\,,
\end{align}
with $T_{c,\mathrm{max}}$, $k_{F,0}$, $\delta k_F$ and $r$ being free parameters.
In Tab.~\ref{tab:fit-para-p-2}, we show the values of these parameters corresponding to the gap models in Fig.~\ref{fig:gap-3p2}.

\section{Thermodynamics of neutron stars}
\label{sec:thermodynamics}

The NS cooling was first studied about half a century ago in Ref.~\cite{1966CaJPh..44.1863T}, and has
been studied over the years along with the development of observational
technology, and also of our understanding in the microphysics inside a NS (See,
e.g., Refs~\cite{Yakovlev:2004iq, Yakovlev:2007vs} for
reviews).\


In the following sections, we review the standard theory of NS cooling, with particular
emphasis on the theory in the NS core, which has direct relevance to our study.
This section presents the basic equation for NS cooling (Sec.~\ref{sec:basic-eqs}), the NS envelope (Sec.~\ref{sec:envelopew}) and specific heat of nucleons and leptons (Sec.~\ref{sec:spec-heat}).
The dominant energy loss process in the core is neutrino emission.
We discuss various neutrino emission processes and their luminosities in Sec.~\ref{sec:neutrino-emis}.


\subsection{Basic equations}
\label{sec:basic-eqs}

\subsubsection{Temperature of a neutron star}
\label{sec:temp-ns}

In the general relativity, we need to be careful of the coordinate system.
We have to distiguish the temperature we observe near the earth, $T^\infty$, from
the temperature locally measured inside the star, $T(r)$, where $r$ denotes the
distance from the NS center.%
\footnote{
  $T^\infty$ is the temperature measured in the Schwarzschild coordinate since
  Eq.~\eqref{eq:schwarzschild} is asymptotically flat.
}
The temperature is obtained by $1/T = \partial S/\partial E$, where $S$
and $E$ are the entropy and energy in a given coordinate system, respectively.
Since the entropy is scalar, temperatures are different by the amount of the
gravitational redshift between different coordinate systems. Threfore, $T^\infty$ and
$T(r)$ are related as
\begin{align}
  \label{eq:temp-redshift}
  T^\infty = e^{\Phi(r)}T(r)\,.
\end{align}

In the following, we use
the superscript $\infty$ to denote quantities measured by the distant observer.
Otherwise they are defined in the local frame
inside the star. This local frame is convenient for the calculation of
thermodynamic quantities because we can make the metric locally flat and thus
use the Lagrangian in the flat space.

\subsubsection{Thermal evolution equations}
\label{sec:thermal-evolution-eqs}

Then we provide the differential equations which govern the thermal evolution of a NS. We assume the star is spherically symmetric.
We do not assume the whole star is in thermal equilibrium. Hence we also consider the heat transport inside the NS.
In the following, we use the unit of $c=1$.

The energy conservation determines the temperature evolution as
\begin{align}
  \label{eq:ene-consv}
  c_V\frac{dT(r)}{dt}e^{\Phi(r)}4\pi r^2 e^{\Lambda(r)}
  =
  - \frac{d(Le^{2\Phi(r)})}{dr}
\end{align}
where $c_V$ is the specific heat and $L$ is the luminosity, energy loss per unit time.
Note that the luminosity is redshifted twice because of the time delay as well as energy redshift.

Neutrinos are produced by the weak interaction, and they are collision-less inside the NS for the temperature of our interest
($T\lesssim 10^{10}\unit{K}$). Therefore, they are emitted from the whole star.
Then the neutrino luminosity $L_\nu$ is written by the neutrino emissivity $Q_\nu$, the
energy loss per unit time and unit volume, as
\begin{align}
  \label{eq:nu-lum-def}
  \frac{d(L_\nu e^{2\Phi(r)})}{dr}
  =
  Q_\nu e^{2\Phi(r)}4\pi r^2 e^{\Lambda(r)}\,.
\end{align}
On the other hand, other particles such as photons and electrons collide with
each other, transporting heat from one place to another.
The heat transport equation is written as~\cite{1983ApJ...272..286G} 
\begin{align}
  \label{eq:heat-cond-eq}
  -\lambda\frac{d(T(r)e^{\Phi(r)})}{dr}
  =
  \frac{L_{d}}{4\pi r^2}e^{\Phi(r)}e^{\Lambda(r)},
\end{align}
where $\lambda$ is thermal conductivity and $L_d$ the luminosity due to the thermal
radiation and conduction. From the definition of thermal conductivity,
$-\lambda d(Te^{\Phi})/dr$ corresponds to the energy flux per
unit area due to the temperature gradient along $r$ direction. Thus this equation is the definition of $L_d$.
Since the total luminosity is decomposed as $L=L_\nu + L_d$,
Eq.~\eqref{eq:ene-consv} is written as
\begin{align}
  \label{eq:en-balance}
  c_V\frac{dT(r)}{dt}e^{\Phi(r)}4\pi r^2e^{\Lambda(r)}
  =
  -Q_\nu e^{2\Phi(r)}4\pi r^2 e^{\Lambda(r)}
  -
  \frac{d(L_de^{2\Phi(r)})}{dr}.
\end{align}

The calculation of thermal evolution goes as follows:
\begin{enumerate}
\item
  Solve TOV equation with an input EOS.
\item
  Using the solution of TOV equation, calculate $c_V$, $\lambda$ and $Q_\nu$ with some microphysics input.
\item
  Solve Eq.~\eqref{eq:heat-cond-eq} and~\eqref{eq:en-balance} for $T(r)$ and
  $L_d(r)$ with an appropriate boundary condition.
\end{enumerate}

\subsubsection{Equations for an isothermal neutron stars}
\label{sec:isothermal-equations}

In principle, we have to solve the coupled equations~\eqref{eq:heat-cond-eq}
and~\eqref{eq:en-balance}. However, it is known that thermal relaxation time
scale is $10^{2-3}\unit{yr}$~\cite{Yakovlev:1999sk, Yakovlev:2000jp,
  Yakovlev:2004iq}, so when 
we are interested only in $t \gg 10^{2-3}\unit{yr}$, the whole star is
approximated as being isothermal (see also Sec.~\ref{sec:minimal-cooling}).
This corresponds to the limit of $\lambda \to
\infty$ in Eq.~\eqref{eq:heat-cond-eq}, so that in an isothermal NS,
\begin{align}
  \label{eq:isothermal}
  T^\infty = T(r)e^{\phi(r)} = \text{ indep. of }r\,
\end{align}
holds. 
Then integrating Eq.~\eqref{eq:en-balance} by $r$ from $0$ to $R$, we
obtain
\begin{align}
  \label{eq:cooling-isothermal}
  C_V\frac{dT^\infty}{dt}
  &=
    -L_\nu^\infty - L_\gamma^\infty\, \text{ (isothermal)},
\end{align}
where we have used $L_d(0) = 0$, $L_d(R) = L_\gamma$ with $L_\gamma$ being
the surface photon luminosity (see the next subsection), and
\begin{align}
  \label{eq:heat-cap}
  C_V &= \int_0^R dr\, 4\pi r^2 e^{\Lambda(r)} c_V
\end{align}
is the total heat capacity.
The redshifted luminosity is calculated as $L_\gamma^\infty = L_\gamma e^{2\Phi(R)}$ and
\begin{align}
  \label{eq:lnu-infty}
  L_\nu^\infty = \int_0^R dr\,4\pi r^2e^{\Lambda(r)} Q_\nu e^{2\Phi(r)}\,.
\end{align}

\subsection{Envelope and surface photon emission}
\label{sec:envelopew}

The surface photon emission is the blackbody radiation from the entire surface.
The luminosity is expressed as
\begin{align}
  \label{eq:surface-photon}
  L_\gamma = 4\pi R^2\sigma_B T_s^4\,,
\end{align}
where $T_s$ is the surface temperature and $\sigma_B = 0.567\unit{erg\,s^{-1}\,m^{-2}\,K^{-4}}$ is the Stefan-Boltzmann constant.

The surface temperature $T_s$ is very different from the temperature in the core or crust because of the thin layer called \textit{envelope}, locating near the surface.
The envelope is the region for the density $\rho \lesssim 10^{10}\unit{g/cm^3}$ and thickness $\lesssim 100\unit{m}$, which consists of the iron and lighter elements.
This layer becomes a thermal insulator which separates the atmosphere from the interior.
The temperature gradient in the envelope is so large that it is convenient to treat this layer separately from the crust and core~\cite{1983ApJ...272..286G}. 
For a nonmagnetic iron envelope, the thermal conduction equation is solved in Ref.~\cite{1983ApJ...272..286G}, which provides an approximated relation between the surface temperature $T_s$ and the internal temperature $T_b$, defined by the outer most part of the crust, as
\begin{align}
  \label{eq:ts-tb-iron}
  T_b \simeq 1.288\times10^8\bcm{T_{s6}^4}{g_{14}}{0.455}\unit{K}\,,
\end{align}
where $T_{s6} = T_s/10^6\unit{K}$, and $g_{14}$ is the surface gravity normalized by $10^{14}\unit{cm\,s^{-2}}$.

\begin{figure}
  \centering
  \includegraphics[width=0.7\linewidth]{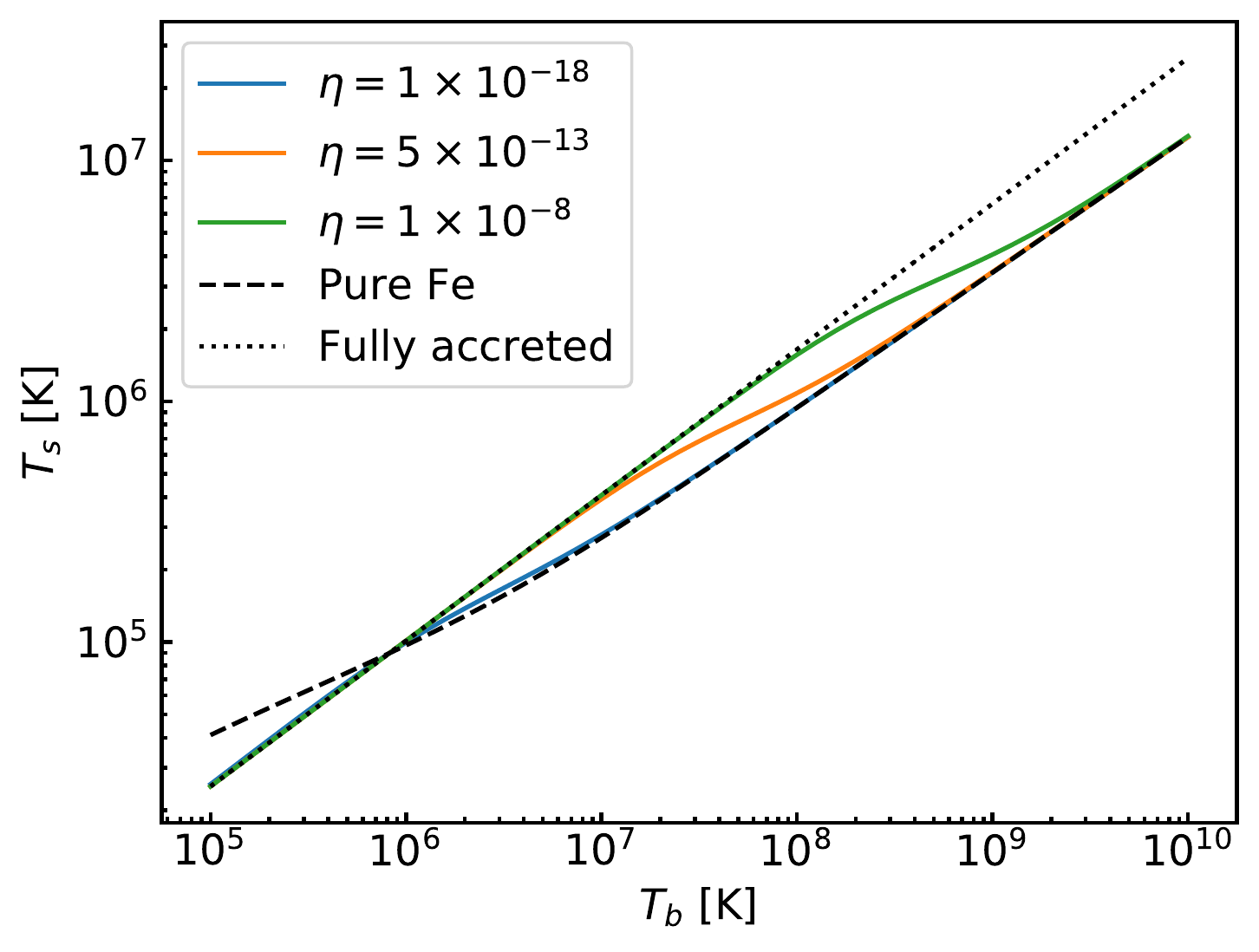}
  \caption{The relation between surface temperature $T_s$ and the internal temperature $T_b$. See the text for detail.}
  \label{fig:ts-tb}
\end{figure}
A more accurate relation, which is also applicable beyond the purely iron envelope, is calculated in Ref.~\cite{Potekhin:1997mn}.
The fitting formula of the $T_s-T_b$ relation is obtained as a function of $\eta = g_{14}^2\Delta M/M$, where $\Delta M$ is the mass of the light element in the envelope coming from the accretion.
The light element changes the conductivity in the envelope and thus changes the relation. 
This formula is calculated as follows:
\begin{itemize}
\item 
  The crude approximation of the purely iron envelope is
  \begin{align}
    \label{eq:ts-tb-crude}
    T_{s6} = T_* = \left( 7T_{b9} g_{14}^{1/2} \right)^{1/2}\,,
  \end{align}
  which roughly equals to Eq.~\eqref{eq:ts-tb-iron}. 
\item
  For the purely iron envelope ($\eta\to0$), a more accurate relation is given by
  \begin{align}
    \label{eq:ts-tb-iron-accurate}
    T_{s6,\mathrm{Fe}}^4 = g_{14}\left( (7\zeta)^{2.25} + (\zeta/3)^{1.25} \right)\,,
  \end{align}
  where $\zeta \equiv T_{b9} - T_*/10^3$.
\item
For a fully accreted envelope ($\eta \sim 10^{-7}$), the fitting formula is obtained as
\begin{align}
  \label{eq:ts-tb-fully-acc}
  T_{s6,a}^4 = g_{14}\left( 18.1T_{b9} \right)^{2.42}\,.
\end{align}
\item
The partially accreted envelope is obtained by interpolating these as follows:
\begin{align}
  \label{eq:ts-tb-partial}
  T_{s6}^4 = \frac{aT_{s6,\mathrm{Fe}}^4 + T_{s6,a}^4}{a+1}\,,
\end{align}
where $a = \left(1.2 + (5.3\times10^{-6}/\eta)^{0.38})\right)  T_{b9}^{5/3}$.
\end{itemize}
Note that these results are obtained in the calculation for $4.7\lesssim \log_{10}T_s \lesssim 6.5$.
Hence the fitting formulae above may not be applicable outside this range.
Figure~\ref{fig:ts-tb} shows $T_s-T_b$ relation calculated by Eq.~\eqref{eq:ts-tb-partial} for $\eta = 1\times10^{-18}$ (blue), $5\times10^{-13}$ (yellow) and $1\times10^{-8}$ (green) for the surface gravity of a $1.4\Msun$ NS and the APR EOS.
The surface temperature is higher for larger $\eta$, and thus for the larger conductivity.
We also show the relation for a purely iron envelope calculated by Eq.~\eqref{eq:ts-tb-iron-accurate} (dashed) and for a fully accreted envelop by Eq.~\eqref{eq:ts-tb-fully-acc} (dotted).
We see that these two lines are not consistent with Eq.~\eqref{eq:ts-tb-partial} outside the applicable range, $4.7\lesssim \log_{10}T_s \lesssim 6.5$.

\subsection{Specific heat}
\label{sec:spec-heat}

The specific heat of Fermi liquid is calculated by%
\begin{align}
  \label{eq:spec-heat}
  c_V
  &=
    2\int \frac{d^3p}{(2\pi\hbar)^3}\,(\varepsilon - \mu)\frac{\partial f}{\partial T}\,,
\end{align}
where the factor 2 takes care of the spin degrees of freedom.
At low temperature, the distribution function is close to the step function.
Hence only the excitation in the vicinity of the Fermi surface determines the specific heat.

\paragraph{Normal fluid}
We first consider the specific heat of the normal fluid such as leptons or nucleons above the critical temperature.
Since we are interested in the temperature smaller than the Fermi momentum, the usual low temperature expansion is applicable: 
\begin{align}
  \label{eq:spec-heat-normal}
  c_V = \frac{m^* p_F}{3\hbar^3} T\,.
\end{align}
Note that the effective mass of leptons is given by $m^*_\ell = \sqrt{m_\ell^2 + p_{F,\ell}^2}$ (see Eq.~\eqref{eq:mstar-lepton}).

\paragraph{Superfluid}

For nucleons in superfluid state, the specific heat receives suppression due to the presence of pairing gap.
Since the specific heat is determined by the excitation around the Fermi surface, we can use Eq.~\eqref{eq:energy-spectrum-gap} as an energy spectrum and extend the energy integration to the infinity.
Then the so called superfluid \textit{reduction factor} is obtained as
\begin{align}
  \label{eq:reduction-cv}
  R
   \equiv \frac{c_V}{c_V|_{\Delta = 0}} 
  = \frac{3}{2\pi^3}\int d\Omega \int_0^\infty dx\, zT\frac{\partial f}{\partial T}\,, 
\end{align}
where $c_V|_{\Delta = 0}$ is the specific heat without pairing (Eq.~\eqref{eq:spec-heat-normal}) and 
\begin{align}
  \label{eq:dimless-vars}
  x \equiv v_F\frac{p-p_F}{T},\quad
  y \equiv \frac{\delta}{T},\quad
  z \equiv \frac{\varepsilon - \mu}{T} = \mathrm{sign}(x)\sqrt{x^2 + y^2}\,.
\end{align}
Using the gap equation~\eqref{eq:gap-eq-2}, one can show
\begin{align}
  \label{eq:gap-t-derv}
  \frac{d\Delta(T)}{dT}
  =
  \frac{v\mathcal{F}(v)}{\mathcal{F}(v)-1}\,,
\end{align}
where $v=\Delta(T)/T$ and
\begin{align}
  \label{eq:f-function}
  \mathcal{F}(v)
  =
  2\lambda\int\frac{d\Omega}{4\pi}F(\theta)
  \int_0^\infty dx\frac{e^z}{(e^z+1)^2}\,.
\end{align}
Then the reduction factor is written as~\cite{1994ARep...38..247L} 
\begin{align}
  \label{eq:reduction-cv-2}
  R
  =
  \mathcal{G}(v)
   + \frac{3}{\pi^2\lambda}\frac{v^2\mathcal{F}^2(v)}{1-\mathcal{F}(v)}\,,
\end{align}
where
\begin{align}
  \label{eq:g-function}
  \mathcal{G}(v)
  =\frac{3}{2\pi^3}\int d\Omega
  \int_0^\infty dx
  \frac{z^2 e^z}{(e^z+1)^2}\,.
\end{align}

The nucleon pairing causes the phase transition to superfluid state.
We can see the signature of the phase transition from the \textit{jump} of the specific heat.
For that purpose, we take $v\to 0$ in Eq.~\eqref{eq:reduction-cv-2}; $\mathcal{G}$ and $\mathcal{F}$ are expanded as
\begin{align}
  \mathcal{F}(v)
  &\simeq
    1 + \frac{7\zeta^\prime(-2)}{\lambda}v^2\times
    \begin{cases}
      1\quad({}^1S_0) \\
      \frac{6}{5}\quad({}^3P_2\,(m_j=0,\pm2))
    \end{cases}
  \label{eq:f-func-asympt}
  \\
  \mathcal{G}(v)
  &\simeq
    1 + \frac{3}{2\pi^2\lambda}v^2
    \label{eq:g-func-asympt}\,.
\end{align}
Using these, we read the reduction factor near the phase transition as
\begin{align}
  \label{eq:reduction-cv-u0}
  R
  &\simeq 1 - \frac{3}{7\pi^2\zeta^\prime(-2)}\times
    \begin{cases}
      1\quad({}^1S_0) \\
      \frac{5}{6}\quad({}^3P_2)
    \end{cases}
  \notag\\
  &\simeq
    \begin{cases}
      2.42613\quad({}^1S_0) \\
      2.18844\quad({}^3P_2)
    \end{cases}
\end{align}
Thus when we decrease the temperature and measure the specific heat, we observe that it jumps by factor $\sim2$ at $T=T_c$.%
\footnote{
  Although we call $R$ reduction factor, it actually enhances the specific heat just below the critical temperature.
  The reduction occurs for $T \ll T_c$.
}

For numerical calculation, we use Eq.~\eqref{eq:spec-heat-normal} for normal matter, and below the critical temperature, we multiply the reduction factor.
The numerical fitting formula for each superfluid type in Tab.~\ref{tab:gap-angle} is given  by~\cite{Yakovlev:1999sk}
\begin{align}
  \label{eq:r-c-fit}
  R_A(v) &= \left(0.4186 + \sqrt{(1.007)^2 + (0.5010v)^2}\right)^{2.5}\exp\left(1.456 - \sqrt{(1.456)^2 + v^2}\right)\,,\\
  R_B(v) &= \left(0.6893 + \sqrt{(0.790)^2 + (0.2824v)^2}\right)^{2}\exp\left(1.934 - \sqrt{(1.934)^2 + v^2}\right)\,,\\
  R_C(v) &= \frac{2.188 - (9.537\times10^{-5}v)^2+(0.1491v)^4}{1 + (0.2846v)^2 + (0.01355v)^4 + (0.1815v)^6}\,.
\end{align}
Note that temperature evolution of $v$ is computed by Eqs.~\eqref{eq:v-tau-a}-\eqref{eq:v-tau-c}.
We can see that for the superfluidity of type A and B, the specific heat is suppressed exponentially at $v\gg1$ due to the small phase space around $\varepsilon\sim \mu$.
On the other hand, the suppression of type C is weak ($R_C \propto 1/v^2$), because the gap amplitude, $\delta\propto\sin\theta$, always vanishes at $\sin\theta=0$.


\section{Neutrino emission}
\label{sec:neutrino-emis}

Inside a NS, neutrinos are produced by the various processes involving the weak interaction.
For the temperature of our interest, $T\lesssim 10^{10}\unit{K}$, a NS is fully transparent for thermally produced neutrinos.
As we will see in the following, for $t\lesssim 10^6\unit{yr}$, neutrino emission is major source of the loss of thermal energy.
In this section, we will review the main neutrino emission processes in NS core: direct/modified Urca process, bremsstrahlung and Cooper pair breaking and formation.
There are a number of sub-dominant processes both in the core and crust; we refer Ref.~\cite{Yakovlev:2000jp} as the comprehensive review. 


\subsection{Direct Urca process}
\label{sec:direct-urca}
The direct Urca process consists of the beta decay of neutrons and the inverse decay:
\begin{align}
  \label{eq:durca-reaction}
  n \to p + \ell + \bar\nu_\ell\,,\quad
  p + \ell \to n + \nu_\ell\,,
\end{align}
where $\ell = e$ and $\mu$.
In this chapter, we assume that the reaction rates of these two processes are the same and the chemical equilibrium is maintained; this is also called \textit{beta equilibrium} and restricts the chemical potentials as
\begin{align}
  \label{eq:beta-eq}
  \mu_n = \mu_p + \mu_\ell\,,
\end{align}
where neutrinos do not have chemical potential since they escape from the star.
Although the beta equilibrium is often assumed in the study of neutron star cooling, it is not correct in general~\cite{Reisenegger:1994be}. 
We will discuss the consequence of the deviation from the beta equilibrium in Chap.~\ref{chap:neutron-star-heating}.

Assuming the beta equilibrium, the emissivity, energy loss rate per unit time and unit volume, is calculated by
\begin{align}
  \label{eq:durca-emissivity}
  Q_{D,\ell}
  =
  &2\int\prod_{i=n,p,\ell,\nu}\frac{d^3p_i}{(2\pi)^3}
    (2\pi)^4\delta^4(p_n - p_p - p_\ell -p_\nu)
    \sum_{\mathrm{spin}}|\mathcal{M}_{\mathrm{Durca}}|^2
    \varepsilon_\nu f_n(1-f_p)(1-f_\ell)\,,
\end{align}
where $f$'s are the Fermi-Dirac distributions, and the prefactor $2$ is the consequence of beta equilibrium.
$\mathcal{M}_{\mathrm{Durca}}$ is the non-relativistic matrix element, whose spin sum is given by
\begin{align}
  \label{eq:non-rel-beta-decay-mat-el}
  \sum_{\mathrm{spin}}|\mathcal{M}_{\mathrm{Durca}}|^2
      \simeq 2G_F^2\cos\theta_c^2(1+3g_A^2)\,,
\end{align}
where $G_F$ is the Fermi constant, $\cos\theta_c$ the Cabibbo angle, $g_A\simeq1.27$ the axial-vector coupling of nucleons and we have dropped the terms proportional to the neutrino momentum because it vanishes after the angular integration.

Let us first calculate the emissivity by neglecting nucleon superfluidity.
The phase space integration is performed by the \textit{phase-space decomposition}.
Since we consider the temperature much below the Fermi momenta of nucleons and leptons, they are Fermi degenerate, and only the excitations around the Fermi surface participate in the reaction due to the Pauli blocking effect.
Thus, we set $p=p_F$ for nucleons and leptons in all smooth functions of the integrand; in this case, $p=p_F$ except for the distribution functions.
The momentum integration for degenerate particles is thus decomposed as
\begin{align}
  \label{eq:integ-decompose}
  p^2dp \simeq p_Fm^*d\varepsilon\,.
\end{align}

The energy conservation and beta equilibrium give $\varepsilon_\nu = p_\nu \sim T$.
Thus neutrino momentum is neglected compared to those of degenerate particles, and the momentum conserving delta function is approximated as $\delta^3(\bm{p}_n - \bm{p}_p-\bm{p}_\ell)$ where $|\bm{p}_i| = p_{F,i}$.
This delta function is satisfied only when these three momenta can form a triangle; the triangle condition
\begin{align}
  \label{eq:triangle-cond}
  p_{F,n} < p_{F,p} + p_{F,\ell}
\end{align}
has to be satisfied for nonzero emissivity.

Taking all these into account, the emissivity is evaluated as~\cite{Lattimer:1991ib} 
\begin{align}
  \label{eq:durca-emis}
  Q_{D,\ell}^{(0)}
  &=
    \frac{457\pi}{10080}G_F^2\cos^2\theta(1+3g_A^2)m_n^*m_p^*m_e^*T^6\Theta(-p_{F,n} + p_{F,p} + p_{F,e})
    \notag
  \\
  &=
    4.001\times10^{27}\cdot
    \frac{m_n^*}{m_n}\cdot\frac{m_p^*}{m_p}\cdot\frac{m^*_\ell}{k_0}\cdot
    T_9^6\,\Theta(-p_{F,n} + p_{F,p} + p_{F,e})
    \unit{erg}\unit{cm^{-3}}\unit{s^{-1}}\,,
\end{align}
where $\Theta$ is step function, $T_9=T/10^9\unit{K}$ and $k_0=1.68\unit{fm^{-1}}$.
The superscript $0$ denotes the emissivity without superfluidity.
The phase space integration is complicated, but we can easily check the temperature dependence, $T^6$, as follows.
The energy integrations of degenerate particles contribute only in the region of $\varepsilon \sim T$, which provides $T$ for each integration.
On the other hand, neutrino energy integration, $\int dp_\nu p^2_\nu$, gives $ T^3$.
There is $\varepsilon_\nu$ in the integrand, giving another $T$, while $T^{-1}$ appears from the energy conserving delta function.
Thus the temperature dependence is evaluated as
\begin{align}
  Q_{D,\ell}^{(0)}\propto T\cdot T\cdot T\cdot T^3\cdot  T\cdot  T^{-1} = T^6.
\end{align}
The emissivity rapidly decreases at low temperature.

\begin{figure}[t]
  \centering
  \begin{minipage}{0.5\linewidth}
    \includegraphics[width=1.0\linewidth]{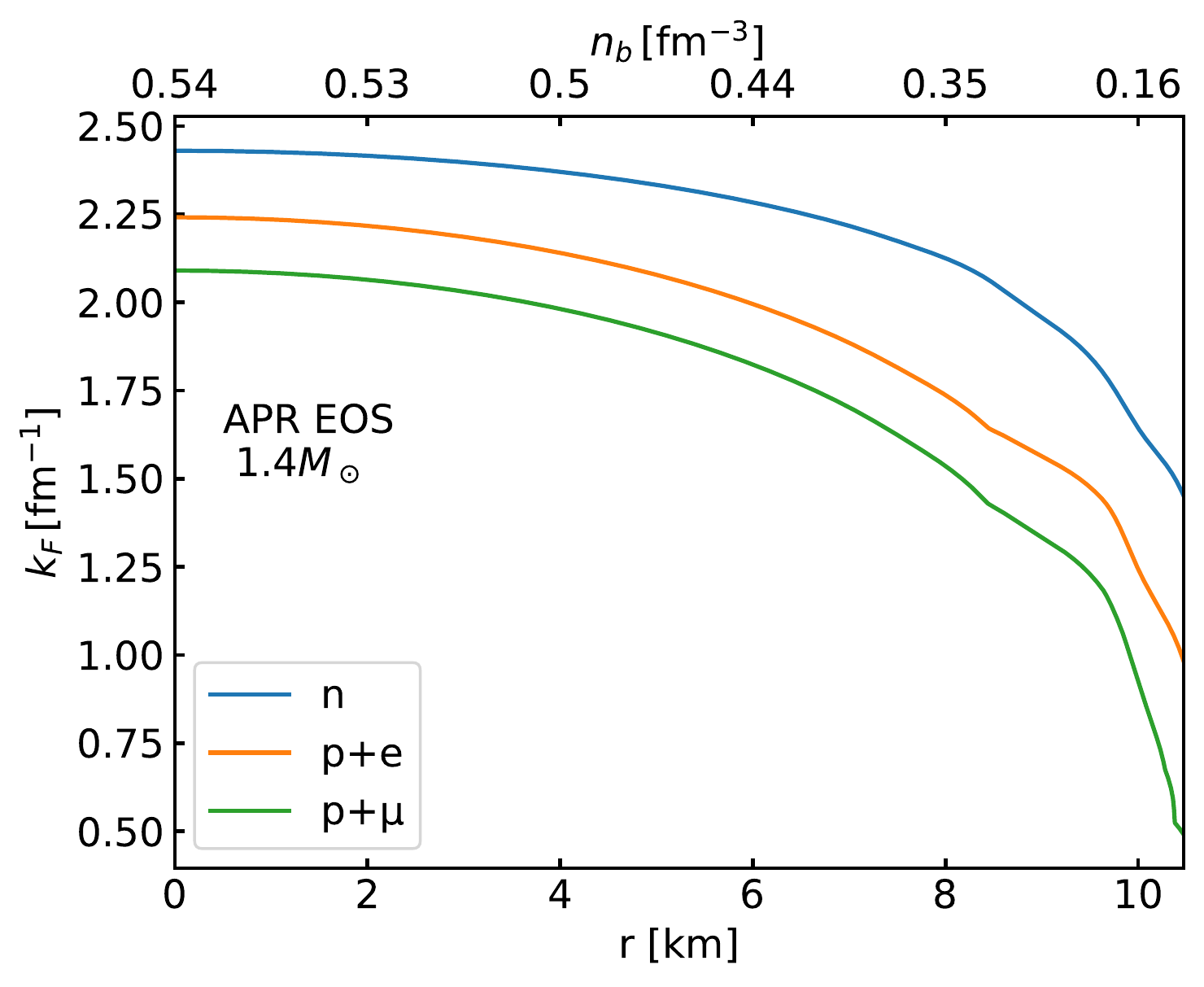}
  \end{minipage}%
  \begin{minipage}{0.5\linewidth}
    \includegraphics[width=1.0\linewidth]{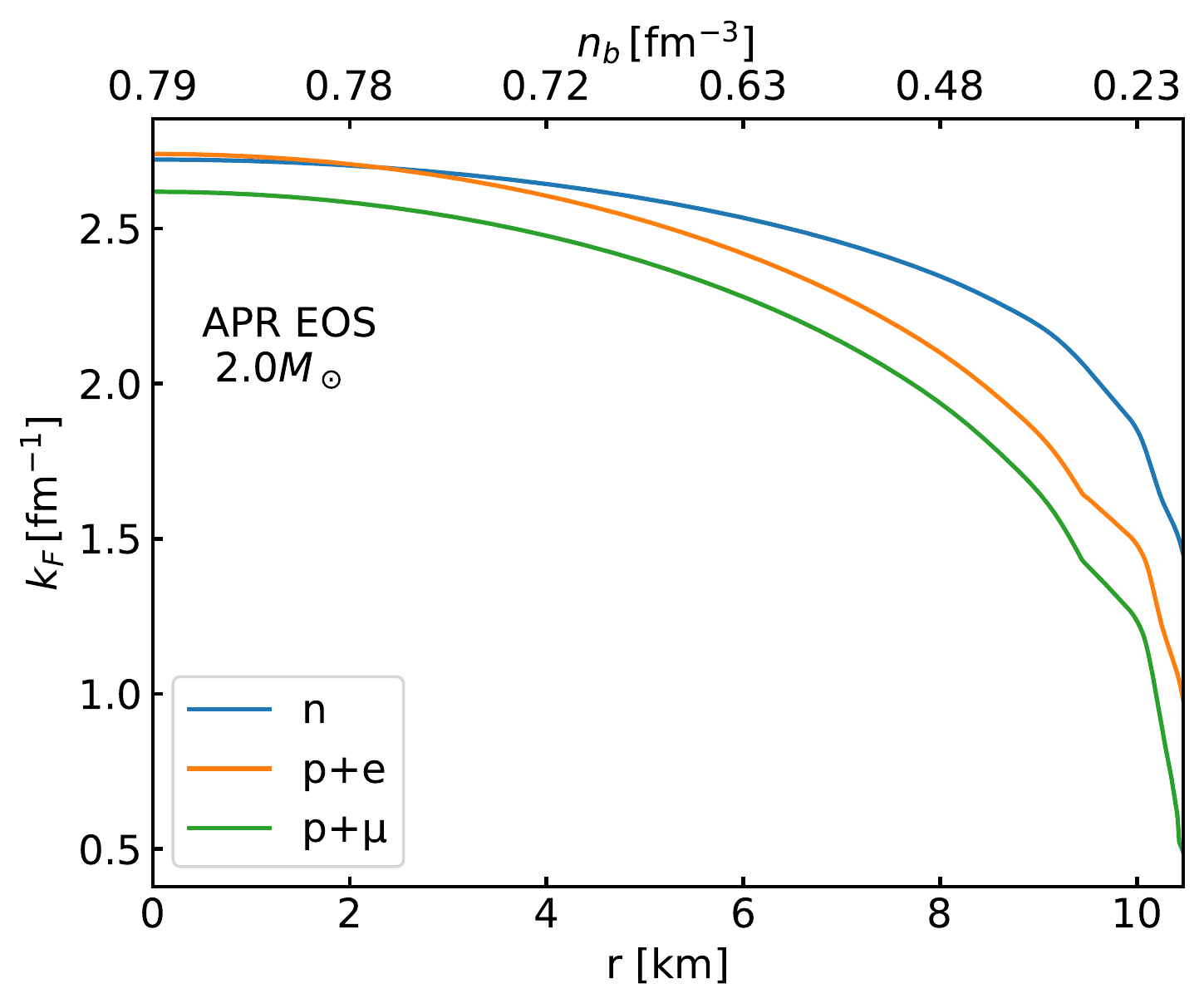}
  \end{minipage}
  \caption{Fermi momenta for $M=1.4\Msun$ (left) and $M=2.0\Msun$ (right) NS calculated by APR EOS~\cite{Akmal:1998cf}.
    The blue, yellow and green lines correspond to $k_{F,n}$, $k_{F,p} + k_{F,e}$ and $k_{F,p} + k_{F,\mu}$, respectively.}
  \label{fig:durca-condition}
\end{figure}
The triangle condition, Eq.~\eqref{eq:triangle-cond}, implies that for the direct Urca process to occur, protons and leptons need to have rather large number density.
Usually this large density is realized only near the center of very heavy NSs.
In Fig.~\ref{fig:durca-condition}, we show the Fermi momentum of nucleons and leptons for $M=1.4\Msun$ (left panel) and $M=2.0\Msun$ (right panel) using the APR EOS.
The blue, yellow and green lines correspond to $k_{F,n}$, $k_{F,p} + k_{F,e}$ and $k_{F,p} + k_{F,\mu}$, respectively.
For $M=2.0\Msun$, only the small region ($r\lesssim 2\unit{km}$) allows the direct Urca process, while it is completely forbidden in $M=1.4\Msun$ NS.
In the case of APR EOS, the direct Urca process is allowed only for $M\gtrsim 1.97\Msun$~\cite{Page:2004fy}.

\subsection{Modified Urca process}
\label{sec:modified-urca}

\begin{figure}[t]
  \centering
  \begin{fmffile}{murca}
    \begin{fmfgraph*}(120,100)
      \fmfstraight
        \fmfleft{i1,i15,i2,i3,i4}
        \fmfright{o1,o15,o2,o3,o4}
        \fmf{fermion,tension=0.6}{i1,v11}
        \fmf{plain,tension=1.3}{v11,v12}
        \fmf{fermion}{v12,o1}
        \fmf{fermion,tension=0.6}{i2,v21}
        \fmf{plain,tension=1.3}{v21,v22}
        \fmf{fermion}{v22,o2}
        \fmffreeze
        \fmf{fermion}{o4,v3,o3}
        \fmf{dashes,label=$\pi$}{v11,v21}
        \fmf{photon,tension=3}{v22,v3}
        \fmf{phantom,tension=0.3}{i4,v3}
        \fmf{phantom}{v12,v22}
      \fmflabel{$n$}{i2}
      \fmflabel{$p$}{o2}
      \fmflabel{$N_1$}{i1}
      \fmflabel{$N_2$}{o1}
      \fmflabel{$\ell$}{o3}
      \fmflabel{$\bar\nu$}{o4}
    \end{fmfgraph*}
  \end{fmffile}
  \caption{Left: One of the diagrams for the modified Urca process.}
  \label{fig:murca}
\end{figure}
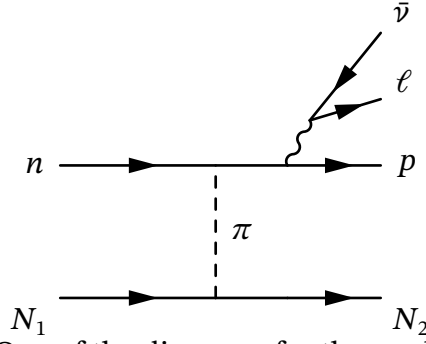

The modified Urca process relaxes the strong threshold of the direct Urca process by adding spectator nucleons.
It consists of the reactions
\begin{align}
  &n + N_1 \to p + N_2 + \ell + \bar\nu_\ell \,,
    \label{eq:murca1} \\
  & p + N_2 + \ell \to n + N_1 + \nu_\ell\,,
    \label{eq:murca2}
\end{align}
where $N_1=N_2=n$ (neutron branch) or $N_1=N_2=p$ (proton branch).
These extra nucleons, $N_1$ and $N_2$, are coupled by nuclear interaction (see Fig.~\ref{fig:murca}).
We again assume the beta equilibrium.
The emissivity of this process is given by
\begin{align}
  \label{eq:murca-emis}
  Q_{M,N\ell}
  &=
    2\int \biggl[\prod_{j=1}^4 \frac{d^3p_j}{(2\pi)^3} \biggr] \frac{d^3p_\ell}{(2\pi)^3}\frac{d^3p_\nu}{(2\pi)^3}\,
    (2\pi)^4\delta^4(P_f - P_i) \cdot \varepsilon_\nu \cdot\frac{1}{2} \sum_{\mathrm{spin}}|\mathcal M_{M,N\ell}|^2
    \notag\\
  &\times f_1f_2(1-f_3)(1-f_4)(1-f_\ell) \,,
\end{align}
where $j=1,2,3,4$ denote the nucleons $n,N_1,p, N_2$, respectively, $\delta^4(P_f - P_i)$ the energy-momentum conserving delta function, and $1/2 \times\sum_{\mathrm{spin}}|\mathcal M_{M,N\ell}|^2$ the matrix element summed over all the particles' spins with the symmetry factor.

The matrix element is calculated in Ref.~\cite{1979ApJ...232..541F} based on the free one-pion exchange, with the angular dependence of the matrix element on the momenta of nucleons and leptons neglected. We also adopt this approximation in the following analysis.

The emissivities are calculated in the same way as the direct Urca process.
Without superfluidity, they are evaluated for the neutron branch as
\begin{align}
  \label{eq:murca-emis-nbr}
  Q_{M,n\ell}^{(0)}
  =
  8.05\times10^{21}v_{F,\ell}
  \bcm{m^*_n}{m_n}{3}\bcm{m^*_p}{m_p}{}\bcm{p_{F,p}}{k_0}{}T_9^8\cdot
  \alpha\cdot\beta \unit{erg}\unit{cm^{-3}}\unit{s^{-1}}\,,
\end{align}
and for the proton branch as
\begin{align}
  \label{eq:murca-emis-pbr}
  Q_{M,p\ell}^{(0)}
  =
  Q_{M,n\ell}^{(0)}\times\bcm{m_p^*}{m_n^*}{2}
  \frac{(p_{F,\ell} + 3p_{F,p} - p_{F,n})^2}{8p_{F,\ell}p_{F,p}}
  \Theta(p_{F,\ell} + 3p_{F,p} - p_{F,n})\,,
\end{align}
where $\alpha$ and $\beta$ are introduced to take care of the correction beyond the one-pion exchange approximation; following Ref.~\cite{1979ApJ...232..541F, 1995A&A...297..717Y}, we use $\alpha = 1.76-0.63(n_0/n_n)^{2/3}$ and $\beta=0.68$.%
\footnote{
  The nuclear potential is attractive at long range due to the pion exchange, whereas at short range it is highly repulsive due probably to the heavier mesons contribution.
  Reference~\cite{1979ApJ...232..541F} introduces the $\alpha$ parameter to incorporate such corrections in the matrix element.
  At the same time the OPE interaction is cut off at that short range, which slightly reduces the contribution from the OPE part.
  This is parametrized by $\beta$, and changes by $\sim 10\%$ depending on how to cut off the potential.
  In addition, recent development in the calculation of the modified Urca process suggests that there may be a significant change in the result if, e.g., the in-medium effects are included \cite{Blaschke:1995va, DehghanNiri:2016cqm, Schmitt:2017efp, Shternin:2018dcn}.
}
Since the extra nucleons provide another $T$ through the phase space integration, the proportionality is $Q_{M,N\ell}^{(0)} \propto T^8$.
Thus the emissivity of the modified Urca is several orders of magnitude smaller than the direct Urca process.
However, the threshold of the reaction is significantly relaxed; the neutron branch does not have the threshold, while the proton branch has only the weak condition, $3p_{F,p} + p_{F,\ell}  > p_{F,n} $.

\subsection{Bremsstrahlung}
\label{sec:brems}

The neutrino bremsstrahlung is another neutrino emission process:
\begin{align}
  \label{eq:neutrino-brems}
  N_1 + N_2 \to N_1 + N_2 + \nu_\ell + \bar\nu_\ell\,,
\end{align}
where $N_{1,2} = n,p$.
This process is expressed by similar diagrams to the modified Urca process (see Fig.~\ref{fig:murca}).
The temperature dependence of the emissivity without superfluidity is estimated as follows: as in the case of the modified Urca process, through the phase space integration, the four external nucleons, the neutrino energy and the energy conserving delta function give $T^4$, $T$ and $T^{-1}$, respectively.
Unlike the modified Urca process, the squared matrix element is not constant but proportional to $\varepsilon_\nu^{-2}$, giving $T^{-2}$, since the internal nucleon propagator is close to the on-shell.
Finally the phase space integration of two neutrinos provides $(T^{3})^2$.
In total, the emissivity is proportional to
\begin{align}
  \label{eq:nu-brems-emis}
  Q_{\nu\text{-brems}}^{(0)}\propto T^4\cdot T \cdot T^{-1}\cdot T^{-2}\cdot (T^{3})^2 = T^8
\end{align}
Although this temperature dependence is the same as the modified Urca process, $Q^{(0)}\propto T^8$, its emissivity is numerically sub-dominant compared to the modified Urca process~\cite{Yakovlev:2000jp}.

\subsection{Superfluid reduction factors for Urca processes}
\label{sec:urca-reduction}
From Sec.~\ref{sec:direct-urca} to~\ref{sec:brems}, we ignore the superfluidity of nucleons.
Once we take it into account, the neutrino emissivity is highly suppressed at low temperature. 
The reason of this suppression is similar to that for the specific heat; the gap in the energy spectrum suppresses the excitation around the Fermi surface.
Following the literatures, we introduce the reduction factor $R$ by
\begin{align}
  \label{eq:reduction-urca-def}
  Q = Q^{(0)}R\,,
\end{align}
where $Q$ ($Q^{(0)}$) is the emissivity of any neutrino emission process with (without) the pairing.
In this subsection, we explore the reduction factors for direct/modified Urca process and the bremsstrahlung.
Note that we still assume that the matter is in beta equilibrium.

\subsubsection{Direct Urca process}
\label{sec:durca-red}

We begin with the direct Urca process. Equation~\eqref{eq:durca-emissivity} is also valid for the superfluid nucleons.
Since the matrix element is constant, the only difference is the energy spectra in the distribution functions.
The reduction factor is written as~\cite{1994AstL...20...43L, Yakovlev:2000jp} 
\begin{align}
  \label{eq:durca-red}
  R_{D,\ell}
  &=
  \frac{5040}{457\pi^6}
  \int_0^1 d\cos\theta_n
  \int_0^\infty dx_\nu x_\nu^3
  \int_{-\infty}^\infty dx_n dx_p dx_\ell\notag\\
  &\times
    \delta(z_n+z_p+x_\ell-x_\nu)
  f(z_n)f(z_p)f(x_\ell)\,,
\end{align}
where $\cos\theta_n$ is the polar angle of the neutron momentum around the quantization axis,
\begin{align}
  \label{eq:dimless-vars-durca-red}
  x_{N,\ell} \equiv \frac{\varepsilon_{N,\ell}-\mu_{N,\ell}}{T}\,, \quad
  x_\nu \equiv \frac{\varepsilon_\nu}{T}\,,\quad
  y_N \equiv \frac{\delta_N}{T}\,,\quad
  z_N \equiv \mathrm{sign}(x_N)\sqrt{x_N^2 + y_N^2}\,,\quad
\end{align}
and $f(z) = 1/(e^z+1)$ with $N=n,p$.
The prefactor is the inverse of
\begin{align}
  \int_0^\infty dx_\nu \, x_\nu^3 \,
  \left[\prod_{j=1}^3 \int_{-\infty}^{\infty} dx_j\, f(x_j)\right]\, \delta \left( \sum_{j=1}^3 x_j -x_\nu \right) 
  = \frac{457 \pi^6}{5040}\,.
\end{align}
Thus $R_{D,\ell}$ is normalized such that it becomes unity without nucleon pairings.
Note that the ${}^1S_0$ pairing does not have angular dependence ($F(\theta) = 1$), while the ${}^3P_2$ pairing depends on the angle (see Tab.~\ref{tab:gap-angle}).
The temperature dependence of $v_N = \Delta_N(T)/T$ is given by Eqs.~\eqref{eq:v-tau-a} - \eqref{eq:v-tau-c}

This integration has been performed numerically.
In the core, protons form ${}^1S_0$ paring and neutrons form ${}^3P_2$ pairing.
The reduction factor is a function of both proton and neutron gaps.
The neutron triplet gap is anisotropic and angular dependence is different for different total angular momentum.
As we discussed in Tab.~\ref{tab:gap-angle}, $m_j = 0$ (type B) and $m_j=\pm2$ (type C) are often studied.
The fitting formulae $R_{D,\ell}(v_p, v_n)$ for both types of neutron pairings are presented in Ref.~\cite{1994AstL...20...43L}.

\subsubsection{Modified Urca process}
\label{sec:murca-red}

The reduction factor of the modified Urca process is obtained in the similar way:
\begin{align}
  R^N_{M,\ell}
  &=
    \frac{120960}{11513\pi^8}
    \frac{1}{A_0^N}
    \int \prod_{j=1}^5\frac{d\Omega_j}{4\pi}
    \delta^3\left(\sum_{j=1}^5\bm p_j\right)
    \int_0^\infty dx_\nu \int_{-\infty}^{\infty}dx_ndx_pdx_{N_1}dx_{N_2}
    \notag\\
  &\times
    x_\nu^3 \cdot 
    f(z_n)f(z_p)f(z_{N_1})f(z_{N_2})f(x_\ell)
  \delta(z_n+z_p+z_{N_1}+z_{N_2}+x_\ell-x_\nu)
    \,,
    \label{eq:murca-r-integ-emis}
\end{align}
where $N_1 = N_2 = N$ takes $p$ (proton branch) or $n$ (neutron branch), and
$j=1,2,\dots,5$ correspond to $n, p, \ell, N_1, N_2$, respectively.
Note that the momentum of a degenerate particle is set as $|\bm{p}_{F,i}| = p_{F,i}$.
The constant factor $A_0^N$ is given by 
\begin{align}
  A_0^N  &\equiv \int \prod_{j=1}^5\frac{d\Omega_j}{4\pi} 
           \delta^3\biggl(\sum_{j=1}^5\bm p_j\biggr) 
           =
           \begin{cases}
             \frac{1}{8\pi p_{F,n}^3} & (N=n) \\
             \frac{(p_{F,\ell} + 3 p_{F,p} - p_{F,n})^2}{64\pi p_{F,n} p_{F,p}^3 p_{F,\ell}} 
             \Theta_p & (N=p)
           \end{cases}
                        ~,
\end{align}
where $\Theta_p = \Theta(3 p_{F,p} + p_{F,\ell}- p_{F,n} )$ and for neutron branch, we neglect the proton and lepton Fermi momenta.
The prefactor of Eq.~\eqref{eq:murca-r-integ-emis} is the inverse of
\begin{align}
  \label{eq:a0n}
  \int_0^\infty dx_\nu \, x_\nu^5 \,
  \left[\prod_j^3 \int_{-\infty}^{\infty} dx_j\, f(x_j)\right]\, \delta \left( \sum_{j=1}^5 x_j -x_\nu \right) 
  = \frac{11513 \pi^8}{120960}\,,
\end{align}
so that the reduction factor is properly normalized.
We note that due to the angular dependence of neutron triplet pairing, angular integrals in Eq.~\eqref{eq:murca-r-integ-emis} differ for different branches.

\begin{figure}
  \centering
  \begin{minipage}{0.5\linewidth}
    \includegraphics[width=1.0\linewidth]{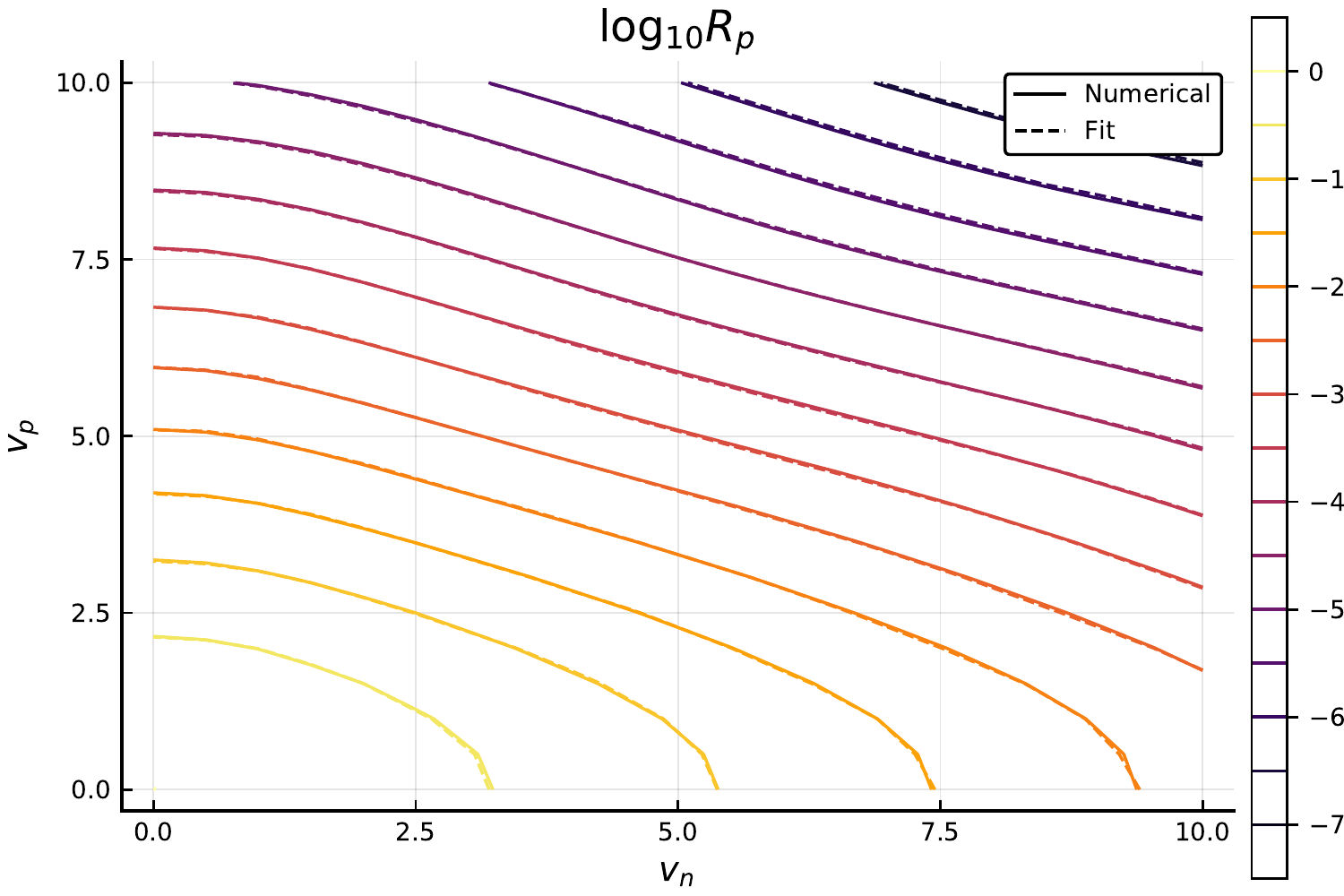}
  \end{minipage}%
  \begin{minipage}{0.5\linewidth}
    \includegraphics[width=1.0\linewidth]{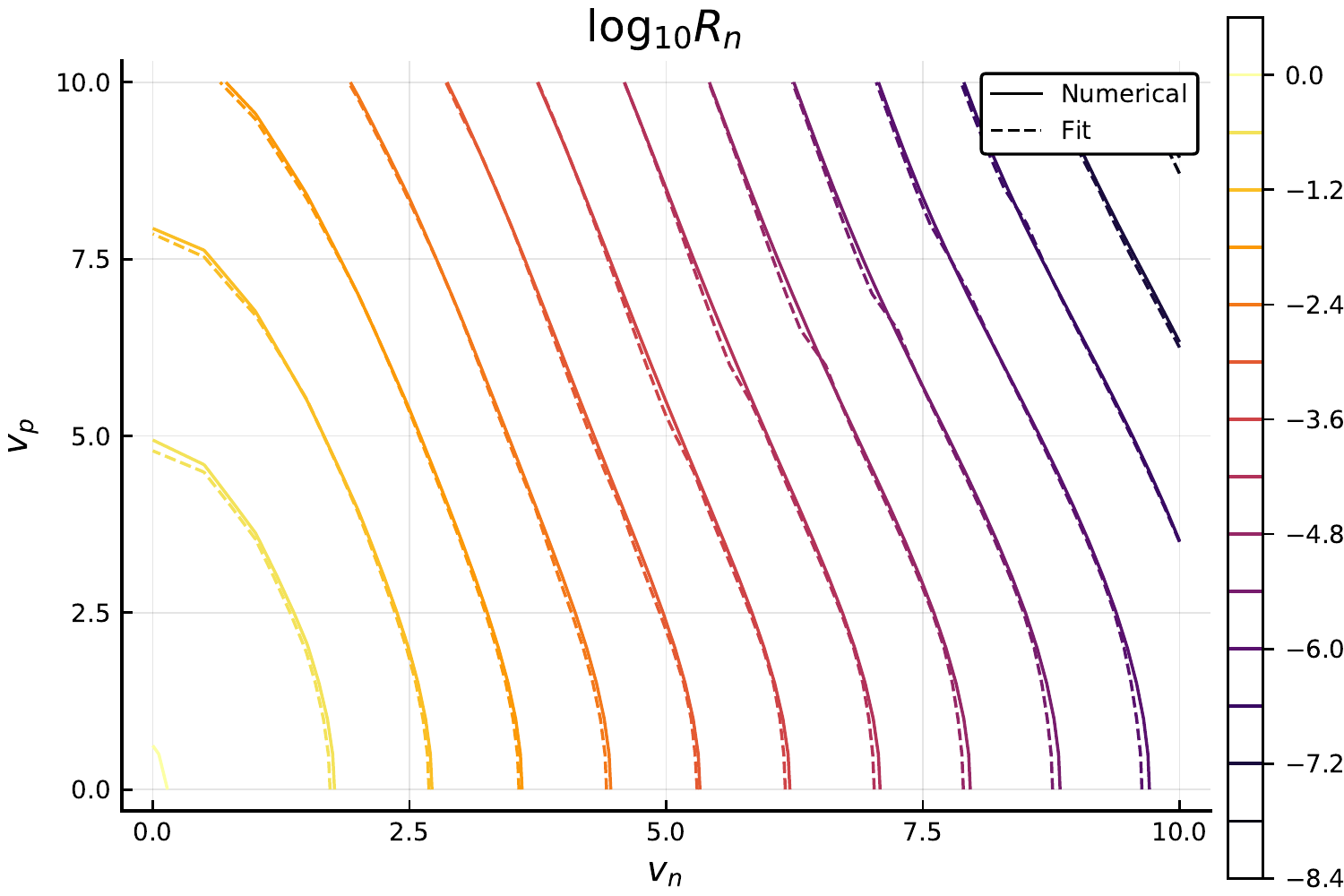}
  \end{minipage}
  \caption{The reduction factor of modified Urca process for proton branch (left) and neutron branch (right).
    We take proton type A and neutron type B pairing (see Tab.~\ref{tab:gap-angle}).
  The solid contours show $\log_{10}R^N_{M,\ell}$ from the numerical integration as a function of dimensionless gap $v_N=\Delta_N(T)/T$, while the dashed contours are obtained from the corresponding fitting formulae presented in Ref.~\cite{Gusakov:2002hh}.}
  \label{fig:murca-red-compare}
\end{figure}
The detailed study of this integration is performed in Refs.~\cite{1995A&A...297..717Y, Gusakov:2002hh}.
In particular, Ref.~\cite{Gusakov:2002hh} analyzed the reduction factor in the presence of both proton singlet and neutron triplet pairings.  
The numerical integration is performed for type A proton and type B or C neutron superfluidity, and the results are summarized in the fitting formulae.
In Fig.~\ref{fig:murca-red-compare}, we compare the numerical integration of Eq.~\eqref{eq:murca-r-integ-emis} (solid contours) to the fitting formulae in Ref.~\cite{Gusakov:2002hh} (dashed contours).
We assume proton type A and neutron type B pairings, and colored contours indicate the $\log_{10}$ values of reduction factor for proton branch (left panel) and neutron branch (right panel). 
We can see that the fitting formulae reproduce the numerical results very well.
We also see that except for the vicinity of the axes ($v_p=0$ or $v_n=0$), the reduction factor of proton (neutron) branch depends on the gap amplitudes mostly through the combination of $3v_p + v_n$ ($v_p + 3v_n$).
This counts the gap of nucleons involved in the process.
Therefore, if the proton gap is larger than the neutron gap, only the neutron branch is important, and vice versa.

\subsubsection{Bremsstrahlung}
\label{sec:brems-reduc}

The neutrino bremsstrahlung also receives the superfluid suppression in its emissivity.
The reduction factors are calculated only when either neutrons or protons form Cooper pairs.
For the case of both singlet proton and triplet neutron superfluidity, only the approximate formulae is available~\cite{Yakovlev:2000jp}. 

\subsection{The Cooper pair breaking and formation}
\label{sec:pbf}

There is another important neutrino emission process so-called the Cooper pair breaking and formation (PBF).
It is associated with the Cooper pair breaking due to the thermal disturbance and its subsequent reformation.
During this reformation, the neutrino anti-neutrino pair is emitted.
The PBF process for singlet pairing is first proposed in Ref.~\cite{Flowers:1976ux}.
This work has been forgotten for almost 20 years and finally incorporated in analyses of thermal evolution in late 90's, when the PBF of neutron triplet pairing is also developed~\cite{Yakovlev:1998wr}.
As we will see, the PBF process usually dominates the neutrino emission luminosity after the nucleon pairing.

In this section, we discuss the PBF process based on the second quantized Hamiltonian.
The relevant weak interaction for non-relativistic nucleons are described by
\begin{align}
  \label{eq:hamiltonian-pbf}
  \hat{\mathcal H}_{\mathrm{int}}
  =
  -\frac{G_F}{2\sqrt 2}\,\left( c_Vl^0J^0 - c_A \bm{l}\cdot\bm{J}\right)\,,
\end{align}
where $l^\mu = \bar\nu\gamma^\mu(1-\gamma_5)\nu$ is the neutrino current for any flavor, $c_V(c_A)$ the vector (axial-vector) coupling constant, whose numerical value is shown later in this subsection.
$J^\mu$ is the nucleon current written in the non-relativistic limit as
\begin{align}
  \label{eq:nucl-current-nr}
  J^0 = \hat\Psi^\dagger \hat\Psi\,,\quad
  \bm{J} = \hat\Psi^\dagger\bm\sigma\hat\Psi\,.
\end{align}
where the 2-component spinor $\hat\Psi$ is the second quantized annihilation operator of nucleons.
Using the annihilation operator corresponding to the plain wave, $\hat{c}_{\bm p\sigma}$, this is expanded in the Schr\"odinger picture as
\begin{align}
  \label{eq:psi-exp}
  \hat\Psi(\bm r)
  =
  \frac{1}{\sqrt V}\sum_{\bm p,\sigma}\hat{c}_{\bm p\sigma}\chi_\sigma e^{i\bm{p}\cdot\bm{r}}\,,
\end{align}
where $\sum_{\bm p} = \int \frac{Vd^3p}{(2\pi)^3}$, $\sigma=\uparrow\downarrow$ denotes spin, and $\chi_\sigma$ is corresponding 2-component spinor which satisfies $\chi^\dagger_\sigma \chi_{\sigma^\prime} = \delta_{\sigma\sigma^\prime}$.

Because of the pairing, the operator $\hat{c}_{\bm p\sigma}$, corresponding to the excitation of individual nucleons, is not the appropriate operator to describe the elementary excitations above the ground state.
Instead the elementary excitations are collective excitations called quasiparticles or quasinucleons created by $\hat\alpha_{\bm p\sigma}^\dagger$.
The Bogoliubov transformation provides the relation between these two operators, which is derived in App.~\ref{chap:gap-equation} for ${}^1S_0$ and ${}^3P_2$ pairing as
\begin{align}
  \label{eq:inverse-bogoliubov}
  \hat{\bm c}_{\bm k} &= U_\lambda(\bm k)\hat{\bm\alpha}_{\bm k} + V_\lambda(\bm k)\hat{\bm\alpha}^\dagger_{-\bm k}\,,
                        \notag\\
  \hat{\bm c}^\dagger_{\bm k} &= U_\lambda(\bm k)\hat{\bm\alpha}_{\bm k}^\dagger + V_\lambda^*(\bm k)\hat{\bm\alpha}_{-\bm k}\,,
\end{align}
where $\hat{\bm{c}}_{\bm k} = (\hat{c}_{\bm k \uparrow}, \hat{c}_{\bm k \downarrow})^\intercal$ and $\hat{\bm{\alpha}}_{\bm k} = (\hat{\alpha}_{\bm k \uparrow}, \hat{\alpha}_{\bm k \downarrow})^\intercal$. 
The $2\times2$ unitary matrices $U_\lambda$ and $V_\lambda$ are 
\begin{align}
  \label{eq:u-mat-1s0}
  U_\lambda(\bm k) = u_{\bm k}\bm{1}\,,
  V_\lambda(\bm k) = v_{\bm k}\Gamma(\bm k)\,,
  \text{ with }
  u_{\bm k}^2 + v_{\bm k}^2 = 1\,,
\end{align}
where for singlet pairing, $\Gamma(\bm k) = i\sigma^2$, 
while for triplet pairing, $\Gamma(\bm k)$ is a symmetric unitary matrix satisfying $\Gamma(-\bm k) = -\Gamma(\bm k)$, and determined by the equilibrium condition along with the gap.
Note that in both cases, $\Gamma^\intercal(-\bm k) = -\Gamma(\bm k)$.

Substituting the Bogoliubov transformation into Eq.~\eqref{eq:hamiltonian-pbf}, we obtain the term proportional to $\hat\alpha\hat\alpha$, which servers as the annihilation of two quasinucleons into a neutrino pair:
\begin{align}
  \label{eq:pbf-qp-diagram}
  \tilde{N}\tilde{N}\to [\tilde{N}\tilde{N}] + \nu\bar\nu\,,
\end{align}
where $\tilde{N}$ denotes quasinucleons, and $[\tilde{N}\tilde{N}]$ the paired state.
In terms of the individual nucleon, denoted by $N$, the vertex Eq.~\eqref{eq:hamiltonian-pbf} corresponds to the bremsstrahlung $N\to N\nu\bar\nu$, and so is forbidden by the kinematics.
The PBF process~\eqref{eq:pbf-qp-diagram} is thus due to the collective effect of the superfluidity, and is allowed only below the critical temperature.

The emissivity of the PBF is obtained in the usual way as
\begin{align}
  \label{eq:pbf-emis}
  Q_{\mathrm{PBF},N}
  &=
    \frac{1}{2}N_\nu \int \frac{d^3p_N}{(2\pi)^3}\frac{d^3p_N^\prime}{(2\pi)^3}
    \frac{d^3p_\nu}{2\varepsilon_\nu(2\pi)^3} \frac{d^3p_\nu^\prime}{2\varepsilon_\nu^\prime(2\pi)^3}
    (2\pi)^4\delta(p_N+p_N^\prime-p_\nu-p_\nu^\prime)
    \notag\\
  &\times(\varepsilon_\nu + \varepsilon_\nu^\prime)
    \sum_{\mathrm{spin}}\left|\mathcal{M}_{\mathrm{PBF}} \right|^2
    f(\varepsilon_N)f(\varepsilon_N^\prime)\,,
\end{align}
where $p_N$ and $p_N^\prime$ are the incoming quasinucleons momenta, $p_\nu$ and $p^\prime_\nu$ are neutrino and anti-neutrino momenta, and $\varepsilon$'s are corresponding energies.
The prefactor $1/2$ excludes the double counting of the quasinucleons, and $N_\nu = 3$ is the number of neutrino flavors.
The matrix element, $\mathcal{M}_\mathrm{PBF}$, is different for different superfluid types and nucleon species.
The emissivity is evaluated in the following form~\cite{Flowers:1976ux, Yakovlev:1998wr}%
\footnote{
  The proportionality for $T$ is estimated in the same way as we did in the Urca processes, except that the momentum conserving delta function in this case also provides $T^{-1}$ because the Cooper pair formation occurs for two nucleons with $\bm p_N + \bm p_N^\prime \simeq 0$.
}
\begin{align}
  \label{eq:pbf-emis-2}
  Q_{\mathrm{PBF},N}
  &=
    \frac{4N_\nu G_F^2m_N^*p_{F,N}}{15\pi^5}T^7
    \cdot a_{N,j}\cdot F_j\left( v_j \right)
    \notag\\
  &=
    3.51\times10^{21}\cdot\frac{m_N^*}{m_N}\frac{p_{F,N}}{m_{N}}T_9^7
    \cdot a_{N,j}\cdot F_j\left( v_j \right)
    \unit{erg\, cm^{-3}\,s^{-1}}\,,
\end{align}
where $a_{N,j}$ involves the vector and axial-vector couplings with $j$ referring to the superfluid type (we will give its explicit expression shortly), and $v_j = \Delta_j(T)/T$.
$F_j$ is the phase space integral involving the nucleon distribution functions and neutrino momenta:
\begin{align}
  \label{eq:pbf-control-integ}
  F_j(v_j)
  &=
    \int\frac{d\Omega}{4\pi}y^2\int_0^\infty dx\,\frac{z^4}{(e^z+1)^2}\,,
\end{align}
where $y=\delta_j/T$ and $z=\sqrt{x^2+y^2}$ provide the dependence on the superfluid type.
Due to the energy gap, this integral has to be performed numerically, and the fitting formulae are obtained as follows~\cite{Yakovlev:1998wr}:
\begin{align}
  F_A(v)
  &=
    (0.602v^2 + 0.5942v^4 + 0.288v^6)
    \left(  0.5547 + \sqrt{(0.4453)^2 + 0.0113v^2}\right)^{1/2}
    \notag\\
    &\times
    \exp\left(-\sqrt{4v^2 + 2.245^2} + 2.245\right)\,,
      \label{eq:pbf-control-a}
  \\
  F_B(v)
  &=
    \frac{1.204v^2 + 3.733v^4 + 0.3191v^6}{1 + 0.3511v^2}
    \left(0.7591 + \sqrt{(0.2409)^2 + 0.3145v^2}\right)^2
    \notag\\
  &\times
    \exp\left( -\sqrt{4v^2 + (0.4616)^2} + 0.4616 \right)\,,
    \label{eq:pbf-control-b}
  \\
  F_C(v)
  &=
    \frac{0.4013v^2 - 0.043v^4 + 0.002172v^6}{1 - 0.2018v^2 + 0.02601v^4 - 0.001477v^6 + 0.0000434v^8}\,.
    \label{eq:pbf-control-c}
\end{align}
These functions indicate that for large $v$, the emissivity is suppressed.
This is because the PBF occurs through the annihilation of excited quasiparticles, whose number density is quite suppressed for $T \ll \Delta$.
This suppression is exponential for type A and B while it is power low for type C, since type C gap has always gap-less direction at $\sin\theta = 0$ (see Tab.~\ref{tab:gap-angle}).
Therefore, the PBF neutrino emission is very efficient for the temperature only slightly below the critical temperature.

To determine the factor $a_{N,j}$, we need to evaluate the matrix element of the process.
This has been done in Refs.~\cite{Flowers:1976ux, 1986ZhETF..90.1505V, Yakovlev:1998wr} using the second quantized Hamiltonian~\eqref{eq:hamiltonian-pbf} and Bogoliubov transformation~\eqref{eq:inverse-bogoliubov}, and the resultant expressions are
\begin{align}
  \label{eq:anj-tree}
  a_{N,A}^{(\mathrm{tree})}
  &=
    c_{V,N}^2
    +
    c_{A,N}^2\cdot v_{F,N}^2
    \bcm{m^*_N}{m_N}{2}
    \left( 1 + \frac{11}{42} \bcm{m^*_N}{m_N}{-2} \right)\,,
    \notag\\
  a_{N,B}^{(\mathrm{tree})}
  &= a_{N,C}^{(\mathrm{tree})} = c_{V,N}^2 + 2c_{A,N}^2\,,
\end{align}
where $N=n,p$.
These correspond to the tree level calculation in a more sophisticated Green's function approach.
While their analyses provide the correct results for the axial vector part (proportional to $c_{A,N}^2$), they are not sufficient for the nucleon vector current (proportional to $c_{V,N}^2$) because such lowest order calculation in the BCS Hamiltonian violates the vector current conservation.
As is well known, the quasiparticle $\tilde N$ is not the eigenstate of charge.%
\footnote{Here the charge refers to the electrical charge or baryon number.}
The corresponding charge conservation or Ward identity is satisfied by including the radiative correction to the vertex, which arises from the collective excitation of the superfluid nucleons~\cite{Nambu:1960tm} (see also Ref.~\cite{PhysRevB.26.4883}).%
\footnote{
  This collective excitation is the Nambu-Goldstone mode of the condensation.
  For the charged current, it is absorbed into the longitudinal mode of photon field, and becomes the plasma oscillation~\cite{Nambu:1960tm}.
  Thus it further modifies the vector current vertex of the proton.
  This modification is, however, $\Order(1)$ for protons in a NS and does not affect the thermal evolution\cite{Kolomeitsev:2008mc}. 
}
Taking this correction properly into account and expanding by $v_F$, Ref.~\cite{Leinson:2006gf} shows that tree level contribution to the PBF from the vector current cancels with the radiative correction up to $\Order(v_F^4)$.

For the singlet pairing, this cancellation provides the significant suppression to the PBF emissivity since at tree level, the leading contribution comes from the vector current.
After the inclusion of the radiative corrections, the vector current contributes only at $\Order(v_F^4)$ while the axial-vector part remains to be $\Order(v_F^2)$.
The resultant factor $a_{N,j}$ is calculated as~\cite{Kolomeitsev:2008mc} (see also Ref.~\cite{Page:2009fu}) 
\begin{align}
  \label{eq:anj-1s0}
  a_{N,A}
  &=
    c_{V,N}^2\cdot \frac{4}{81}
    v_{F,N}^4
    +
    c_{A,N}^2\cdot v_{F,N}^2
    \bcm{m^*_N}{m_N}{2}
    \left( 1 + \frac{11}{42} \bcm{m^*_N}{m_N}{-2} \right)\,.
\end{align}
For the triplet pairing, since the leading order term comes from both the vector and axial-vector currents, the effect of the radiative correction is moderate.
The vector current contribution again receives the strong suppression, and hence we only consider the axial-vector contribution, which results in~\cite{Leinson:2006gh, Page:2009fu} 
\begin{align}
  \label{eq:anj-triplet}
  a_{N,B} = a_{N,C} = 2c_{A,N}^2\,.
\end{align}

The PBF neutrino emission of ${}^1S_0$ proton superfluid is described by Eq.~\eqref{eq:anj-1s0} with the coupling constants being $c_{V,p} = 4\sin^2\theta_W-1\simeq -0.08$ and $c_{A,p} = -g_A \simeq -1.27$, where $\theta_W$ is the Weinberg angle.
Thus the proton PBF is dominated by the axial-vector contribution.
The neutrons in the crust also emit neutrinos through PBF of ${}^1S_0$ pairing, whose emissivity is calculated with $c_{V,n}=1$ and $c_{A,n}=g_A$.
The main contribution is also the axial-part, which is suppressed by $v_{F,n}^2$.
On the other hand, neutrons in the core form ${}^3P_2$ pairing, and its $a_{N,j}$ factor is given by Eq.~\eqref{eq:anj-triplet}, which does not receive the suppression of the powers of $v_{F,n}$.
Therefore, the triplet pairing in the core dominates the neutron PBF.

\section{Standard theory of cooling}
\label{sec:standard-cooling}

In this section, we integrate all the ingredients discussed in the previous sections.
In Sec.~\ref{sec:basic-eqs}, we presented the thermal evolution equations
\begin{align}
  &-\lambda\frac{d(T(r)e^{\Phi(r)})}{dr}
  =
  \frac{L_{d}}{4\pi r^2}e^{\Phi(r)}e^{\Lambda(r)}\,,
  \tag{\ref{eq:heat-cond-eq}}\\
  &c_V\frac{dT(r)}{dt}e^{\Phi(r)}4\pi r^2e^{\Lambda(r)}
  =
  -Q_\nu e^{2\Phi(r)}4\pi r^2 e^{\Lambda}
  -
  \frac{d(L_de^{2\Phi(r)})}{dr}.
  \tag{\ref{eq:en-balance}}
\end{align}
These are solved with microphysics inputs, specifically the specific heat discussed in Sec.~\ref{sec:spec-heat} and neutrino emissivity $Q_\nu$ in Sec.~\ref{sec:neutrino-emis}.
The resultant internal temperature is converted to the surface effective temperature $T_s$ by the $T_s-T_b$ relation in Sec.~\ref{sec:envelopew}. 

Following Refs.~\cite{Page:2004fy, Gusakov:2004se, Page:2009fu}, we introduce the classification of \textit{minimal cooling} and \textit{enhanced cooling}.
The minimal cooling does not consider the direct Urca process. 
With this assumption, the main neutrino emission source is modified Urca and PBF processes.
Meanwhile, the enhanced cooling incorporates the direct Urca process, which enhances the neutrino emission rate drastically, and thus provide the colder NS than the minimal cooling.
Although we do not consider the exotic particle such as hyperons or quarks, their emergence also tends to enhance the cooling and such a situation is also classified as the enhanced cooling. 

\subsection{Minimal cooling}
\label{sec:minimal-cooling}

In minimal cooling, we do not consider the direct Urca process which could greatly enhances the neutrino emission rate.
Using the APR EOS, this is equivalent to considering NSs lighter than $1.97\Msun$ (see Sec.~\ref{sec:direct-urca}).
In such a case, the major uncertainties of theoretical predictions come from the uncertainties of superfluid gaps and the amount of light elements in the envelope. 

\begin{figure}
  \centering
  \begin{minipage}{0.5\linewidth}
    \includegraphics[width=1.0\linewidth]{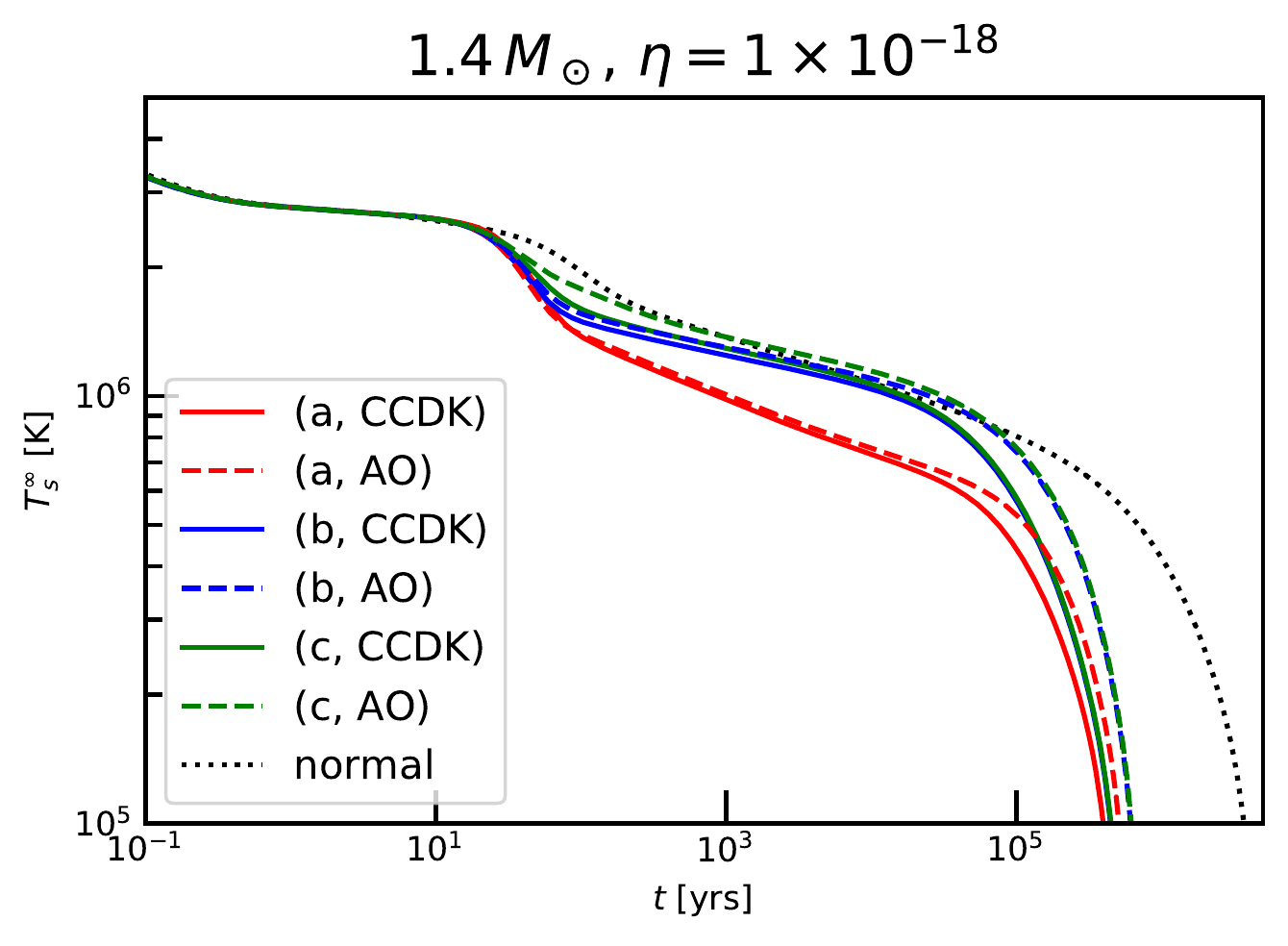}
  \end{minipage}%
  \begin{minipage}{0.5\linewidth}
    \includegraphics[width=1.0\linewidth]{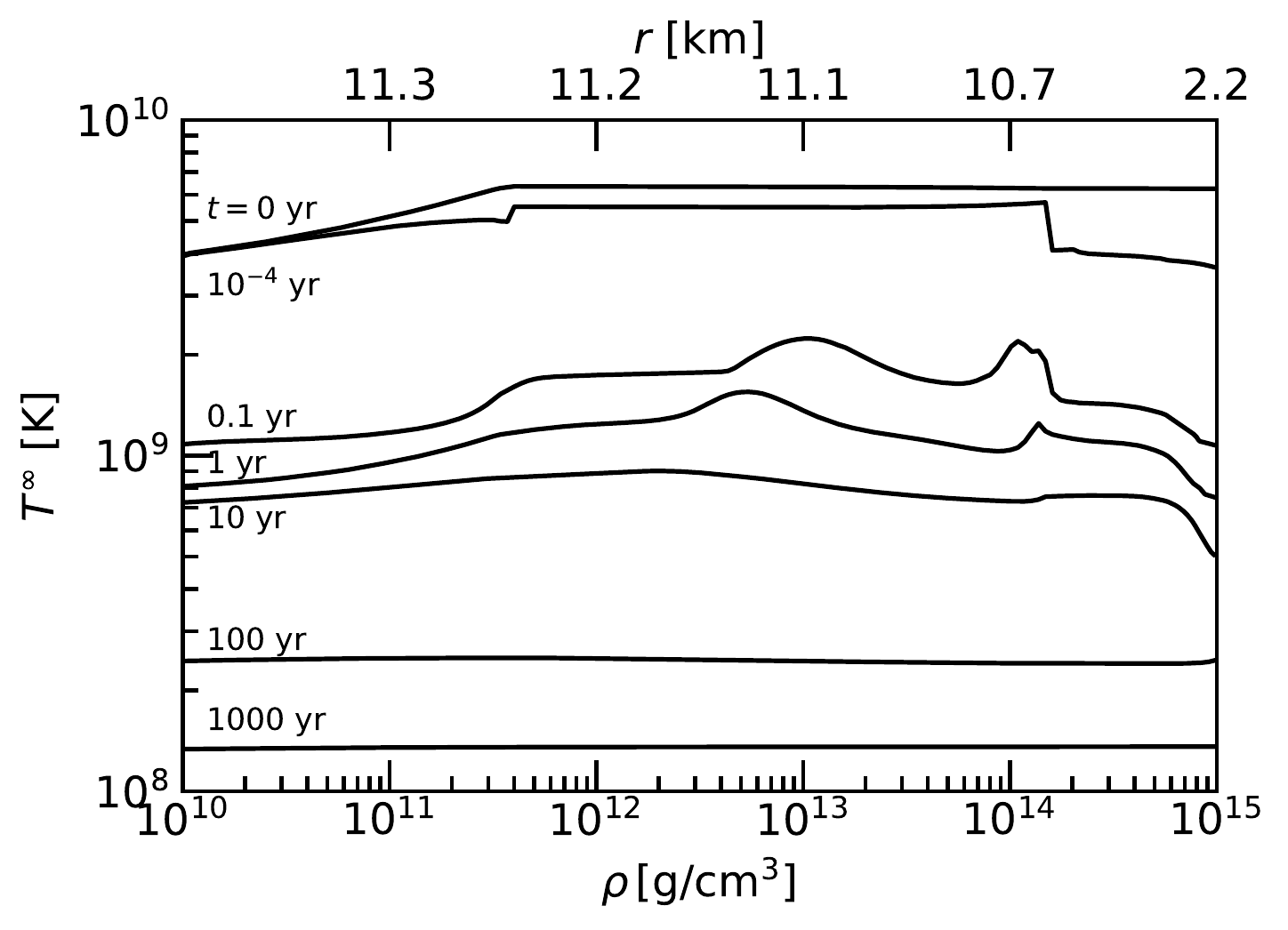}
  \end{minipage}%
  \caption{Left: Surface temperatures of a $1.4\Msun$ NS with $\eta = 1\times 10^{-18}$ calculated by \texttt{NSCool}~\cite{NSCool}.
    We use ``a'' (red), ``b'' (blue) and ``c'' (green) models for neutron ${}^3P_2$ gap, and CCDK (solid) or AO (dashed) model for proton ${}^1S_0$ gap. The results without superfluidity is also shown (dashed).
    Right: The redshifted internal temperature as a function of mass density inside the NS corresponding to the case of (a, CCDK) gap models in the left panel.
}
  \label{fig:ts-sf-compare}
\end{figure}
The left panel of Fig.~\ref{fig:ts-sf-compare} shows the thermal evolution of a $1.4\Msun$ NS with different gap models (colored lines) compared with that without nucleon superfluidity (black dotted line).
The redshifted surface temperature, $T_s^\infty = T_se^{\Phi(R)}$, is shown.
These are calculated by the public code \texttt{NSCool}~\cite{NSCool}, which incorporates all the relevant microphysics including the conductivity $\lambda$, and solves the coupled equations~\eqref{eq:heat-cond-eq} and~\eqref{eq:en-balance}.%
\footnote{
  In the presence of both proton and neutron pairings, \texttt{NSCool} adopts the simplified prescription for the reduction factor of modified Urca process (so called similarity criteria~\cite{Yakovlev:2000jp, Yakovlev:1999sk}).
  It provides a crude approximation of a more accurate expressions in Ref.~\cite{Gusakov:2002hh}, which are plotted in Fig.~\ref{fig:murca-red-compare}.
}
\footnote{
  In \texttt{NSCool}, the vector current contribution to the PBF (see Eqs.~\eqref{eq:anj-1s0} and~\eqref{eq:anj-triplet}) is set to be zero.
  This does not change the resultant temperature since the axial vector part always dominates the emissivity.
}
In Fig.~\ref{fig:ts-sf-compare}, we use ``a'' (red), ``b'' (blue) and ``c'' (green) models for neutron ${}^3P_2$ gap,\footnote{We assume $m_j=0$ (type B) pairing throughout this section.} and CCDK (solid) or AO (dashed) model for proton ${}^1S_0$ gap.
The SFB model~\cite{Schwenk:2002fq} is used for neutron ${}^1S_0$ gap in the crust.
The neutron gap amplitude becomes larger from ``a'' to ``c''. The proton ``CCDK'' gap is large while ``AO'' is small (See Fig.~\ref{fig:gap-3p2}).
The light element amount in the envelope is fixed at $\eta =1\times 10^{-18}$.

We first note that for $t\lesssim 100-200\unit{yr}$, the thermal equilibrium is not achieved.
In the right panel of Fig.~\ref{fig:ts-sf-compare}, we show the redshifted internal temperature, $T^\infty = T(r)e^{\Phi(r)}$, as a function of the mass density $\rho(r)$.
The line of $t=0\unit{yr}$ is the initial condition.
We can see that the temperature profile is quite density dependent at $t \lesssim 10\unit{yr}$; the core ($\rho \gtrsim 10^{14}\unit{g/cm^3}$) is kept to be colder than the crust ($\rho \lesssim 10^{14}\unit{g/cm^3}$).
The left panel shows the sharp decline of $T_s^\infty$ at $t=10-100\unit{yr}$, which corresponds to the thermalization of the core and the crust.
Thus the thermalization time scale is about $10-100\unit{yr}$ in a typical NS, and after that $T^\infty$ becomes constant.

The cooling at $t \lesssim 10^5\unit{yr}$ is dominated by the neutrino emission.
In the left panel of Fig.~\ref{fig:ts-sf-compare}, we see the inclusion of the nucleon pairing tends to enhance the neutrino emission at $t \sim 10^{2-3}\unit{yr}$.
The quantitative effect is sensitive to the choice of nucleon gap models.
At $t\sim 10^5\unit{yr}$, the surface photon luminosity becomes comparable with the neutrino luminosity, and the NS undergoes the photon cooling stage.
After the photon emission dominates the cooling, the surface temperature sharply decreases.

\begin{figure}[t]
  \centering
  \begin{minipage}{0.5\linewidth}
   \includegraphics[width=1.0\linewidth]{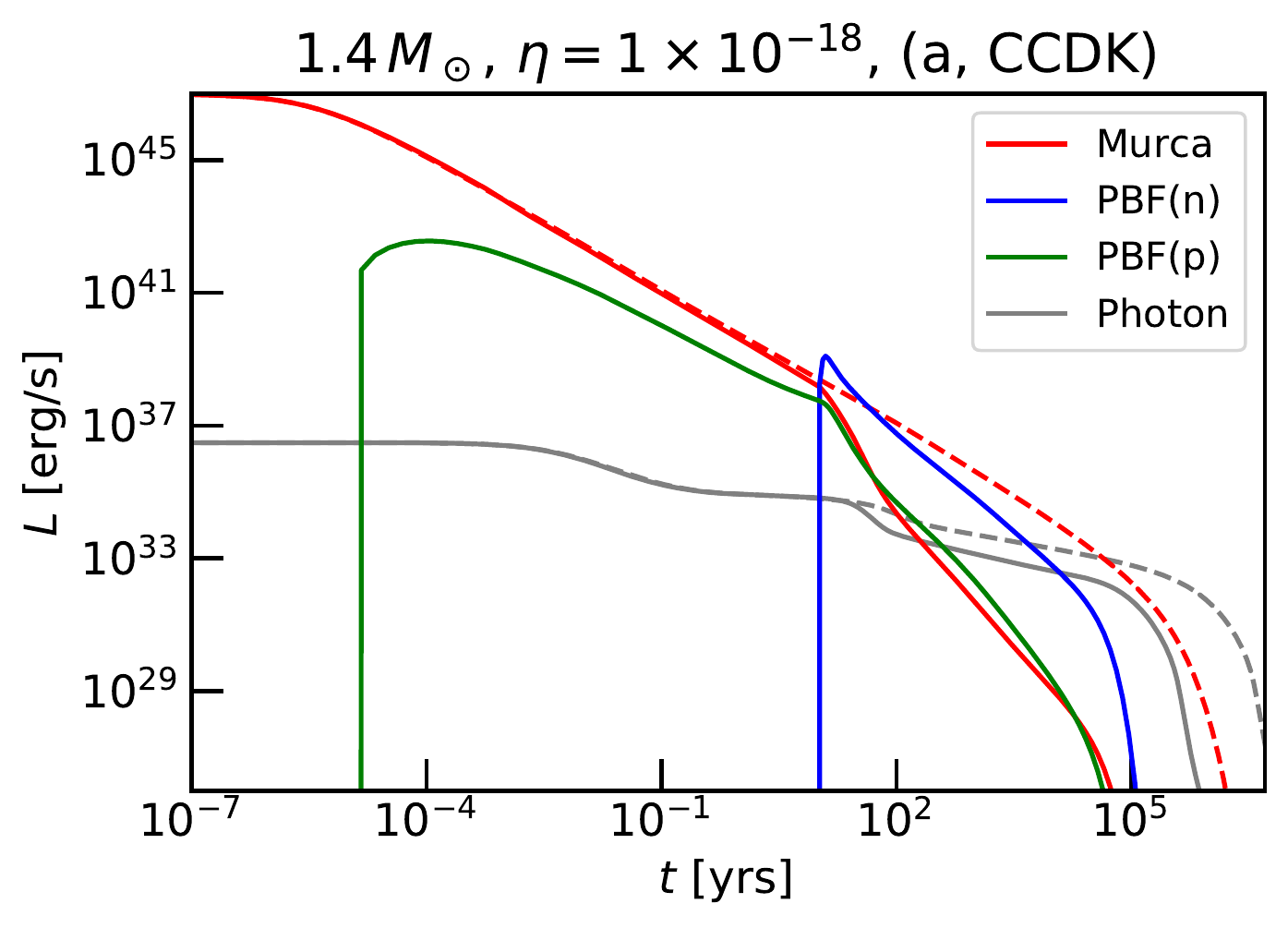} 
  \end{minipage}%
  \begin{minipage}{0.5\linewidth}
    \includegraphics[width=1.0\linewidth]{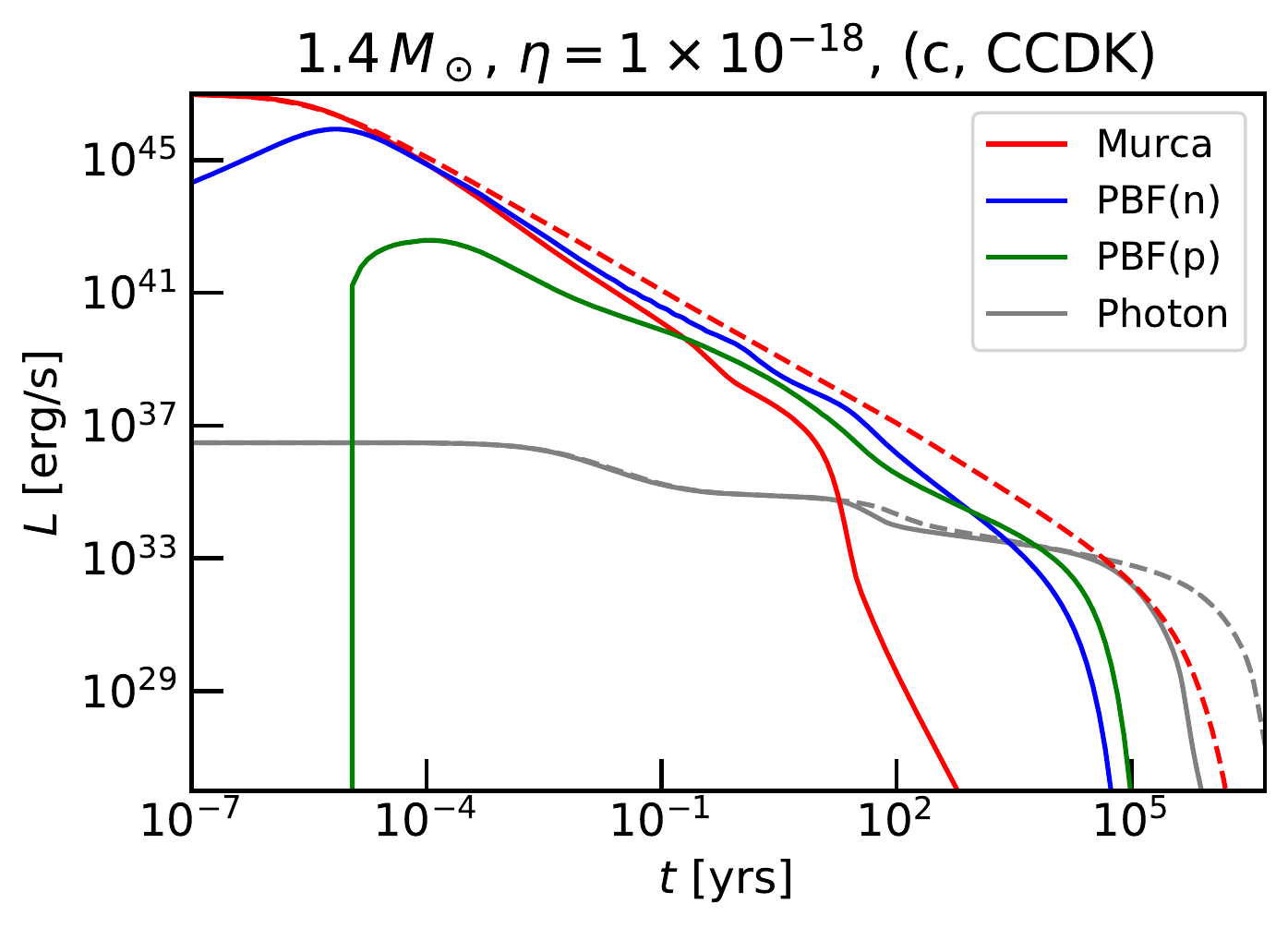} 
  \end{minipage}
  \begin{minipage}{0.5\linewidth}
    \includegraphics[width=1.0\linewidth]{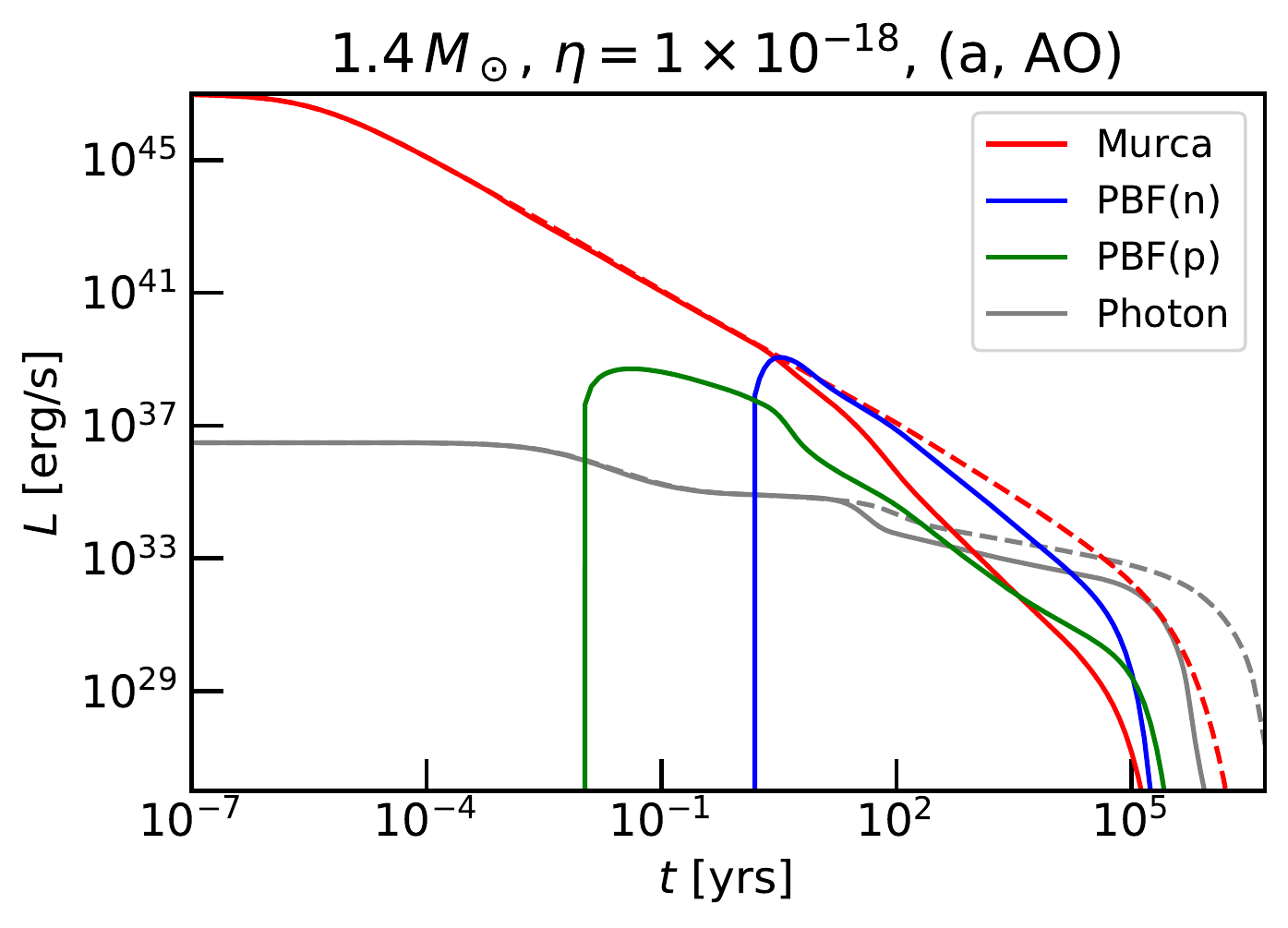} 
  \end{minipage}%
  \begin{minipage}{0.5\linewidth}
    \includegraphics[width=1.0\linewidth]{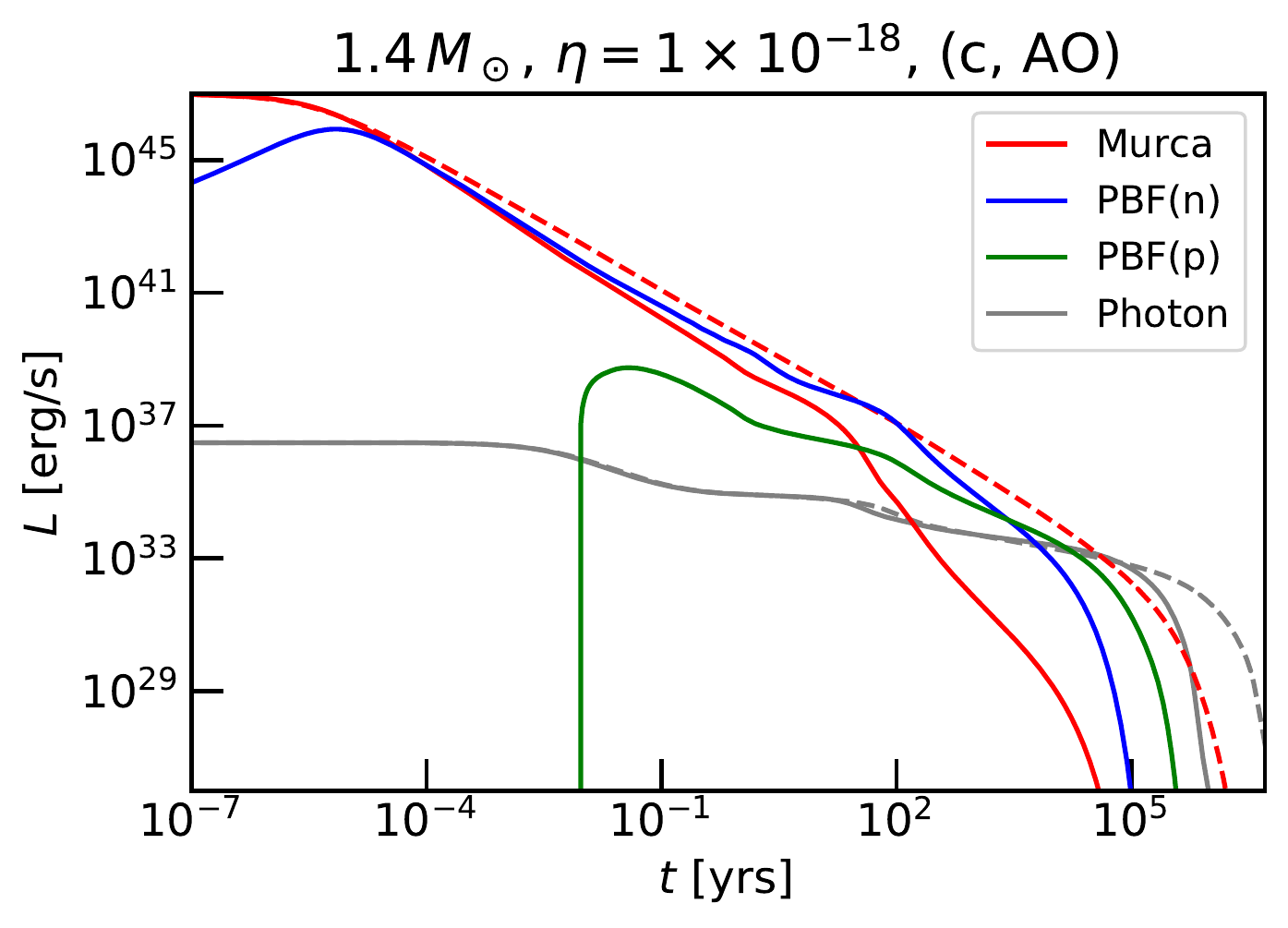} 
  \end{minipage}
  \caption{Neutrino emission luminosity from modified Urca (red), neutron ${}^3P_2$ PBF (blue), and proton ${}^1S_0$ PBF (green) processes.
    The surface photon luminosity is also shown (gray).
    Solid lines are the results computed with the gap models indicated on each panel.
  The luminosity without nucleon superfluidity is shown by dashed lines for comparison.}
  \label{fig:lum-sf-compare}
\end{figure}
For the closer look at the effect of the pairing gap, we show in Fig.~\ref{fig:lum-sf-compare} the luminosity of each neutrino emission process (colored lines) as well as surface photon one (gray lines).
Each panel shows the specific combination of neutron and proton gaps, and the corresponding luminosities of normal nucleons are shown by dashed lines.
We can see that the modified Urca process (red) dominates the neutrino emission at early stage, while the PBF of neutron (blue) or proton (green) dominates at later stage.
For neutron ``a'' gap (left two panels), the beginning of the neutron triplet pairing, characterized by the sharp increase of its PBF emissivity, dramatically suppresses the modified Urca process.
For neutron ``c'' gap (right two panels), the modified Urca process is suppressed after the proton PBF occurs.
The PBF process is most effective around the phase transition, and for $T\ll T_c\sim\Delta$, the PBF, as well as Urca processes, is Boltzmann suppressed by $f\sim e^{-\Delta/T}$, which results in the slow temperature decrease for $10^2\unit{yr} \lesssim t \lesssim 10^5\unit{yr}$ in the left panel of Fig.~\ref{fig:ts-sf-compare}.

\begin{figure}[t]
  \centering
  \begin{minipage}{0.5\linewidth}
    \includegraphics[width=1.0\linewidth]{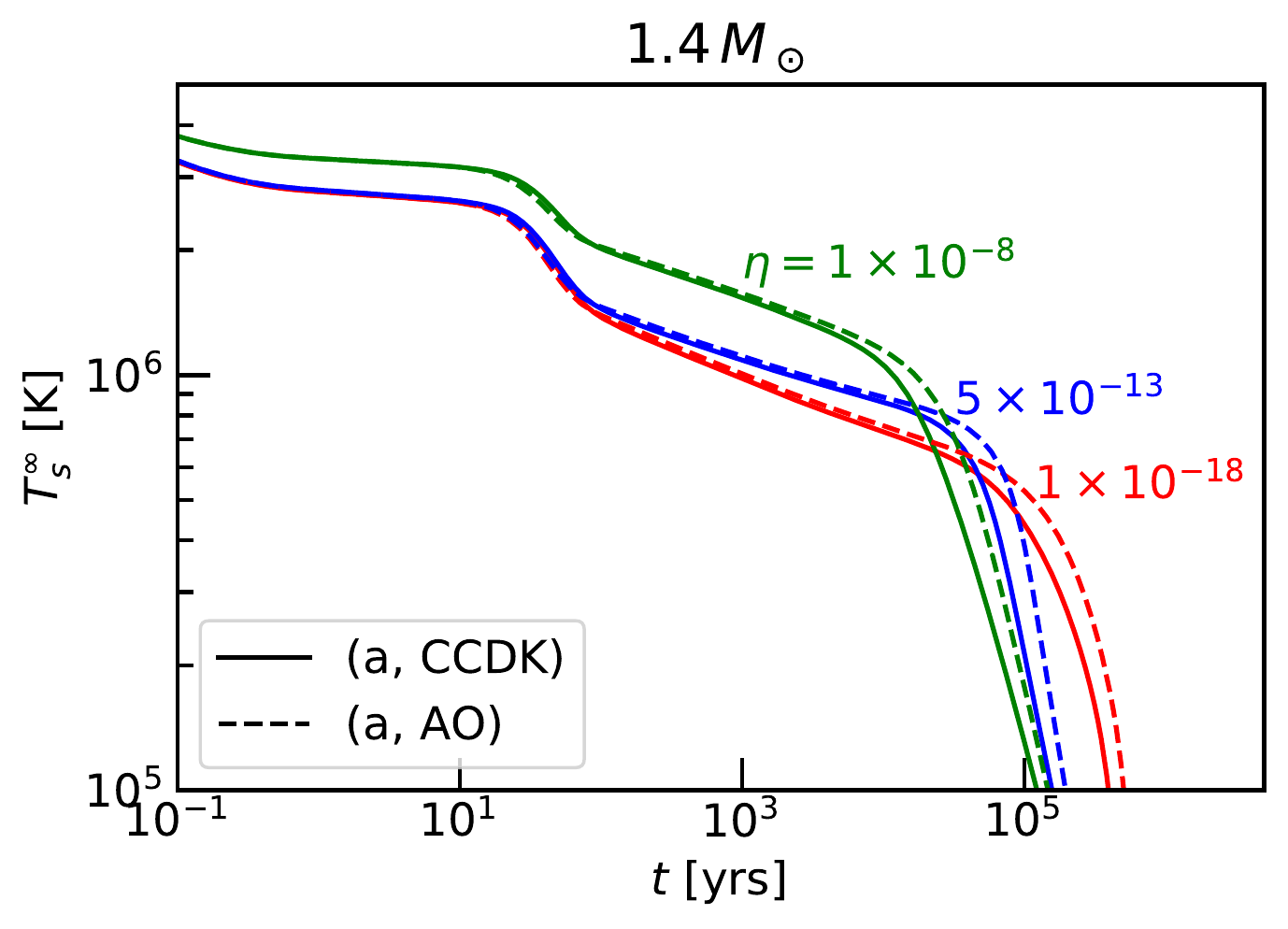}
  \end{minipage}%
  \begin{minipage}{0.5\linewidth}
    \includegraphics[width=1.0\linewidth]{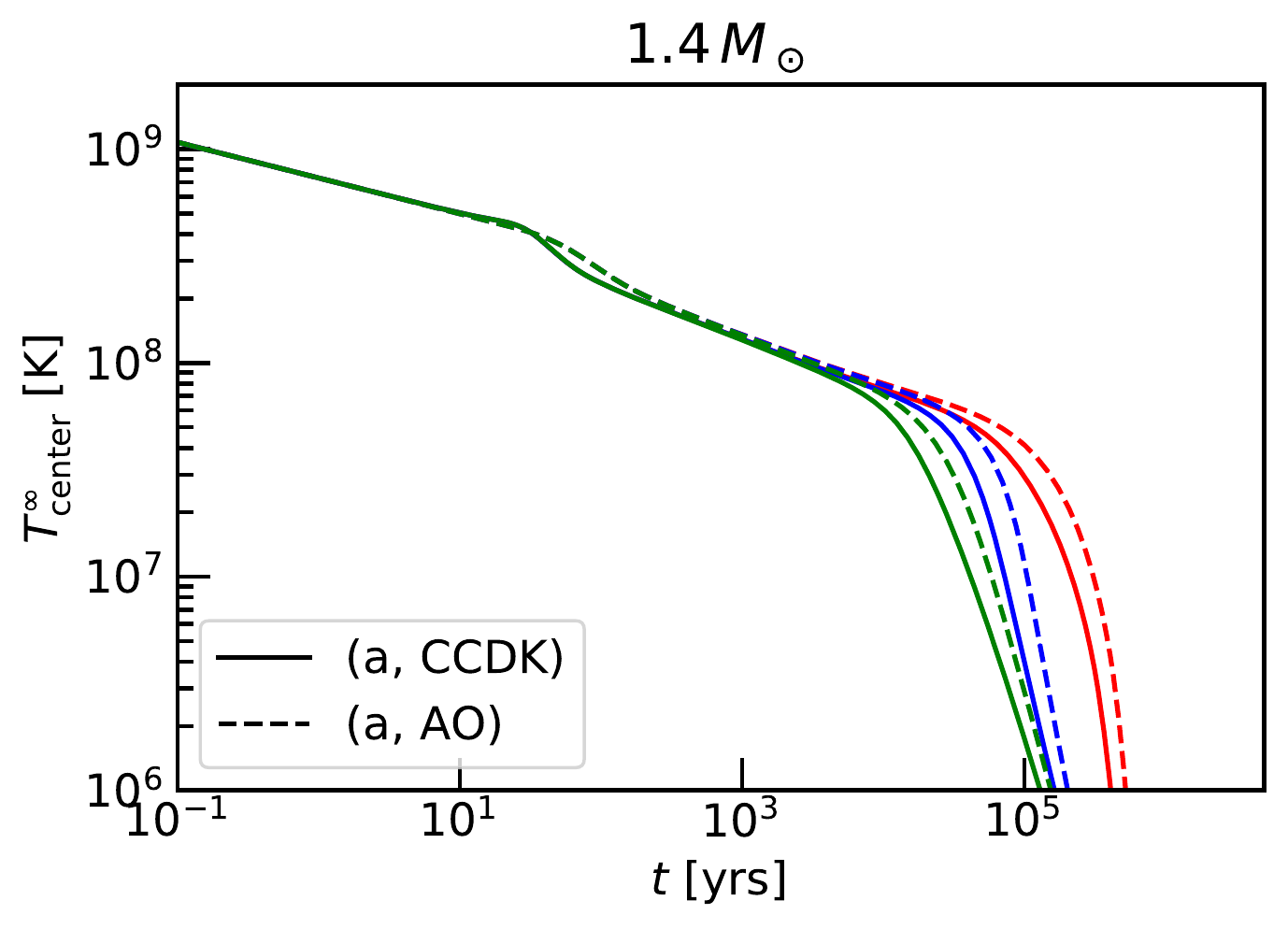}
  \end{minipage}
  \caption{The thermal evolution with different envelope parameters $\eta$.
    The left panel shows the surface temperature, and the right panel shows the redshifted temperature at the center, $T(0)e^{\Phi(0)}$.
    The light element in the envelope is taken as $\eta=1\times10^{-18}$ (red), $5\times10^{13}$ (blue) and $1\times10^{-8}$.
    Gap models are chosen as (a, CCDK) (solid) and (a, AO) (dashed).
  }
  \label{fig:ts-envelope}
\end{figure}
The amount of the light element affects the surface temperature.
Figure.~\ref{fig:ts-envelope} shows the redshifted surface temperature (left) and the temperature at NS center (right).
The red lines corresponds to the case of $\eta=1\times10^{-18}$, which is close to the purely iron envelope.
The green lines show $\eta=1\times10^{-8}$, and this case is the same as fully accreted envelope for $T_b \lesssim 10^8\unit{K}$, as we see in Fig.~\ref{fig:ts-tb}.
During neutrino emission stage ($t\lesssim 10^5\unit{yr}$), the surface temperature is higher for larger $\eta$.
This is because the internal temperature is almost the same, as shown in the right panel, but the $T_s-T_b$ relation provides the higher surface temperature for larger $\eta$.
Since the higher surface temperature generates larger photon luminosity, the photon cooling begins earlier for larger $\eta$.


\subsection{Minimal cooling vs. observation}
\label{sec:minimal-cool-vs-obs}

We now compare predictions in the minimal cooling paradigm with the observed surface temperatures of young and middle-aged pulsars.
We first review how to determine the age and surface temperature of observed NSs, and then discuss the compatibility of these observations with minimal cooling.
In Tab.~\ref{tab:psr-temp-1}, we collect the ordinary pulsars whose surface temperatures are observed.
The young NS in the supernova remnant of Cas A is not shown since it is the main subject in Chap.~\ref{chap:limit-axion-decay}, and will be discussed in detail there.
There are also several middle-aged and old pulsars which are not shown in the table.
They are discussed in Chap.~\ref{chap:neutron-star-heating} in detail because their temperatures tend to be higher than the prediction of the minimal cooling, and are important for the study of internal heating.
\begin{table}
  \centering
  \caption{Observational data used in this section. $t_{\text{sd}}$, $t_{\text{kin}}$, 
  $T_s^\infty$, and $P$ denote the spin-down age, kinematic age, 
  effective surface temperature, and period of neutron stars, respectively. 
  The sixth column shows the atmosphere model used in the estimation of 
  the surface temperature, where H, BB, C, and PL indicate hydrogen,  
  blackbody, carbon, and power-law, respectively, while M represents a magnetized NS hydrogen atmosphere model, such as NSA \cite{1995ASIC..450...71P} and NSMAX \cite{Ho:2008bq}. Data are taken from ATNF Pulsar
  Catalogue \cite{Manchester:2004bp, atnf} unless other references are shown explicitly.}
  \vspace{3mm}
    \begin{adjustbox}{width=\textwidth}
  \begin{tabular}{lccccc}\toprule
    Name & $\log_{10}t_{\mathrm{sd}}$& $\log_{10}t_{\mathrm{kin}}$ & $\log_{10}T_s^\infty$ &$P$   &Atmos. \\ 
    &[yr]&[yr] &[K] &[s] &model \\\midrule
    PSR J2043+2740  &$6.1$&&$5.64(8)$ \cite{Beloin:2016zop}&$0.096$&H\\
    PSR B1055-52 &$5.7$&&$5.88 (8)$ \cite{Beloin:2016zop}&$0.20$&BB\\
    PSR J0357+3205 &$5.7$&&$5.62^{+0.09}_{-0.08}$ \cite{Kirichenko:2014ona}&$0.44$&M+PL\\
    PSR J1741-2054 &$5.6$&&$5.85^{+0.03}_{-0.02}$ \cite{Auchettl:2015wca}&$0.41$&BB+PL\\
    PSR J0633+1748 &$5.5$&&$5.71(1)$ \cite{Mori:2014gaa}&$0.24$&BB+PL\\
    PSR J1740+1000 &$5.1$&&$6.04(1)$ \cite{2012Sci...337..946K}&$0.15$&BB\\
    PSR B0656+14  &$5.0$&&$5.81(1)$ \cite{DeLuca:2004ck}&$0.38$&BB+PL\\
    PSR B2334+61 &$4.6$&&$5.5-5.9$ \cite{McGowan:2005kt}&$0.50$&M\\
    PSR J0538+2817 &$5.8$& $4.3-4.8$ \cite{Ng:2006vh}&  $6.02(2)$ \cite{Ng:2006vh}&$0.14$&H\\
    XMMU J1732-344 &&$4.0-4.6$ \cite{Tian:2008tr, Klochkov:2014ola} &$6.25(1)$ \cite{Klochkov:2014ola}&&C\\
    PSR B1706-44   &$4.2$ & & $5.8^{+0.13}_{-0.13}$ \cite{McGowan:2003sy}&$0.10$&M+PL\\
    PSR B0833-45 (Vela) & $4.05$ & $3.97^{+0.23}_{-0.24}$ \cite{tsurutaThermalEvolutionHyperonMixed2009}& $5.83(2)$ \cite{Pavlov:2001hp}&$0.089$ &H\\ 
    PSR J1357-6429 &$3.9$&&$5.88^{+0.03}_{-0.04}$ \cite{Zavlin:2007nx} &$0.17$&H\\
    RX J0822-4247 &$3.9$&$3.57^{+0.04}_{-0.05}$ \cite{1999ApJ...525..959Z} &$6.24(4)$ \cite{1999ApJ...525..959Z}&$0.075$ \cite{1999ApJ...525..959Z}&H\\
    PSR J1119-6127 &$3.2$&&$6.09(8)$ \cite{SafiHarb:2008mj}&$0.41$&NSA+PL\\
    \bottomrule
  \end{tabular}
\end{adjustbox}
  \label{tab:psr-temp-1}
\end{table}

\subsubsection{Spin-down age}
\label{sec:spin-down-age}

In many cases, a NS is found as a pulsar, which emits radio pulses with very precise period.
The periods of pulsars are measured with great precision.
A typical period is $P\sim 1\unit{s}$, and it is gradually increasing.
This increase of the spin period means that the rotational energy is dissipating, which is called \textit{spin-down}.
The time derivative of the period, $\dot{P}$, is also measured very precisely.
Although the mechanism of the spin-down is not completely understood, it is qualitatively explained by the magnetic dipole radiation.
The energy loss rate by the magnetic dipole radiation is~\cite{Shapiro:1983du}
\begin{align}
  \label{eq:dipole-radiation-rate}
  \dot{E}
  =
  -\frac{2B_s^2R^6\Omega^4}{3}\sin^2\alpha\,,
\end{align}
where $B_s$ is a dipole magnetic field at the magnetic equator, $\Omega = 2\pi/P$ the angular velocity, and $\alpha$ the angle between rotation axis and magnetic axis. 
Once we equate this to the total loss of the rotational energy $\dot{E}=I\Omega\dot\Omega$, where $I$ is the moment of inertia, we obtain the evolution of angular velocity as
\begin{align}
  \label{eq:spin-down-n3}
  \dot{\Omega} = -k\Omega^n\,,\quad
  \text{with }
  k = \frac{2B_s^2R^6}{3I}\sin^2\alpha\,,
\end{align}
with $n=3$.
Note that $k$ is estimated by the current period and its derivative as $k = P\dot{P}/(4\pi)$.
Thus the combination $P\dot{P}$ is constant if the dipole radiation model is correct.
From the measurement of $P$ and $\dot P$, we can estimate the dipole magnetic field.
For the canonical NS of $R=10\unit{km}$ and $I=10^{45}\unit{g\,cm^2}$ with $\sin\alpha=1$, 
\begin{align}
  \label{eq:bs}
  B_s \sim 3.2\times10^{19}\left( \frac{P\dot{P}}{s} \right)^{1/2}\unit{G}\,.
\end{align}
A typical ordinary pulsar of $P=1\unit{s}$ and $\dot{P} = 10^{-15}$ has a magnetic field of $B_s\sim 10^{12}\unit{G}$.

The solution of Eq.~\eqref{eq:spin-down-n3} is given by 
\begin{align}
  \label{eq:omega-sol-n3}
  \Omega(t)
  =
  \frac{2\pi}{\sqrt{P_0^2 + 2P\dot{P}t}}\,,
\end{align}
where $P_0$ is the initial period.
If $P_0\ll P$, the pulsar age is estimated by the following \textit{spin-down age}:
\begin{align}
  \label{eq:spin-down-age}
  t_{\mathrm{sd}} = \frac{P}{2\dot P}\,.
\end{align}
Therefore one can estimate an age of a pulsar simply through the current $P$ and $\dot{P}$, both of which are precisely measured for many pulsars.
In the second column of Tab.~\ref{tab:psr-temp-1}, we list the spin-down ages.

We note that the spin-down age is just a crude estimation of the true age.
First of all, the assumption of the energy loss purely from dipole radiation may not be valid for many pulsars.
If we assume the general power-law behavior for the spin-down and hence Eq.~\eqref{eq:spin-down-n3} with arbitrary $n$, we can estimate $n$ by
\begin{align}
  \label{eq:n-estimator}
  n = \frac{\Omega\ddot\Omega}{\dot\Omega^2}\,.
\end{align}
The pure magnetic dipole model predicts $n=3$, but the measurements of the breaking index of eight pulsars show $n<3$~\cite{Lyne:2014qqa}; for instance the Crab pulsar has $n\simeq2.5$~\cite{Lyne:2014qqa}.
This implies that there are other sources of rotational energy loss, and/or the assumption of constant magnetic field is not valid (see Ref.~\cite{Vigano:2013lea} for the pulsar evolution under the decaying magnetic field).
Moreover, the assumption of $P_0 \ll P$ is also difficult to justify since the estimation of $P_0$ is not available for most pulsars, and even if it exists, the uncertainty is typically very large (see the discussion in Chap.~\ref{chap:neutron-star-heating} and \ref{chap:dm-heating-vs-roto} and references there).
Nevertheless this spin-down age roughly agrees with other age estimation discussed below, and hence is widely used.

\subsubsection{Kinematic age}
\label{sec:kinematic-age}

Another estimation of NS age is to use its kinetic property.
NSs are believed to be born by the supernova.
If we can measure a NS with the remnant of supernova, and if we can measure the velocity of these remnant around the star, we can estimate the age simply by the kinematics.
The age determined in this way is called \textit{kinematic age}, denoted by $t_{\mathrm{kin}}$.
Unfortunately, only a small number of NSs are observed with associated supernova remnant. 
Since it does not assume the property of spin-down, the kinematics age is often considered as a more reliable estimation.
Kinematic ages are shown in the third column of Tab.~\ref{tab:psr-temp-1}, where it is available only for a few pulsars.
In this subsection, we use the kinematic age if available, and otherwise the spin-down age.
In this subsection, we assume the factor 3 uncertainty for the spin-down age following Ref.~\cite{Page:2004fy}.

\subsubsection{Surface temperature}
\label{sec:surface-temperature}
The surface temperature, shown in the fourth column of Tab.~\ref{tab:psr-temp-1}, is estimated from the fit of the observed spectrum with atmosphere models.
Calculating the spectrum in an atmosphere model involves several parameters: chemical composition of the atmosphere, surface temperature $T_s$, the NS radius $R$ and the surface gravity.
This temperature $T_s$ corresponds to the temperature on the outer-most layer of the envelope (see Sec.~\ref{sec:envelopew}).
Since the observed flux suffers the absorption by the interstellar medium, we also need the parameter for its column density.

The chemical composition of the atmosphere changes the spectrum due to the difference in the conductivity.
Generically speaking, the light-element atmosphere (such as hydrogen or carbon) fits the thermal emission of young pulsars ($t \lesssim 10^5\unit{yr}$) well, while the heavy element model or blackbody offers a good fit for a middle-aged or old pulsar.
It is often possible to fit an observed spectrum with different atmosphere models since the range of observed wavelength is limited.
Usually a light-element atmosphere provides a lower temperature and larger radius than a heavy-element by a factor of a few.
In addition, if a NS has a relatively large magnetic field, a magnetized NS hydrogen atmosphere model such as NSA \cite{1995ASIC..450...71P} and NSMAX \cite{Ho:2008bq} may improve its spectrum fit.

In the sixth column in Tab.~\ref{tab:psr-temp-1}, we show the atmosphere model used in the evaluation of the surface temperature of each NS; H, BB, C, PL, and M represent the hydrogen, blackbody, carbon, power-law, and magnetized NS hydrogen atmosphere models, respectively.
For some NSs in the table, there are several atmosphere models that can fit the observed spectrum. In such cases, we choose the model which is considered to give the best fit and/or yields a NS radius of a plausible size ($\sim 10$~km). Moreover, there are some cases where the use of two or more blackbody components improves the fit due to the presence of hot spots, though we do not show this explicitly in the table.
In these cases, we use the temperature of the component with a NS radius that is consistent with the typical NS radius. 
Note, however, that the fits with atmosphere models are often performed with the mass, radius, or distance of the NS being fixed, though these quantities are not precisely known for most of the NSs---if this is the case, additional systematic uncertainty might be present. For more details, see the references cited in the fourth column of the table.
\subsubsection{Comparison with observations}
\label{sec:comparison}

\begin{figure}[t]
  \centering
  \begin{minipage}{0.5\linewidth}
    \includegraphics[width=1.0\linewidth]{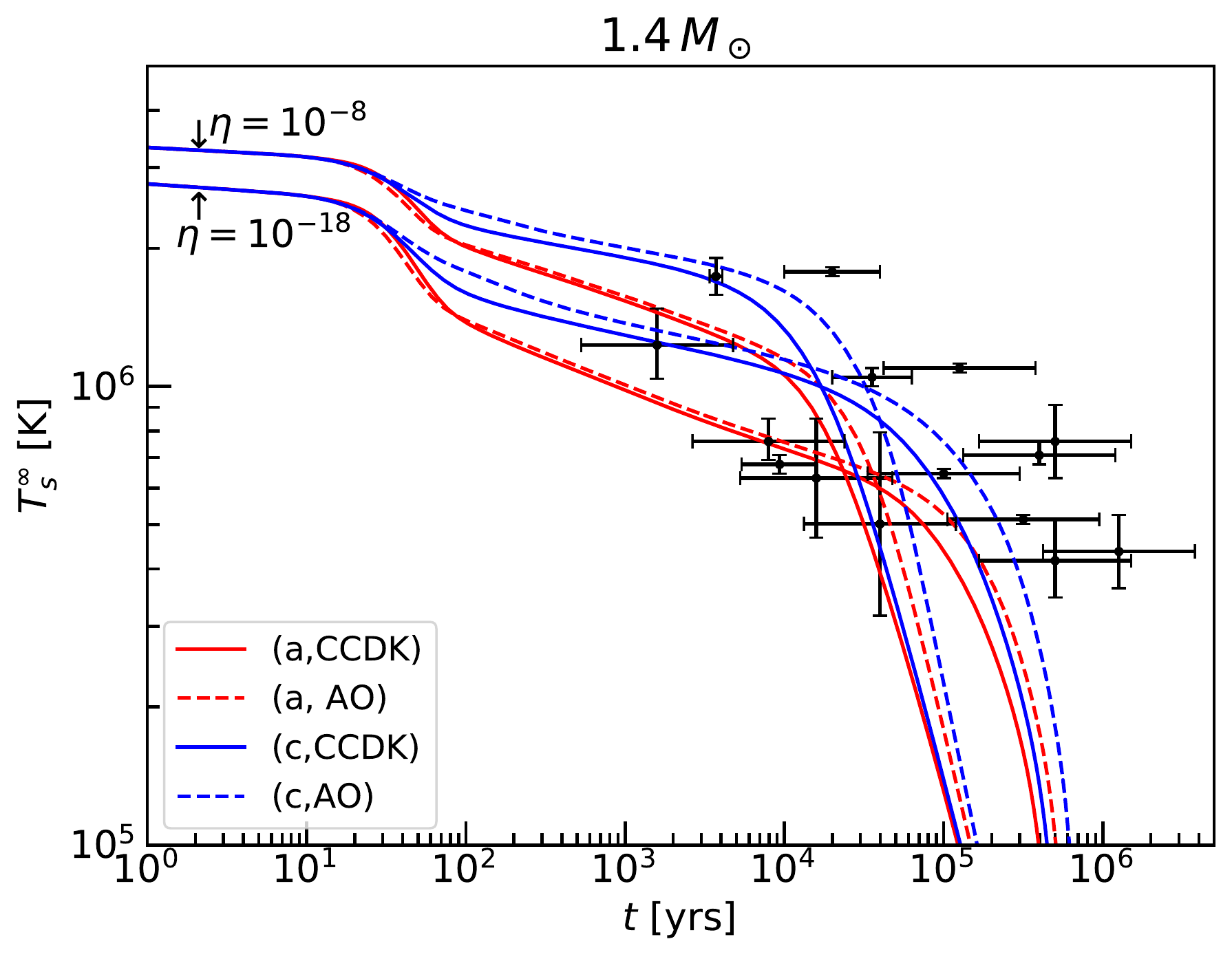}
  \end{minipage}%
  \begin{minipage}{0.5\linewidth}
    \includegraphics[width=1.0\linewidth]{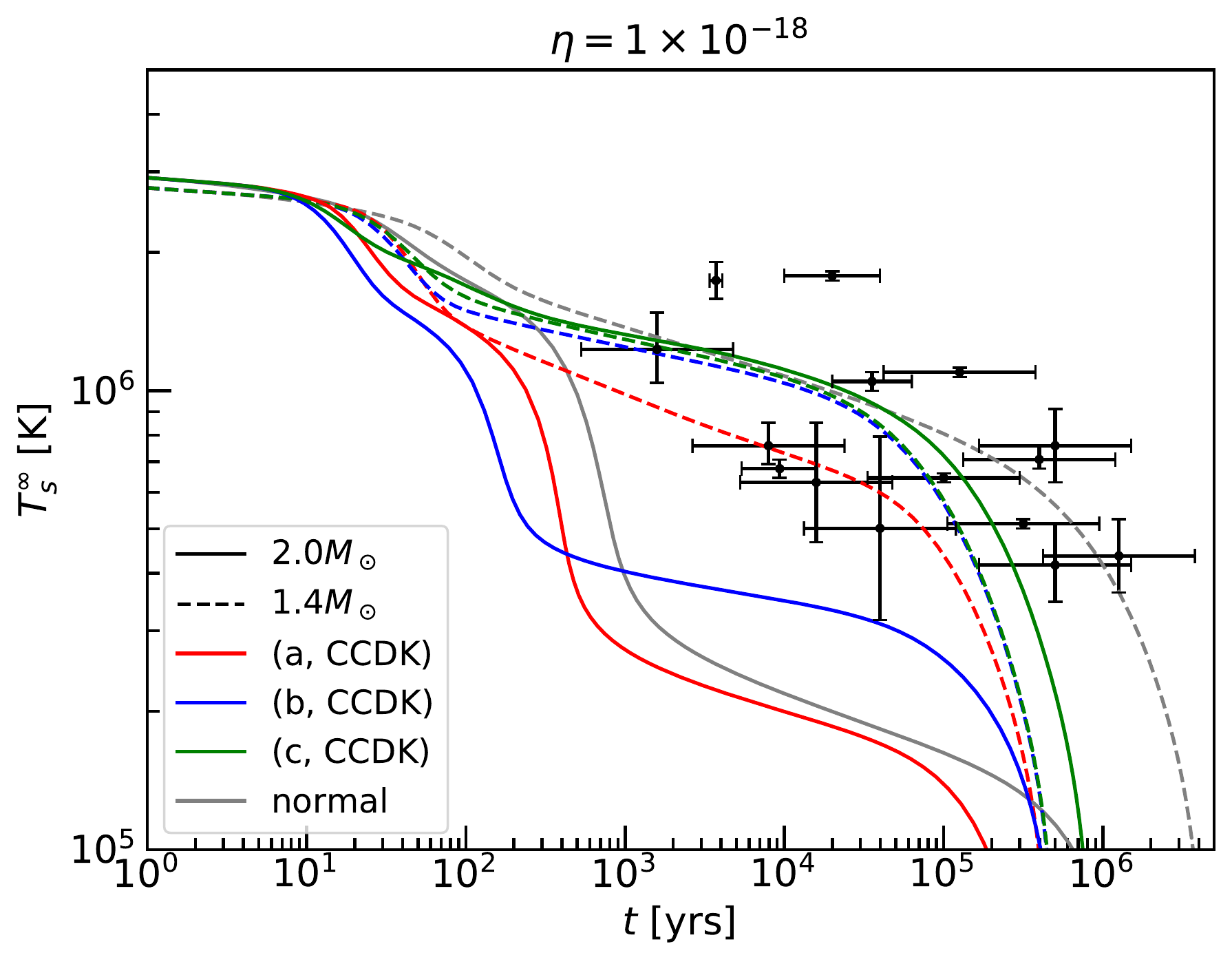}
  \end{minipage}
  \caption{Left: The comparison of minimal cooling and observations. The red (blue) lines are the case of neutron ``a'' (``c'') gap model, and solid (dashed) lines are proton CCDK (AO) gap model. The envelope parameter is indicated. The data in Tab.~\ref{tab:psr-temp-1} is plotted by the black points with error bars.
    Right: The effect of direct Urca process. Evolution of a $2.0\Msun$ ($1.4\Msun$) NS is shown by solid (dashed) line. The neutron ``a'' (red), ``b'' (blue) or ``c'' (green), and proton CCDK gap models are used.
  The cooling curves without nucleon superfluidity are shown by gray lines.}
  \label{fig:obs-and-cooling}
\end{figure}
Now we compare the data in Tab.~\ref{tab:psr-temp-1} with the predictions of minimal cooling.
In the left panel of Fig.~\ref{fig:obs-and-cooling}, we show the cooling curves of different gap models and envelope parameter.
The mass is fixed at $1.4\Msun$ and the APR EOS is used.
The neutron gap ``a'' (``c'') is taken as small (large) gap model and shown in the red (blue) line.
Similarly, the proton CCDK (AO) gap is taken as large (small) gap model, shown in the solid (dashed) line.
The envelope parameter is chosen to be $\eta=10^{-18}$ or $\eta=10^{-8}$.
This range of $\eta$ covers the uncertainty in the amount of light element for $10^5\unit{yr}\lesssim T_s \lesssim 10^6\unit{yr}$, from the purely iron to fully accreted envelope (see Fig.~\ref{fig:ts-tb}).
The points correspond to the observed ages and the surface temperatures with uncertainties.
We can see that most of the observations are compatible with the prediction of minimal cooling within the uncertainties.
The several pulsars of $t\sim10^6\unit{yr}$ is slightly hotter than the theoretical prediction.
Although this discrepancy can be  due to the uncertainties in theory and/or observations, they may be explained by the heating mechanism such as rotochemical heating, which we will discuss in Chap.~\ref{chap:neutron-star-heating}.

\subsection{Enhanced cooling}
\label{sec:enhanced-cooling}


So far we have used the NS having the canonical mass $M=1.4\Msun$.
As we have discussed in Sec.~\ref{sec:direct-urca}, a heavier NS has more concentration of protons, electrons and muons near the center, and once their Fermi momenta satisfy $k_{F,p} + k_{F,\ell} > k_{F,n}$, the direct Urca process is allowed.
Since the emissivity of the direct Urca process for the normal nucleons is proportional to $T^6$, it is much more powerful than the modified Urca process, whose emissivity is proportional to $T^8$.
In the right panel of Fig.~\ref{fig:obs-and-cooling}, we compare the predictions of $2.0\Msun$ NSs (solid) to $1.4\Msun$ NSs (dashed).
The gap models are taken as ``a'' (red), ``b'' (blue) and ``c'' (green) for neutrons, and CCDK for protons.
For the neutron ``a'' or ``b'' gap model, the surface temperature of a $2.0\Msun$ NS is much lower than that of $1.4\Msun$ because of the strong direct Urca process, and is not consistent with the observations.
Thus this enhanced cooling should not have occurred in these pulsars.
On the other hand, neutron ``c'' gap model does not make large difference between $1.4\Msun$ and $2.0\Msun$.
In this case, the direct Urca process still operates near the center but the neutron triplet gap there is so large that the emissivity is highly suppressed.

If a NS much colder than the minimal cooling prediction is observed in the future, and if it is sufficiently heavy, its temperature will be explained by the enhanced cooling.

\chapter{Limit on the axion decay constant from the cooling neutron star in Cassiopeia A}
\label{chap:limit-axion-decay}

This chapter presents one of the main results of the dissertation, based on the author's work.~\cite{Hamaguchi:2018oqw}.
In Sec.~\ref{sec:standard-cooling}, we compare predictions of the minimal cooling to several young and middle-aged NSs.
In fact, there is another important NS to test the theoretical prediction: the NS located at the center of the supernova remnant Cassiopeia A (Cas A).
Since the surface temperature of the Cas A NS has been observed for about ten years, its cooling rate, as well as the temperature itself, is available.
Therefore, we can constrain models of NS cooling in a more stringent way.
In particular, this stringent constraint is useful to probe the physics beyond the standard model; if new particles couple to nucleons, their emission can be an extra source of thermal energy release, and alter the prediction of the standard cooling.

In this chapter, we use the cooling of the Cas A NS to constrain the axion, one of the well-motivated candidates of BSM particles.
It is found that the success of the standard cooling for the Cas A NS is spoiled if the axion-nucleon interaction becomes sufficiently large. 
We obtain a lower limit on the axion decay constant, $f_a\gtrsim (5-7)\times 10^8$ GeV, if the star has an envelope with a thin carbon layer.
It turns out that this is as strong as existing limits imposed by other astrophysical observations such as SN1987A.

\section{Standard NS Cooling and Cas A NS}

\subsection{The neutron star in the Cassiopeia A}
\label{sec:cas-a}

The Cassiopeia A (Cas A) is a supernova remnant in the Cassiopeia
constellation. This supernova may be identical to the \textit{3 Cassiopeiae}
recorded by John Flamsteed on August 16, 1680
\cite{1980JHA....11....1A, 1980Obs...100....3K, 1980Natur.285..132H},
which is consistent with the supernova explosion date estimated from the
remnant expansion: $1681\pm 19$ \cite{Fesen:2006zma}. In 1999, the
\textit{Chandra} X-ray observatory discovered a hot point-like source in
the center of the supernova remnant \cite{1999IAUC.7246....1T}, which is
now identified as a NS. 

\begin{figure}[t]
  \centering
  \includegraphics[width=0.6\linewidth]{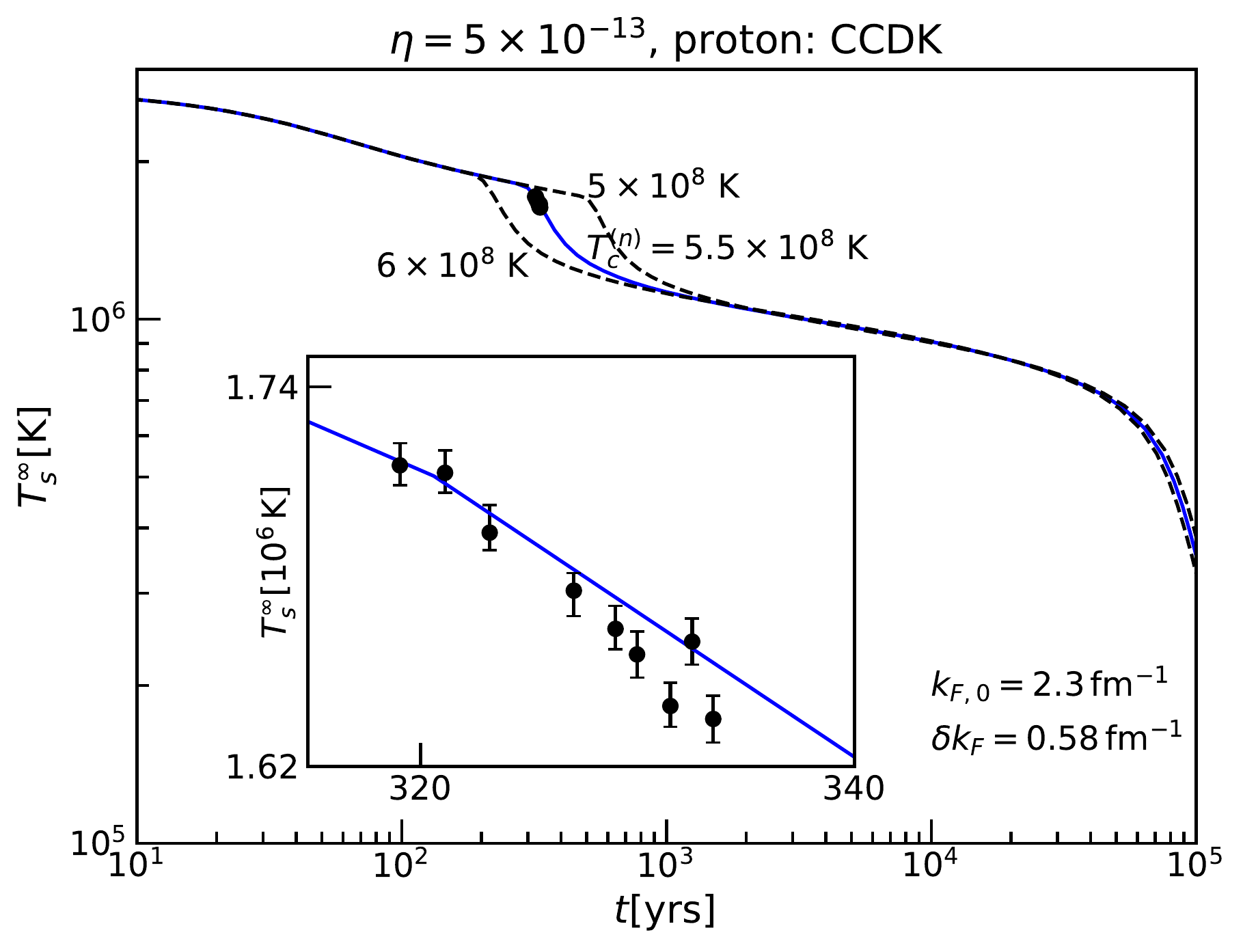}
  \caption{Minimal cooling vs. Cas A NS. The black points are the measured surface temperatures of Cas A NS taken from Ref.~\cite{Ho:2014pta}.
    These are explained by the minimal cooling with the neutron triplet gap of $T_{c}^{(n)}\simeq 5.5\times10^8\unit{K}$ shown by the blue line.
    Dotted lines are the cooling curves with slightly different $T_{c}^{(n)}$, showing that the prediction around Cas A age is very sensitive to the gap parameters.}
  \label{fig:cas-a-minimal}
\end{figure}

Given the distance to Cas A, $d = 3.4^{+0.3}_{-0.1}$~kpc
\cite{Reed:1995zz}, the NS radius can be determined by measuring
the X-ray spectrum thermally emitted from the NS. With the black-body
and hydrogen atmosphere models, a rather small radius of the X-ray emission
area was obtained---about 0.5 and 2 km, respectively \cite{Pavlov:1999tca, Chakrabarty:2000ps, Pavlov:2003eg}. This implied a
hot spot on the NS surface. On the other hand, a lack of the observation of
pulsations in the X-ray flux \cite{Murray:2001fy, Mereghetti:2001cf}
indicates that the X-ray emission comes from the whole surface, which is
incompatible with the above observation. This contradiction was resolved 
by Heinke and Ho, who found that a carbon atmosphere model with low
magnetic field gave a good fit to the X-ray spectrum with a
typical size of the NS radius ($\gtrsim 10$~km)
\cite{Ho:2009mm}. Moreover, they observed that the surface temperature
of the NS evaluated with the carbon atmosphere model was decreasing over
the years, which provided the first direct observation of NS cooling
\cite{Heinke:2010cr}.
The Cas A NS cooling data collected so far \cite{Ho:2014pta} is plotted in Fig.~\ref{fig:cas-a-minimal} by the black points. 
It clearly shows that the temperature is decreasing at a constant rate. 
The authors in Ref.~\cite{Ho:2014pta} performed a fit of this X-ray spectrum using a non-magnetic carbon atmosphere model~\cite{Ho:2009mm} and obtained $M \simeq (1.4 \pm 0.3) M_{\odot}$.

\subsection{Cas A NS in the minimal cooling}
\label{sec:cas-a-minimal}


In the standard NS cooling scenario, as we have discussed in Sec.~\ref{sec:standard-cooling}, the NS cooling proceeds via the emission of neutrinos and photons.
The former dominates over the latter in the earlier epoch ($t \lesssim 10^{5}$~years).
Various processes participate in neutrino emission, such as the direct Urca process, the modified Urca process, bremsstrahlung, and the PBF process.
The direct Urca process is the ``fast'' process; if it occurs, a NS cools quite rapidly (see Sec.~\ref{sec:enhanced-cooling}).
However, this process can occur only at very high density regions \cite{Lattimer:1991ib}, which can be achieved only for a heavy NS (see Sec.~\ref{sec:direct-urca}).
For instance, the APR EOS \cite{Akmal:1998cf} allows the direct Urca process only for $M \gtrsim 1.97M_{\odot}$, which is well above the Cas A NS mass estimated in Ref.~\cite{Ho:2014pta}.
Thus, we can safely assume that the fast process never occurs in the Cas A NS, as in the minimal cooling paradigm.
In this case, the neutrino emission proceeds through the ``slow'' processes such as the modified Urca and bremsstrahlung processes. 
In the absence of nucleon pairings, the neutrino luminosity $L_\nu$ caused by these processes is expressed as $L_\nu \simeq L_9 T_9^8$ with the internal temperature $T_9 \equiv T/(10^9~\text{K})$ and the coefficient $L_9 \sim 10^{40}~\text{erg}\cdot\text{s}^{-1}$ \cite{Yakovlev:2004iq,
  Page:2005fq}.

In general, a NS is brought into an isothermal state with relaxation time $\sim 10$--100~years \cite{1994ApJ...425..802L,
  Gnedin:2000me} (see also the right panel of Fig.~\ref{fig:ts-sf-compare}), and in fact the Cas A NS is very likely to be thermally relaxed \cite{Yakovlev:2010ed}.
As we discussed in Sec.~\ref{sec:basic-eqs}, the thermal evolution in this case is determined by
\begin{equation}
 C \frac{dT}{dt} = -L_\nu -L_{\text{cool}} ~,
\label{eq:balanceeq}
\end{equation}
where $L_{\text{cool}}$ denotes the luminosity caused by potential extra cooling sources, and we neglect the redshift factor for simplicity.
We have dropped the photon luminosity since this is much smaller than $L_\nu$ for a young NS like the Cas A NS (see Fig.~\ref{fig:lum-sf-compare}).
The heat capacity $C$ has temperature dependence of the form $C = C_9 T_9$ with $C_9 \sim 10^{39}~\text{erg} \cdot \text{K}^{-1}$ \cite{Yakovlev:2004iq,
Page:2005fq}. 
If $L_{\text{cool}} = 0$, Eq.~\eqref{eq:balanceeq} leads to 
\begin{equation}
 T_9 = \biggl(\frac{C_9 \cdot 10^9~\text{K}}{6L_9
  t}\biggr)^{\frac{1}{6}} 
\sim  \biggl(\frac{1~\text{year}}{ t}\biggr)^{\frac{1}{6}}
~,
\label{eq:t9}
\end{equation}
where we have assumed that the initial temperature is much larger than
that at the time of interest.

As we have discussed in Sec.~\ref{sec:envelopew}, the NS surface is insulated from the hot interior by its envelope. 
For a non-magnetic iron envelope at temperatures as high as the Cas A NS temperature, the relation between the surface temperature $T_s$ and the internal temperature $T$ is approximated by \cite{1983ApJ...272..286G}
\begin{equation}
  \tag{\ref{eq:ts-tb-iron}}
 T_9 \simeq 0.1288 \times \biggl(\frac{T_{s6}^4}{g_{14}}\biggr)^{0.455} ~.
\end{equation}
Note that in the subsequent numerical analysis, we use a more accurate relation, Eq.~\eqref{eq:ts-tb-partial}.

The cooling rate of the Cas A NS observed in Ref.~\cite{Heinke:2010cr,Ho:2014pta}
was about 3--4\% in ten years around $t \simeq 320$~years.\footnote{See also \cite{Elshamouty:2013nfa,Posselt:2013xva} for possible uncertainties.} On the other hand, 
from Eq.~\eqref{eq:t9} and Eq.~\eqref{eq:ts-tb-iron}, we find that the surface
temperature goes as $T_s \propto t^{0.09}$,
which results in only $0.3$\% decrease in temperature in ten years.
Hence, the slow neutrino emission cannot explain the observed rapid cooling of the Cas A NS. 
 
This difficulty can be resolved with the help of superfluidity in the NS.
As we have seen in Sec.~\ref{sec:neutrino-emis}, the onset of Cooper pairing of nucleons triggers the rapid PBF emission of neutrino while suppresses other emission processes which these nucleons participate in~\cite{Flowers:1976ux, Voskresensky:1987hm}.
The PBF lasts only for a short time---to explain the rapid cooling of the Cas A NS by this PBF process, therefore, the phase transition of the neutron triplet pairing should occur just before $t\simeq 320$~years.
This condition implies that the critical temperature of this phase transition, $T_c^{(n)}$, should agree to the internal temperature around this time; thus, $T_c^{(n)}$ is fixed via Eq.~\eqref{eq:t9}.
One also finds that the resultant cooling rate increases as $T_c^{(n)}$ gets larger, which is achieved with a smaller $L_9$ according to Eq.~\eqref{eq:t9}.
The reduction in $L_\nu$ can be realized again with the aid of Cooper pairing---with proton pairings formed, the neutrino emission processes which contain protons are suppressed by the proton gap, which then results in a small $L_9$.
A small $L_9$ also ensures that the NS was not overcooled by the time of observation.

Indeed, the authors in Refs.~\cite{Page:2010aw, Shternin:2010qi} found that the rapid cooling of the Cas A NS can be explained in the minimal cooling scenario with an appropriate choice of $T_c^{(n)}$ and a sufficiently large proton pairing gap.
For instance, it is shown in Ref.~\cite{Page:2010aw} that the observed data points are
fitted quite well 
for $T_c^{(n)} \simeq 5.5 \times 10^8$~K and $\Delta M = 5 \times 10^{-13} M_{\odot}$, where the CCDK model for proton gap \cite{Chen:1993bam} and the APR EOS \cite{Akmal:1998cf} are adopted.
In Fig.~\ref{fig:cas-a-minimal}, we show the cooling curve using similar setup in Ref.~\cite{Page:2010aw}.
We take CCDK gap model for proton pairing.
For neutron triplet pairing, we use phenomenological formula \eqref{eq:p-gap-mod-gauss} with $r=0$, i.e., the Gaussian with three parameters.
The cooling curve fits the observation well for $k_{F,0}=2.3\unit{fm^{-1}}$, $\delta k_F=0.58\unit{fm^{-1}}$ and $T_{c,\mathrm{max}}^{(n)}=5.5\times10^8\unit{K}$ (blue line).
If we change $T_{c,\mathrm{max}}^{(n)}$ only by $10\%$, the theoretical prediction becomes very bad (dotted lines).
In Refs.~\cite{Page:2012se,Shternin:2010qi}, it was shown  that the temperature observations of other NSs can also be fitted by cooling curves with pairing models required by the rapid cooling of the Cas A NS.
Because of its simplicity, the minimal cooling scenario is a very promising candidate of the correct NS cooling model.

In the rest of this section, we will consider the compatibility between the minimal NS cooling model and axion models by looking for the highest axion decay constant $f_a$ with which the rapid Cas A NS cooling cannot be fitted by any pairing model.
This serves as a lower bound of $f_a$ under the assumption that the minimal NS cooling model describes the cooling of NS correctly.

\section{Axion Emission from Neutron Stars}

The discussion in the last section would be changed if there is an additional cooling source, i.e., if $L_{\text{cool}} \neq 0$ in Eq.~\eqref{eq:balanceeq}.
In this case, the temperature at $t\simeq 320$~years is predicted to be lower than that in the minimal cooling scenario.
However, 
the observed surface temperature of the Cas A NS, $T_s \simeq 2 \times 10^6$~K, implies $T \simeq 4\times 10^8$~K (see Eq.~\eqref{eq:ts-tb-iron}), and $T_c^{(n)}$ needs to be larger than this value in order for the PBF process to operate at $t\simeq 320$ years.
In other words, if we fix $T_c^{(n)} \simeq 5.5 \times 10^8$~K in the case of $L_{\text{cool}} \neq 0$, the rapid cooling due to the PBF process has occurred much before $t \simeq 320$~years---then, the rapid cooling would have already ceased and/or the present surface temperature would be much lower than the observed value.
Accordingly, we may obtain a constraint on extra cooling sources from the Cas A NS cooling data.

\subsection{Axion}
\label{sec:axion-review}

To discuss this possibility, in this work, we take axion \cite{Weinberg:1977ma, Wilczek:1977pj} as a concrete example for a
cooling source.
They are emitted out of NSs through their couplings to nucleons. 

Axion is a well-motivated candidate of particles in BSM.
It is a pseudoscalar, and dynamically explains the so-called strong $CP$ problem: $CP$ conservation in the strong interaction~\cite{Peccei:1977hh, Peccei:1977ur}.
The common feature of axion $a$ is that it couples to gluons through the anomalous triangle diagrams as
\begin{align}
  \label{eq:lag-agg}
  \mathcal{L}_{agg}
  =
  \frac{\alpha_S}{8\pi}\frac{a}{f_a}G_{\mu\nu}^a\tilde{G}^{a\mu\nu}\,,
\end{align}
where $\alpha_S$ is the strong coupling constant, $f_a$ the axion decay constant, $G^a_{\mu\nu}$ the gluon field strength and $\tilde{G}^{a\mu\nu}=\epsilon^{\mu\nu\alpha\beta}G^a_{\alpha\beta}/2$ its dual.
The axion mass is $m_a = \Order(m_\pi f_\pi/f_a)$~\cite{diCortona:2015ldu}, where $m_\pi$ and $f_\pi$ are pion mass and decay constant, respectively.
Considering the previously obtained constraints, e.g. $f_a\gtrsim 10^8\GeV$ from SN1987A~\cite{Raffelt:2006cw}, the axion mass of our interest is $m_a \lesssim \mathrm{meV}$.
Therefore, we can treat axions as massless particles inside NSs.

The axion couplings to quarks depend on models, and so do the couplings to nucleons.
We write the axion-nucleon couplings in the form
\begin{equation}
\mathcal{L}_{aNN} = \sum_{N =p,n} \frac{C_N}{2 f_a} \bar{N} \gamma^\mu
  \gamma_5 N \partial_\mu a ~.
\label{eq:axion_nucleon}
\end{equation}
The coefficients $C_N$ are expressed in terms of the axion-quark couplings $C_q$ (having the same form as in Eq.~\eqref{eq:axion_nucleon}), the quark masses $m_q$, and the spin fractions $\Delta q^{(N)}$ defined by $2 s_\mu^{(N)} \Delta q^{(N)} \equiv \langle N| \bar{q} \gamma_\mu \gamma_5 q |N\rangle$ with $s_\mu^{(N)}$ the spin of the nucleon $N$.
At the leading order in the strong coupling constant $\alpha_s$, we have $C_N = \sum_{q} (C_q-m_*/m_q) \Delta q^{(N)}$ with $m_* \equiv m_um_dm_s/(m_u m_d + m_d m_s + m_u m_s) $.
QCD corrections to this formula are discussed in Ref.~\cite{diCortona:2015ldu}; for instance, in the case of the KSVZ axion ($C_q = 0$)~\cite{Kim:1979if, Shifman:1979if}, we have
\begin{align}
  \label{eq:cncp-ksvz}
  C_p = -0.47(3),\quad
  C_n =-0.02(3),
  \quad(\text{KSVZ})
\end{align}
while for the DFSZ axion~\cite{Zhitnitsky:1980tq, Dine:1981rt} ($C_{u,c,t} = \cos^2 \beta/3$ and $C_{d,s,b} = \sin^2 \beta/3$ with $\tan\beta$ the ratio of the vacuum expectation values of the two doublet Higgs fields, $\tan \beta \equiv \langle H_u\rangle /\langle H_d\rangle$),
\begin{align}
  \label{eq:cncp-dfsz}
  C_p = -0.182(25) - 0.435 \sin^2 \beta,\quad
  C_n = -0.160(25) + 0.414 \sin^2 \beta, \quad
  (\text{DFSZ})
\end{align}
where $\Delta u^{(p)} = \Delta d^{(n)} = 0.897(27)$, $\Delta d^{(p)} = \Delta u^{(n)} = -0.376(27)$, and $\Delta s^{(p)} = \Delta s^{(n)} = -0.026(4)$ are used~\cite{Patrignani:2016xqp}. 

\subsection{Axion emissivty}
\label{sec:axion-emissivity}

The axion-nucleon couplings induce axion emission via the PBF and bremsstrahlung processes~\cite{Leinson:2014ioa, Sedrakian:2015krq, Iwamoto:1984ir, Nakagawa:1987pga, Nakagawa:1988rhp, Iwamoto:1992jp, Umeda:1997da, Paul:2018msp}.
The axion PBF
\begin{align}
  \label{eq:axion-pbf-diag}
  \tilde{N}\tilde{N} \to [\tilde{N}\tilde{N}] + a
\end{align}
occurs after nucleons form Cooper pairs, similar to the neutrino PBF~\eqref{eq:pbf-qp-diagram}.
The emissivity reads~\cite{Leinson:2014ioa, Sedrakian:2015krq}
\begin{align}
  \label{eq:axion-pbf-emis}
  Q_{a,\mathrm{PBF}}
  &=
    \frac{2C_N^2m_N^*p_{F,N}}{3\pi^3f_a^2}T^5F_2(\tau)
    \times
    \begin{cases}
      v_{F,N}^2&({}^1S_0) \\
      1 &({}^3P_2)
    \end{cases}
\end{align}
where $\tau = T/T_{c}^{(N)}$ and
\begin{align}
  \label{eq:phasespaec-axion-pbf}
  F_2(\tau)
  =
  \int \frac{d\Omega}{4\pi}\,
  y^2\int_0^\infty\,dx\frac{z^2}{\left( e^z +1 \right)^2}
\end{align}
with $y = \delta/T$ and $z=\sqrt{x^2+y^2}$.
Another important process is axion bremsstrahlung:
\begin{align}
  \label{eq:axion-brems}
  N_1 + N_2 \to N_1 + N_2 + a\,,
\end{align}
where $N_{1,2} = n,p$.
The bremsstrahlung consists of $nn$, $np$ and $pp$ branches.
The emissivities are calculated as~\cite{Iwamoto:1992jp} 
\begin{align}
   Q_{aNN}^{(0)}
  &=
    \frac{31}{3780\pi}
    \left( \frac{C_Nm_N}{f_a} \right)^2
    \left( \frac{f}{m_\pi} \right)^4
    m_N^{*2}p_{F,N}T^6F(x)\,,\quad(N=n,p)
    \label{eq:axion-brems-nn-pp}\\
  Q_{anp}^{(0)}
  &=
    \frac{31}{5670\pi}  \left( \frac{f}{m_\pi} \right)^4
    m_N^{2}p_{F,p}T^6
    \notag\\
    &\times
   \left[  
    \frac{1}{2}(g^2+h^2)F(y)
      + (g^2+h^2)G(y)\right.
      \notag\\
  &+\left.\left( g^2+\frac{h^2}{2} \right)
  \left( F\left( \frac{2xy}{x+y} \right)+ F\left( \frac{2xy}{-x+y} \right) 
    +\frac{y}{x}\left( F\left( \frac{2xy}{x+y}\right) - F\left( \frac{2xy}{-x+y} \right) \right)\right)
      \right]\,,
    \label{eq:axio-brems-np}
\end{align}
where $f\simeq1$ is the parameter for the nucleon-pion interaction, $x=m_\pi/(2p_{F,n})$, $y=m_\pi/(2p_{F,p})$, and
\begin{align}
  g &= \frac{C_pm_p}{f_a} + \frac{C_nm_n}{f_a}\,,
      \label{eq:axion-g} \\
  h &= \frac{C_pm_p}{f_a} - \frac{C_nm_n}{f_a}\,,
      \label{eq:axion-h} \\
  F(x) &=
         1 - \frac{3}{2}\arctan(x^{-1}) + \frac{x^2}{2(1+x^2)}\,,
         \label{eq:axion-brems-f}\\
  G(y) &= 1-y\arctan(y^{-1})\,.
           \label{eq:axion-brems-g}
\end{align}
For the superfluid nucleons, as we discussed in Sec.~\ref{sec:urca-reduction}, we multiply the superfluid reduction factor.
We use the same reduction factor as the neutrino bremsstrahlung processes.


We have modified the public code \texttt{NSCool} \cite{NSCool} to implement these processes and use it to compute the luminosity of axion and its effect on the NS cooling curves.
We adopt the APR EOS~\cite{Akmal:1998cf} and fix the NS mass to be $M = 1.4 M_\odot$ in this work.

\begin{figure}[t]
  \centering
  \begin{minipage}{0.5\linewidth}
    \includegraphics[clip, width = 1.0\textwidth]{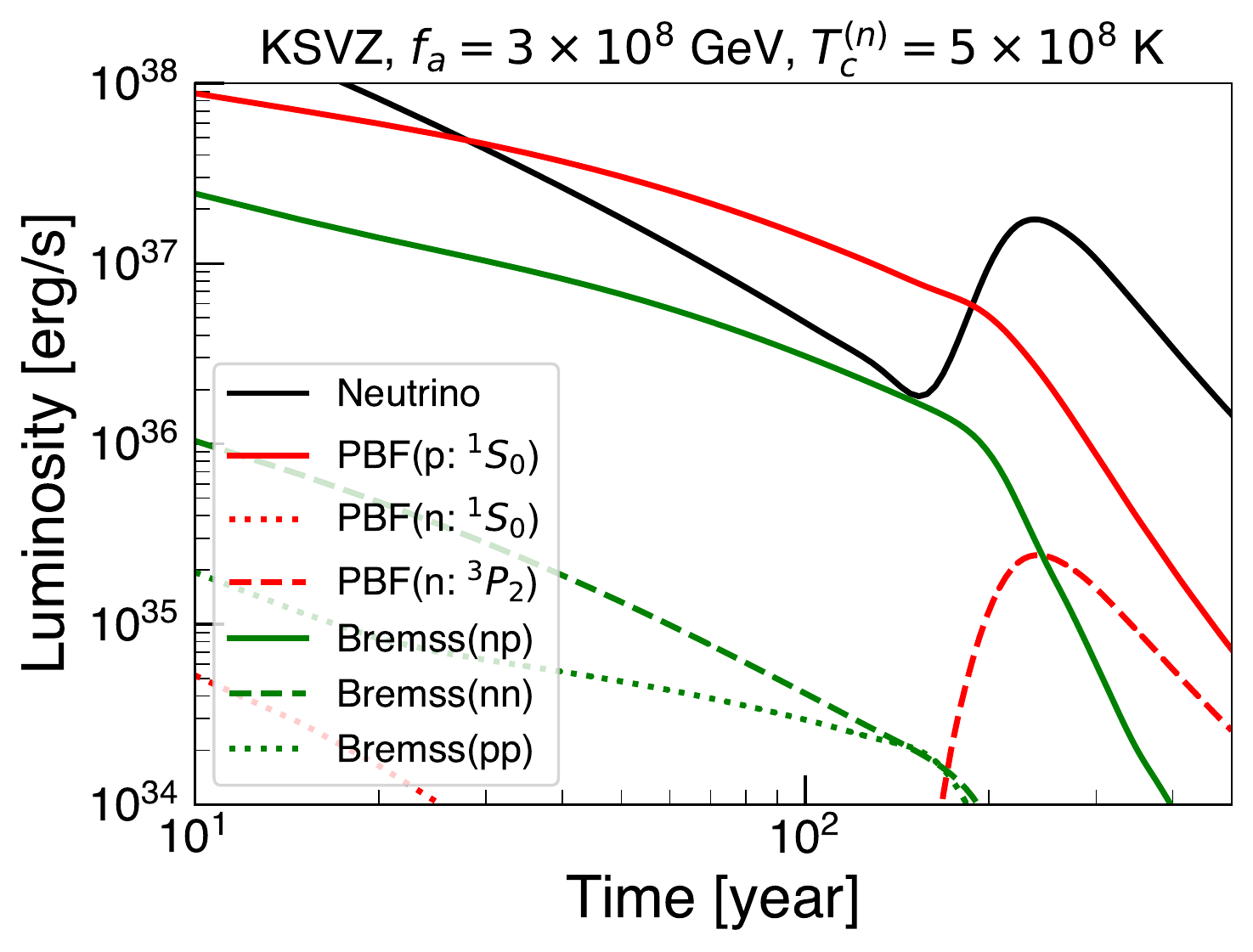}
  \end{minipage}%
  \begin{minipage}{0.5\linewidth}
    \includegraphics[clip, width = 1.0\textwidth]{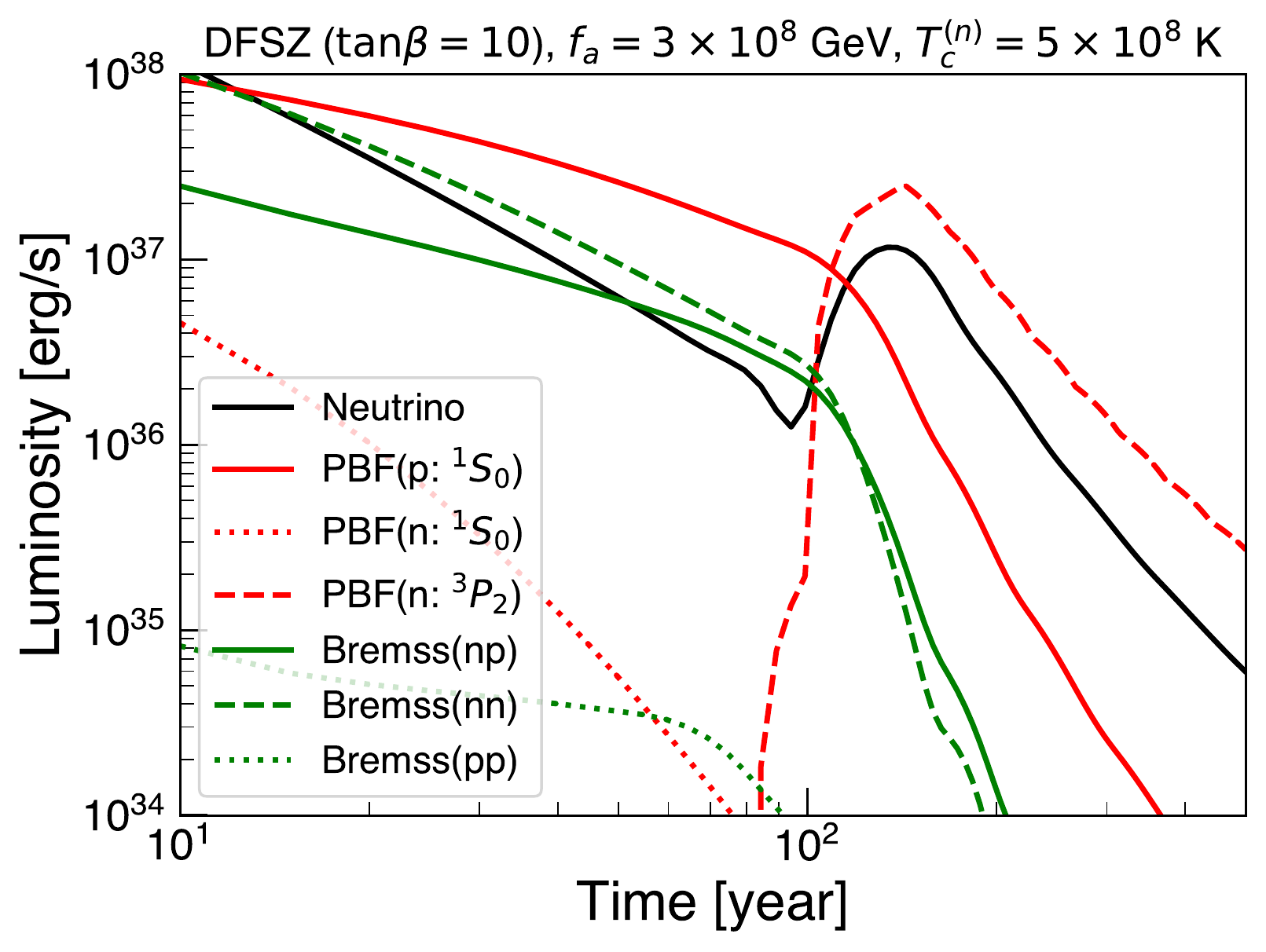}
  \end{minipage} 
  \caption{
    Luminosity of each axion emission (red and green) and the total 
    neutrino emission (black) processes as a function of time.
    The left panel shows the result of KSVZ model, and the right panel shows that of the DFSZ model with $\tan\beta=10$.
  }
  \label{fig:Lumiksvz}
\end{figure}

In Fig.~\ref{fig:Lumiksvz}, we show luminosities of various axion emission processes in the KSVZ (left) and DFSZ (right) model with $f_a=3\times 10^8$~GeV as functions of time (red and green).
For comparison, we also show the total luminosity of neutrino emission (black). 
We use the SFB model~\cite{Schwenk:2002fq} for the gap of singlet neutron pairings.
Our analysis is insensitive to this choice.
For the proton singlet pairings, the CCDK model~\cite{Chen:1993bam} is chosen because it has the largest gap in the NS core among those presented in Ref.~\cite{Page:2004fy}---this results in a strong suppression of neutrino emission before the onset of the neutron triplet Cooper pairing and thus improves the fit onto the observed Cas A NS data \cite{Page:2010aw} as we discussed above. 
Note that a large gap for the proton singlet pairing also suppresses the axion emission, and therefore the CCDK model gives a conservative bound.
On the contrary, there are large uncertainties in choosing a model of neutron
triplet pairings.
In this chapter, we assume the triplet pairing of $m_j=0$ state (see Tab.~\ref{tab:gap-angle}), and model the gap with a Gaussian shape, i.e., we use Eq.~\eqref{eq:p-gap-mod-gauss} with $r=0$.
In Fig.~\ref{fig:Lumiksvz}, we take $T_{c,\mathrm{max}}^{(n)} = 5 \times 10^8$~K.
The instantaneous increase in luminosity at $t \simeq 300$ years for the neutrino emission, as well as the axion emission in the PBF process, is due to the formation of neutron triplet pairings.
As we see from the left panel, the axion emission in the KSVZ model via the proton PBF and proton-neutron bremsstrahlung processes is as strong as the neutrino emission before the formation of neutron triplet parings.
In particular, the emission via the proton PBF dominates over other axion emission processes in this case because it is suppressed by less powers of $T_{\text{core}}$ resulting from a smaller number of states involved in the process.%
\footnote{
  For the axion PBF process, the phase space integration of nucleon and axion gives $T^2$ and $T^3$, respectively, while the energy-momentum conserving delta function gives $T^{-2}$, as we saw in the neutrino PBF.
  The axion energy term results in $T$. The interaction vertex contains the derivative coupling, which provides another $T^2$ from the squared matrix element. Finally, another $T^{-1}$ comes from the normalization of axion state. In total, we obtain $Q_{a,\mathrm{PBF}}\propto T^5$.
  We can estimate the proportionality of the axion bremsstrahlung in the same way. 
}
This allows us to set stringent bounds even on the KSVZ model where $|C_n|$ is vanishingly small.
If $|C_n|$ is sizable as in the DFSZ model, the neutron bremsstrahlung process is also significant (see the right panel).

\subsection{Related works}
\label{sec:related-work-axion}

There are several studies that discuss their effects on the Cas A NS cooling. 
A detailed study on the axion emission processes and their consequences on
NS cooling was performed in Ref.~\cite{Sedrakian:2015krq}, where
predicted cooling curves were compared with the temperature data of
young NSs including the Cas A NS. However, only the average
temperature of the Cas A NS was concerned and no attempt was made to fit
the slope of its cooling curve. In Ref.~\cite{Leinson:2014ioa},
the axion-neutron PBF process was utilized to enhance the cooling rate so that
the slope of the Cas A NS cooling curve was reproduced. This
analysis focused on the time around which the neutron
superfluid transition was supposed to occur, with the
axion-proton PBF and axion bremsstrahlung processes neglected; 
especially, there was no discussion on the
temperature evolution at $t\lesssim 300$~years.

In this chapter, we study the Cas A NS cooling including all the relevant processes of axion emission, and taking account of the temperature evolution in the whole lifetime of the Cas A NS.

\section{Limit on axion decay constant}

\subsection{Constraints from the temperature}
\label{sec:constraint-temp}

\begin{figure}[t]
  \centering
  \begin{minipage}{0.5\linewidth}
    \includegraphics[clip, width = 1.0\textwidth]{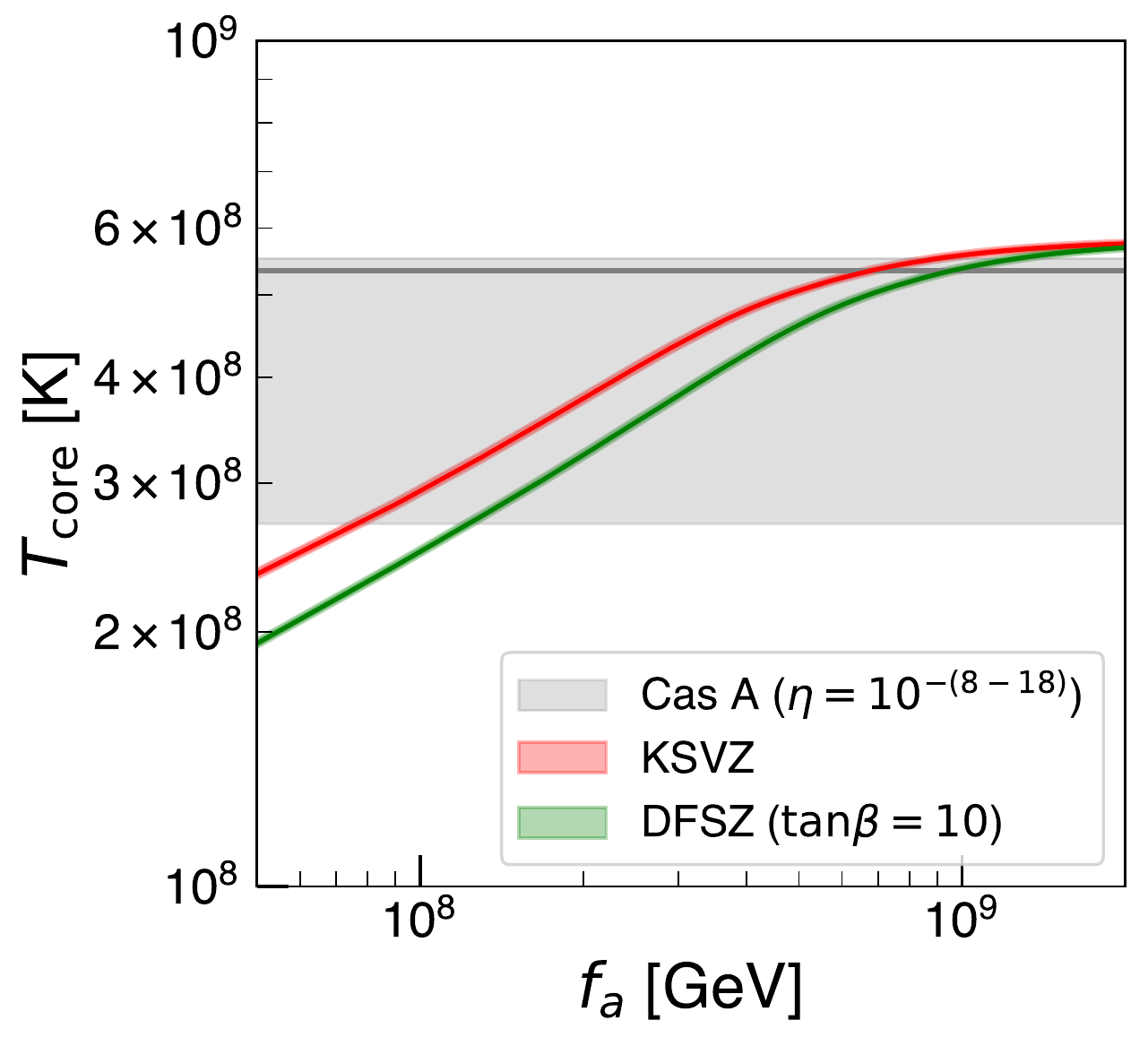}
  \end{minipage}%
  \begin{minipage}{0.5\linewidth}
    \includegraphics[clip, width = 1.0\textwidth]{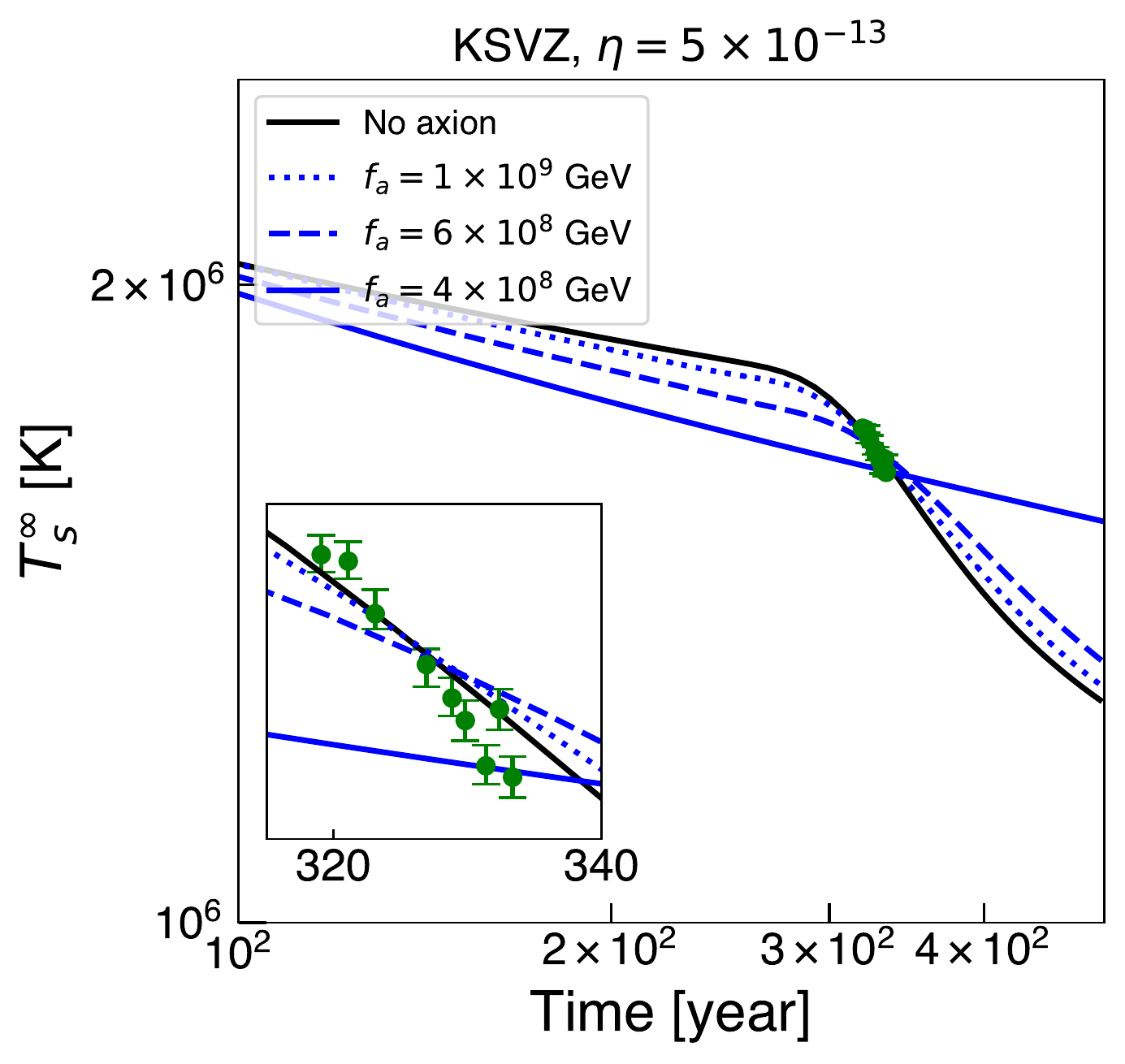}
  \end{minipage}
  \caption{
    Left: The core temperature $T_{\mathrm{core}}$ at $t = 300$--338~years against $f_a$, without neutron triplet superfluidity. The gray shaded region reflects the uncertainty of $T_{\mathrm{core}}$ due to $\eta$ and the grey solid line corresponds to $\eta = 5\times 10^{-13}$.
    Right: Cooling curves compared to observed data.
  }
  \label{fig:tvsfa}
\end{figure}

Now let us study the effect of the axion emission processes on the NS temperature evolution.
In the left panel of Fig.~\ref{fig:tvsfa}, we show the core temperature $T_{\mathrm{core}}$ at the time of the Cas A NS age on January 30, 2000 ($t= 300$--338 years) as functions of $f_a$ for the KSVZ and DFSZ ($\tan\beta = 10$) models in the red and green bands, respectively, with the bands reflecting the uncertainty in the NS age.
Since we are interested in the drop in the temperature before the onset of neutron triplet pairings, we have switched off the neutron triplet superfluidity in this plot. The Cas A NS core temperature $T_{\mathrm{core, A}}$ inferred from the observation for the envelope model Ref.~\cite{Potekhin:1997mn} with $\eta = 5\times 10^{-13}$ is shown in the gray line, while its uncertainty is estimated by varying $\eta = 10^{-(8-18)}$ (gray band)~\cite{Ho:2014pta}.
We find that the predicted core temperature falls below $T_{\mathrm{core, A}} \simeq 5 \times 10^8$~K for $f_a = (\text{a few}) \times 10^{8}$~GeV, and thus $f_a$ smaller than this value is disfavored.
We also note that the bound derived in this manner has a large uncertainty due to the ignorance of the envelope parameter $\eta$.

\subsection{Constraints from the cooling rate}
\label{sec:constraint-cas-a-fit}


The right panel of Fig.~\ref{fig:tvsfa} shows the best-fit curves of the red-shifted surface temperature $T_s^\infty$ for several values of $f_a$ in the KSVZ model (blue lines), as well as that obtained in the minimal cooling scenario (black line).
For each curve, we vary the neutron triplet gap parameters and the Cas A NS age to fit the observed data shown in the green points \cite{Ho:2014pta}, where the envelope parameter is fixed to be $\eta = 5\times 10^{-13}$ as in \cite{Page:2010aw}.
We find that as $f_a$ gets smaller, the NS temperature at $t \simeq 320$~years gets
lower, which then requires a smaller value of $T^{(n)}_c$ and results in a shallower slope.
As a result, the fit gets considerably worse for a smaller $f_a$. 

Finally, we show the lower bound on $f_a$ obtained from our attempt to fit the observed data.
If the Cas A NS has an iron envelope with a thin carbon layer ($\eta=5\times 10^{-13}$ as in \cite{Page:2010aw}), 
\begin{align}
f_a&\gtrsim 5~(7)\times10^8~\mathrm{GeV}\quad &\text{KSVZ~(DFSZ)}~,
\label{eq:limitonfa}
\end{align}
where we take $\tan\beta=10$ for the DFSZ model.
The bound on the DFSZ model is comparable to the one on the KSVZ model with $C_n\simeq 0$ because of the large luminosity from proton PBF and the proton-neutron bremsstrahlung as shown in Fig.~\ref{fig:Lumiksvz}.
For general couplings, the limit can be roughly estimated by
\begin{align}
f_a&\gtrsim \sqrt{0.9~C_p^2+1.4~C_n^2}
\times10^9~\mathrm{GeV}~.
\end{align}
As a comparison, the bound derived from SN1987A is $f_a\gtrsim 4\times 10^8\unit{GeV}$ \cite{Raffelt:2006cw} for the KSVZ model, comparable to the bounds from the Cas A NS obtained above.

\subsection{Uncertainties from the envelope}
\label{sec:uncertainty-envelope}

\begin{figure}
\centering
{\includegraphics[clip, width = 0.6 \textwidth]{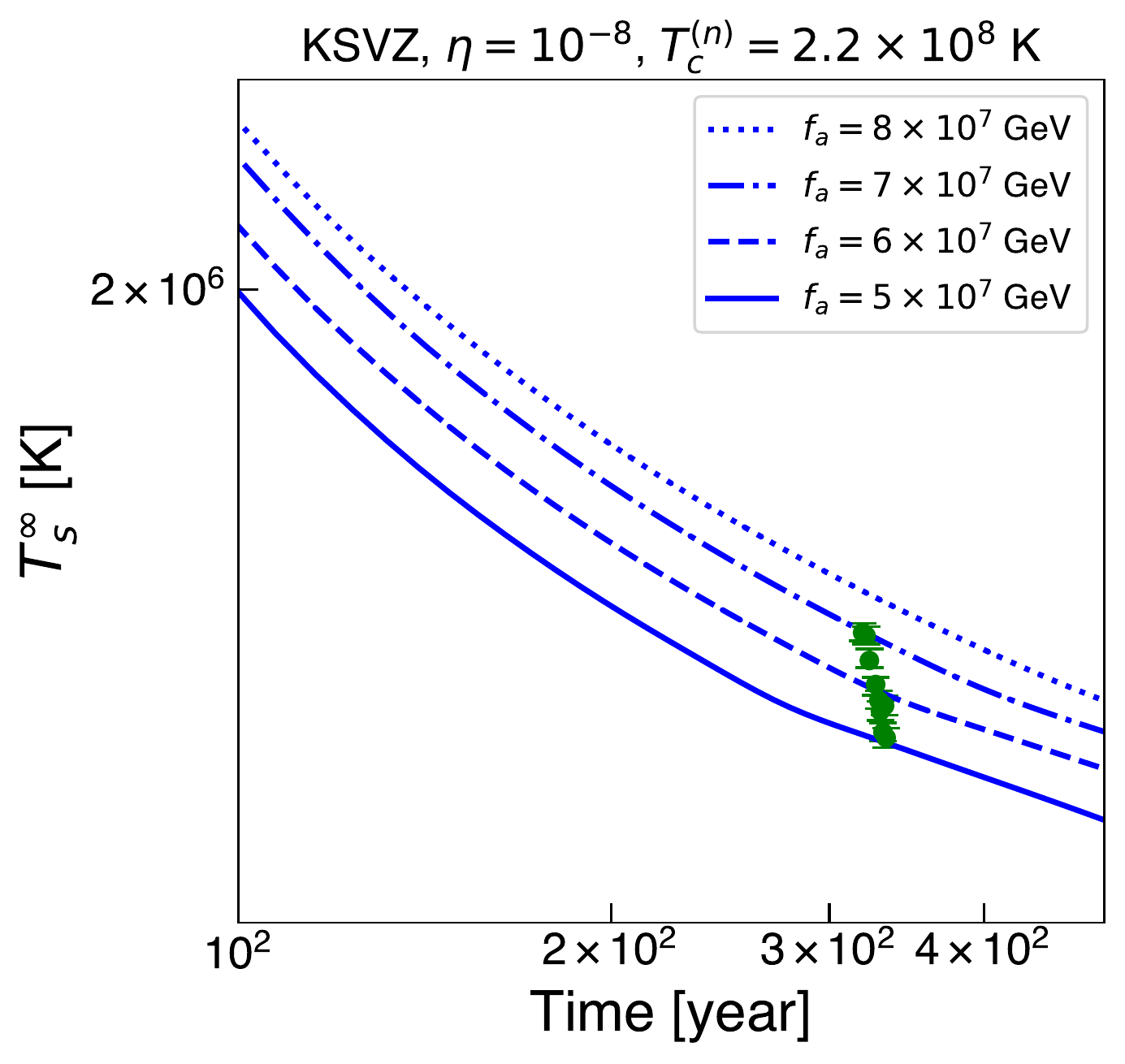}} 
\caption{
Cooling curves of the KSVZ model for $\eta=10^{-8}$.
}
\label{fig:cooling_hieta}
\end{figure}

The bounds in the last section are derived with $\eta = 5\times10^{-13}$.
If the envelope is maximally carbon-rich ($\eta=10^{-8}$) instead,
naively, the bound will be weakened by an $\Order(1)$ factor as shown in the left panel of Fig.~\ref{fig:tvsfa}. 
However, as we increase $\eta$ and hence the thermal conductivity of the envelope, the same observed surface temperature corresponds to a lower core temperature in the NS.
This reduces drastically the neutrino luminosity from the neutron PBF that scales as $T_{\mathrm{core}}^7$ for $T_{\mathrm{core}} \simeq T_c^{(n)}$, making it harder to fit the rapid cooling slope alone.
An axion emission may help to cool the NS, but in the KSVZ model the neutrino emission dominates over axion in the neutron PBF process and hence it can be incompatible with the observed rapid cooling.
In Fig.~\ref{fig:cooling_hieta}, we plot the cooling curves of the KSVZ model for a NS with $\eta=10^{-8}$.
Due to the large $\eta$, neutrino emission cannot cool the NS to its observed temperature and a sizable axion emission from proton with $f_a\lesssim 8\times10^7$~GeV is needed. 
The neutron triplet pairing temperature is set to $T_c^{(n)}=2.2\times10^8\unit{K}$ so that the phase transition occurs shortly before the observation.
However, we cannot see any rapid cooling drop in the curves because the neutrino PBF luminosity is suppressed as described above.
For a moderate $\eta$ ($\sim 10^{-10}$), on the other hand, we find that the slope of the cooling data constrains $f_a\gtrsim 5\times 10^8\unit{GeV}$, which is similar to that for $\eta = 5\times 10^{-13}$ given in Eq.~\eqref{eq:limitonfa}. 
Thus the limit on the KSVZ model is rather stringent even if we allow $\eta$ to vary.

For the DFSZ model, $|C_n|$ is non-vanishing in general so the axion emission during the neutron triplet-pairing phase transition can rapidly cool the Cas A NS via the PBF process even when $\eta = 10^{-8}$.
According to Fig.~\ref{fig:tvsfa}, $f_a\sim 10^8$~GeV is needed to reproduce the observed rapid cooling in this case.
We note in passing that such a stellar cooling source may be favored by several astrophysical observations~\cite{Giannotti:2015kwo, Giannotti:2017hny}, for which our new limits (or a favored value of $f_a$ in the case of the DFSZ model for a large $\eta$) may have important implications. 

\subsection{Mean free path}
\label{sec:mfp}

To conclude this section, we point out that if $f_a$ is too small the axion may have a short mean free path in the NS and thus avoid all the limits set above.
For the purpose of qualitative estimation, we only consider the partial axion decay rate by the inverse proton PBF $a\rightarrow\tilde{p} + \tilde{p}$ that is important to both the KSVZ and DFSZ models.
Here, $ \tilde{p}$ is the quasi-particle excitation inside a medium of proton Cooper pairs.
A more careful evaluation is beyond the scope of this paper.
The matrix element of the related process is given in Ref.~\cite{Keller:2012yr}, which leads to
\begin{equation}
\Gamma_{a\rightarrow\tilde{p}\tilde{p}}
\sim 
\frac{m_p^* p_{F} v^2_{F} T}{3\pi f_a^2}\left(\frac{C_p}{2}\right)^2~.
\end{equation}
For $p_F\sim100$~MeV, $m_p^*\sim1$~GeV, $T\sim \Delta_p\sim 1~$MeV, we need 
\begin{equation}
f_a\gtrsim \left(\frac{C_p}{2}\right)\times 10^6\unit{GeV}~,
\end{equation}
for the mean free path of axion $l_a=1/\Gamma_{a\rightarrow\tilde{p}\tilde{p}}$ to be larger than $\sim 10\unit{km}$, the radius of the Cas A NS.

\section{Summary and Discussion}

We have discussed the implications of the Cas A NS rapid cooling for axion models.
It is found that the requirement of fitting the slope of the Cas A NS cooling curve in accordance with the temperature evolution from its birth results in a limit of $f_a \gtrsim 5\times 10^8$~GeV if the envelope only has a thin layer of carbon. 

This limit is stronger than those obtained in the previous studies.
For instance, Ref.~\cite{Sedrakian:2015krq} sets $f_a \gtrsim (5-10)\times 10^7 \GeV$ for the KSVZ model without taking the cooling rate of the Cas A NS into account.
In Ref.~\cite{Leinson:2014ioa}, it is argued that the rapid cooling of the Cas A NS can be explained with $|C_n|/f_a \simeq \sqrt{0.16}/(10^9~\mathrm{GeV})$ in the KSVZ model---this corresponds to $f_a \simeq 5\times 10^7$~GeV for $C_n = -0.02$, which is actually in tension with the observation as shown in Fig.~\ref{fig:tvsfa}. 

Finally, some remarks on the uncertainties of our analysis are in order.
First of all, lower cooling rates of the Cas A NS have been reported and the actual cooling rate is still in dispute~\cite{Elshamouty:2013nfa,Posselt:2013xva}.
In the worst scenario where the NS is found to cool slowly by future observations, our strong limit on the KSVZ model will no longer hold.
However, the conservative limits obtained directly from Fig.~\ref{fig:tvsfa} by assuming maximal $\eta$ are hardly affected since they do not rely on the cooling rate and depend dominantly on the emission from proton.
Other mechanisms~\cite{Yang:2011yg, Negreiros:2011ak, Blaschke:2011gc, Noda:2011ag, Sedrakian:2013xgk, Blaschke:2013vma, Bonanno:2013oua, Leinson:2014cja, Taranto:2015ubs, Noda:2015pvn, Grigorian:2016leu} such as an extended thermal relaxation time have also been proposed to explain the rapid cooling rate and more study on neutron star physics is needed to test them against the minimal cooling scenario that we base our work on.

Apart from the observation and theoretical uncertainties stated above, the limit on $f_a$ obtained in this work also suffers from the ignorance of the envelope parameter $\eta$.
While a maximal $\eta$ parameter is inconsistent with the KSVZ axion model, it is compatible with the DFSZ model with $f_a \sim 10^8$~GeV.
Further cooling data of the Cas A NS%
\footnote{Recently, the rapid cooling rate of the Cas A NS is confirmed again in Ref.~\cite{Wijngaarden:2019tht} with additional temperature data.
  On the other hand, Ref.~\cite{Posselt:2018xaf} performs an independent analysis and reports a lower cooling rate, which is consistent with no rapid cooling at $3\,\sigma$.
The actual cooling rate of the Cas A NS is still in dispute, and we have to wait for further data and/or analyses.},
as well as additional observations of direct cooling curves of other NSs, are of great importance to verify the rapid cooling of the Cas A NS and to test the present scenario against potential alternative explanations of the rapid cooling of the Cas A NS, and may allow us to obtain a more robust limit on the axion.

\chapter{Rotochemical heating of neutron stars with proton and  neutron pairings}
\label{chap:neutron-star-heating}

In Chap.~\ref{chap:neutron-star-cooling}, we have seen that the minimal cooling paradigm explains many observations of neutron star surface temperature, in particular with the help of nucleon superfluidity.
The minimal cooling predicts that the neutrino emission rate gets highly suppressed as the NS cools down, and eventually the surface photon emission dominates the cooling, which then results in a rapid decrease in the NS surface temperature to $T_s^\infty \lesssim 10^4\unit{K}$ at $t\gtrsim 1~\unit{Myr}$.
This is a generic consequence of the cooling theory.

This minimal cooling theory is build upon the assumption that nucleons and leptons (electrons and muons) in the NS core are in chemical (or beta) equilibrium through the Urca reactions. This assumption, however, turns out to be invalid for (especially old) NSs.
As the rotation rate of a pulsar decreases, the centrifugal force decreases~\cite{Reisenegger:1994be}. 
This reduction makes the NS continuously contract, which perturbs the local number density of each particle species away from the equilibrium value. On the other hand, the timescale of the Urca reaction is typically much longer than that of the NS contraction (especially for old NSs), and thus the beta equilibrium cannot be maintained. This has a significant impact on the cooling of NSs, since the imbalance in the chemical potentials of nucleons and leptons, which quantifies the degree of the departure from beta equilibrium, is partially converted to the heat inside the NS~\cite{Reisenegger:1994be, 1992A&A...262..131H, 1993A&A...271..187G}. This heating mechanism due to the non-equilibrium Urca process is sometimes called the \textit{rotochemical heating}.
In the presence of the rotochemical heating, the NS surface temperature reaches $T_s^\infty \sim 10^5 \unit{K}$ at $t \gtrsim \unit{Gyr}$~\cite{Fernandez:2005cg}, which is in contrast to the prediction of the minimal cooling with the beta equilibrium.

Intriguingly, recent observations of old pulsars suggest the presence of old ``warm'' NSs; several observed ordinary and millisecond pulsars (MSPs) show the temperature of $T_s^\infty \sim 10^5\unit{K}$ at $t>10^7\unit{yr}$~\cite{Kargaltsev:2003eb, Durant:2011je, Rangelov:2016syg, Mignani:2008jr, Pavlov:2017eeu}.
On the other hand, the observation of the ordinary old pulsar J2144-3933 imposes an upper limit on its surface temperature, $T_s^\infty < 4.2 \times 10^4\unit{K}$ \cite{Guillot:2019ugf}, giving an evidence for the presence of an old ``cold'' NS.
It is quite important to study if we can explain these observations by means of the minimal cooling with the non-equilibrium beta processes, i.e., with the rotochemical heating.

To that end, it is necessary to include the non-equilibrium effect into the minimal cooling paradigm in a consistent manner.
As we have seen in Sec.~\ref{sec:nucleon-pairing}, both the neutrons and protons form the pairing in the NS core, and their pairing gaps depend on the nucleon density.
The previous studies~\cite{Reisenegger:1996ir, Villain:2005ns, Petrovich:2009yh, Pi:2009eq, Gonzalez-Jimenez:2014iia} have revealed the roles of pairing gaps in the rotochemical heating.
In particular, the superfluid gap provides the threshold of the heating.
However, the numerical study including both proton and neutron parings with density-dependent gaps is not performed yet.
For a more realistic calculation of the rotochemical heating, we need to include all of these effects simultaneously.
In this chapter, we include both the proton singlet and neutron triplet pairings into the calculation of the NS thermal evolution in the presence of the non-equilibrium beta processes, with their density- and temperature-dependence fully taken into account. 
We then compare the results to the latest observational data of the NS surface temperature. 
We find that the heating with both nucleon pairings can be stronger than that only with neutron pairing~\cite{Gonzalez-Jimenez:2014iia}, which turns out to be advantageous in explaining the old warm MSPs. 
Meanwhile, it is found that the same setup can also account for the temperature data of the ordinary classical pulsars if the diversity of their initial spin periods is taken into account. 
We also discuss the compatibility of our scenario with the so-called X-ray dim Isolated Neutron Stars (XDINSs).

This chapter is organized as follows.
After reviewing the rotochemical heating in Sec.~\ref{sec:rotochemical-heating}, we summarize the observational data of NS surface temperatures in Sec.~\ref{sec:observations-old-hot}.
We then give our numerical analysis in Sec.~\ref{sec:role-non-equilibrium} and devote Sec.~\ref{sec:conclusion-roto} to discussion.
In appendix~\ref{sec:phase-space}, we summarize the formulas for phase space factors.
This chapter is based on the author's work~\cite{Yanagi:2019vrr}.

\section{Rotochemical heating}
\label{sec:rotochemical-heating}

The rotochemical heating occurs in the very same setup as in the minimal cooling.
The only difference is whether the beta equilibrium of the Urca processes is assumed or not.
In the case of an actual pulsar, its rotational rate keeps decreasing, which results in a continuous reduction of the centrifugal force.
Consequently, NSs contract continuously and the equilibrium number densities of nucleons and charged leptons change at all times.
The number densities of these particles in a NS follow the equilibrium values if the Urca reactions are fast enough.
It however turns out that the typical timescale of these reactions is much longer than that of the NS contraction, especially for an old NS.
As a result, the Urca reactions are no longer in beta equilibrium.
The departure from the beta equilibrium leads to an imbalance in the chemical potentials of nucleons and leptons, which keeps increasing over the time. The energy stored in the chemical imbalance is released partly via the neutrino emission and partly as a heat, generating a non-zero heating luminosity, $L_H^\infty$.
We emphasize that this heating effect due to the non-equilibrium beta reactions is an inevitable consequence for actual rotating NSs and thus needs to be included into the calculation of the temperature evolution of NSs.
In this section, we review the rotochemical heating~\cite{Reisenegger:1994be, 1992A&A...262..131H, 1993A&A...271..187G, Fernandez:2005cg}.
In Sec.~\ref{sec:eq-roto}, we derive the general thermal evolution equation beyond the beta equilibrium.
We then discuss the heating rate from the non-equilibrium modified Urca process.

\subsection{Equations for rotochemical heating}
\label{sec:eq-roto}
The local temperature $T(r)$ in general depends on the position, especially for a very young NS.
It is however known~\cite{Yakovlev:1999sk, Yakovlev:2000jp, Yakovlev:2004iq} that the typical timescale of thermal relaxation in a NS is $\sim 10^{2-3}\unit{yr}$ (see also the right panel of Fig.~\ref{fig:ts-sf-compare}), and thus a NS with the age $t\gtrsim 10^4\unit{yr}$ can safely be regarded as isothermal.
Since our main focus is on old NSs, in this chapter, we assume that NSs have already reached an isothermal state.
In this case, the red-shifted internal temperature defined by $T^\infty \equiv T(r) e^{\Phi(r)}$ is constant throughout the NS core.

As we discussed above, the Urca processes in spin-down pulsars are out of equilibrium, which leads to imbalance in the chemical potentials of nucleons and leptons. We denote this imbalance by 
\begin{equation}
  \eta_\ell \equiv \mu_n - \mu_p - \mu_\ell ~,
\end{equation}
for $\ell = e, \mu$.
This parameter quantifies the departure from the beta equilibrium and is equal to the amount of the energy released in each reaction.
It is discussed in Ref.~\cite{Reisenegger:1996ir} that the diffusion timescale is shorter than the time scale of its evolution, and thus we can safely assume the diffusive equilibrium in the stellar core; $\mu_i^\infty = \mu_ie^{\Phi(r)}$ and thus $\eta_\ell^\infty = \eta_\ell e^{\Phi(r)}$ are constant throughout the NS core.
In the following, we will discuss the evolution of the temperature and chemical imbalance.

\subsubsection{Evolution of temperature}

We first see the temperature evolution with the imbalance of the chemical potentials.
As in the minimal cooling, the thermal energy of a NS is released by the neutrino and photon emissions.
With the relation of the thermodynamics, the energy decrease for small time interval $dt$ is written as
\begin{align}
  \label{eq:de-noneq}
  dE^\infty
  =
  T^\infty dS + \sum_{i=n,p,e,\mu}\mu_i^\infty dN_i
  =
  -(L_\nu^\infty + L_\gamma^\infty)dt\,,
\end{align}
where $dS$ is the change of the entropy, and $dN_i$ is the change of number of each particle species.
The entropy term is rewritten as $T^\infty dS = CdT^\infty$, so that we obtain the thermal evolution equation
\begin{align}
  \label{eq:time-evl}
  C\frac{dT^\infty}{dt} &= -L_\nu^\infty -L_\gamma^\infty + L_H^\infty,
\end{align}
with the heating luminosity being
\begin{align}
  \label{eq:lum-h-general}
  L_H^\infty
  =
  -\sum_{i=n,p,e,\mu}\mu_i^\infty\frac{dN_i}{dt}\,.
\end{align}
The particle number $N_i$ changes by the charged current interactions, i.e., the Urca processes.
Let us concentrate on the modified Urca process since it is the dominant charged current process in the minimal cooling.
In this case, the heating luminosity is put into
\begin{align}
  L_H^\infty &= \sum_{\ell=e,\mu}\sum_{N=n,p} \int dV\, \eta_{\ell} \cdot \Delta\Gamma_{M,N\ell} \, e^{2\Phi(r)}
               \label{eq:hating-rate}\,,
\end{align}
where  $dV=4\pi r^2 e^{\Lambda(r)}dr$, and $\Delta \Gamma_{M,N\ell}$ is the difference between the reaction rates of the processes \eqref{eq:murca1} and \eqref{eq:murca2}.
Note that it can be easily generalized to other weak reactions such as direct Urca process.
These equations clearly show that the non-zero heating luminosity arises when the beta equilibrium is not maintained.

Note that in the limit of $L_\nu^\infty \to 0$ and $L_\gamma^\infty \to 0$, the NS is regarded as the thermally isolated system and Eq.~\eqref{eq:de-noneq} reads $T^\infty dS = L_H^\infty dt$.
The total entropy of an isolated system never decreases, which ensures the positivity of Eq.~\eqref{eq:lum-h-general}.

\subsubsection{Evolution of chemical imbalance}
To solve the thermal evolution, we need equations governing the evolution of $\eta_\ell$. 
The spin-down perturbs the chemical potential of each particle species from the equilibrium, which is denoted by $\delta\mu_i^\infty = \mu_i^\infty - \mu_{i,\mathrm{eq}}^\infty$.%
\footnote{Thus in practice, the deviation from $\beta$ equilibrium is calculated by $\eta^\infty = \delta\mu_n^\infty - \delta\mu_p^\infty - \delta\mu_\ell^\infty$.}
We assume that the departure from equilibrium is small: $|\delta\mu_i^\infty| \ll \mu_i^\infty$.
Then $\delta\mu_i^\infty$ is related to the deviation of $N_i$ as
\begin{align}
  \label{eq:delta-mu-rel}
  \delta N_i
  =
  \int dV \,\delta n_i
  \equiv
  \sum_j B_{ij}\delta\mu_j^\infty\,,
\end{align}
where $\delta n_i$ denotes the deviation from the equilibrium of the local number density.
Note that $B_{ij}$ is evaluated in equilibrium.
By inverting the matrix $B_{ij}$, we can write $\delta \mu_i^\infty$ by the linear combination of $\delta N_i$.
First such calculation was performed in Ref.~\cite{Fernandez:2005cg}, in which the authors neglected the electrostatic potential, and hence the local charge neutrality is violated in $B_{ij}$.
After that the same authors included the electrostatic correction, by which the matrix $B_{ij}$ realize the identity $B_{pj} = B_{ej} + B_{\mu j}$ so that it respects the local charge neutrality~\cite{Reisenegger:2006ky}. 
Due to this identity, however, $B_{ij}$ cannot be inverted, and one has to invert its submatrix.
The results are summarized as~\cite{Reisenegger:2006ky}
\begin{align}
  \label{eq:eta-dn}
  \eta^\infty_e
  =
  -Z_{npe}\delta N_e - Z_{np}\delta N_\mu\,,\quad
  \eta_\mu^\infty
  =
  -Z_{np}\delta N_e - Z_{np\mu}\delta N_\mu\,,
\end{align}
where $Z$'s are composed by the inverse of the submatrix of $B_{ij}$, and determined by EOSs.
We can find their analytic expression and numerical value in Refs.~\cite{Fernandez:2005cg, Reisenegger:2006ky}. 
We note that the effect of including electrostatic potential on $Z$'s is numerically small~\cite{Reisenegger:2006ky}.

The change of particle number is caused by the Urca processes.
Thus
\begin{align}
  \label{eq:ndot}
  \dot N_n &= -\sum_{\ell=e,\mu} \sum_{N=n,p}\int dV\,\Delta\Gamma_{M,N\ell}e^{\Phi(r)}\,,\notag\\
  \dot N_p &= \sum_{\ell=e,\mu} \sum_{N=n,p}\int dV\,\Delta\Gamma_{M,N\ell}e^{\Phi(r)}\,,\notag\\
  \dot N_\ell &=  \sum_{N=n,p}\int dV\,\Delta\Gamma_{M,N\ell}e^{\Phi(r)}\,.
\end{align}
We decompose $N_i$ into the equilibrium component and the deviation from it: $N_i = N_i^{\mathrm{eq}} + \delta N_i$.
The equilibrium number is determined by the hydrostatic equilibrium, and thus it is determined by the angular momentum $\Omega$ at the moment of $t$.
We parameterize its evolution as follows:
\begin{align}
  \label{eq:neq-dot}
  \dot{N}_i^{\mathrm{eq}} = 2\Omega\dot\Omega\frac{\partial N_i^{\mathrm{eq}}}{\partial \Omega^2}
  \equiv 2\Omega\dot\Omega I_{\Omega,i}\,.
\end{align}
In Ref.~\cite{Fernandez:2005cg}, $I_{\Omega,i}$ is calculated to the lowest order of $\Omega^2$: this approximation is valid as long as $R\Omega \ll c$.
In practice, expanding the metric by $\Omega^2$, we obtain equations for hydrostatic equilibrium, and as in the case of the TOV equation, we need an input EOS to solve the equations.
In beta equilibrium, thermodynamic quantities other than $P$ and $\rho$ are determined by equilibrium condition (see, e.g., the determination of proton fraction, Eq.~\eqref{eq:eq-yp}): thus such quantities are constant on a surface of constant pressure.
Without beta equilibrium, however, these quantities can vary on this surface.
Here, to avoid such complexity, we further assume that on a surface of constant pressure, the other thermodynamic quantities are also constant~\cite{Fernandez:2005cg}.%
\footnote{
In Ref.~\cite{Fernandez:2005cg}, it is shown that this assumption holds in a uniformly-rotating, perfect-fluid NS in hydrostatic equilibrium.
}
In the following numerical analysis, we use the numerical results of $I_{\Omega,i}$ in Ref.~\cite{Fernandez:2005cg}. 

Substituting Eqs.~\eqref{eq:ndot} and \eqref{eq:neq-dot} into Eq.~\eqref{eq:eta-dn}, we obtain the evolution of the chemical imbalance:
\begin{align}
  \frac{d\eta_e^\infty}{dt}
  &=
    -\sum_{N=n,p}\int dV \,(Z_{npe}\Delta\Gamma_{M,Ne} + Z_{np}\Delta\Gamma_{M,N\mu})\, e^{\Phi(r)} +2W_{npe}\Omega\dot\Omega\,,
    \label{eq:roto-diff-eta-e}\\
  \frac{d\eta_\mu^\infty}{dt}
  &=
    -\sum_{N=n,p}\int dV \,(Z_{np}\Delta\Gamma_{M,Ne} + Z_{np\mu}\Delta\Gamma_{M,N\mu})\,e^{\Phi(r)} +2W_{np\mu}\Omega\dot\Omega\,,
    \label{eq:roto-diff-eta-mu}
\end{align}
where
\begin{align}
  \label{eq:wnpell}
  W_{npe} = Z_{npe}I_{\Omega,e} + Z_{np}I_{\Omega,\mu}\,,\quad
  W_{np\mu} = Z_{np}I_{\Omega,e} + Z_{np\mu}I_{\Omega,\mu}\,.
\end{align}
The first terms correspond to the equilibration by the modified Urca process, and the second term to the effect of spin-down which perturbs the system away from the equilibrium.

Finally, we need to specify the evolution of $\Omega(t)$.
In this work, we assume the purely magnetic dipole radiation for the spin-down, i.e., Eq.~\eqref{eq:spin-down-n3} with the braking index $n=3$:
\begin{align}
  \tag{\ref{eq:omega-sol-n3}}
  \Omega(t)
  =
  \frac{2\pi}{\sqrt{P_0^2 + 2P\dot{P}t}}\,.
\end{align}
We again note that $P$ and $\dot{P}$ are the present values, and $P\dot{P}$ is constant.

\subsection{Heating rate}
\label{sec:heating-rate}

In this subsection, we discuss $L_H^\infty$ and $\Delta\Gamma_{M,N\ell}$ of the modified Urca process.
The neutrino emissivity is similar to Eq.~\eqref{eq:murca-emis}.
Here we need to relax the condition of the beta equilibrium.
The difference is only in the distribution functions, and the resultant expression of the emissivity reads
\begin{align}
  \label{eq:murca-emis-noneq}
  Q_{M,N\ell}
  &=
    \int \biggl[\prod_{j=1}^4 \frac{d^3p_j}{(2\pi)^3} \biggr] \frac{d^3p_\ell}{(2\pi)^3}\frac{d^3p_\nu}{(2\pi)^3}\,
    (2\pi)^4\delta^4(P_f - P_i) \cdot \epsilon_\nu \cdot\frac{1}{2} \sum_{\mathrm{spin}}|\mathcal M_{M,N\ell}|^2
    \notag\\
  &\times \left[f_1f_2(1-f_3)(1-f_4)(1-f_\ell) + (1-f_1)(1-f_2)f_3f_4f_\ell \right]\,,
\end{align}
where again $j=1,2,3,4$ denote the nucleons $n,N_1,p, N_2$, respectively, $\delta^4(P_f - P_i)$ the energy-momentum conserving delta function, $1/2 \times\sum_{\mathrm{spin}}|\mathcal M_{M,N\ell}|^2$ the matrix element summed over all the particles' spins with the symmetry factor, and $f$'s the Fermi-Dirac distribution functions.
Similarly the reaction rate is
\begin{align}
  \label{eq:murca-rate-noneq}
  \Delta\Gamma_{M,N\ell}
  &=
    \int \biggl[\prod_{j=1}^4 \frac{d^3p_j}{(2\pi)^3} \biggr] \frac{d^3p_\ell}{(2\pi)^3}\frac{d^3p_\nu}{(2\pi)^3}\,
    (2\pi)^4\delta^4(P_f - P_i) \cdot\frac{1}{2} \sum_{\mathrm{spin}}|\mathcal M_{M,N\ell}|^2
    \notag\\
  &\times \left[f_1f_2(1-f_3)(1-f_4)(1-f_\ell) - (1-f_1)(1-f_2)f_3f_4f_\ell \right]\,,
\end{align}
Following Refs.~\cite{Petrovich:2009yh, Gonzalez-Jimenez:2014iia}, we factorize the emissivity and reaction rate as
\begin{equation}
  \label{eq:q-gamma-i-factor}
  Q_{M,N\ell} = Q_{M,N\ell}^{(0)} \, I^N_{M,\epsilon}\,,\quad
  \Delta\Gamma_{M, N\ell} = \frac{Q_{M,N\ell}^{(0)}}{T(r)}\, I^{N}_{M,\Gamma}\,,
\end{equation}
where $Q_{M,N\ell}^{(0)}$ is the equilibrium emissivity without superfluidity given by Eqs.~\eqref{eq:murca-emis-nbr} and~\eqref{eq:murca-emis-pbr}.
The phase space factor for the emissivity, $I^N_{M,\epsilon}$, is equivalent to the reduction factor $R$ when the beta equilibrium is maintained.

Now let us give more concrete expressions for the phase space factors $I^N_{M,\epsilon}$ and $I^N_{M,\Gamma}$.
They are analogous to the superfluid reduction factor and written as\footnote{
We note that $I^N_{M,\Gamma}$ in Eq.~\eqref{eq:i-integ-gamma} has the opposite sign to the corresponding phase space integrals given in Refs.~\cite{Petrovich:2009yh} and \cite{Gonzalez-Jimenez:2014iia}, while it is consistent with those in Ref.~\cite{Villain:2005ns}.
}
\begin{align}
  I^N_{M,\epsilon}
  &=
    \frac{60480}{11513\pi^8}
    \frac{1}{A_0^N}
    \int \prod_{j=1}^5\frac{d\Omega_j}{4\pi}
    \int_0^\infty dx_\nu \int_{-\infty}^{\infty}dx_ndx_pdx_{N_1}dx_{N_2} \,
    x_\nu^3 \cdot 
    f(z_n)f(z_p)f(z_{N_1})f(z_{N_2})\notag\\
  &\times\left[f(x_\nu - \xi_\ell -z_n-z_p-z_{N_1}-z_{N_2})
    + f(x_\nu + \xi_\ell -z_n-z_p-z_{N_1}-z_{N_2})\right]
    \delta^3\biggl(\sum_{j=1}^5\bm p_j\biggr)
    \,,
    \label{eq:i-integ-emis}
    \\
    I^N_{M,\Gamma}
  &=
    \frac{60480}{11513\pi^8}
    \frac{1}{A_0^N}
    \int \prod_{j=1}^5\frac{d\Omega_j}{4\pi}
     \int_0^\infty dx_\nu \int_{-\infty}^{\infty}dx_ndx_pdx_{N_1}dx_{N_2} \,
     x_\nu^2 \cdot 
    f(z_n)f(z_p)f(z_{N_1})f(z_{N_2})\notag\\
  &\times\left[f(x_\nu - \xi_\ell -z_n-z_p-z_{N_1}-z_{N_2})
    - f(x_\nu + \xi_\ell -z_n-z_p-z_{N_1}-z_{N_2})\right]
    \delta^3\biggl(\sum_{j=1}^5\bm p_j\biggr)
    \,,
    \label{eq:i-integ-gamma}
\end{align}
where we have defined 
\begin{equation}
    x_N \equiv \frac{\epsilon_N - \mu_N}{T}~, \quad 
    x_\nu \equiv \frac{\epsilon_\nu}{T}~, \quad 
    y_N \equiv \frac{\delta_N}{T} ~, \quad 
    z_N \equiv \mathrm{sign}(x_N)\sqrt{x_N^2 + y_N^2} ~, \quad 
    \xi_\ell \equiv \frac{\eta_\ell}{T} ~,
\end{equation}
and $f(x) = 1/(e^x + 1)$.
The factor $A_0^N$ is given by Eq.~\eqref{eq:a0n}.
As can be seen from the definition in Eq.~\eqref{eq:q-gamma-i-factor}, we have $I^N_{M,\epsilon}=1$ and $I^N_{M,\Gamma}=0$ in the limit of $\delta_n=\delta_p=\eta_\ell = 0$, i.e., for a non-superfluid NS in beta equilibrium.
The angular integral in the above equations is non-trivial in the presence of a neutron triplet pairing since its gap amplitude depends on the direction of $\bm p_{F,n}$ (see Sec.~\ref{sec:nucleon-pairing}); otherwise it is just reduced to the factor $A_0^N$. 

For superfluid matter, the phase space factors provide a threshold for the modified Urca process: $\eta_\ell \gtrsim 3\Delta_n + \Delta_p$ for the neutron branch and $\eta_\ell \gtrsim \Delta_n + 3\Delta_p$ for the proton branch~\cite{Reisenegger:1996ir}.
Hence, $\Delta_{\mathrm{th}} = \mathrm{min}\{3\Delta_n + \Delta_p, \Delta_n + 3\Delta_p\}$ is the threshold of the rotochemical heating---for $\eta_\ell \lesssim \Delta_{\mathrm{th}}$, heating does not occur because the modified Urca reaction is suppressed ($\Delta\Gamma_{M,N\ell}\simeq0$).%
\footnote{It is seen by taking $T\to0$ in the distribution functions. See also Eqs.~\eqref{eq:itildenmep} and~\eqref{eq:itildenmgam} in App.~\ref{sec:phase-space}.}
For a very young NS, the Urca reaction is fast enough so that the chemical equilibrium is maintained, i.e., $\eta_\ell = 0$. Later, the NS departs from beta equilibrium due to the spin-down and $\eta_\ell$ monotonically increases until it exceeds $\Delta_{\mathrm{th}}$, after which the accumulated $\eta_\ell$ is converted to heat.\footnote{We however note that if $\Delta_{\mathrm{th}}$ is large and/or the increase rate of $\eta_\ell$ is small, $\eta_\ell$ may never exceed the threshold and thus rotochemical heating is always ineffective.
}
Therefore, the rotochemical heating is efficient usually at late times, when $T \ll \Delta_N$ and $T \ll \eta_\ell$. In such a situation, we can safely exploit the zero temperature approximation in the calculation of the phase space factors \cite{Petrovich:2009yh}.

A numerical calculation in Ref.~\cite{Fernandez:2005cg} shows that the late-time heating indeed occurs in a non-superfluid NS.
In this case, we have analytical expressions for the phase space factors~\cite{Reisenegger:1994be}:
\begin{align}
  I^N_{M,\varepsilon} = F_M(\xi)
  &=
    1 + \frac{22020\xi^2}{11513\pi^2} + \frac{5670\xi^4}{11513\pi^4}
    + \frac{420\xi^6}{11513\pi^6} + \frac{9\xi^8}{11513\pi^8}\,,
    \label{eq:i-epsilon-normal}\\
  I^N_{M,\Gamma} = H_M(\xi)
  &=
    \frac{14680\xi}{11513\pi^2} + \frac{7560\xi^3}{11513\pi^4}
 + \frac{840\xi^5}{11513\pi^6} + \frac{24\xi^7}{11513\pi^8}\,.
    \label{eq:i-gamma-normal}
\end{align}
From these equations, we can see the effects of the non-equilibrium ($\eta_\ell\neq0$); $I^N_{M,\varepsilon} > 1$ enhances the neutrino emission, while $I^N_{M,\Gamma} \neq 0$ generates non-zero $L_H^\infty$.
Such analytical expressions for the neutron (proton) branch can also be obtained for the case where only protons (neutrons) form a constant paring gap in the limit of $T\to 0$~\cite{Petrovich:2009yh}. As for the numerical evaluation of the phase space factors, Refs.~\cite{Villain:2005ns, Pi:2009eq} give the results for the case in which either proton or neutron has a non-zero gap. Reference~\cite{Petrovich:2009yh} also performs the numerical computation of the phase space factors in the presence of the neutron and proton singlet uniform pairings.
In Ref.~\cite{Gonzalez-Jimenez:2014iia}, neutron triplet pairings whose gap has density and temperature dependence are considered, but the effect of proton superfluidity is neglected.
In this dissertation, we include the effect of both the singlet proton and triplet neutron pairing gaps with taking account of their density and temperature dependence. For the calculation of the phase space factors, we use the zero temperature approximation as in Ref.~\cite{Petrovich:2009yh}; see App.~\ref{sec:phase-space} for more details.

\section{Observations of neutron star temperatures}
\label{sec:observations-old-hot}

\begin{table}
  \centering
  \caption{Observational data used in this work. $t_{\text{sd}}$, $t_{\text{kin}}$, 
  $T_s^\infty$, and $P$ denote the spin-down age, kinematic age, 
  effective surface temperature, and period of neutron stars, respectively. 
  The sixth column shows the atmosphere model used in the estimation of 
  the surface temperature, where H, BB, C, and PL indicate hydrogen,  
  blackbody, carbon, and power-law, respectively, while M represents a magnetized NS hydrogen atmosphere model, such as NSA \cite{1995ASIC..450...71P} and NSMAX \cite{Ho:2008bq}. Data are taken from ATNF Pulsar
  Catalogue \cite{Manchester:2004bp, atnf} unless other references are shown explicitly.
Several middle-aged pulsars overlap with those in Tab.~\ref{tab:psr-temp-1}.}
  \vspace{3mm}
  {
    \begin{adjustbox}{width=\textwidth}
  \begin{tabular}{lccccc}\toprule
    Name & $\log_{10}t_{\mathrm{sd}}$& $\log_{10}t_{\mathrm{kin}}$ & $\log_{10}T_s^\infty$ &$P$   &Atmos. \\ 
    &[yr]&[yr] &[K] &[s] &model \\\midrule
    PSR J2124-3358  & $10.0^{+0.2}_{-0.1}$ \cite{Rangelov:2016syg}&&$4.7-5.3$ \cite{Rangelov:2016syg} &$4.9\times 10^{-3}$& BB+PL \\
    PSR J0437-4715  & $9.83^{+0.01}_{-0.02}$ \cite{Durant:2011je}&& $5.1-5.5$ \cite{Durant:2011je}&$5.8\times 10^{-3}$& BB \\
 \hline
    PSR J2144-3933  &$8.5$&&$<4.6$ \cite{Guillot:2019ugf}&$8.5$& BB\\
    PSR J0108-1431  &$8.3$ \cite{2009ApJ...701.1243D}&&$5.0-5.7$ \cite{Mignani:2008jr}&$0.81$& BB  \\
    PSR B0950+08  &$7.2$&&$5.0-5.5$ \cite{Pavlov:2017eeu}&$0.25$& BB+PL  \\
    RX J2143.0+0654  &$6.6$&&$6.082 (2)$ \cite{Kaplan:2009au} &$9.4$&BB\\
    RX J0806.4-4123 &$6.5$&&$6.005(5)$ \cite{Kaplan:2009ce} &$11.4$& BB \\
    PSR B1929+10 &$6.5$&&$<5.7$ \cite{Becker:2005dk} &$0.23$& BB+PL \\
    RX J0420.0-5022 &$6.3$&&$5.742(3)$ \cite{Kaplan:2011xd}&$3.5$& BB\\
    PSR J2043+2740  &$6.1$&&$5.64(8)$ \cite{Beloin:2016zop}&$0.096$&H\\
    RX J1605.3+3249 &$4.5$ \cite{Pires:2014qza}&$5.6-6.6$ \cite{Tetzlaff:2012rz}&$5.88(1)$ \cite{Pires:2019qsk}&$3.4$ \cite{Pires:2014qza}& BB \\
    RX J0720.4-3125 &$6.3$& $5.8-6.0$ \cite{Tetzlaff:2011kh} &$5.5-5.7$ \cite{Kaplan:2003hj}&$8.4$&BB\\
    RX J1308.6+2127 &$6.2$& $5.5-6.2$~\cite{Motch:2009nq}&$6.07(1)$ \cite{Schwope:2006ra}&$10.3$& BB \\
    PSR B1055-52 &$5.7$&&$5.88 (8)$ \cite{Beloin:2016zop}&$0.20$&BB\\
    PSR J0357+3205 &$5.7$&&$5.62^{+0.09}_{-0.08}$ \cite{Kirichenko:2014ona}&$0.44$&M+PL\\
    RX J1856.5-3754 &$6.6$&$5.66(5)$ \cite{Tetzlaff:2011kh} &$5.6-5.7$ \cite{Sartore:2012fk}&$7.1$&BB \\
    PSR J1741-2054 &$5.6$&&$5.85^{+0.03}_{-0.02}$ \cite{Auchettl:2015wca}&$0.41$&BB+PL\\
    PSR J0633+1748 &$5.5$&&$5.71(1)$ \cite{Mori:2014gaa}&$0.24$&BB+PL\\
    PSR J1740+1000 &$5.1$&&$6.04(1)$ \cite{2012Sci...337..946K}&$0.15$&BB\\
    PSR B0656+14  &$5.0$&&$5.81(1)$ \cite{DeLuca:2004ck}&$0.38$&BB+PL\\
    PSR B2334+61 &$4.6$&&$5.5-5.9$ \cite{McGowan:2005kt}&$0.50$&M\\
    PSR J0538+2817 &$5.8$& $4.3-4.8$ \cite{Ng:2006vh}&  $6.02(2)$ \cite{Ng:2006vh}&$0.14$&H\\
    XMMU J1732-344 &&$4.0-4.6$ \cite{Tian:2008tr, Klochkov:2014ola} &$6.25(1)$ \cite{Klochkov:2014ola}&&C\\
    PSR B1706-44   &$4.2$ & & $5.8^{+0.13}_{-0.13}$ \cite{McGowan:2003sy}&$0.10$&M+PL\\
    \bottomrule
  \end{tabular}
\end{adjustbox}
  }
  \label{tab:psr-temp}
\end{table}

Before going to our numerical study, we summarize the current status of the NS temperature observations (see Sec.~\ref{sec:minimal-cool-vs-obs} for how to determine the age and temperature).
We focus on isolated NSs with the age $t>10^4\unit{yr}$, for which we can safely assume that the thermal and diffusion relaxation in the NS core has already been completed. 
In Tab.~\ref{tab:psr-temp}, we list the NSs whose surface temperature is measured, together with two NSs (PSR J2144-3933 and PSR B1929+10) for which only the upper bound on the surface temperature is obtained.%
\footnote{Several middle-aged pulsars overlap with those in Tab.~\ref{tab:psr-temp-1}.}
For most of the NSs in the table, only the spin-down age $t_{\mathrm{sd}}=P/(2\dot P)$ can be used for the estimation of their age, while in some cases, the kinematic age $t_{\mathrm{kin}}$, which is derived from the motion of the supernova remnant, is also available. 
Here we use the kinematic age if available, and the spin-down age otherwise.%
\footnote{We note that once we adopt the pure dipole radiation model, and fix the initial period and the value of $P \dot{P}$ (hence the dipole magnetic field), the NS age is unambiguiously determined from the observed value of the NS period through Eq.~\eqref{eq:omega-sol-n3}. In particular, the spin-down age should agree to the real pulsar age without uncertainty if $P_0 \ll P$. Thus in this chapter we do not consider the uncertainty of the spin-down age.}
The values of $t_{\mathrm{sd}}$ and $P$ are taken from ATNF Pulsar Catalogue \cite{Manchester:2004bp, atnf} unless other references are shown explicitly.

The first two pulsars in Tab.~\ref{tab:psr-temp}, PSR J2124-3358 and J0437-4715, are classified into the MSPs.
They have small $P$ and $\dot P$, and hence a small dipole magnetic field, compared to the ordinary (classical) pulsars.
PSR J0437-4715 is the closest millisecond pulsar at present.
Its rotational period, mass, and distance 
are estimated to be $5.76$~ms, 
$1.44 \pm 0.07~M_{\odot}$, and $d= 156.79 \pm 0.25$~pc, 
respectively \cite{Reardon:2015kba}.
This pulsar is in a binary system accompanied with a white 
dwarf. The spin-down age of PSR J0437-4715 is estimated in 
Ref.~\cite{Durant:2011je} to be $t_{\text{sd}} = (6.7\pm 0.2)
\times 10^9$~years with the Shklovskii correction 
\cite{1970SvA....13..562S} included. This is in a 
good agreement with
the estimated age of the white dwarf, $t_{\text{WD}} = 
(6.0\pm 0.5) \times 10^9$~years. 
In Ref.~\cite{Durant:2011je}, it is found 
that a fit with the Rayleigh-Jeans law in the far UV range 
is consistent with a blackbody emission from the whole
NS surface with a temperature of 
$T_s^\infty = (1.25-3.5) \times 10^5$~K. As argued in 
Ref.~\cite{Durant:2011je}, it is unlikely for the observed
surface temperature to be due to heat flow coming from
the magnetosphere regions. 
Since the minimal cooling theory predicts $T_s^\infty \ll 10^3\unit{K}$ for $t\sim 10^{10}\unit{yr}$, the observed surface temperature requires late time heating.
PSR J2124-3358 is an isolated millisecond pulsar with a period of $4.93$~ms \cite{Reardon:2015kba}. Its spin-down age, 
after corrected by the Shklovskii effect, is 
$11^{+6}_{-3}\times 10^9$~years \cite{Rangelov:2016syg} for 
the distance $d = 410^{+90}_{-70}$~pc. Its surface temperature is obtained with a blackbody plus power-law fit to be $(0.5-2.1) \times 10^5$~K \cite{Rangelov:2016syg}, with the radius fixed to be 12~km. This is also well above the cooling theory prediction.

We also have examples of old warm ordinary pulsars: PSR J0108-1431 and B0950+08. PSR J0108-1431 is an old NS with the spin-down age of $2.0 \times 10^8$~years 
\cite{2009ApJ...701.1243D}, where the Shklovskii correction 
is taken into account. The analysis in Ref.~\cite{Mignani:2008jr} with a  Rayleigh-Jeans spectrum fit shows that its surface
temperature is $T_s^\infty = (7-10) \times 10^4 \,
(d_{130}/R_{13})^2$~K, where $d_{130}$ is the distance 
in units of 130~pc and $R_{13}$ is the apparent radius 
in units of 13~km. Within the error of the distance, 
$d = 210^{+ 90}_{-50}$~pc \cite{2012ApJ...755...39V}, the 
maximum (minimum) temperature is estimated as $T_s^\infty 
= 5.3 \times 10^5$~K ($1.1 \times 10^5$~K) for a radius 
of 13~km. PSR B0950+08 has the spin-down age of $1.75 \times 10^{7}$~years. Its surface temperature is obtained with a power-low plus blackbody spectrum fit
in Ref.~\cite{Pavlov:2017eeu} as $(1-3) \times 10^5$~K, with other parameters such as the pulsar
radius varied in a plausible range.

Seven middle aged pulsars, RX J2143.0+0654, J0806.4-4123, J0420.0-5022, J1308.6+2127, J0720.4-3125, J1856.5-3754, and J1605.3+3249 are classified into the X-ray Dim Isolated Neutron Stars (XDINSs), which are also dubbed as the Magnificent Seven. 
They exhibit thermal X-ray emission without any signature of magnetospheric activity, and have a rather long spin period.
See 
Refs.~\cite{2007Ap&SS.308..181H, vanKerkwijk:2006nr, Kaplan:2008qn} for reviews of XDINSs. 
Their surface temperatures are found to be $\sim 10^6\unit{K}$, which are again higher than the prediction of the cooling theory.
The inferred dipole magnetic fields of these NSs are relatively large: $B\sim 10^{13} - 10^{14}\unit{G}$.
As we will discuss in the following sections, in these NSs, a different type of heating mechanism due to the magnetic field decay \cite{Pons:2008fd, Vigano:2013lea} may operate.

Contrary to the above examples, PSR J2144-3933 is an old ``cool'' NS. This is one of the slowest pulsars, having $P=8.51$~s, and its spin-down age is $t_{\mathrm{sd}} = 333\unit{Myr}$ with the Shklovskii correction taken into account. Assuming the blackbody spectrum, the authors in Ref.~\cite{Guillot:2019ugf} obtained an upper limit on the surface temperature of J2144-3933: $T_s^\infty < 4.2\times 10^4\unit{K}$. This is the lowest limit on the surface temperature of NSs for the moment. 

In the next section, we discuss if the minimal cooling setup with the non-equilibrium beta processes is compatible with the observed surface temperatures in Tab.~\ref{tab:psr-temp}.

\section{Results}
\label{sec:role-non-equilibrium}




Now we show the results of our numerical analysis, where we follow the thermal evolution of NSs with the effect of the non-equilibrium beta reactions included. We then compare the results with the observed surface temperatures given in Tab.~\ref{tab:psr-temp}.

\subsection{Physical input}
\label{sec:phys-input}

We perform the numerical study in the framework of the minimal cooling with the non-equilibrium beta process discussed in Sec.~\ref{sec:rotochemical-heating}. The following inputs are common to all of the analyses: 
\begin{itemize}
\item
  APR EOS~\cite{Akmal:1998cf}. 
\item
 Initial condition: $T^\infty = 10^{10}\unit{K}$ and $\eta_e^\infty =\eta_\mu^\infty =0$. 
\item
  Protons and neutrons form singlet ($^1S_0$) and triplet ($^3P_2\,(m_j=0)$) pairings in the core, respectively. 
\item
  The pulsar braking index $n=3$, i.e., $\Omega(t)$ obeys Eq.~\eqref{eq:omega-sol-n3}.
\item
  We use Eq.~\eqref{eq:ts-tb-partial} for the relation between $T^\infty$ and $T_s$.
\end{itemize}
For the superfluid gap models, we use the CCDK and AO models for proton and the ``a'', ``b'' and ``c'' models for neutron, which are shown in Fig.~\ref{fig:gap-3p2}. The values of $Z_{np}$, $Z_{np\ell}$, and $W_{np\ell}$ in Eqs.~\eqref{eq:roto-diff-eta-e} and \eqref{eq:roto-diff-eta-mu} are read from Fig.~3 in Ref.~\cite{Fernandez:2005cg}, which are summarized in Tab.~\ref{tab:const}.
The numerical values of EOS and the solution of TOV equation are taken from \texttt{NSCool}~\cite{NSCool}.

\begin{table}
  \centering
  \begin{tabular}{lccccc}\toprule
    $M$ &$Z_{npe}$&$Z_{np\mu}$&$Z_{np}$&$W_{npe}$&$W_{np\mu}$ \\\relax
    [$M_\odot$] &[$10^{-61}\unit{erg}$]&[$10^{-61}\unit{erg}$]&[$10^{-61}\unit{erg}$]&[$10^{-13}\unit{erg}\unit{s^2}$]&[$10^{-13}\unit{erg}\unit{s^2}$]\\ \midrule
    $1.4$ & $10$ & $12$ & $4$ & $-1.5$ & $-2$ \\
    $1.8$ & $6$ & $7$ & $2$ & $-1.4$ & $-1.8$ \\\bottomrule
  \end{tabular}
  \caption{The values of $Z_{np}$, $Z_{np\ell}$, and $W_{np\ell}$ in Eqs.~\eqref{eq:roto-diff-eta-e} and \eqref{eq:roto-diff-eta-mu}, which are taken from Ref.~\cite{Fernandez:2005cg}.}
  \label{tab:const}
\end{table}

We divide the NSs listed in Tab.~\ref{tab:psr-temp} into two categories: MSPs and the others.
The latter contains ordinary pulsars and XDINSs. We exploit a representative parameter set for each category as follows:

\paragraph{Millisecond pulsars}

MSPs have much smaller $P$ and $\dot P$ than ordinary pulsars.
With MSP J0437-4715 in mind, we use the following parameters for this category: 
\begin{itemize}
\item
  $M = 1.4\,M_\odot$.
\item
  $P=5.8\unit{ms}$.
\item
  $\dot P = 5.7\times10^{-20}$.
\item
  $\Delta M/M = 10^{-7}$.
\end{itemize}
We also note that the values of $P$ and $\dot{P}$ of J2124-3358, $P = 4.9\unit{ms}$ and $\dot{P} = 2.1 \times 10^{-20}$, are fairly close to those of J0437-4715, while its mass is unknown. We have fixed the amount of the light elements in the envelope, $\Delta M/M = 10^{-7}$, as it turns out that the result is almost independent of this choice for old NSs such as J0437-4715 and J2124-3358.

\paragraph{Ordinary pulsars and XDINSs}

For ordinary pulsars and XDINSs, we use
\begin{itemize}
  \item
$M = 1.4\,M_\odot$ or $1.8\,M_\odot$.
\item
  $P=1\unit{s}$.
\item
  $\dot P = 1\times10^{-15}$.
\item
  $\Delta M/M = 10^{-7}$ or $10^{-15}$.
\end{itemize}
Note that $P$ and $\dot{P}$ affect the rotochemical heating only through Eq.~\eqref{eq:omega-sol-n3}, and thus the result depends only on the combination $P\dot P$.
Ordinary pulsars have $P\dot P \sim 10^{-17} - 10^{-13}$, corresponding to $B\sim 10^{11} - 10^{13}\unit{G}$.
The dependence of the thermal evolution on $P\dot P$ is weaker than that on gap models and $P_0$, and thus we fix it to be $P\dot P = 1\times10^{-15} \unit{s}$ in the following analysis.

\vspace{5mm}
Once we fix the NS parameters as above, the time evolution of the NS surface temperature depends only on the nucleon gap models and the initial period $P_0$. As we see in Sec.~\ref{sec:heating-rate}, the heating rate depends on the nucleon pairing gaps via the phase space factors. On the other hand, the initial period $P_0$ affects the time evolution of the NS angular velocity $\Omega(t)$ in Eq.~\eqref{eq:omega-sol-n3}, through which the accumulation rate of the chemical imbalance is modified. We will study these effect in the following subsections.

\subsection{Millisecond pulsars}
\label{sec:millisecond-pulsars}

\begin{figure}[t]
  \centering
  \begin{minipage}{0.5\linewidth}
    \includegraphics[width=1.0\linewidth]{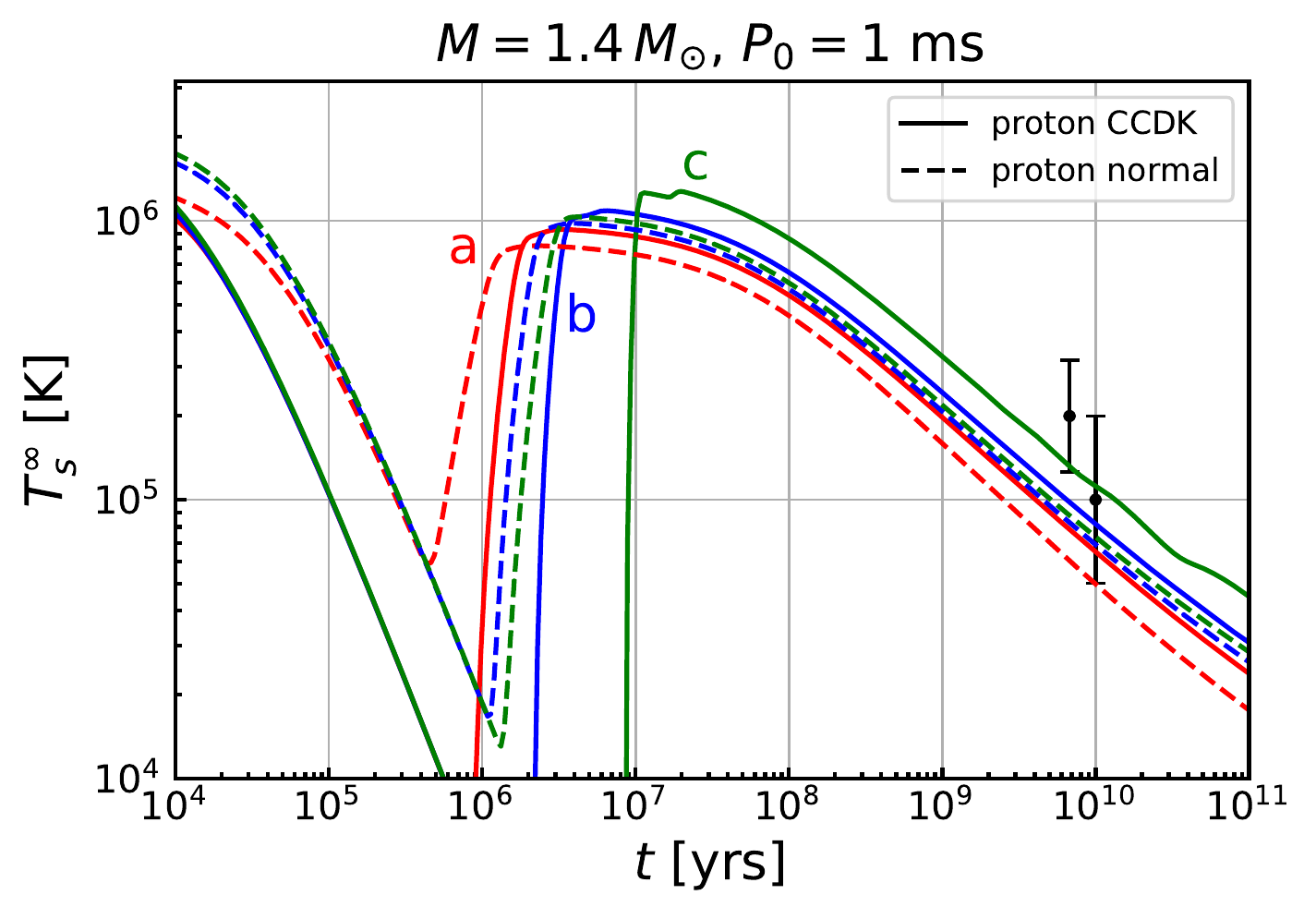}
  \end{minipage}%
  \begin{minipage}{0.5\linewidth}
    \includegraphics[width=1.0\linewidth]{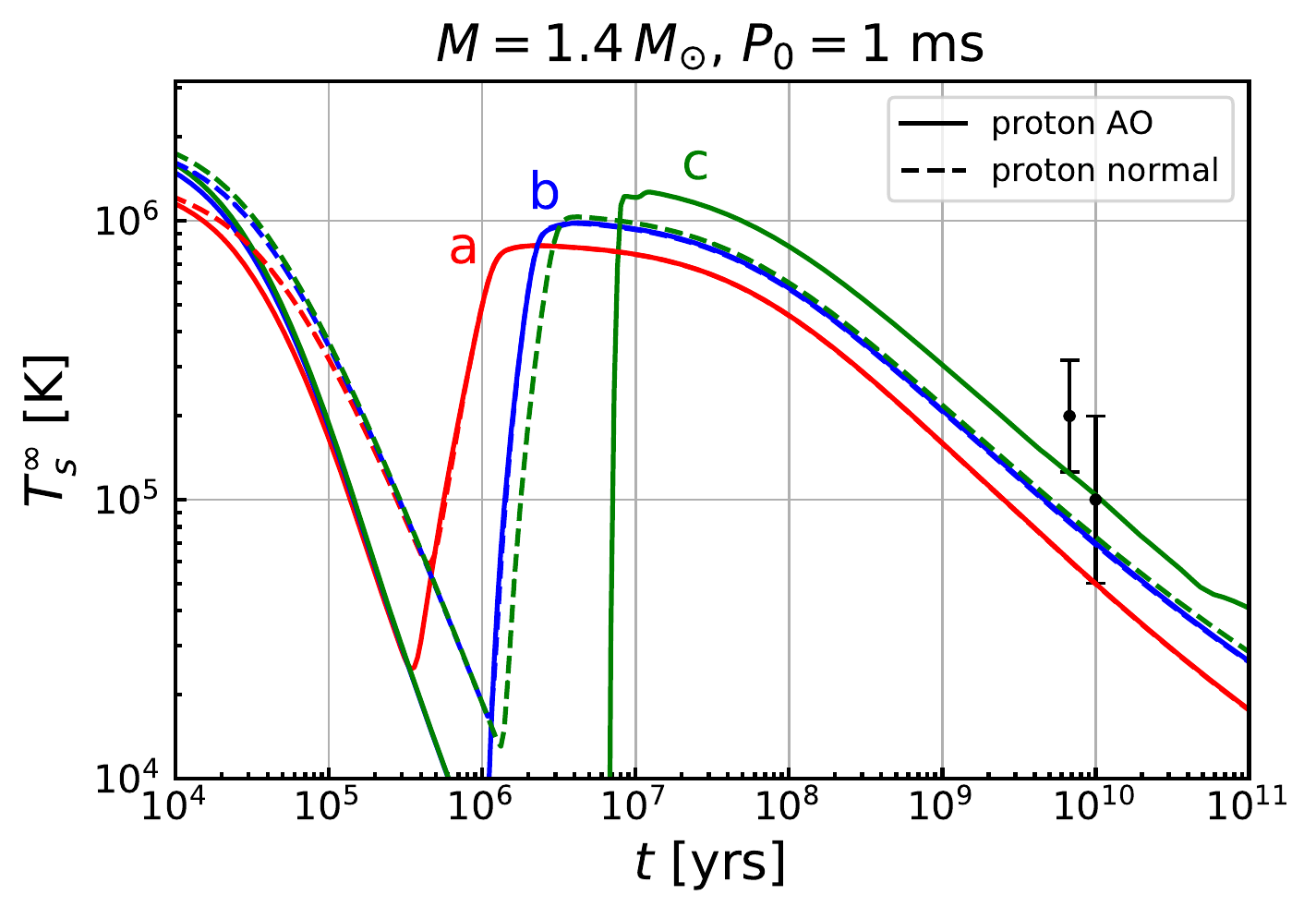}
  \end{minipage}
  \begin{minipage}{0.5\linewidth}
    \includegraphics[width=1.0\linewidth]{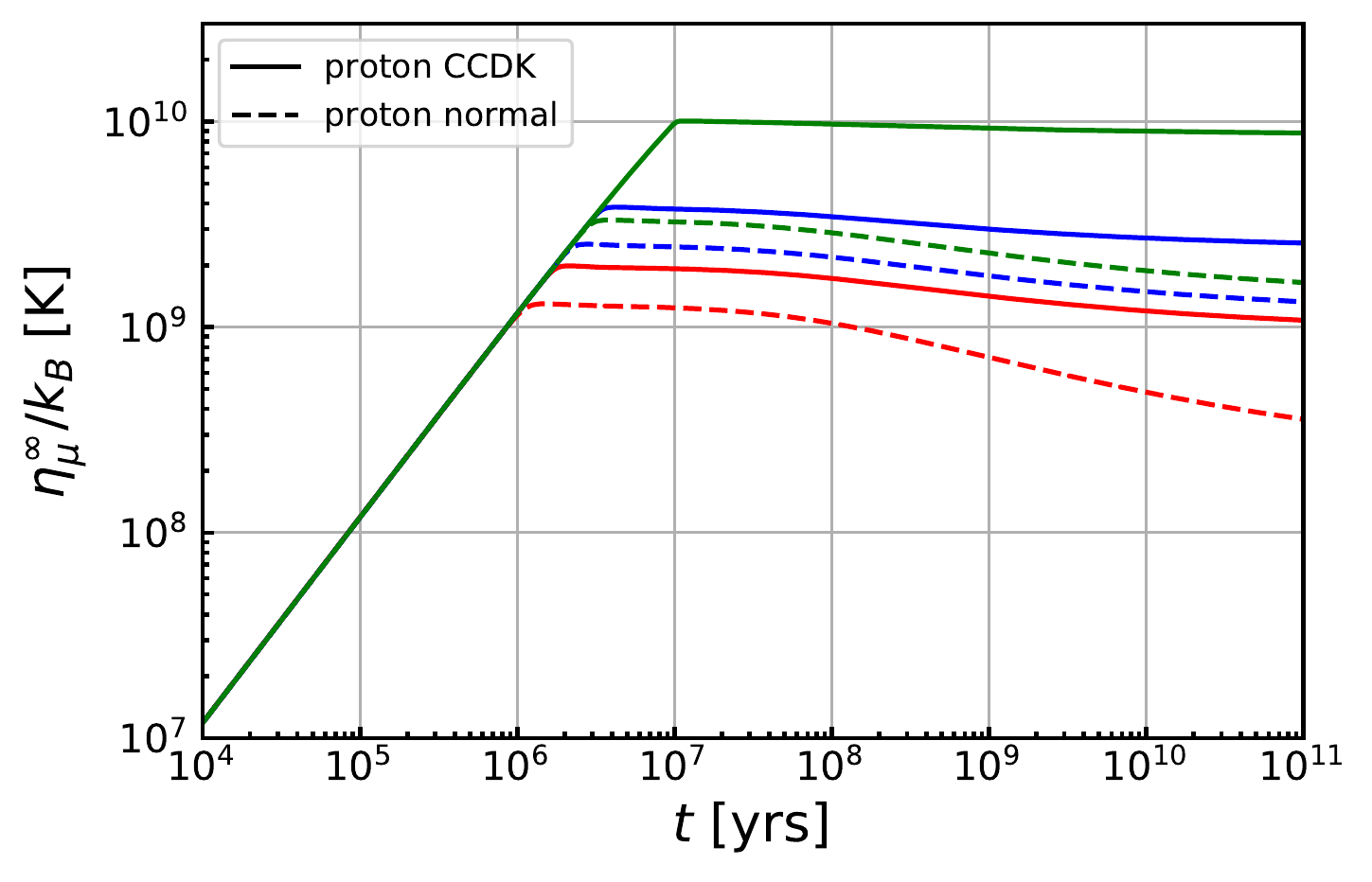}
  \end{minipage}%
  \begin{minipage}{0.5\linewidth}
    \includegraphics[width=1.0\linewidth]{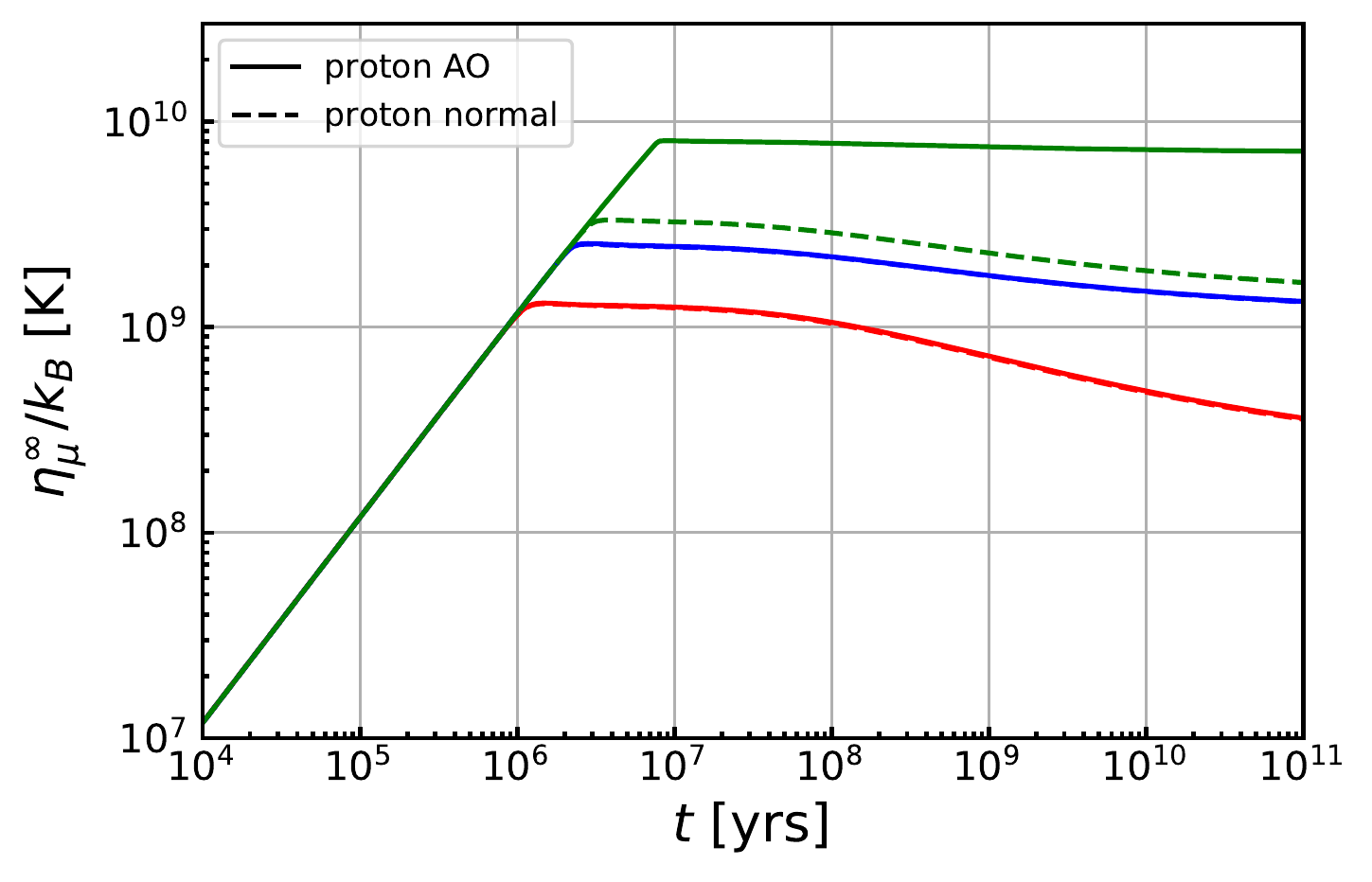}
  \end{minipage}
  \caption{Top two panels show the time evolution of the redshifted surface temperature $T_s^\infty$ for the MSP category with $P_0 = 1\unit{ms}$. We use the CCDK (AO) model for proton pairing in the left (right) panel. The red, blue, and green lines correspond to the ``a'', ``b'', and ``c'' models for neutron pairing, respectively. The solid (dashed) lines are for the case with (without) proton superfluidity. The observed surface temperatures of J0437-4715 and J2124-3358 are also shown by the black points with black solid lines indicating the uncertainty.
  The bottom two panels show the evolution of $\eta_\mu^\infty$ corresponding to each line in the upper panels.}
  \label{fig:msp}
\end{figure}
We first compute the evolution of the redshifted surface temperature $T_s^\infty$ for the MSP category.
The resultant temperature (chemical imbalance) evolution is shown in the top (bottom) panels in Fig.~\ref{fig:msp}, where the initial period is taken to be $P_0=1\unit{ms}$. We use the CCDK (AO) model for proton pairing in the left (right) panel. The red, blue, and green lines correspond to the ``a'', ``b'', and ``c'' models for neutron pairing, respectively. The solid (dashed) lines are for the case with (without) proton superfluidity. The observed surface temperatures of J0437-4715 and J2124-3358 are also shown by the black points with black solid lines indicating the uncertainty.%

As seen in the bottom panels in Fig.~\ref{fig:msp}, the chemical imbalance $\eta_\ell^\infty$ monotonically increases until $t=10^{6-7}\unit{yr}$ because equilibration by the modified Urca process is highly suppressed. 
In this case, the evolution of $\eta_\ell^\infty$ (Eqs.~\eqref{eq:roto-diff-eta-e} and~\eqref{eq:roto-diff-eta-mu}) is solved by
\begin{align}
  \label{eq:eta-sol-early}
  \eta_\ell^\infty
  \simeq
  W_{np\ell}(\Omega(t)^2 - \Omega(0)^2)
  =
  \frac{4\pi^2|W_{np\ell}|}{P_0^2}\cdot\frac{t/t_c}{1+t/t_c}\,,
\end{align}
where we have defined $t_c \equiv P_0^2/2P\dot{P}$.
At $t = 10^{6-7}\unit{yr}$, the imbalance becomes large enough for the modified Urca process to occur efficiently.
As a result, the surface temperatures quickly rise to $T_s^\infty \sim 10^6\unit{K}$ by the rotochemical heating, while the  growth of $\eta_\ell$ stops.
Then the system reaches a quasi-steady state, where the increase in $\eta_\ell^\infty$ due to spin-down is compensated by the consumption via the Urca processes and the heating rate balances with the photon cooling rate. During this stage, $T_s^\infty$ gradually decreases due to the decline of the term $|\Omega\dot\Omega|$.

As we discussed in Sec.~\ref{sec:heating-rate}, in this work we include the effect of both proton and neutron superfluidity simultaneously. We can see the consequence of this simultaneous inclusion by comparing the solid and dashed lines for each case.
Let us first study the case with the proton CCDK for $P_0 = 1\unit{ms}$, i.e., the left panels in Fig.~\ref{fig:msp}.
In this case, for all neutron gap models, the heating effect starts to be visible at a later time in the presence of proton superconductivity.
This is because the additional contribution from the proton pairing gap increases the threshold of rotochemical heating $\Delta_{\mathrm{th}}$, and thus an extra amount of $\eta_\ell^\infty$ needs to be accumulated.
This results in a delay in the onset of rotochemical heating.
Moreover, a larger value of $\Delta_{\mathrm{th}}$ leads to a larger value of $\eta_\ell^\infty$ eventually, as seen in the bottom panel.
This then results in a higher $T_s^\infty$ at late times, since the heating power is proportional to $\eta_\ell$ as in Eq.~\eqref{eq:hating-rate}.
This feature can also be seen for every neutron gap model in the left panel in Fig.~\ref{fig:msp}.

Next, we examine the cases with the AO proton gap model shown in the right panels in Fig.~\ref{fig:msp}. For this proton paring gap, we do not see enhancement in $T_s^\infty$ due to the proton superfluidity for the neutron gaps ``a'' and ``b''. As we see in the left panel in Fig.~\ref{fig:gap-3p2}, the size of the AO proton gap is smaller than the CCDK gap, and even vanishes deep inside the NS core. For this reason, the rotochemical threshold $\Delta_{\mathrm{th}}$ is determined almost solely by the neutron gap, which makes the effect of proton superconductivity invisible. 
For the neutron ``c'' gap, on the other hand, the gap amplitude is very large near the NS center, and hence heating is ineffective there. Instead, the rotochemical heating mainly occurs in the intermediate region where the AO proton gap is sizable, which makes the effect of proton gap manifest. This observation indicates that it is crucial to take account of the density dependence of the nucleon pairing gaps for the evaluation of the rotochemical heating effect.

The results in Fig.~\ref{fig:msp} show that the observed surface temperatures of J0437-4715 and J2124-3358 can be explained by the heating effect of non-equilibrium beta reactions, especially for  moderate/large nucleon gaps. In particular, for the neutron ``c'' gap model, the simultaneous inclusion of proton superconductivity improves the fit considerably such that the predicted thermal evolution is totally consistent with the observed temperatures.

\subsection{Ordinary pulsars and XDINSs}
\label{sec:ordinary-pulsars}

\begin{figure}
  \centering
  \begin{minipage}{0.5\linewidth}
    \includegraphics[width=1.0\linewidth]{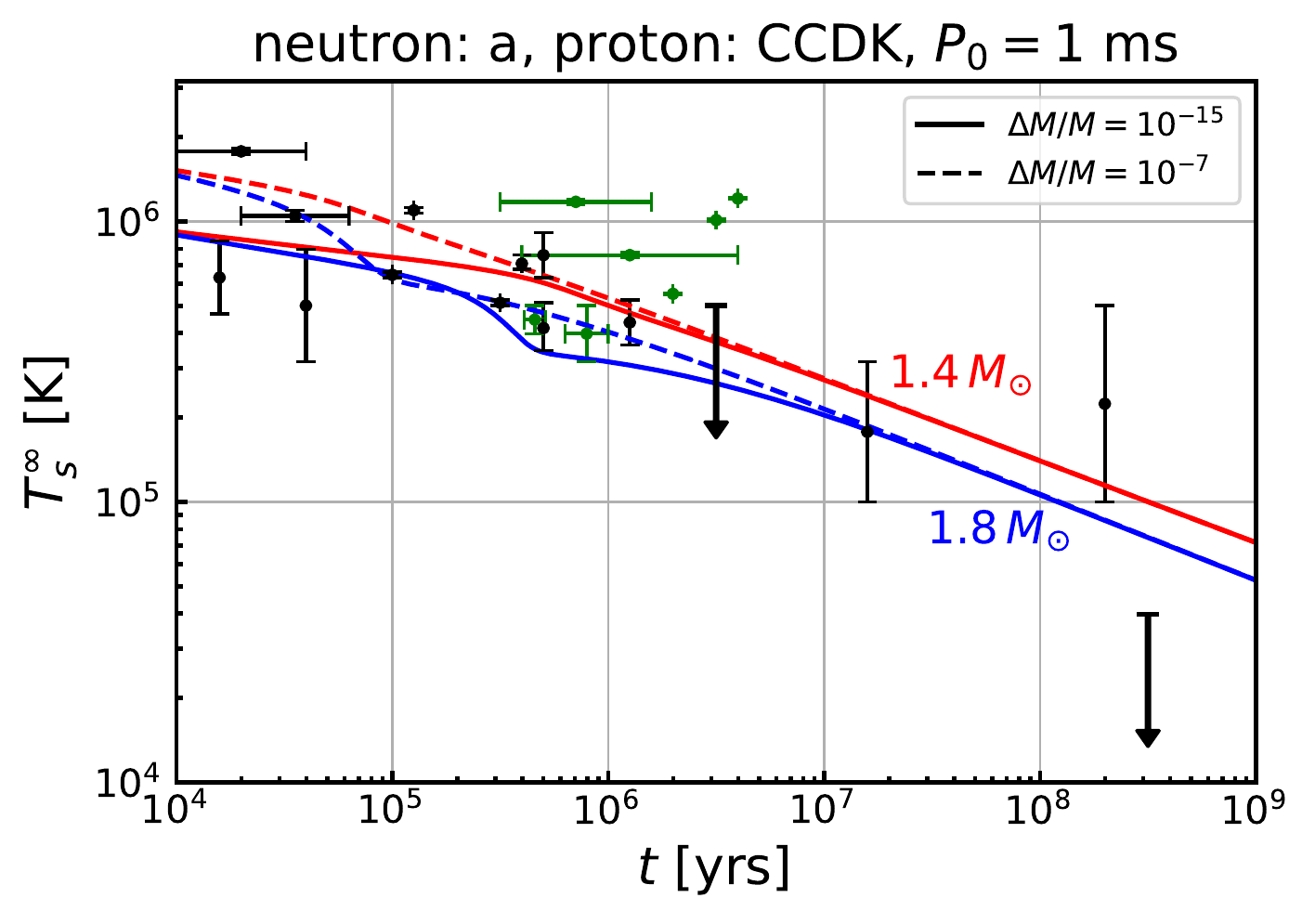}
  \end{minipage}%
  \begin{minipage}{0.5\linewidth}
    \includegraphics[width=1.0\linewidth]{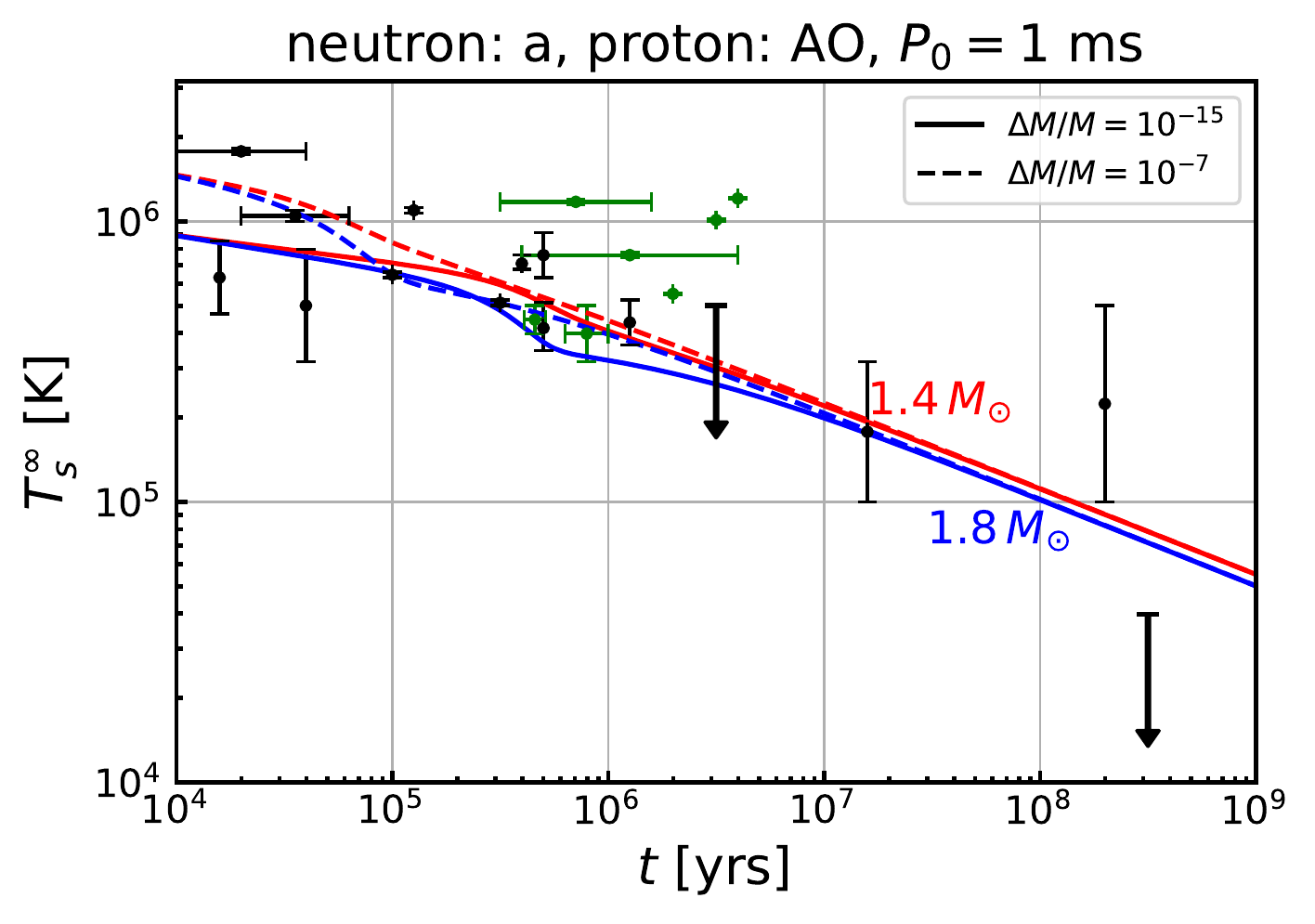}
  \end{minipage}
  \begin{minipage}{0.5\linewidth}
    \includegraphics[width=1.0\linewidth]{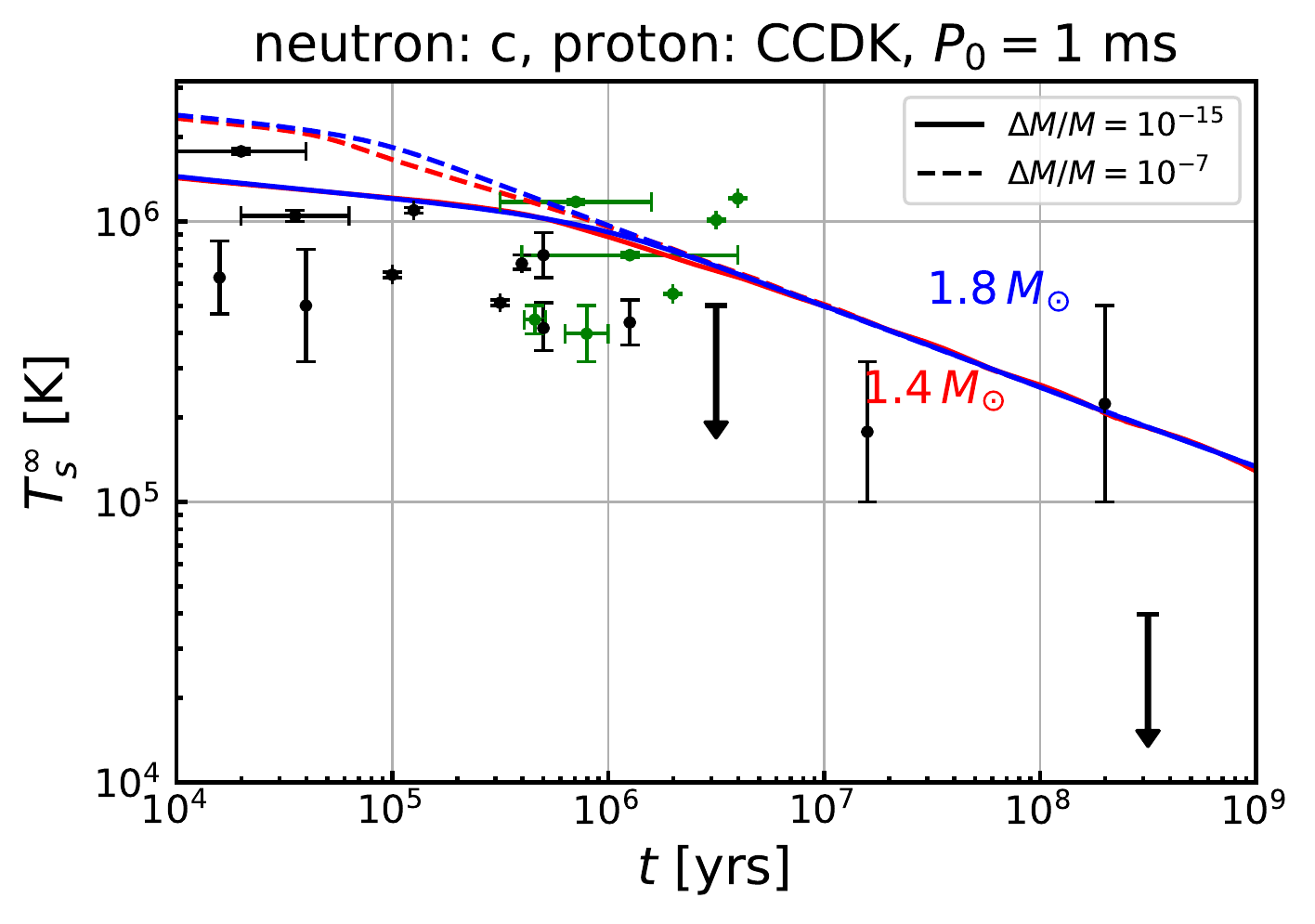}
  \end{minipage}%
  \begin{minipage}{0.5\linewidth}
    \includegraphics[width=1.0\linewidth]{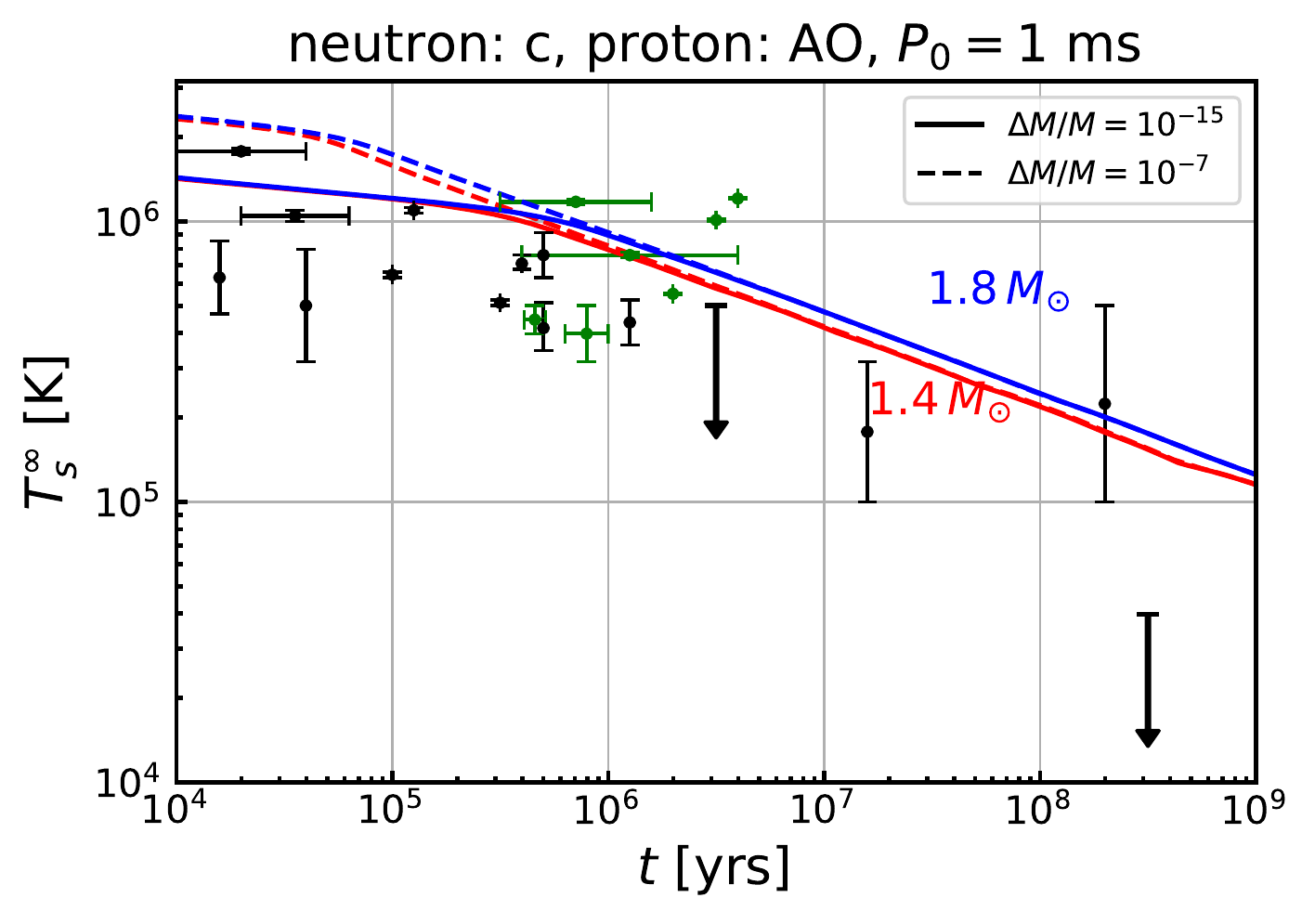}
  \end{minipage}
  \caption{Time evolution of the surface temperature $T_s^\infty$ for ordinary pulsars and XDINSs with $P_0=1\unit{ms}$. We use the neutron gap ``a'' (``c'') in the upper (lower) panels and the CCDK (AO) proton gap in the left (right) panels. The red and blue lines correspond to $M=1.4\,M_\odot$ and $1.8\,M_\odot$, while the solid and dashed lines represent the cases for the heavy ($\Delta M/M = 10^{-15}$) and light ($\Delta M/M = 10^{-7}$) element envelope models, respectively. We also plot the observed surface temperatures of ordinary pulsars and XDINSs with black and green points, respectively, with the horizontal (vertical) lines indicating the uncertainty in the age ($T_s^\infty$).}
  \label{fig:cp-1}
\end{figure}
Next we discuss the second category, which is comprised of ordinary pulsars and XDINSs. In Fig.~\ref{fig:cp-1}, we show the time evolution of the surface temperature $T_s^\infty$ for this category with $P_0=1\unit{ms}$. We use the neutron gap ``a'' (``c'') in the upper (lower) panels and the CCDK (AO) proton gap in the left (right) panels. The red and blue lines correspond to $M=1.4\,M_\odot$ and $1.8\,M_\odot$, while the solid and dashed lines represent the cases for the heavy ($\Delta M/M = 10^{-15}$) and light ($\Delta M/M = 10^{-7}$) element envelope models, respectively. It is found that another choice of the envelope parameter $\Delta M/M$ just falls in between these two cases.
We also plot the observed surface temperatures of ordinary pulsars and XDINSs with black and green points, respectively, with the horizontal (vertical) lines indicating the uncertainty in the kinematic age ($T_s^\infty$). The bars with down arrows correspond to the upper limits on $T_s^\infty$ of J2144-3933 and B1929+10.

For the cases shown in Fig.~\ref{fig:cp-1}, the rotochemical heating begins earlier than $10^4\unit{yr}$.
The difference from the MSP category is due to the larger $P\dot P$ of the ordinary pulsars and XDINSs. 
The predicted temperatures tend to be higher for the light element envelope than the heavy element one at earlier times, but this difference disappears at later times. In addition, the evolution curves depend on the NS mass. This dependence is, however, rather non-trivial compared with that observed in Ref.~\cite{Gonzalez-Jimenez:2014iia}, where only the neutron triplet paring is taken into account and the temperature always gets higher for a lighter NS mass. This complexity is caused by the non-trivial dependence of the rotochemical threshold on the matter density, as both proton and neutron pairing gaps contribute to the threshold. 

It is found from the top two panels that the small neutron gap ``a'' offers the predictions consistent with most of the surface temperatures of the ordinary pulsars. 
We also find that three hot XDINSs, having relatively high temperatures $T_s^\infty \sim 10^6\unit{K}$ at $t\sim 10^6\unit{yr}$, are located above the predictions.
The bottom two panels, on the other hand, show that the large neutron gap ``c'' gives higher temperatures than the ``a'' model, and can be consistent only with a part of the NS temperatures shown in the figure. The hot XDINSs are still above the prediction, while several young pulsars ($t\sim 10^{4-5}\unit{yr}$ and $T_s^\infty < 10^6\unit{K}$) are below the thermal evolution curves.

\begin{figure}[t]
  \centering
  \begin{minipage}{0.5\linewidth}
    \includegraphics[width=1.0\linewidth]{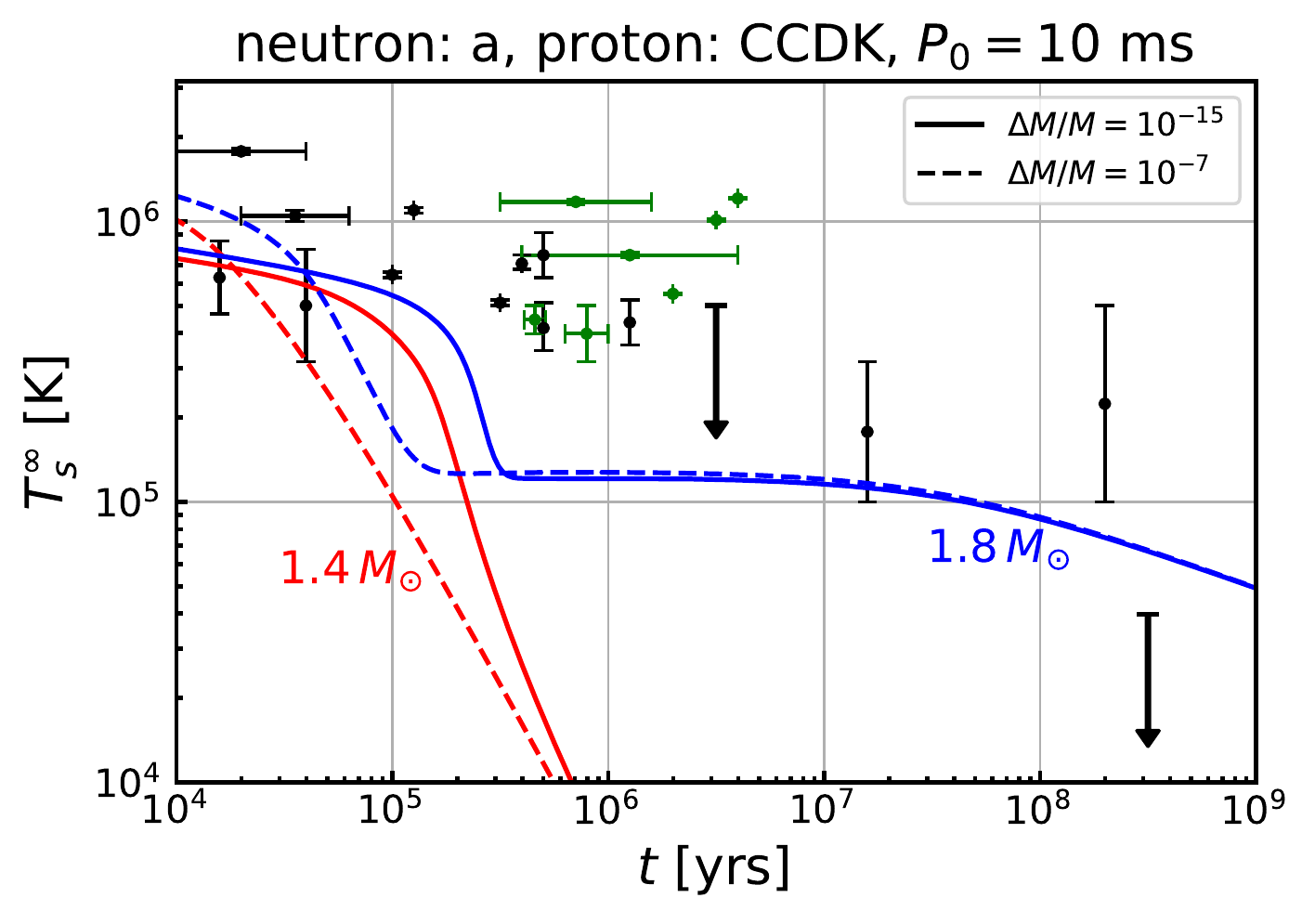}
  \end{minipage}%
  \begin{minipage}{0.5\linewidth}
    \includegraphics[width=1.0\linewidth]{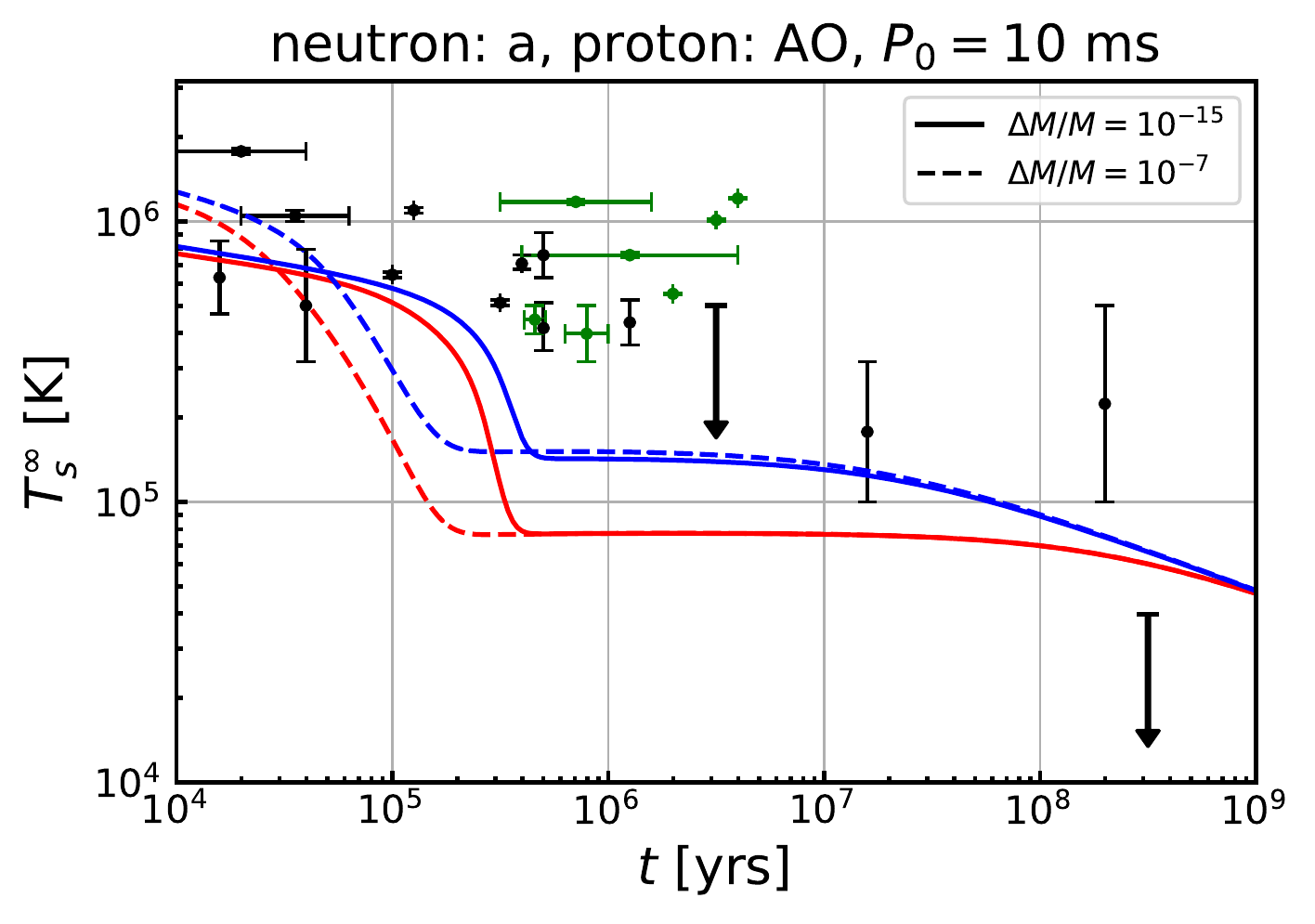}
  \end{minipage}
  \begin{minipage}{0.5\linewidth}
    \includegraphics[width=1.0\linewidth]{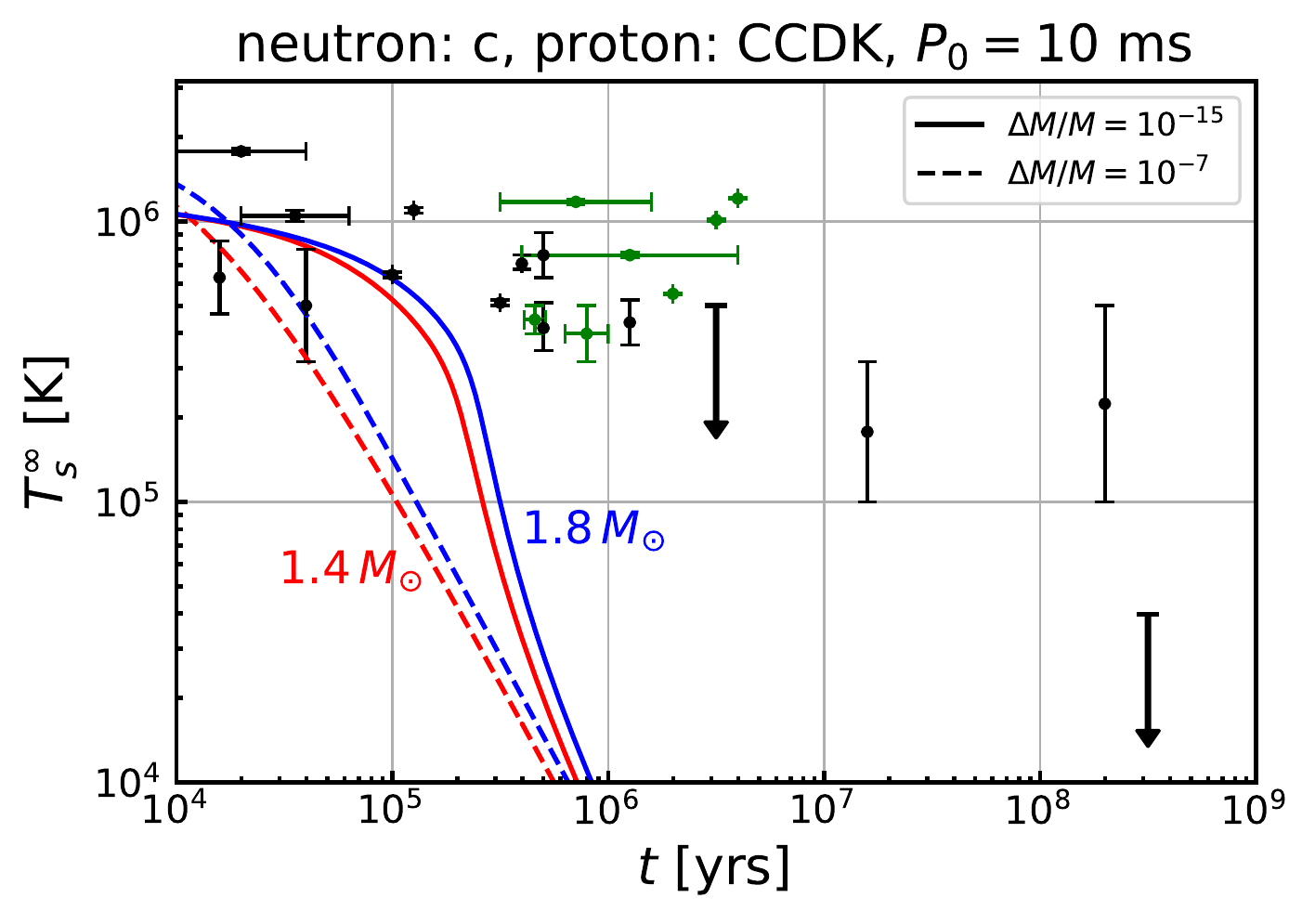}
  \end{minipage}%
  \begin{minipage}{0.5\linewidth}
    \includegraphics[width=1.0\linewidth]{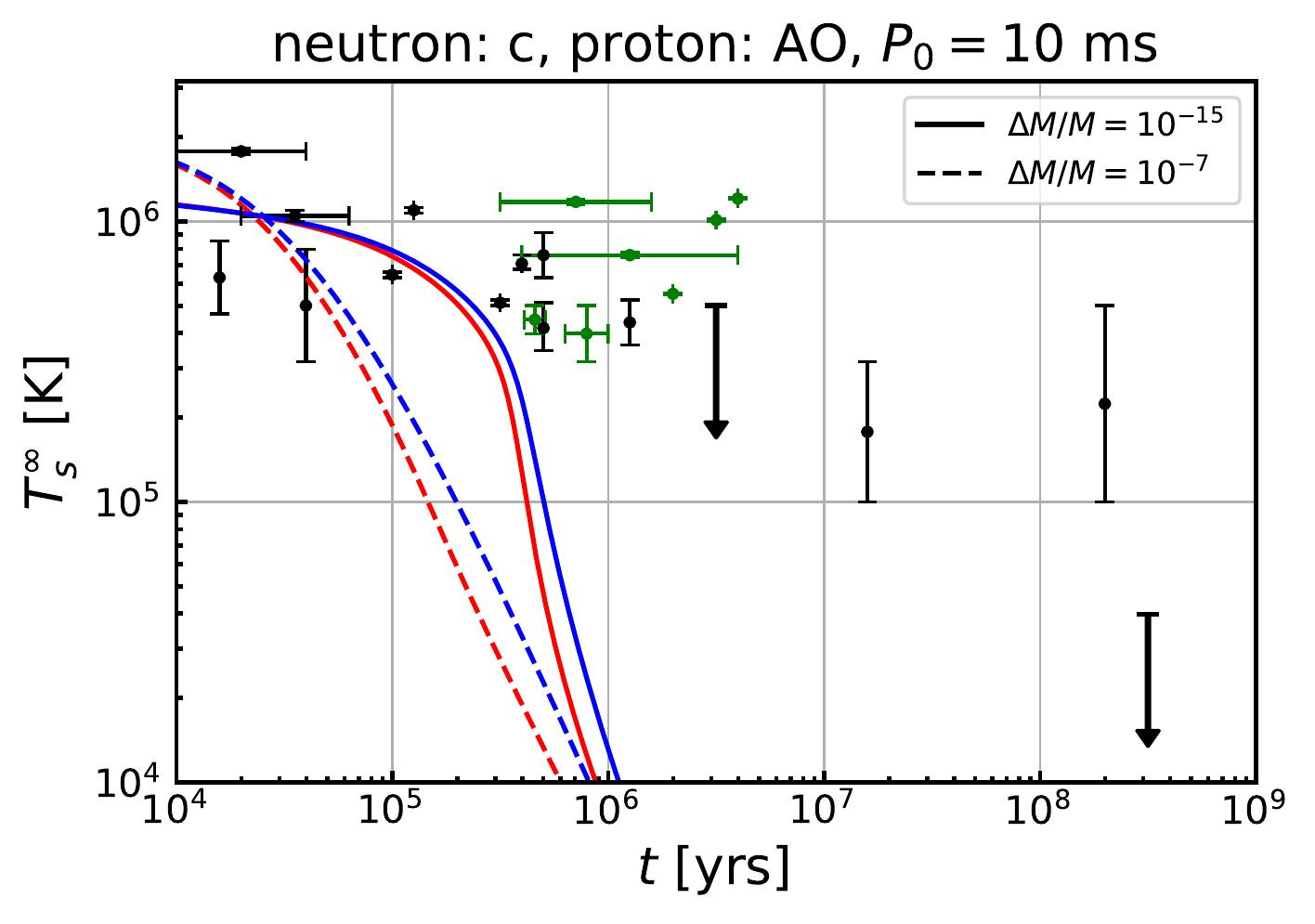}
  \end{minipage}
  \caption{The same as the left panels in Fig.~\ref{fig:cp-1} for $P_0=10\unit{ms}$, where the meaning of the lines is the same as in the figure. }
  \label{fig:cp-2}
\end{figure}
A different choice of $P_0$ may change the evolution of $T_s^\infty$ via the modification of the increase rate in $\eta_\ell^\infty$.
From Eq.~\eqref{eq:eta-sol-early}, the maximum amount of the imbalance is read by taking $t\gg t_c$ as
\begin{align}
  \label{eq:max-eta}
  \eta_\ell^{\infty,\,\mathrm{max}}
  =
  \frac{4\pi^2|W_{np\ell}|}{P_0^2}
  \simeq
  3\times 10^{10}\unit{K}\times \bcm{1\unit{ms}}{P_0}{2}\,,
\end{align}
where we use $|W_{np\ell}| \sim 10^{-13}\unit{erg\,s^2}$.
In general, a larger initial period makes the departure from the beta equilibrium smaller and hence the heating effect milder. To see this, in Fig.~\ref{fig:cp-2}, we show the time evolution of the surface temperature for $P_0=10\unit{ms}$ for the CCDK (left) and AO (right) proton gap models and the ``a'' (top) or ``c'' (bottom) neutron gap models. It is found that this slightly larger initial period strongly suppresses the heating effect. 
This can be seen in the bottom panels, where the heating does not occur for both of the NS masses due to the large neutron gap of ``c''. For the neutron gap ``a'', on the other hand, the internal heating can occur when the nucleon pairing gaps are sufficiently suppressed in the NS center so that even a small amount of the chemical imbalance can trigger the rotochemical heating. The predictions in all of these cases are consistent with the surface temperatures of relatively young pulsars ($t\sim 10^4 - 10^6\unit{yr}$) of $10^5\unit{K}<T_s^\infty < 10^6\unit{K}$. The surface temperature of B0950+08 may be explained for some cases. 
The upper limit on the surface temperature of J2144-3933 can also be satisfied.

We note that, for most of the NS parameter choices, at later times, the intensity of the rotochemical heating has little dependence on $P\dot{P}$ and thus on its dipole magnetic field, as discussed in detail in App.~\ref{chap:magnetic-field}.
This is because for a sufficiently old pulsar, the spin-down effect is very weak and the evolution of $\eta_\ell^\infty$ is solely determined by the equilibration caused by the modified Urca process.  
This observation, therefore, indicates that we cannot explain the low surface temperature of J2144-3933 by changing the magnetic field while keeping $P_0$ small.

In summary, the observed surface temperatures of ordinary pulsars are compatible with the rotochemical heating for both small and large neutron gaps. For a small neutron gap, the predicted curves of the time evolution of $T_s^\infty$ tend to be in good agreement with most of the observations for $P_0 = 1\unit{ms}$. In the case of a large neutron gap, on the other hand, the middle-aged and old pulsars can still be explained with $P_0 = 1\unit{ms}$, while it is required to assume a larger initial period for young pulsars; for an initial period of $1\unit{ms} < P_0 < 10\unit{ms}$, the heating can still occur but predicted temperatures tend to be lower than those for $P_0 = 1\unit{ms}$, with which we can explain all of the data below the curves shown in Fig.~\ref{fig:cp-1}. In particular, the J2144-3933 limit can be satisfied for any nucleon gap models if we take a sufficiently large initial period. It is intriguing to note that the J2144-3933 has one of the longest periods observed so far, as mentioned in Sec.~\ref{sec:observations-old-hot}. This observation may suggest that the initial period of this NS is also relatively long so that the rotochemical heating has never operated and the NS has been cooled down below the limit $T_s^\infty < 4.2\times 10^4\unit{K}$.

On the other hand, the observed temperatures of some of the XDINSs may be higher than the prediction of the rotochemical heating. This discrepancy could be merely due to the errors of their age and temperature; in particular, the temperatures shown in the Tab.~\ref{tab:psr-temp} may suffer from large systematic uncertainty because of our ignorance of their masses, radii, and distances. We also note that the temperature data of the three especially hot XDINS---J2143+0654, RX J0806.4-4123, and RX J1308.6+2127---used in this analysis are obtained using a spectrum fit with one blackbody component plus Gaussian absorption lines. If there is a hot spot in these NSs, however, such a fit tends to give a higher temperature and a smaller NS radius than the actual ones. In fact, the inferred radii of J2143+0654 and RX J0806.4-4123, $R = 3.10(4)\unit{km}$ for $d = 500\unit{pc}$ \cite{Kaplan:2009au} and $R=2.39(15)\unit{km}$ for $d = 250\unit{pc}$ \cite{Kaplan:2009ce}, respectively, are considerably smaller than the typical size of the NS radius $\sim 10-15\unit{km}$. In addition, these NSs exhibit X-ray pulsations \cite{Pires:2014qza}, which indicate that the temperature distribution on their surface is inhomogeneous. It is indeed found that a spectrum analysis based on a two-temperature blackbody model gives a larger radius and a lower surface temperature for these XDINSs \cite{Yoneyama:2018dnh}, with which the discrepancy between the prediction and observation is significantly reduced. 
Another potential explanation for the discrepancy is that the XDINSs have undergone the magnetic field decay, which makes the spin-down age differ from the actual age. 
Moreover, the magnetic field decay itself can be another source of heating~\cite{2012MNRAS.422.2878D, Vigano:2013lea}. If this is the case, the inclusion of only the rotochemical heating effect may be insufficient to explain the temperatures of the XDINSs.

\section{Summary and discussion}
\label{sec:conclusion-roto}

We have studied the non-equilibrium beta process in the minimal cooling scenario, which gives rise to the late time heating in NSs. Extending the previous works, we have included the singlet proton and triplet neutron pairing gaps simultaneously in the calculation of the rate and emissivity of the process, with their density dependence taken into account. We then compare the time evolution of the NS surface temperature predicted in this framework with the latest observations of the NS temperatures, especially with the recent data of the old warm NSs. It is found that the simultaneous inclusion of both proton and neutron gaps is advantageous for the explanation of the old warm NSs, since it increases the threshold of rotochemical heating and thus enhances the heating effect. We find that the observed surface temperatures of warm MSPs, J2124-3358 and J0437-4715, are explained for various choices of nucleon gap models. The same setup can also explain the temperatures of ordinary pulsars by choosing the initial rotational period of each NS accordingly. In particular, with $P_0 = 10\unit{ms}$ or larger, the upper limit on the surface temperature of J2144-3933 can be satisfied.

To explain the observation, we require $P_0\lesssim 10\unit{ms}$ for old warm pulsars and $P_0 \gtrsim 10\unit{ms}$ for old cold one. This assumption is reasonable for hot MSPs since their current periods are also as small as $\Order(1)\unit{ms}$. On the other hand, for ordinary pulsars, several recent studies suggest that they are born with $P_0 = \Order(10-100)\unit{ms}$~\citep{2012Ap&SS.341..457P, 2013MNRAS.430.2281N, Igoshev:2013rqf, FaucherGiguere:2005ny, 2010MNRAS.401.2675P, Gullon:2014dva, Gullon:2015zca, Muller:2018utr}, 
which is apparently in tension with the requirement of $P_0\lesssim 10\unit{ms}$. 
Nevertheless, the initial period of a NS highly depends on the detail of the supernova process where the NS was created. Given that a fully satisfactory simulation for supernova explosion process has not yet been available, we regard the issue of the NS initial period as an open question, expecting future simulations and astrophysical observations to answer this problem.

As we have seen in Sec.~\ref{sec:ordinary-pulsars}, the surface temperatures of the three XDINSs J2143+0654, RX J0806.4-4123, and RX J1308.6+2127 are higher than the prediction of the rotochemical heating. This discrepancy may be due to large magnetic fields of these XDINSs. In fact,
XDINSs commonly have absorption feature in the X-ray spectrum~\cite{Borghese:2015iqa, Borghese:2017vxn},
which is interpreted as the proton cyclotron resonance or atomic transition.
Both require the magnetic field larger than about $10^{13}\unit{G}$, and such a strong magnetic field can affect the thermal evolution of XDINSs.
If this is the case, we need a more involved analysis for these XDINSs including the magnetic field evolution.

Regarding the ambiguity in the choice of pairing gaps, there is a hint from the observation.
As we have seen in Chap.~\ref{chap:limit-axion-decay}, the cooling rate of the Cas A NS requires a large proton gap such as the CCDK model and a small neutron gap with the critical temperature of $T_c \simeq 5\times 10^8\unit{K}$, which is about a factor of 2 smaller than that for the ``a'' model.
The time evolution of $T_s^\infty$ for these gaps is close to that for the proton CCDK + neutron ``a'' model, and thus is consistent with the observations of ordinary pulsars as shown in Fig.~\ref{fig:cp-1}. This setup can also explain the surface temperature of the MSP J2124-3358, while the predicted value of $T_s^\infty$ for J0437-4715 is slightly below the observed one. Although a large neutron gap such as ``c'' model is favored by the observation of these MSPs, it is not consistent with the Cas A NS observation.

Finally, we emphasize that the non-equilibrium beta process discussed in this chapter is not an adhoc assumption but an inevitable consequence of rotating NSs, and therefore the heating mechanism based on this process should always be taken into account. 
In this sense, the minimal cooling plus rotochemical heating is \textit{the minimal} scenario for the NS thermal evolution. Intriguingly, we have found that this ``minimal'' setup is compatible with the observed surface temperatures of NSs for the moment, without relying on exotic physics. Further developments in the evaluation of nucleon pairing gaps, as well as additional data of NS surface temperatures, allow us to test this minimal scenario in the future.


\chapter{DM heating vs. rotochemical heating in old neutron stars}
\label{chap:dm-heating-vs-roto}

In the previous chapter, we have seen that the rotochemical heating is inevitable consequence for pulsars, and it can indeed explains the observed warm pulsars.
This also has an impact of searches for new physics; among them, the detection of DM signature through the observation of NS surface temperature offers a distinct strategy for testing DM models~\cite{Kouvaris:2007ay, Bertone:2007ae, Kouvaris:2010vv, deLavallaz:2010wp}.
DM particles are trapped by the gravitational potential of a NS after they have lost their kinetic energy through the scattering with the NS matter.
These DM particles eventually annihilate and heat the NS. 
Without any other heating source, this heating effect balances with the energy loss due to the photon
emission from the NS surface, and its surface temperature $T_s$ is kept
constant at $T_s \simeq 2 \times 10^3$~K. This consequence is in
stark contrast to the prediction in the standard NS cooling theory which have discussed in Sec.~\ref{sec:standard-cooling}; NSs cool down to $T_s < 10^3$~K for the NS age $t \gtrsim 5 \times 10^6$~years. This implies that we can in principle test this DM heating
scenario by measuring the surface temperature of old NSs. See
Refs.~\cite{Bramante:2017xlb, Baryakhtar:2017dbj, Raj:2017wrv,
Chen:2018ohx, Bell:2018pkk, Camargo:2019wou, Bell:2019pyc, Acevedo:2019agu} for recent
studies on the DM heating.

As we have shown in the previous section, however, the rotochemical heating can raise the NS surface temperature up to $T_s\simeq 10^6$~K for $t \simeq 10^{6-7}$~years, and this prediction is much higher than that of DM heating.
If the rotochemical heating operates in a NS, it may conceal the DM heating effect. 
Given these observations, can we still expect to detect the signature of the DM heating in old NSs?
This is the issue we address here. 

This chapter is organized as follows. In Sec.~\ref{sec:wimp-dm} and \ref{sec:dm-heating}, we briefly review the WIMP DMs and the heating of NSs by their accretion, respectively.
Then we show our numerical results in Sec.~\ref{sec:comparison-dm-roto}.
We conclude the chapter in Sec.~\ref{sec:conclusion}.
This chapter is based on the author's work~\cite{Hamaguchi:2019oev}. 

\section{WIMP dark matter and DM direct detection}
\label{sec:wimp-dm}

Although its existence is well established, we know little about the nature of DMs.
Among many proposed candidates, WIMP DM is the most popular because it naturally appears in a well-motivated model such as the supersymmetric SM, which solves the hierarchy problem and realizes the gauge coupling unification.
A typical WIMP has a mass of $m\sim 100\GeV - 1\TeV$.

The cosmological abundance of WIMPs is determined by \textit{freeze-out} mechanism.
In the early universe, WIMPs are in thermal equilibrium with the SM particles through the weak interaction; the annihilation and creation rapidly occurs in the thermal plasma.
As the universe expands, its number density and hence annihilation rate decrease, and at the same time the creation rate is suppressed due to the decrease of the plasma temperature.
Finally, its comoving density is frozen at $T \sim m/20$, which is called freeze-out.
This production mechanism is another attractive feature since the DM abundance is independent of the initial condition of the universe.%
\footnote{We assume that the reheating temperature is sufficiently high.}

Intensive experimental searches have been conducted to discover WIMPs.
Unfortunately we have no WIMP signal yet, and limits on the WIMP parameter space are becoming more and more stringent.
In particular, such a stringent constraint is given by the DM direct detection experiments, which try to detect rare scattering between WIMPs and ordinary matters in the huge tanks filled by the target materials such as liquid xenon.
For instance, the strongest bound on the spin-independent scattering with nuclei comes from the XENON1T experiments, which reports $\sigma_{\mathrm{SI}} \lesssim 10^{-46} - 10^{-45}\unit{cm^2}$ for $m=100\GeV - 1\TeV$~\cite{Aprile:2018dbl}. 
The future upgrade of these experiments will further constrain the parameter space.

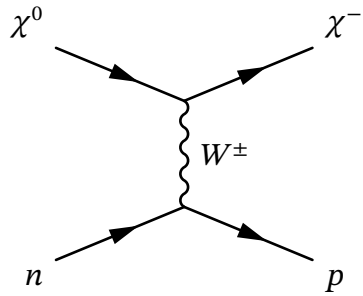
\begin{figure}
  \centering
  \begin{fmffile}{diagram}
    \begin{fmfgraph*}(120,80)
      \fmfleft{i1,i2}
      \fmfright{o1,o2}
      \fmf{fermion}{i1,v1}
      \fmf{fermion}{v1,o1}
      \fmf{fermion}{i2,v2}
      \fmf{fermion}{v2,o2}
      \fmf{photon, label=$W^\pm$}{v1,v2}
      \fmflabel{$\chi^0$}{i2}
      \fmflabel{$n$}{i1}  
      \fmflabel{$\chi^-$}{o2}
      \fmflabel{$p$}{o1}  
    \end{fmfgraph*}
  \end{fmffile}
  \label{fig:inel-diagram}
  \caption{Diagram of inelastic scattering.}
\end{figure}
Nevertheless it is also important to consider another direction to search WIMP DMs because there are models that are difficult to constrain by the direct detection experiments.
For instance, let us consider the WIMP DM that was originally in the electroweak multiplet.
After the electroweak symmetry breaking, it decouples to neutral ($\chi^0$) and charged ($\chi^\pm$) components with the mass difference of $\Delta m = m_{\chi^\pm} - m_{\chi^0} = \Order(100)\MeV$, and only $\chi^0$ remains as the DM in the present universe.
This class of DM includes the wino/Higgsino DM in the supersymmetric model, and so-called minimal DM~\cite{Cirelli:2007xd, Cirelli:2009uv, Farina:2009ez, Farina:2013mla, Nagata:2014aoa}.%
\footnote{
  The Higgsino DM also accompanies a heavier neutral component, which has similar mass difference.
}
The tree level scattering with nucleons occurs in an inelastic channel through the W boson exchange: $\chi^0+p \to \chi^+ + n$ and $\chi^0+n \to \chi^- + p$ (see Fig.~\ref{fig:inel-diagram}).
In order for such inelastic scattering to occur, the scattering energy must be larger than the excitation energy, $\Delta m$.
On the terrestrial experiments using heavy nucleus as the target, the typical DM velocity is $v_{\mathrm{DM}}\sim 10^{-3}$, so that the typical scattering energy is $\Delta E \sim 100 \keV$, which is much smaller than $\Delta m$.
Therefore, the inelastic scattering is highly suppressed on the earth.
In such a case, the leading contribution comes from the elastic scattering induced by the radiative correction.
Because of the loop suppression, the elastic scattering cross section is as small as $10^{-(46-48)}\mathrm{cm^2}$~\cite{Hisano:2011cs, Hisano:2015rsa}.
As a result, constraints on the electroweak DM are fairly weak.


\section{Dark matter heating}
\label{sec:dm-heating}

The thermal evolution of NSs can be complementary to the direct detection experiments.
The accretion and annihilation of WIMP DM in a NS can be another heating
source \cite{Kouvaris:2007ay, Bertone:2007ae, Kouvaris:2010vv,
  deLavallaz:2010wp}.
Let us first derive the condition in which a DM particle accretes onto a NS.
The DMs near the NS are in the effective potential
\begin{align}
  \label{eq:effective-potential}
  V_{\mathrm{eff}}(r)
  =
  -\frac{GMm_{\mathrm{DM}}}{r}
  +\frac{L^2}{2m_{\mathrm{DM}} r^2}
  -\frac{r_cL^2}{2m_{\mathrm{DM}} r^3}\,,
\end{align}
where $L$ is the angular momentum and $r_c=2GM$ the Schwarzschild radius.
Using the impact parameter $b$, the angular momentum is written as $L=m_{\mathrm{DM}} v_{\mathrm{DM}}b$, where $v_{\mathrm{DM}}$ is the DM velocity far from the NS.
From the energy conservation, the closest distance $r_{\mathrm{min}}$ is determined by
\begin{align}
  \label{eq:rmin}
  \frac{m_{\mathrm{DM}} v_{\mathrm{DM}}^2}{2} = V_{\mathrm{eff}}(r_{\mathrm{min}})\,,
\end{align}
If $r_{\mathrm{min}}\leq R$, the DM arrives at the NS.
Thus one can determine the maximum impact parameter, $b_{\mathrm{max}}$, by solving
\begin{align}
  \label{eq:bmax}
  \frac{m_{\mathrm{DM}} v_{\mathrm{DM}}^2}{2} =\left. V_{\mathrm{eff}}(R)\right|_{b=b_{\mathrm{max}}}\,.
\end{align}
The DMs with $b\leq b_{\mathrm{max}}$ fall into the NS.
The solution of Eq.~\eqref{eq:bmax} is
\begin{align}
  \label{eq:bmax-sol}
  b_{\mathrm{max}}
  =
  R\frac{v_{\mathrm{esc}}}{v_{\mathrm{DM}}}e^{-\Phi(R)}\left( 1+ \frac{v_{\mathrm{DM}}^2}{v_{\mathrm{esc}}^2} \right)\,,
\end{align}
where $v_{\mathrm{esc}}\equiv \sqrt{2GM/R}$ is the escape velocity.
Near the earth, DM velocity is $v_{\mathrm{DM}} \simeq 10^{-3}$, while the escape velocity of a typical NS ($M=1.4\Msun$ and $R=10\unit{km}$) is $v_{\mathrm{esc}} \sim 0.6$.
Thus we obtain $b_{\mathrm{max}} \sim 10^3R$; the NS accumulates the DMs in a broader region than its size.
The DM flux is expressed as
\begin{align}
  \label{eq:dm-flux}
  \dot{N}
  \simeq
  \pi b_{\mathrm{max}}^2
  v_{\mathrm{DM}} \frac{\rho_{\mathrm{DM}}}{m_{\mathrm{DM}}}\,.
\end{align}
We use $\rho_{\mathrm{DM}} = 0.42~\mathrm{GeV} \cdot \mathrm{cm}^{-3}$ and $v_{\mathrm{DM}} = 230~\mathrm{km}\cdot \mathrm{s}^{-1}$ \cite{Pato:2015dua} in what follows. 
A more accurate expression of $\dot{N}$ is given in Ref.~\cite{Kouvaris:2007ay}, which we use in the following numerical analysis.


In order for a DM to be captured by the NS, the initial kinetic energy far away from the star must be lost by the scattering with nucleons.
The recoil energy of this scattering is written as
\begin{align}
  \label{eq:recoil-en}
  \Delta E =
  \frac{m_Nm_{\mathrm{DM}}^2\gamma^2v_{\mathrm{esc}}^2}{m_N^2+m_{\mathrm{DM}}^2+2\gamma m_{\mathrm{DM}} m_N}
  (1-\cos\theta_{\mathrm{CM}})\,,
\end{align}
where $m_N$ is the nucleon mass, $\theta_{\mathrm{CM}}$ the scattering angle in the center of mass frame, and $\gamma=1/\sqrt{1-v_{\mathrm{esc}}^2}\simeq\Order(1)$.
Since the initial kinetic energy is $\sim v_{\mathrm{DM}}^2m_{\mathrm{DM}}\sim 10^{-6}m_{\mathrm{DM}}$, it is found that an electroweak/TeV-scale WIMP DM is captured in NSs after one scattering.
If the DM-nucleon scattering cross section is larger than $\sigma_{\mathrm{crit}} \simeq R^2 m_N/M\sim 10^{-45}~\mathrm{cm}^2$, the mean free path of the DM inside the NS is smaller than the radius $R$~\cite{Kouvaris:2007ay}, so that all the DM coming to the NS are captured.
After the capture, the rest of its kinetic energy is soon lost by successive scatterings with the NS matter.
DM particles then accumulate in the NS core and eventually annihilate.
As shown in Ref.~\cite{Kouvaris:2010vv}, for a typical WIMP, its annihilation and capture rates become in equilibrium in old NSs. As a result, the contribution of the DM heating to the luminosity $L_H^\infty$ in Eq.~\eqref{eq:time-evl} is computed as
\begin{equation}
 L_H^\infty|_{\mathrm{DM}} = e^{2\Phi(R)} \dot{N} m_{\mathrm{DM}}\left[
\chi + (\gamma -1)
\right] ~,
\label{eq:lhdm}
\end{equation}
where $\chi$ is the fraction of the
annihilation energy transferred to heat \cite{Kouvaris:2010vv}. In what
follows, we take $\chi = 1$ unless otherwise noted. The first term in Eq.~\eqref{eq:lhdm} represents the
heat from the DM annihilation, while the second term corresponds to the
deposit of the kinetic energy of the incoming WIMP DM
\cite{Baryakhtar:2017dbj}. 

We note that if the DM is lighter than $1\GeV$, the Pauli blocking effect decreases the scattering rate; the momentum transfer becomes insufficient to excite the nucleons above the Fermi surface~\cite{Baryakhtar:2017dbj}.
Meanwhile, the multiple scattering is necessary to capture DM heavier than $1\mathrm{PeV}$, which also decrease the capture probability~\cite{Bramante:2017xlb}. 
Thus the DM heating effect is most efficient for $1\GeV \lesssim m_{\mathrm{DM}} \lesssim 1\mathrm{PeV}$.
Nevertheless the DM heating is better at low mass region than the direct detection experiments because there is no sharp threshold due to the detector material.

If we neglect the rotochemical heating, the DM heating balances with the
cooling due to the photon emission at late times, i.e., 
$L_H^\infty|_{\mathrm{DM}}
\simeq L_\gamma^\infty$. This condition fixes the NS surface temperature ($M=1.4\Msun$ and $R=10\unit{km}$)
to be
\begin{align}
  \label{eq:ts-dm-estimate}
  T_s^\infty \sim 2200\unit{K}\,,
\end{align}
which has been regarded as a smoking-gun
signature of the DM heating \cite{Kouvaris:2007ay, Bertone:2007ae, Kouvaris:2010vv,
  deLavallaz:2010wp}.
Note that the dependence on the DM parameter is $T_s\propto \rho_{\mathrm{DM}}^{1/4}v_{\mathrm{DM}}^{-1/4}$, and hence the prediction is not very sensitive to the choice of $\rho_{DM}$ and $v_{\mathrm{DM}}$.
In the next section, we study if this signature can still be seen even in the presence of the rotochemical heating.

Before closing this section, let us see the significance of the DM heating in the case of the electroweak DM discussed in the previous section.
Due to the strong gravity, the DM kinetic energy is so large near the NS surface that, unlike in the terrestrial experiments, the inelastic scattering is not suppressed; the typical recoil energy given by Eq.~\eqref{eq:recoil-en} is as large as $\Delta m$.
Since it is induced by the tree-level exchange of the $W$ boson, its cross section is much larger than the critical value $\sigma_{\mathrm{crit}}  \sim 10^{-45}~\mathrm{cm}^2$ \cite{Baryakhtar:2017dbj}.
Hence, we can
directly use the results in the next section for this class of DM candidates.\footnote{There
is a small difference since the annihilation of
these DM candidates can generate neutrinos in the final state and thus
the parameter $\chi$ in Eq.~\eqref{eq:lhdm} is smaller than unity. This
difference only results in an $\mathcal{O} (1)$\% change in the
late-time temperature, which is in effect negligible in the present
discussion. }
Notice that this constraint is independent of the DM masses since they are predicted to be 1--10~TeV~\cite{Farina:2013mla}.


\section{Results}
\label{sec:comparison-dm-roto}

Now we examine the time evolution of the NS temperature by including both DM and rotochemical heating.
We first consider a NS which models a typical ordinary pulsar, where we fix $M = 1.4 M_\odot$, $P = 1$~s, $\dot{P} = 1 \times 10^{-15}$, and $\Delta M/M = 1 \times 10^{-15}$ (see also Sec.~\ref{sec:phys-input}).
The initial values of $T^\infty$ and $\eta_\ell^\infty$ are taken to be $T^\infty = 10^{10}$~K and $\eta_\ell^\infty = 0$, respectively.
We find that the following results have little dependence on the choice of these parameters.

\begin{figure}
  \centering
  \begin{minipage}{0.5\linewidth}
    \includegraphics[clip, width = 1.0\textwidth]{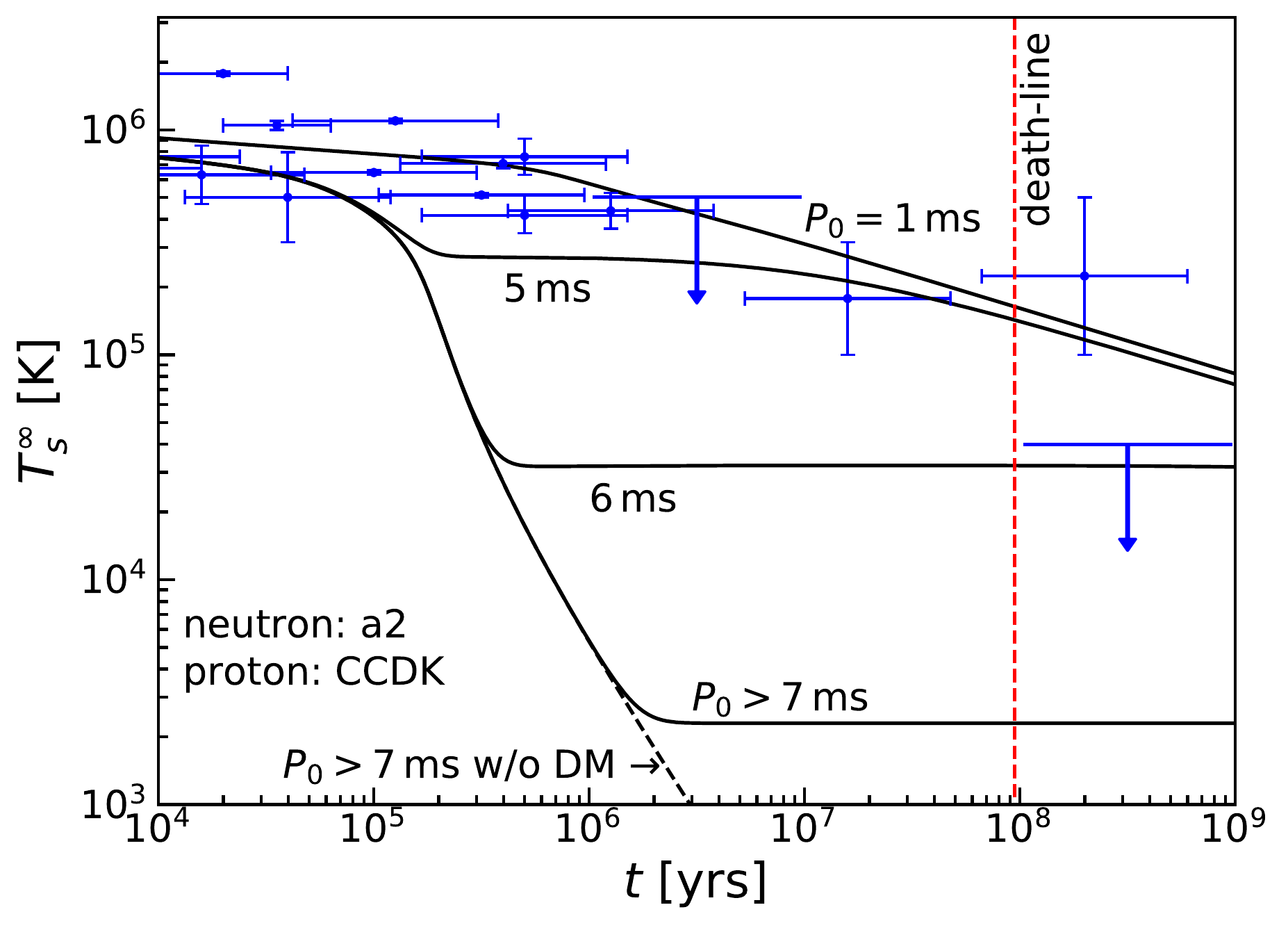}
  \end{minipage}%
  \begin{minipage}{0.5\linewidth}
    \includegraphics[clip, width = 1.0\textwidth]{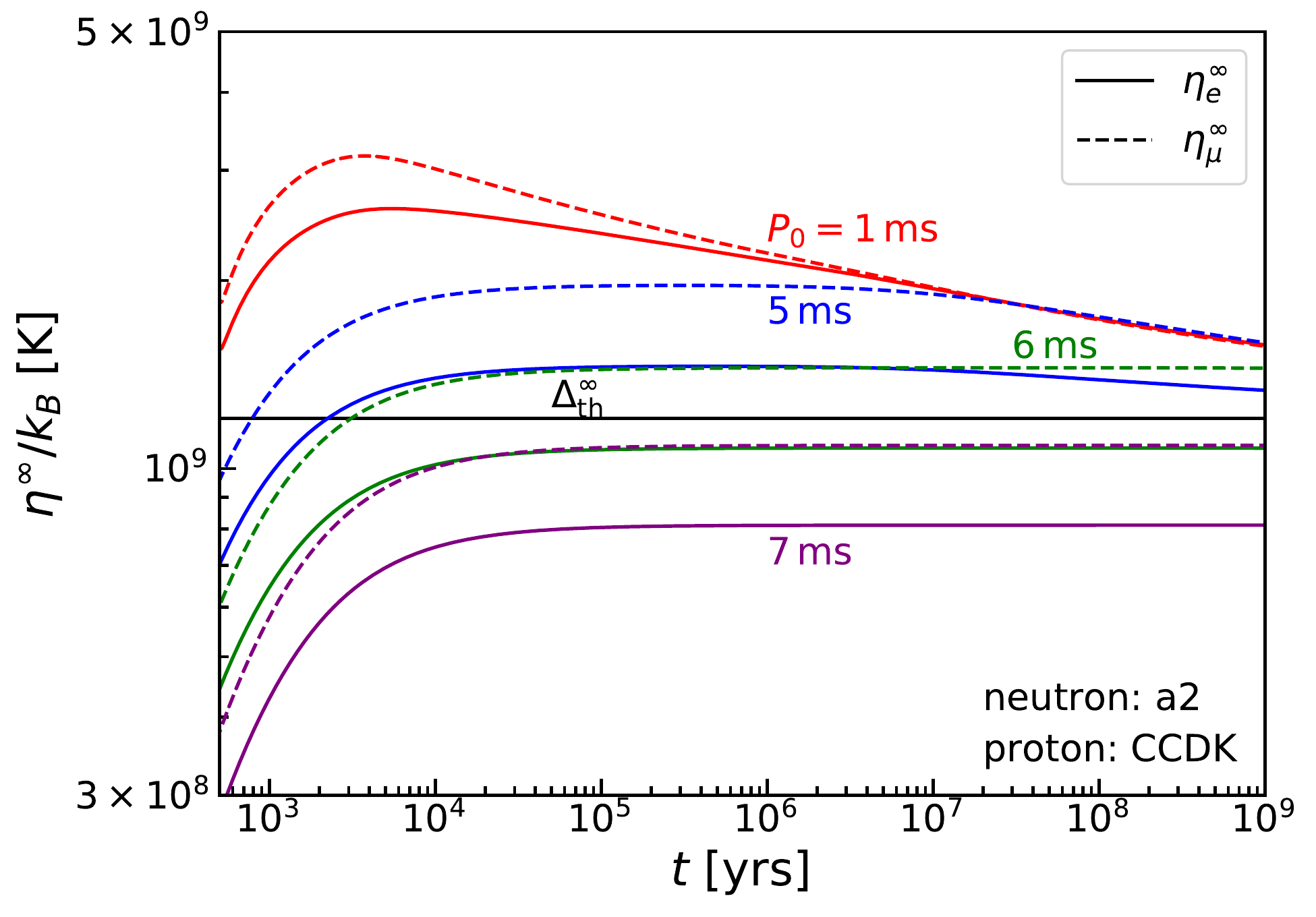}
  \end{minipage}
\caption{
Left: The time evolution of $T_s^\infty$ for different values of $P_0$.
For $P_0=1$, 5, and 6~ms, the time evolution with and without DM heating are indistinguishable and the lines overlap. For $P_0>7$~ms, the solid (dashed) line represents the case with (without) DM heating. The blue crosses
 show the temperature data of ordinary pulsars \cite{Yanagi:2019vrr}.
 The down arrows indicate the upper limits on $T_s^\infty$.
 The
 vertical red dashed line shows the death-line \cite{1992A&A...254..198B}.
 Right: The time evolution of $\eta_\ell^\infty$ for different values of
 $P_0$ corresponding to the left panel. The horizontal line shows
 $\Delta_{\mathrm{th}}^\infty$ defined in the text.
}
\label{fig:a2_CCDK_P0}
\end{figure}

In the left panel of Fig.~\ref{fig:a2_CCDK_P0}, we show the time evolution of $T_s^\infty$ for different values of $P_0$ in the black solid
lines. 
For $P_0=1$, 5, and 6~ms, the time evolutions with and without DM heating are indistinguishable and the lines overlap.
For $P_0>7$~ms, the solid (dashed) line represents the case with (without) DM heating.
We use the CCDK \cite{Chen:1993bam} and ``a2'' \cite{Page:2013hxa} models for the proton and neutron gaps, respectively.\footnote{We assume the neutron triplet pairing of $m_j=0$ state.}
We also show the observed temperatures of old ordinary pulsars with blue crosses, where the lines indicate the uncertainties; we take this data from Tab.~\ref{tab:psr-temp}.
This figure shows that for $P_0 = 1$~ms the surface temperature remains as high as ${\cal O}(10^5)$~K for $t \gtrsim 10^6$~years since the rotochemical heating is quite effective.
The temperature curve in this case is consistent with most of the observed temperatures, but the DM heating effect is completely hidden by the rotochemical heating effect.
For a larger $P_0$, $T_s^\infty$ at late times gets lower, and for $P_0 > 7$~ms, it becomes independent of the initial period.
In the right panel, we show the corresponding evolution of $\eta_\ell^\infty$ with the rotochemical threshold.
The minimal value of $\Delta_{\mathrm{th}} = \mathrm{min}\{3\Delta_n + \Delta_p, \Delta_n + 3\Delta_p\}$ in the core multiplied by a red-shiftfactor at the position is shown.
For $P_0 > 7$~ms, the rotochemical heating is ineffective since $\eta_\ell$ does not exceed the rotochemical threshold.
Thus, the late-time temperature is determined by the DM heating, with $T_s^\infty \simeq 2 \times 10^3$~K.
Notice that a NS cools down to this temperature before it reaches the conventional death-line \cite{Ruderman:1975ju, 1992A&A...254..198B},\footnote{We however note that
the theoretical estimation of the death-line suffers from
huge uncertainty, and thus one should not take this bound too
seriously. Indeed, as can be seen from
Fig.~\ref{fig:a2_CCDK_P0}, J2144-3933, \textit{e.g.}, is located beyond
the conventional death-line, though its pulsation is detected
\cite{1999Natur.400..848Y}. For more discussions on the death-line, see
Refs.~\cite{1993ApJ...402..264C, Zhang:2000rd, Zhang:2002uh, Zhou:2017hfm}. } 
$B_s/P^2 = 0.17 \times 10^{12}~\mathrm{G} \cdot \mathrm{s}^{-2}$,
shown by the
red dashed line in the left panel of Fig.~\ref{fig:a2_CCDK_P0}.
Therefore, it is possible to detect the DM heating effect via the temperature observation of ordinary pulsars if their initial period is sufficiently large.

\begin{figure}
\centering
\begin{minipage}{0.5\linewidth}
  \includegraphics[clip, width = 1.0 \textwidth]{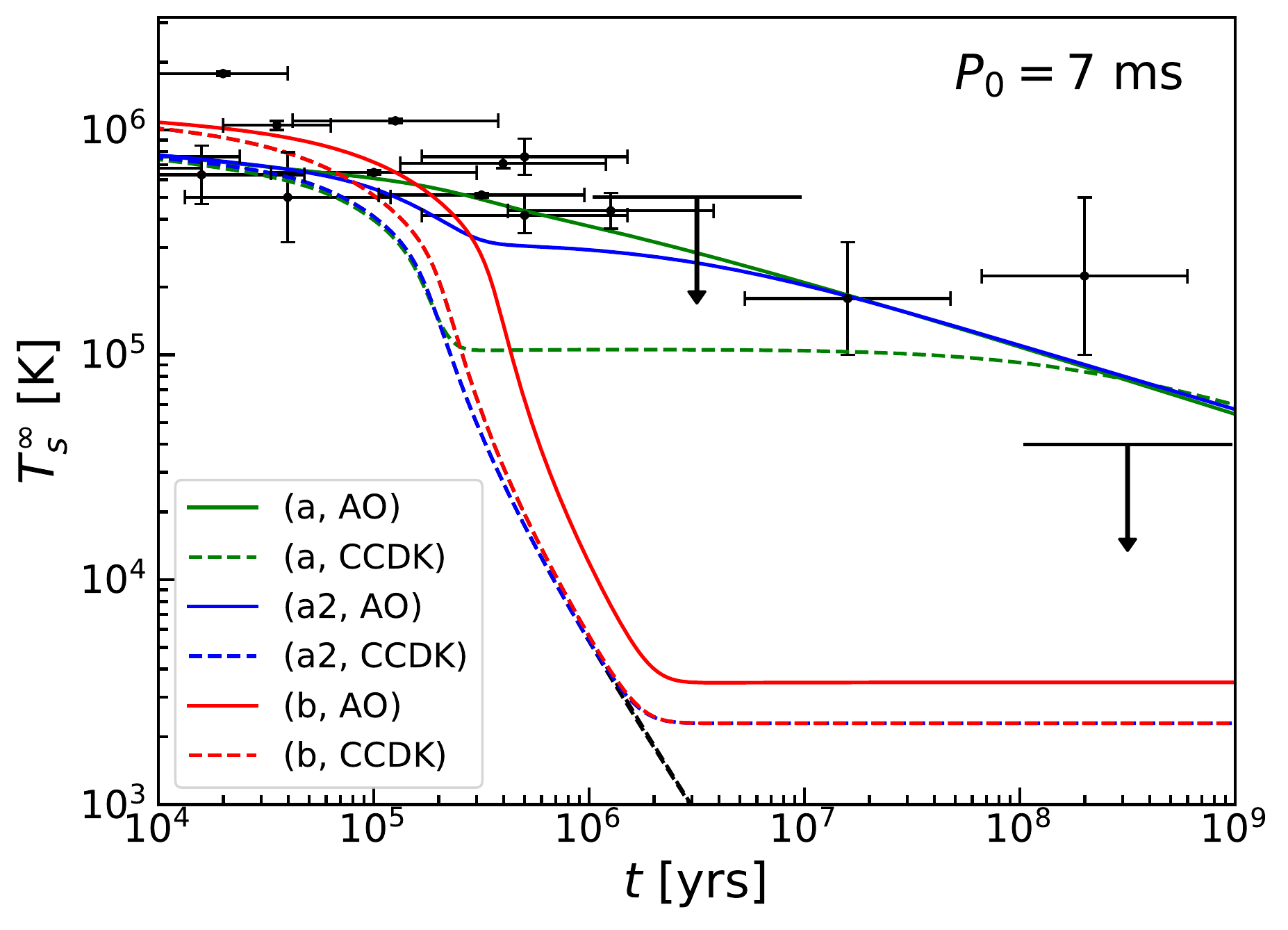}
\end{minipage}%
\begin{minipage}{0.5\linewidth}
  \includegraphics[clip, width = 1.0 \textwidth]{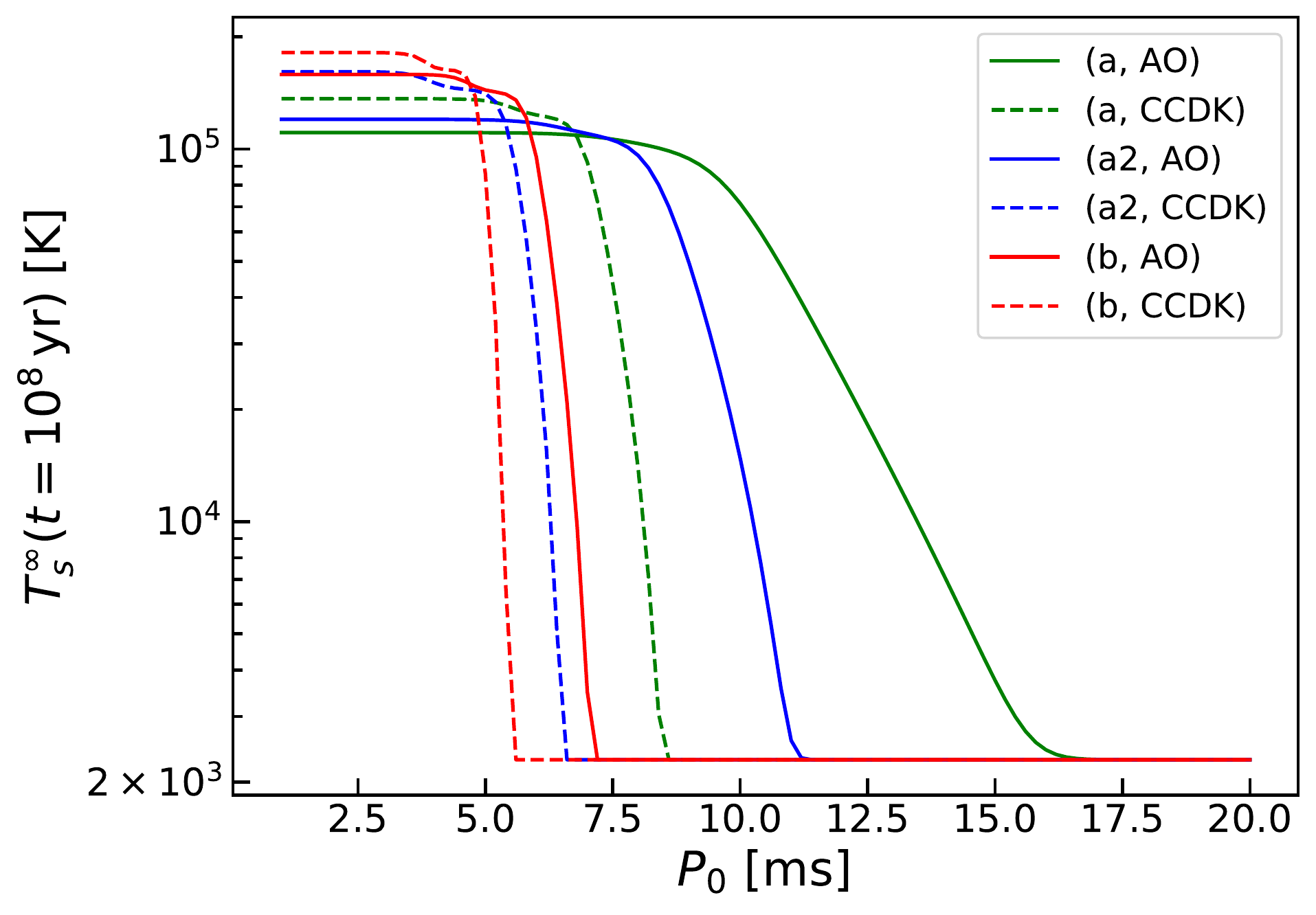}
\end{minipage}
\caption{
Left: The time evolution of $T_s^\infty$ for different nucleon pairing gaps with $P_0 = 7$~ms. 
The green, blue and red lines correspond to 
the ``a'', ``a2'' and ``b'' models for the neutron gap, while the solid and dashed lines show the cases for the AO and CCDK models of the
proton gap, respectively.
For the (a2, CCDK) and (b, CCDK) gap models, the time evolution without DM are also shown in black dashed lines. In the other gap models, the time evolutions with and without DM are almost the same. The black crosses show the temperature data of ordinary pulsars.
}
\label{fig:CP_10ms}
\end{figure}

We note, however, that the lower limit on $P_0$ 
for the condition that the DM heating effect is detectable highly
depends on the nucleon pairing gaps. To see this, in the left panel of 
Fig.~\ref{fig:CP_10ms}, we show the time evolution of $T_s^\infty$ for
different choices of nucleon pairing gaps, with $P_0 = 7$~ms.
The green, blue and red lines correspond to the ``a'', ``a2'' and ``b'' models in Fig.~\ref{fig:gap-3p2} for the
neutron gap, respectively.
The solid and dashed lines show the cases for the
AO \cite{Amundsen:1984qq} and CCDK \cite{Chen:1993bam} models of the
proton gap, respectively. As we see, for the AO model, whose gap
is smaller than that of the CCDK model, the late-time temperature is
predicted to be higher than $T_s^\infty \simeq 2 \times 10^3$~K; in this
case, $\Delta_{\mathrm{th}}^\infty$ is rather small, and thus $\eta_\ell$ can overcome
the rotochemical threshold at late times even for $P_0 = 7$~ms, making
the rotochemical heating operative. On the other hand, for the CCDK proton and "a2" or "b" neutron pairings, the rotochemical heating is
ineffective and thus we can see the DM heating effect at late times.
The right panel of Fig.~\ref{fig:CP_10ms} shows the redshifted surface temperature at $t=10^8\unit{yr}$ for different choices of gap models and $P_0$.
For $P_0 \gtrsim 20\unit{ms}$, the surface temperatures converge to $T_s^\infty \sim 2000\unit{K}$ for any choice of gap; if $P_0$ is smaller, it can be larger than this value due to the strong rotochemical heating, so that the DM heating is concealed.
The results shown in Fig.~\ref{fig:CP_10ms} demonstrate that it is
crucial to take account of both proton and neutron pairings in
order to evaluate the effect of non-equilibrium beta processes
appropriately. 

As seen in the right panel of Fig.~\ref{fig:CP_10ms}, the rotochemical heating does not operate for
any choice of pairing gaps if the initial period is as large as $10-100\unit{ms}$%
\footnote{Since the pairing gap has density dependence, the critical value of $P_0$, above which the rotochemical heating is ineffective, depends also on other star parameters such as NS mass. We find the critical $P_0$ is at most $\Order(10)\unit{ms}$.}---in this case, the late-time surface temperature is always
determined by the DM heating. It is intriguing that recent
studies suggest that the initial period distribution extends to well beyond 100~ms; it is independently estimated from the kinematic age of several tens of observed NSs~\cite{2012Ap&SS.341..457P, 2013MNRAS.430.2281N,  Igoshev:2013rqf}, the population synthesis of pulsars~\cite{FaucherGiguere:2005ny, Popov:2009jn, Gullon:2014dva, Gullon:2015zca}, or the supernova simulation for proto-NSs~\cite{Muller:2018utr}. Hence, we
expect that there are quite a few ordinary pulsars that can be a probe
of the DM heating in future observations.  

Finally, let us consider MSPs. 
In this case, $|\Omega \dot{\Omega}|$ is
much larger than that for ordinary pulsars, and thus $\eta_\ell$ 
can always exceed the rotochemical threshold at late
times. Therefore, the rotochemical heating is highly effective for MSPs.
Although this feature is advantageous for explaining the
old warm MSPs such as J0437-4715 and J2124-3358 (see Sec.~\ref{sec:millisecond-pulsars}), this makes MSPs inappropriate for testing the
DM heating scenario.



\section{Summary and discussion}
\label{sec:conclusion}

We have studied the time evolution of NS surface
temperature, taking account of both the rotochemical and DM heating
effects. We have found that for ordinary pulsars the DM heating effect
can still be observed even with the rotochemical heating
if the initial period of NSs is relatively large, since in
this case the chemical imbalance does not overcome the threshold
$\Delta_{\mathrm{th}}^\infty$ and thus the rotochemical heating is
ineffective. 
The rotochemical heating operates if the initial period is as small as $\Order(1)\unit{ms}$. Thus for MSPs, the DM heating is
always concealed by the rotochemical heating. 

The surface temperature at late times depends not only on the initial
period but also on the choice of the nucleon pairing gaps, as shown in
Fig.~\ref{fig:CP_10ms}. Depending on these unknown quantities, the
rotochemical heating effect may mimic the DM heating effect in old
ordinary pulsars. For instance, the late-time temperature for the proton AO and neutron "b" gaps in the left panel of Fig.~\ref{fig:CP_10ms} is kept at a few thousand K due to the rotochemical heating.
To distinguish these two heating effects, therefore,
it is necessary to improve our knowledge on nucleon pairing gaps as well
as to evaluate the initial period of pulsars accurately. We note in
passing that it is possible to estimate the initial period of a pulsar
if, for instance, the pulsar is associated with a supernova and its age
is computed from the motion of the supernova remnant, as is performed in Ref.~\cite{2012Ap&SS.341..457P},

In any case, in the presence of both the rotochemical and DM heating effects,
the late-time temperature is bounded below,
i.e., $T_s^\infty \gtrsim 2 \times 10^3$~K, which is determined
by the DM heating and thus independent of the initial period and pairing
gaps. As a consequence, an observation of a NS with a surface
temperature that is sufficiently below this lower bound readily excludes
the DM heating caused by typical WIMPs, and thus can severely constrain
such DM models.
For instance, as we have discussed in Sec.~\ref{sec:wimp-dm} and \ref{sec:dm-heating}, the DM heating is good at constraining the electroweak DM that accompanies the charged components with the mass difference $\Delta m \lesssim \Order(100)\MeV$.

The detectability of the DM heating is worth mentioning.
First of all, we need to find a nearby isolated pulsar for the subsequent flux measurement.
As we discussed in the previous section, the DM heating is visible before the death of the active radio emission, so this would be done by ordinary radio telescopes.
The surface temperature of $T_s\sim 2000  \unit{K}$, corresponding to $\lambda \sim \unit{\mu m}$ in wavelength, requires infrared telescopes.
The future infrared telescopes such as the James Webb Space Telescope (JWST), the Thirty Meter Telescope (TMT) and the European Extremely Large Telescope (E-ELT) have sensitivity to this wavelength, and we could detect the DM heating with the exposure time of $t \sim 10^5(d/10\unit{pc})^4\unit{s}$, where $d$ is the distance to the pulsar~\cite{Baryakhtar:2017dbj}.
The population of such close pulsars ($d \lesssim 10\unit{pc}$) is another question.
Perhaps we expect a few nearby NSs~\cite{1993ApJ...403..690B}.
We leave the detailed investigation on the future observational prospect as a future work.

Finally, we note that there are other heating mechanisms proposed in the
literature \cite{Gonzalez:2010ta}, such as the vortex creep heating
\cite{1984ApJ...276..325A, 1989ApJ...346..808S, 1991ApJ...381L..47V,
1993ApJ...408..186U, VanRiper:1994vp, Larson:1998it} and
rotationally-induced deep crustal heating \cite{Gusakov:2015kaa}. These
heating mechanisms may also compete with the rotochemical and DM heating
effects, and therefore the consequence drawn in this letter may be
altered if they are also included. We will study the implications of
these heating mechanisms for the DM heating in the future.

\chapter{Conclusion}
\label{chap:conclusion}

In this dissertation, we have shown how the thermal evolution of NSs is used to constrain/probe the physics beyond the SM.
In Chap.~\ref{chap:limit-axion-decay} we have studied the constraint on the axion models from the cooling rate of the Cas A NS.
We have considered the cooling by axion emissions, and obtained the limit $f_a \gtrsim 10^8\GeV$, which is as strong as the one from SN1987A.
This bound will be improved if, for instance, we obtain more data of young NSs including the new Cas A NS data, or we understand the neutron triplet pairing better.
In Chap.~\ref{chap:dm-heating-vs-roto}, we have investigated the theoretical condition to probe DMs through the heating of old NSs.
The DM capture leads to the heating of NSs, but it may be concealed by the rotochemical heating, which is inherent in pulsars. 
We have shown that even in the presence of rotochemical heating, the DM heating is visible at old NSs if their initial periods are $P_0=\Order(10-100)\unit{ms}$.
We expect that many ordinary pulsars are suitable to look for this DM heating.
However, the better understanding of the pulsar initial period, as well as of the nucleon pairing, is necessary to confirm the evidence of the DM heating.

As stated above, one of the big uncertainties comes from our limited knowledge of the nucleon pairing gap, in particular for neutron triplet pairing.
Currently, the gap amplitude of the neutron triplet pairing is essentially unknown in the NS core.
As we have seen in Chap.~\ref{chap:limit-axion-decay}, the observed cooling rate of the Cas A NS favors a small gap of the neutron triplet pairing ($T_c \sim 5\times 10^8\unit{K}$).
With a larger gap, the neutron PBF would have begun much earlier than the age of the Cas A NS.
In Chap.~\ref{chap:neutron-star-heating}, we have shown that the prediction of the rotochemical heating is quite dependent on the gap models.
In particular, the warm MSPs are better fitted with a large neutron gap such as ``c'' model.
Considering the uncertainties in theory and observations, we cannot tell whether or not this requirement is compatible with the Cas A NS.
Better understanding of nucleon pairing is awaited to explore a unified explanation for the Cas A NS and the old warm NSs.

Confronting current no discovery of new particles in the experiments such as the collider or DM direct searches, we need to develop new ways to probe the physics beyond the SM.
Thermal evolution of NSs can be one of such probes.
Future progresses of both NS observation and theory will be helpful to make theoretical predictions more rigid.
We hope that this dissertation serves as a path toward that.

\appendix

\chapter{Phase space factors}
\label{sec:phase-space}

In this appendix, we give a detailed discussion on the phase space integrals in Eqs.~\eqref{eq:i-integ-emis}  and~\eqref{eq:i-integ-gamma}. While the nucleon pairing is negligible, we use the formulas in Ref.~\cite{Reisenegger:1994be}. In the numerical calculation, we neglect the pairing if $3v_n+v_p < 1$ for the neutron branch and if $v_n+3v_p < 1$ for the proton branch.
If the pairing operates but $\xi_\ell < 1$, we neglect non-equilibrium effects. We use the results in Ref.~\cite{Gusakov:2002hh} for $I^N_{M,\epsilon}$, and set $I^N_{M,\Gamma}=0$. If the pairing operates and $\xi_\ell \geq 1$, we numerically perform the integrals using the zero temperature approximation; we show the detail of this calculation in this appendix. 
We have checked that the result scarcely depends on the choice of the threshold value of pairing and $\xi_\ell$.

\section{Energy integral}
\label{sec:energy-integral}

We define the energy integral part of the phase space factors by\footnote{$\tilde I^N_{M,\Gamma}$ in Eq.~\eqref{eq:itildenmgamdef} has the opposite sign to $I^N_{M,\Gamma}$ in Ref.~\cite{Petrovich:2009yh}. }
\begin{align}
  \tilde I^N_{M,\epsilon}
  &= \frac{60480}{11513\pi^8}\int_0^\infty dx_\nu \int_{-\infty}^{\infty} dx_ndx_pdx_{N_1}dx_{N_2} x_\nu^3
    f(z_n)f(z_p)f(z_{N_1})f(z_{N_2})\notag\\
  &\times\left[f(x_\nu - \xi_\ell -z_n-z_p-z_{N_1}-z_{N_2})
    + f(x_\nu + \xi_\ell -z_n-z_p-z_{N_1}-z_{N_2})\right]\,,
  \\
  \tilde I^N_{M,\Gamma}
  &= \frac{60480}{11513\pi^8}\int_0^\infty dx_\nu \int_{-\infty}^{\infty}dx_ndx_pdx_{N_1}dx_{N_2} x_\nu^2
    f(z_n)f(z_p)f(z_{N_1})f(z_{N_2})\notag\\
  &\times\left[f(x_\nu - \xi_\ell -z_n-z_p-z_{N_1}-z_{N_2})
    - f(x_\nu + \xi_\ell -z_n-z_p-z_{N_1}-z_{N_2})\right]\,.
    \label{eq:itildenmgamdef}
\end{align}
Following the argument in Ref.~\cite{Petrovich:2009yh}, we use the zero temperature approximation in this calculation; namely, 
we replace the Fermi-Dirac distribution $f(x)$ by the step function $\Theta(-x)$. The phase space factors are then written in terms of the following integral expressions:
\begin{align}
  \tilde I^N_{M,\epsilon}
  &=
    \frac{60480}{11513\pi^8}
    \frac{\xi_\ell^8}{4}
    \Theta(1-r_n-r_p-r_{N_1}-r_{N_2})
    \notag\\
    &\times
    \int_{r_n}^{1-r_{N_1}-r_{N_2}-r_p}du_n
    \int_{r_{N_1}}^{1-u_n-r_{N_2}-r_p}du_{N_1}
      \int_{r_{N_2}}^{1-u_n-u_{N_1}-r_p}du_{N_2} 
      \notag\\
  &\times
  \frac{u_{n}}{\sqrt{u_{n}^2-r_{n}^2}}
  \frac{u_{N_1}}{\sqrt{u_{N_1}^2-r_{N_1}^2}}
      \frac{u_{N_2}}{\sqrt{u_{N_2}^2-r_{N_2}^2}}
    K_\epsilon(u_n+u_{N_1}+u_{N_2}, r_p) \, ,
    \label{eq:itildenmep}
\\
  \tilde I^N_{M,\Gamma}
  &=
    \frac{60480}{11513\pi^8}
    \frac{\xi_\ell^7}{3}
    \Theta(1-r_n-r_p-r_{N_1}-r_{N_2})
    \notag\\
    &\times
    \int_{r_n}^{1-r_{N_1}-r_{N_2}-r_p}du_n
    \int_{r_{N_1}}^{1-u_n-r_{N_2}-r_p}du_{N_1}
      \int_{r_{N_2}}^{1-u_n-u_{N_1}-r_p}du_{N_2}
      \notag\\
  &\times
  \frac{u_{n}}{\sqrt{u_{n}^2-r_{n}^2}}
  \frac{u_{N_1}}{\sqrt{u_{N_1}^2-r_{N_1}^2}}
      \frac{u_{N_2}}{\sqrt{u_{N_2}^2-r_{N_2}^2}}
    K_\Gamma(u_n+u_{N_1}+u_{N_2}, r_p)\,,
    \label{eq:itildenmgam}
\end{align}
where $r_N=v_N/\xi$, $u_N=|z_N|/\xi$, and 
\begin{align}
  K_\epsilon(u,r)
  &= -\frac{1}{2}r^2(1-u)\left(3r^2 +4(1-u)^2\right)\ln\left[\frac{1-u+\sqrt{(1-u)^2-r^2}}{r}\right]
    \notag\\
  &+\frac{1}{30}\left(16r^4 + 83r^2(1-u)^2+6(1-u)^4\right)\sqrt{(1-u)^2-r^2}\,,
  \label{eq:kepdef}
  \\
  K_\Gamma(u,r)
  &= -\frac{3}{8}r^2\left(r^2 +4(1-u)^2\right)\ln\left[\frac{1-u+\sqrt{(1-u)^2-r^2}}{r}\right]
    \notag\\
  &+\frac{1}{8}(1-u)\left(13r^2 + 2(1-u)^2\right)\sqrt{(1-u)^2-r^2}\,.
\end{align}
Notice that the integral region of Eqs.~\eqref{eq:itildenmep} and \eqref{eq:itildenmgam} is different from that in Eqs.~(B.12) and (B.13) in Ref.~\cite{Petrovich:2009yh}. We also find that the sign of the first term in Eq.~\eqref{eq:kepdef} is opposite to that in Eq.~(B.11) in Ref.~\cite{Petrovich:2009yh}. 

\section{Angular integral}
\label{sec:angular-integral}

Next, we perform the angular integration.
Due to the different angular dependence of proton and neutron pairing gaps, we need to separately treat the integrals for the proton and neutron branches.

\paragraph{Proton branch}

For the proton branch, we can carry out the angular integration of the momenta of the protons and leptons trivially since the proton singlet gap is isotropic. The integration with respect to the neutron momentum direction is then reduced to a simple average as 
\begin{align}
  I^p_{M,\epsilon} &= \int\frac{d\Omega_n}{4\pi} \tilde I^p_{M,\epsilon}\,,
  \nonumber\\
  I^p_{M,\Gamma} &= \int\frac{d\Omega_n}{4\pi} \tilde I^p_{M,\Gamma}\,,
\end{align}
where only $v_n$ is dependent on $\Omega_n$.

\paragraph{Neutron branch}

The neutron branch involves three neutrons, $n$, $n_1$ and $n_2$, so the angular integral is generically complicated. 
Here we neglect the proton and lepton momenta in the momentum conserving delta function.
Then, the three neutron momenta form an equilateral triangle, and we can put the angular integral into
\begin{align}
  I^n_{M,\epsilon}
  &=
    \int_0^1d\cos\theta_n\int_0^{2\pi}\frac{d\varphi_{n_1}}{2\pi}
    \tilde I^n_{M,\epsilon}\,,
  \nonumber\\
  I^n_{M,\Gamma}
  &=
    \int_0^1d\cos\theta_n\int_0^{2\pi}\frac{d\varphi_{n_1}}{2\pi}
    \tilde I^n_{M,\Gamma}\,,
    \label{eq:angintn}
\end{align}
with $\cos\theta_n$ the polar angle of $n$ around the quantization axis and $\varphi_{n_1}$ the azimuthal angle of $n_1$ around $\bm p_n$.
The relative angles among the momenta of $n$, $n_1$ and $n_2$ are fixed because of the momentum conservation. The polar angles of $\bm{p}_{n_1}$ and $\bm{p}_{n_2}$ with respect to the quantization axis are written as
\begin{align}
  \cos\theta_{n_1}
  &=
    -\frac{\sqrt 3}{2}\sin\theta_n\cos\varphi_{n_1} -\frac{1}{2}\cos\theta_n\,,
  \\
  \cos\theta_{n_2}
  &=
    +\frac{\sqrt 3}{2}\sin\theta_n\cos\varphi_{n_1} -\frac{1}{2}\cos\theta_n\,,
\end{align}
respectively. The integrands in Eq.~\eqref{eq:angintn} depend on 
$\cos\theta_{n}$, $\cos\theta_{n_1}$, and $\cos\theta_{n_2}$ through the triplet neutron pairing gap $\Delta_n \propto \sqrt{1+3\cos^2\theta }$.

\chapter{Rotochemical heating with different magnetic field for ordinary pulsars}
\label{chap:magnetic-field}

In Sec.~\ref{sec:role-non-equilibrium}, we fix the spin-down rate $P\dot P =
10^{-15}\unit{s}$, i.e., the dipole magnetic field $B \sim 10^{12}\unit{G}$, for ordinary pulsars, though it generically takes a value in a rather broad range: $P\dot P = 10^{-17} - 10^{-13}\unit{s}$,
corresponding to $B \sim 10^{11-13}\unit{G}$. Since this value controls the evolution of angular velocity (see Eq.~\eqref{eq:omega-sol-n3}), a change in $P\dot{P}$ may affect the intensity of the rotochemical heating. In this appendix, we study the effects of varying $P\dot{P}$ on  the temperature evolution of ordinary pulsars. We fix the other NS parameters to be $M = 1.4\,M_\odot$ and $\Delta M / M = 10^{-15}$ in the following analysis.

\begin{figure}
  \centering
  \begin{minipage}{0.5\linewidth}
    \includegraphics[width=1.0\linewidth]{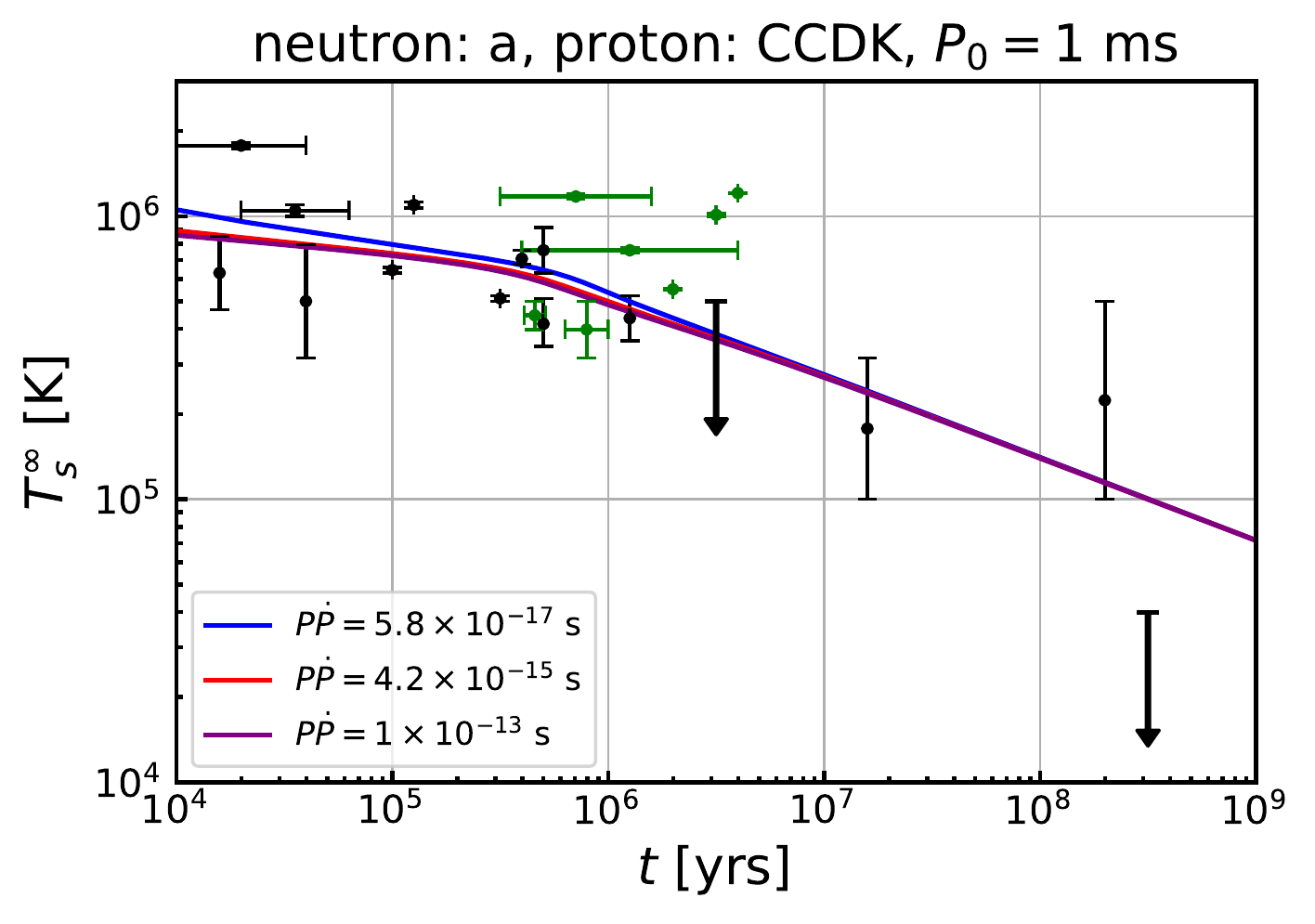}
  \end{minipage}%
  \begin{minipage}{0.5\linewidth}
    \includegraphics[width=1.0\linewidth]{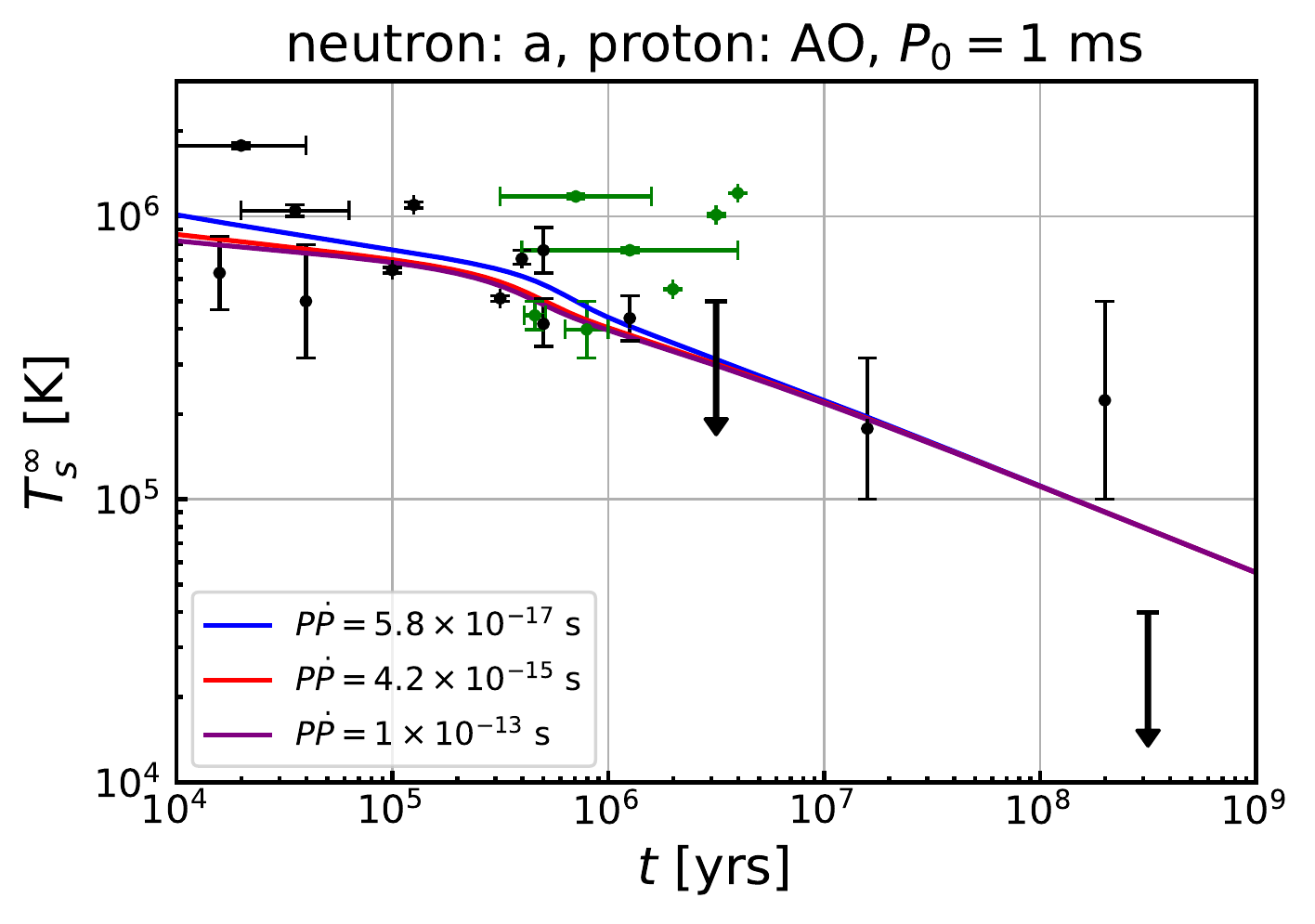}
  \end{minipage}
  \begin{minipage}{0.5\linewidth}
    \includegraphics[width=1.0\linewidth]{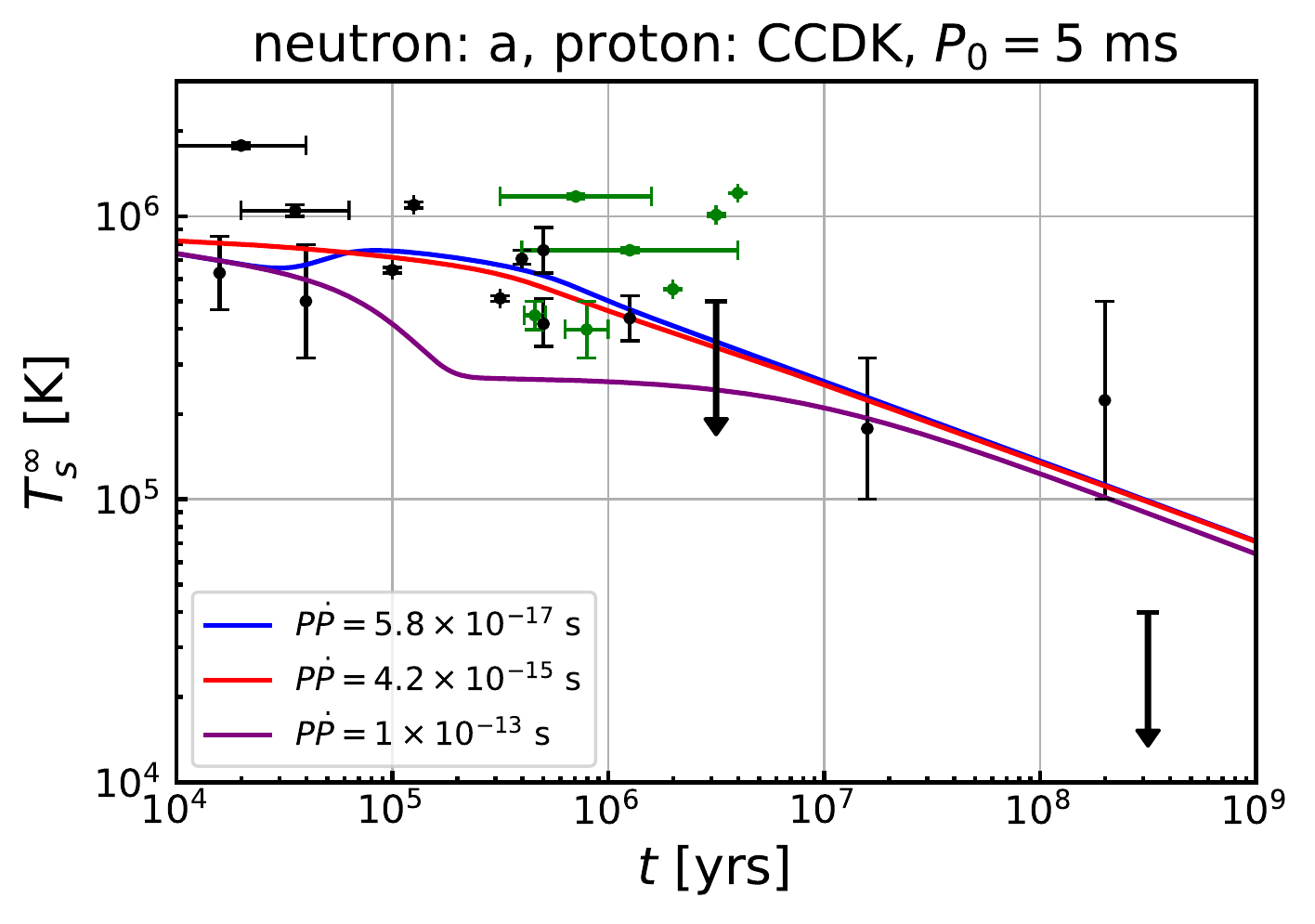}
  \end{minipage}%
 \begin{minipage}{0.5\linewidth}
   \includegraphics[width=1.0\linewidth]{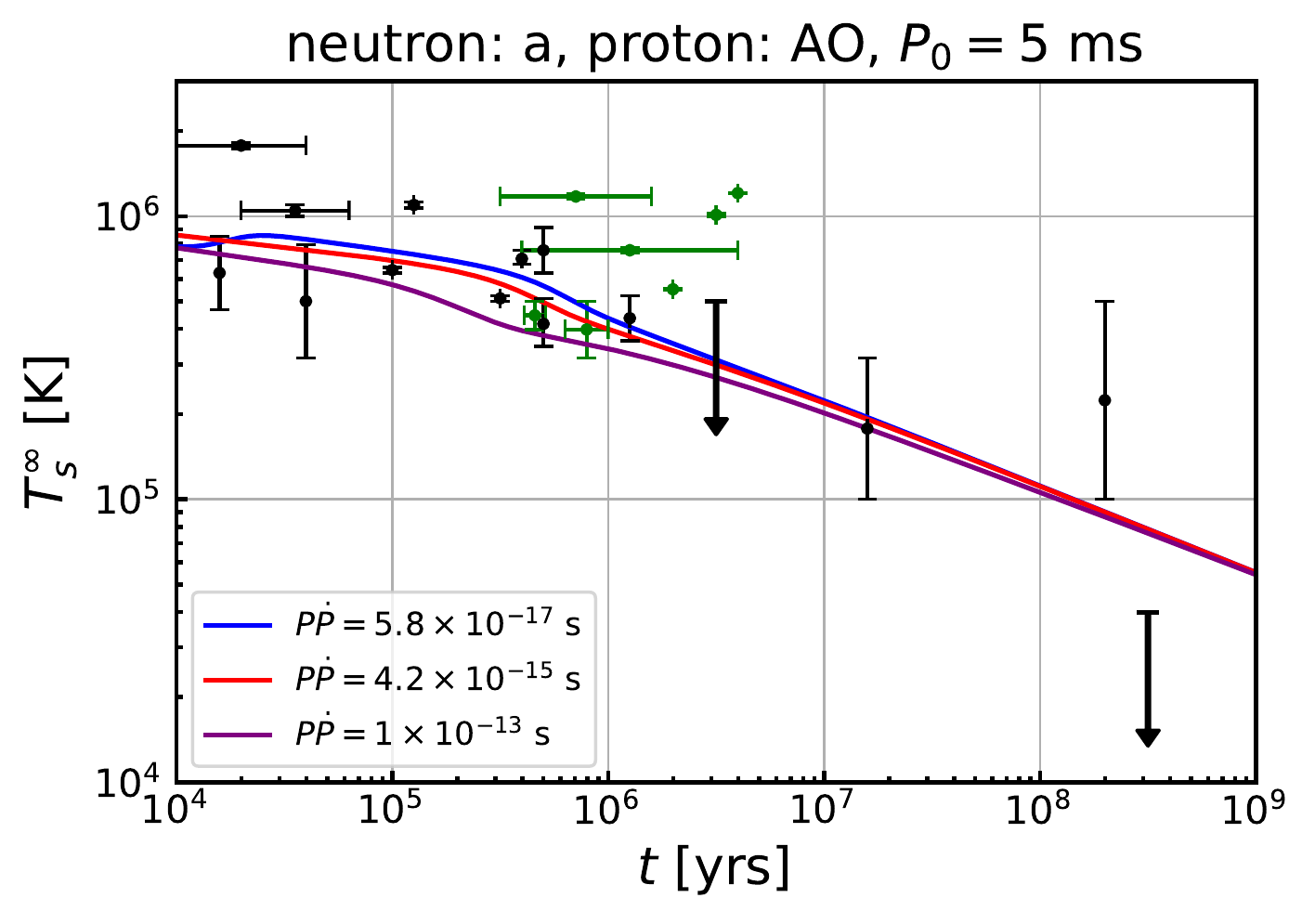}
  \end{minipage}
  \begin{minipage}{0.5\linewidth}
    \includegraphics[width=1.0\linewidth]{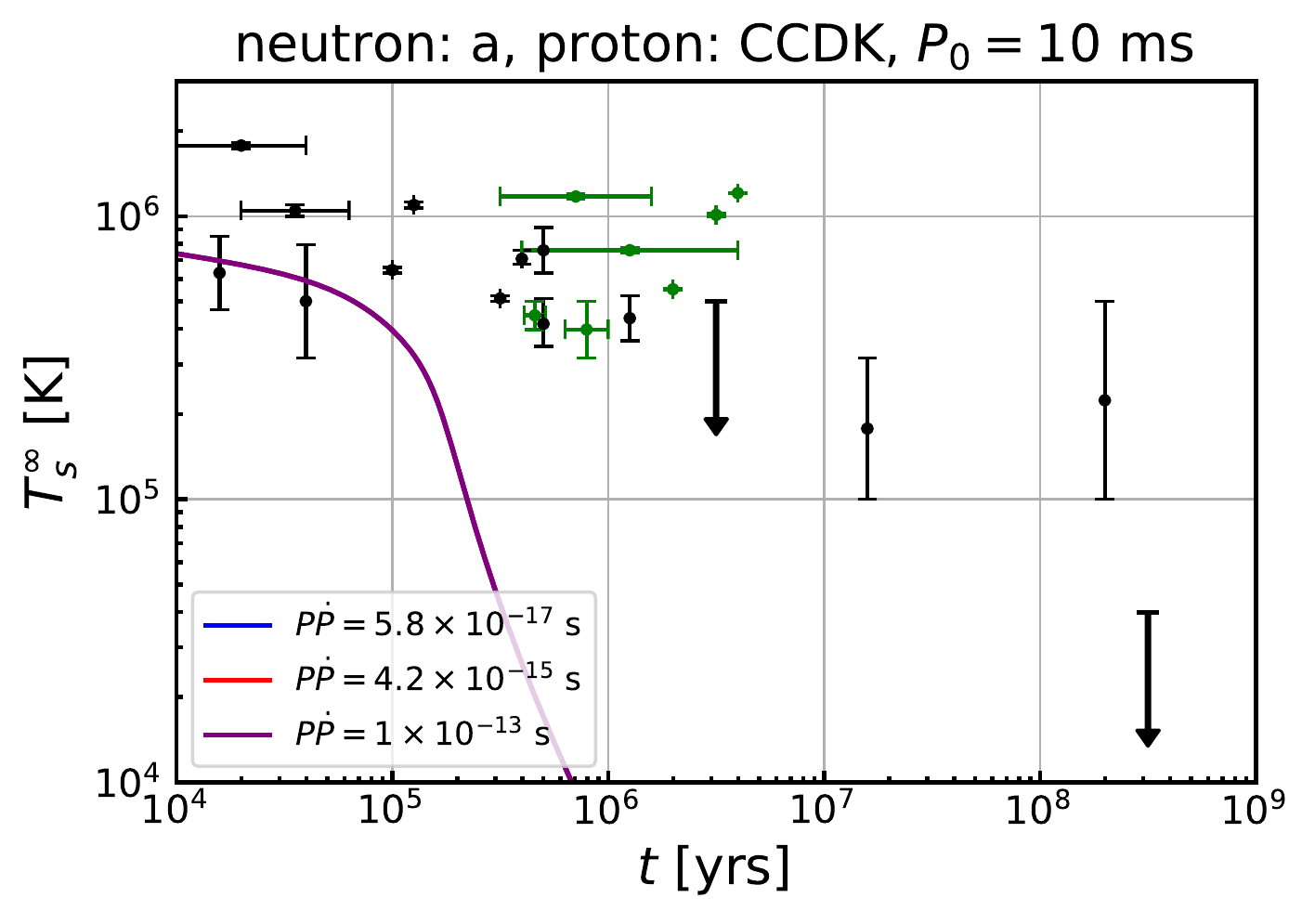}
  \end{minipage}%
  \begin{minipage}{0.5\linewidth}
    \includegraphics[width=1.0\linewidth]{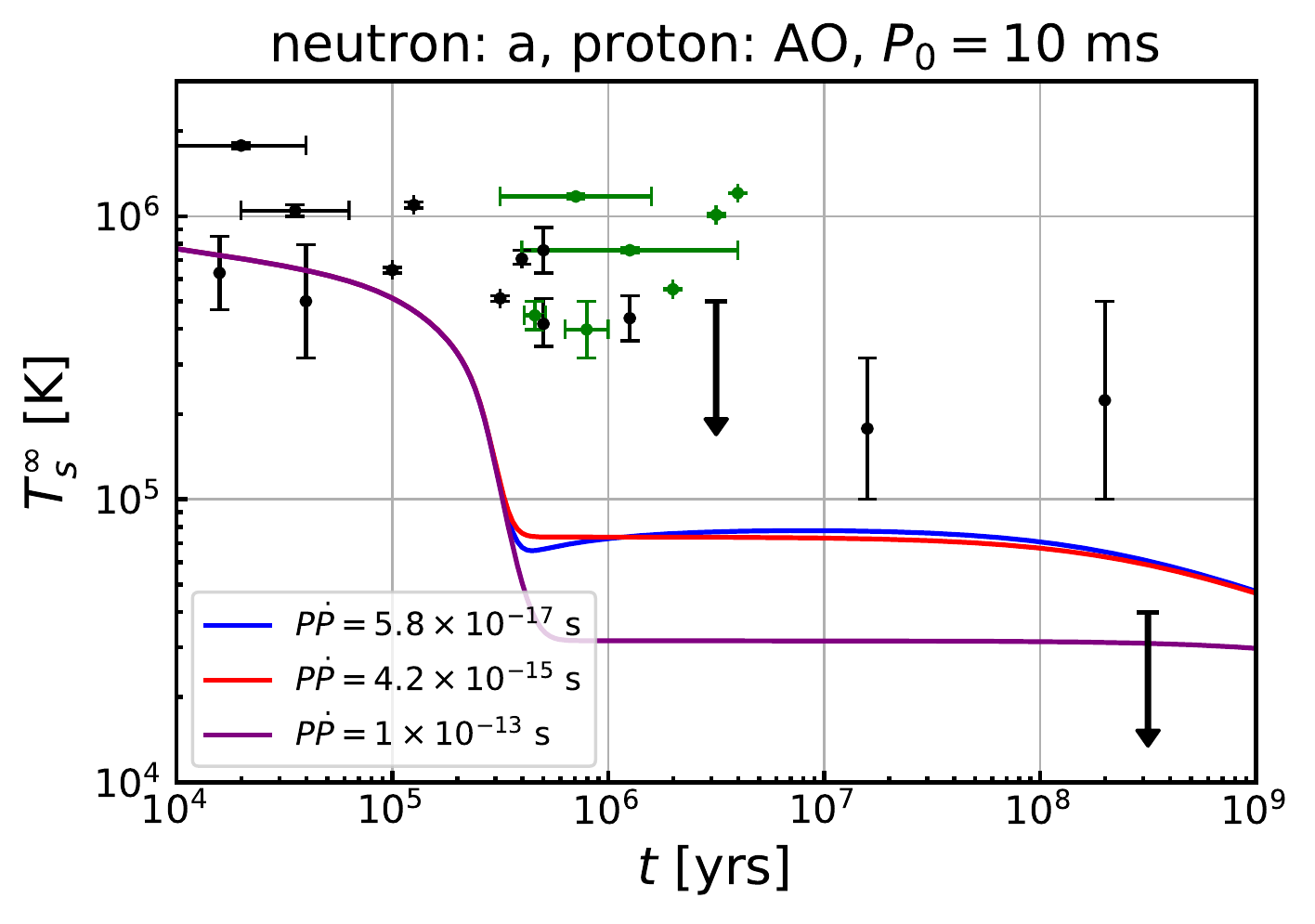}
  \end{minipage}
  \caption{Thermal evolution of ordinary pulsars for different values of $P\dot{P}$.
    We use the neutron ``a'' and proton CCDK (AO) gap models in the left (right) panels.
    The top, middle, and bottom panels show the cases of $P_0=1\unit{ms}$, $5\unit{ms}$, and $10\unit{ms}$, respectively.
    In each panel, we take three values of $P\dot{P}$: $P\dot{P} =5.8\times 10^{-17}\unit{s}$ (blue), $4.2\times10^{-15}\unit{s}$ (red) and $1\times10^{-13}\unit{s}$ (purple).
  The points with error bars are the same as Fig.~\ref{fig:cp-1}, showing the observed surface temperatures.}
  \label{fig:cp-3}
\end{figure}
Figure~\ref{fig:cp-3} shows the thermal evolution for $P\dot{P} = 5.8\times 10^{-17}\unit{s}$ (blue), $4.2\times10^{-15}\unit{s}$ (red), and $1\times10^{-13}\unit{s}$ (purple), corresponding to $B \sim 10^{11}\unit{G}$, $10^{12}\unit{G}$, and $10^{13}\unit{G}$, respectively. Note that $5.8\times10^{-17}\unit{s}$ is the observed value of $P\dot{P}$ for B0950+08 and $4.2\times10^{-15}\unit{s}$ is that for J2144-3933. We use the neutron ``a'' and proton CCDK (left panels) or AO (right panels) gap models.
The top, middle, and bottom panels show the cases of $P_0 = 1\unit{ms}$,
$5\unit{ms}$, and $10\unit{ms}$, respectively. For $P_0 =1\unit{ms}$
and $5\unit{ms}$, we see that the surface temperature is lower for a larger
$P\dot{P}$ at middle age. This is because with a larger $P\dot{P}$, rotochemical
heating begins earlier, resulting in a smaller $\eta_\ell$ at middle age. For $t \gtrsim 10^8\unit{yr}$, however, the surface temperatures of different $P\dot{P}$ converge to each other since the spin-down effect becomes so small that the evolution of $\eta_\ell$ is
mainly determined by the modified Urca process.
For $P_0 = 10\unit{ms}$, the rotochemical heating never occurs
for the proton CCDK gap model (bottom left). In the case of the proton AO model, on the other hand, the heating effect is small but still visible (bottom right).
Moreover, the difference of surface temperatures at $t\gtrsim 10^6\unit{yr}$ is
larger than that for $P_0 = 1\unit{ms}$ or $P_0 = 5\unit{ms}$.
In this case, the accumulated $\eta_\ell$ is close to the rotochemical
threshold, and hence the heating luminosity is very sensitive to the difference in the $\eta_\ell$, making the temperature evolution highly dependent on other NS parameters such as $P\dot{P}$.

Notice that although a large $P\dot{P}$ (and thus a large dipole magnetic field) tends to
predict low surface temperature, it does not help explain the low
temperature of J2144-3933 compared to that of J0108-1431 and B0950+08, as indicated by the red and blue curves in Fig.~\ref{fig:cp-3}, which are always close to each other.
Consequently, the different temperatures of these NSs should be attributed to the difference in $P_0$.

\chapter{Gap equation}
\label{chap:gap-equation}

In this appendix, we review the derivation of the gap equation beyond the
$s$-wave pairing. Following Ref.~\cite{1970PThPh..44..905T}, we introduce the
generalized Bogoliubov transformation. In Ref.~\cite{1970PThPh..44..905T}, gap
equation is derived at $T=0$ by the diagonalization of the Hamiltonian. Here we
choose the different path, and minimize thermodynamic energy, which directly
provides us the gap equation at $T\neq 0$.

\section{Potential}
\label{sec:potential}

The Cooper pairing occurs if the interaction is attractive on the Fermi surface.
Thus we need to know the potential between nucleons. In principle, nucleon
interaction consists of not only pair interaction but also three-body or
higher-order ones. Such higher order correction can be important for heavy NSs
where hyperons may appear. But these beyond-pair interactions have large
uncertainties. Thus we focus on the pair interaction in this appendix.

In a typical situation inside a NS, nucleons are non-relativistic or
semi-relativistic, so the interaction which depends on their momenta is
suppressed. Thus we write the potential as~\cite{10.1143/PTPS.39.23,
  Tamagaki:1968zz}
\begin{align}
  \label{eq:potential-general}
  V(\bm r)
  =
  V_C(r)
  +S_{12}V_T(r)
  +(\bm{L}\cdot\bm{S})V_{LS}(r)\,,
\end{align}
where $S_{12} = 3(\bm{\sigma}_1\cdot \hat{r})(\bm{\sigma}_2\cdot \hat{r}) -
\bm{\sigma}_1\cdot\bm{\sigma}_2$ with $\bm{\sigma}_{1,2}$ the spin of a nucleon,
$\bm{L} = \bm{r}\times\bm{p}$ the angular momentum and $\bm{S} = \bm{\sigma}_1 +
\bm{\sigma}_2$ the total spin. Note that $\bm r$ and $\bm p$ are relative
coordinate and relative momentum of pair nucleons, respectively. There are higher order
terms of the order $\bm{L}^2$, which are treated as the corrections to
Eq.~\eqref{eq:potential-general} when we consider the eigenstates of $J^2 = (\bm{L}+\bm{S})^2$,
$L^2$ and $S^2$~\cite{Tamagaki:1968zz}.

The first term in Eq.~\eqref{eq:potential-general} is scalar force, while the
second one is tensor force. This tensor force mixes the different angular
momentum state. Hence the eigenstates are not necessarily the eigenstate of
$\hat{L}^2$. The third term is spin-orbit interaction which serves as an
attractive force for the spin triplet pairing.

As in the ordinary quantum mechanics, the energy eigenstate is an
eigenstate of the total angular momentum $\hat{J}^2$ and its $J_z$. For
later convenience, we decompose it into the fixed orbital angular momentum
state. Thus we assume the eigenstate of the form 
\begin{align}
  \label{eq:eigenstate}
  \Ket{\Psi_\lambda}
=
\sum_\ell\Ket{\psi_{\lambda\ell}}
\Ket{\mathcal{Y}^{m_j}_{s\ell j}}\,,
\end{align}
where $\lambda$ is the label of energy, $s$, $\ell$, $j$ and $m_j$ are
labels of eigenvalues of $\hat{S}^2$, $\hat{L}^2$, $\hat{J}^2$ and $\hat{J}_3$
respectively, and 
\begin{align}
  \label{eq:y-func}
  \Ket{\mathcal{Y}^{m_j}_{s\ell j}}
=
\sum_{m_\ell}\Ket{s\ell m_s m_\ell}\Braket{s\ell m_s m_\ell|j m_j}\,,
\end{align}
with $m_s$ ($m_\ell$) being the eigenvalue of $\hat{S}_3$ ($\hat{L}_3$). The
coordinate representation of $\Ket{\mathcal{Y}^{m_j}_{s\ell j}}$ is
\begin{align}
  \label{eq:y-func-coord}
  \mathcal{Y}^{m_j}_{s\ell j}(\hat{r}, \sigma_1, \sigma_2)
=
  \Braket{\hat{r},\sigma_1,\sigma_2|\mathcal{Y}^{m_j}_{s\ell j}}
=
\sum_{m_\ell}
  \Braket{s\ell m_s m_\ell|j,m_j}Y_\ell^{m_\ell}(\hat{r})
\Braket{\frac{1}{2}\frac{1}{2}\sigma_1\sigma_2|sm_s}\,,
\end{align}
where $Y_\ell^{m_\ell}(\hat{r})$ is the spherical harmonics and $\sigma_i=\pm 1/2$
($i=1,2$) is the eigenvalue of $\hat{\sigma}_{i,z}$. Note that
$\Braket{\mathcal{Y}^{m_j}_{s\ell j}|\mathcal{Y}^{m_j}_{s\ell j}} =
\sum_{\sigma_1,\sigma_2}\int d\hat{r}|\mathcal{Y}^{m_j}_{s\ell j}(\hat{r},
\sigma_1, \sigma_2)|^2 = 1$.

Substituting these expressions into the Schr\"odinger equation, and taking the
inner product with $\Ket{\mathcal{Y}^{m_j}_{s\ell j}}$, we obtain
\begin{align}
  \label{eq:schrodinger-radial}
  \frac{1}{m}
  \left[
  \frac{1}{r^2}\frac{d}{dr}\left( r^2\frac{d}{dr} \right) - \frac{\ell(\ell+1)}{r^2} + E_\lambda
  \right]
  \psi_{\lambda\ell}(r)
  +
  \sum_{\ell^\prime}V_{j\ell\ell^\prime}(r) \psi_{\lambda\ell^\prime}(r)
  =
  0\,,
\end{align}
where
\begin{align}
  \label{eq:potential-ellell}
  V_{j\ell\ell^\prime}(\bm r)
  =
  \sum_{\sigma_1,\sigma_2}\int d\hat{r}\,
  \mathcal{Y}^{m_j*}_{s\ell j}(\hat{r},\sigma_1,\sigma_2)
  V(\bm r)
  \mathcal{Y}^{m_j}_{s\ell^\prime j}(\hat{r},\sigma_1,\sigma_2)\,.
\end{align}
Note that the spin $s$ is chosen according to even/odd of $\ell$ and $\ell^\prime$.

\subsection{\texorpdfstring{$s$}{s}-wave}
\label{sec:s-wave}

For the ${}^1S_0$ state, the angular part becomes
$\mathcal{Y}^0_{000}(\hat{r},\sigma_1, \sigma_2) =
\Braket{\frac{1}{2}\frac{1}{2}\sigma_1\sigma_2|0,0}/\sqrt{4\pi}$. Hence only the
scalar part of the potential contributes as
\begin{align}
  \label{eq:potential-s}
  V_{000}(r)
  =
  V_C(r)\,.
\end{align}
As we see in Fig.~\ref{fig:phase-shift}, this potential is attractive at long range, and
becomes repulsive at short distance.

\subsection{\texorpdfstring{$p$}{p}-wave}
\label{sec:p-wave}

For the ${}^3P_j$ states, we easily evaluate the scalar and spin-orbit part of the
potential as
\begin{align}
  \label{eq:potential-p-c-ls}
  V_{j11}^{C+LS}(r)
  =
  \begin{cases}
    V_C(r) - 2V_{LS}(r) & (j=0)
    \\
    V_C(r) - V_{LS}(r) & (j=1)
    \\
    V_C(r) + 2V_{LS}(r) & (j=2)
  \end{cases}\,.
\end{align}
The scalar interaction $V_C$ becomes repulsive at short range, whereas the spin-orbit one $V_{LS}$ becomes attractive~\cite{Wiringa:1994wb}.
Thus, Eq~\eqref{eq:potential-p-c-ls} shows only $j=2$ state can condensate. 

The tensor interaction is more complicated. In particular, the mixing term from
$\ell=3$ appears for ${}^3P_2$ states. Thus ${}^3P_2$ state is actually ${}^3P_2 +
{}^3F_2$ sate. This mixing interaction is, however, not large for the
energy range of our interest~\cite{1970PThPh..44..905T}. Therefore we can safely
ignore it.

\section{Derivation of gap equation}
\label{sec:derv-gap-eq}
Given the expressions of the potential, we can
perform the second quantization and derive the gap equation. As we discussed in Sec.~\ref{sec:potential}, the
mixing of different orbital angular momentum states is negligible. Thus we
assume $V_{j\ell\ell^\prime} = 0$ for $\ell \neq \ell^\prime$.

\subsection{Second quantization}
\label{sec:app-second-quantization}

We begin with the following second quantized Hamiltonian for pair interaction between normal
quasiparticles
\begin{align}
  \label{eq:hamiltonian}
  \hat H
  = \sum_{\bm k \sigma}\eta_k\hat{c}_{\bm k \sigma}^\dagger \hat{c}_{\bm k \sigma}
  +\frac{1}{2}\sum_{\bm k, \bm k^\prime}\sum_{\mathrm{spin}}
  \Braket{\bm{k}^\prime \sigma_1^\prime, -\bm{k}^\prime\sigma_2^\prime | \hat{V} |\bm{k}\sigma_1, -\bm{k}\sigma_2}
  \hat c_{\bm{k}^\prime \sigma_1^\prime}^\dagger \hat c_{-\bm{k}^\prime \sigma_2^\prime}^\dagger
  \hat c_{-\bm{k} \sigma_2} \hat c_{\bm{k} \sigma_1}\,,
\end{align}
where $\eta_k = \hbar^2k^2/2m^* - \mu$.%
\footnote{
  The ``Hamiltonian'' here is in fact $\hat H - \mu \hat N$.
  We call both $\hat H$ and $\hat{H} - \mu \hat{N}$ Hamiltonian. 
}
$\hat c_{\bm k\sigma}$ ($\hat c_{\bm k\sigma}^\dagger$) is the
annihilation (creation) operator which obeys the usual anti-commutation relation
\begin{align}
  \label{eq:com-rel}
  \{\hat c_{\bm k\sigma},\hat c_{\bm k^\prime\sigma^\prime}^\dagger \}
  =
  \delta_{\bm k,\bm k^\prime}\delta_{\sigma,\sigma^\prime}\,,\\
  \{\hat c_{\bm k\sigma},\hat c_{\bm k^\prime\sigma^\prime} \}
  =
  \{\hat c_{\bm k\sigma}^\dagger,\hat c_{\bm k^\prime\sigma^\prime}^\dagger \}
  =
  0\,.
\end{align}

For the following discussion, it is convenient to slightly modify the definition
of the quasiparticles (see also the end of Sec. 1 of Ref.~\cite{lifshitz2013statistical}).
Suppose that we excite a quasiparticle inside the Fermi sphere with
momentum $p < p_F$. The excited state has momentum $p^\prime > p_F$, and the
total energy increases by the sum of $v_F(p_F-p)$ and $v_F(p^\prime - p_F)$.
We can regard this process as the pair creation of a quasiparticle outside the
Fermi sphere and a \textit{hole} inside the sphere; the former has momentum
$p^\prime > p_F$, while the latter has $p < p_F$. The energy spectrum is
$\varepsilon = v_F(p-p_F)$ for quasiparticles and  $\varepsilon = v_F(p_F-p)$
for holes.%
\footnote{
  To treat quasiparticles and holes explicitly, one may introduce new creation/annihilation operators by
  \begin{align}
    \label{eq:qsqp-op}
    \hat{a}_{\bm{k}\sigma} =
    \begin{cases}
      \hat{c}_{\bm{k}\sigma} \quad k \geq k_F \\
      \hat{c}^\dagger_{\bm{k}\sigma} \quad k < k_F
    \end{cases}
  \end{align}
  Then the free energy spectrum becomes $\varepsilon = |p^2/2m - \mu|$.
}

This redefinition of quasiparticles is advantageous to focus on the process
around the Fermi surface because by the previous definition, the quasiparticles
deep inside the Fermi sphere, which do not participate in interactions, come
into the discussion. By the new definition, the number of quasiparticles is zero
at $T=0$, and is created by thermal fluctuation at $T\neq 0$.
The chemical potentials of quasiparticles and holes are zero due to the number non-conservation.
Hence their equilibrium distribution function is
\begin{align}
  \label{eq:dist-qpsh}
  f = \frac{1}{e^{\varepsilon/T} + 1}.
\end{align}

In the following, we focus on quasiparticle degrees of freedom: $k \geq k_F$.
The equations for the holes are obtained by $\eta_k \to \mu - \hbar^2k^2/2m^*$, or one may explicitly treat both degrees of freedom by substituting $\eta_k \to |\hbar^2k^2/2m^* - \mu|$.

We first introduce the fermion pair operator
\begin{align}
  \label{eq:fermion-pair-op}
  \hat b^\dagger_{\lambda m_j}(k)
  \equiv
  \frac{1}{\sqrt 2}
  \sum_{\sigma_1, \sigma_2}
  \Braket{\frac{1}{2}\frac{1}{2}\sigma_1\sigma_2|s m_s}
  \Braket{s\ell m_s m_\ell|j m_j}
  \int d\hat{k}Y_{\ell}^{m_\ell}(\hat{k})
  \hat c_{\bm k\sigma_1}^\dagger\hat c_{-\bm k\sigma_2}^\dagger\,,
\end{align}
where $\lambda = (j \ell s)$ specifies the total, orbital, and coupled spin angular
momenta, respectively, and $m$'s are their $z$-components. The brackets are
Clebsch-Gordan coefficients and $Y_\ell^{m_\ell}$
is the spherical harmonics. Note that spin angular
momentum takes either $s=0$ (singlet) or $s=1$ (triplet), and $m_j = m_s +
m_\ell$ and $m_s=\sigma_1+\sigma_2$ by the angular momentum conservation.

We rewrite the Hamiltonian using $b_{\lambda m_j}$. By the property of
Clebsch-Gordan coefficients and spherical harmonics, one can write the pair of
creation operators as
\begin{align}
  \label{eq:cc-b}
  \hat c_{\bm k\sigma_1}^\dagger\hat c_{-\bm k\sigma_2}^\dagger
  =
  \sqrt{2}\sum_{s,\ell,j,m_j}
  \Braket{\frac{1}{2}\frac{1}{2}\sigma_1\sigma_2|s m_s}
  \Braket{s\ell m_s m_\ell|j m_j}
  \hat b^\dagger_{\lambda m_j}(k)
  Y_\ell^{m_j*}(\hat k)\,.
\end{align}
It is convenient to rewrite the potential term by the spherical basis. 
The
matrix element of the pair interaction potential is put into
\begin{align}
  \label{eq:pot-spherical}
  \Braket{\bm{k}^\prime \sigma_1^\prime, -\bm{k}^\prime\sigma_2^\prime | \hat{V} |\bm{k}\sigma_1, -\bm{k}\sigma_2}
  =
  \delta_{\sigma_1\sigma_1^\prime}\delta_{\sigma_2\sigma_2^\prime}
  \Omega^{-1}\int d^3r e^{-i\bm{k}\cdot\bm{r}}e^{i\bm{k}^\prime\cdot\bm{r}}
  V(\bm r)
\end{align}
where $\Omega$ is the spatial volume and the potential is given by
Eq.~\eqref{eq:potential-general}.
Using the well-known decomposition of the plane wave to the spherical harmonics,
\begin{align}
  \label{eq:plane-spherical}
  e^{i\bm{k}\cdot\bm{r}}
  =
  4\pi\sum_{\ell,m_\ell}
  i^\ell j_\ell(kr)Y_\ell^{m_\ell}(\hat{r})Y_\ell^{m_\ell*}(\hat{k})\,,
\end{align}
where $j_\ell$ is the spherical Bessel function of the first kind, and neglecting the mixing of
different $\ell$'s, the second quantized
Hamiltonian is written as
\begin{align}
  \label{eq:hamiltonian-bb}
  \hat{H}
  = \sum_{\bm k \sigma}\eta_k\hat c_{\bm k \sigma}^\dagger\hat c_{\bm k, \sigma}
  +
  \frac{(4\pi)^2}{\Omega}
  \sum_{k,k^\prime}\sum_{j\ell}
  \Braket{k^\prime | V_\lambda | k}
  \sum_{m_j}\hat b^\dagger_{\lambda m_j}(k^\prime)\hat b_{\lambda m_j}(k)\,,
\end{align}
where $\sum_k = \frac{\Omega}{(2\pi)^3} \int dk k^2$, $\lambda = (j\ell s)$, and
\begin{align}
  \label{eq:potential-kk}
  \Braket{k^\prime | V_\lambda | k}
  =
  \int_0^\infty dr r^2 j_\ell(kr)j_\ell(k^\prime r)V_{j\ell\ell}(r)\,.
\end{align}

\subsection{Genelarized Bogoliubov transformation}
\label{sec:bog-trsf}
Let $\Ket{\Phi_0}$ be the ground state of the Fermi gas, i.e., $\Ket{\Phi_0}$ is the ground state for $V
\to 0$. The true ground state of the quasiparticles, $\Ket{\Psi_0}$, can be
related to $\Ket{\Phi_0}$ by an unitary operator as
\begin{align}
  \label{eq:bog-trasf}
  \Ket{\Psi_0} = e^{i\hat{S}}\Ket{\Phi_0}\,.
\end{align}
The Hermitian operator $\hat S$ is bosonic, and expanded as
\begin{align}
  \label{eq:S-exp}
  i\hat{S}
  &=
    \sum_{k,m_j}\left( \phi_{\lambda m_j}(k)\hat{b}^\dagger(k)_{\lambda m_j} - \mathrm{h.c.} \right)
   =
    \frac{1}{2}\sum_{\bm{k},\sigma_1,\sigma_2}\theta_\lambda(\bm{k}, \sigma_1, \sigma_2)
    \hat{c}^\dagger_{\bm{k}\sigma_1}\hat{c}^\dagger_{-\bm{k}\sigma_2} - \mathrm{h.c.}\,,
\end{align}
where
\begin{align}
  \label{eq:theta-def}
  \theta_\lambda(\bm{k}, \sigma_1, \sigma_2)
  =
  \sqrt{2}\sum_{m_j}
  \Braket{\frac{1}{2}\frac{1}{2}\sigma_1\sigma_2|s m_s}
  \Braket{s\ell m_s m_\ell|j m_j}
  Y_{\ell}^{m_\ell}(\hat{k})
  \phi_{\lambda m_j}(k)\,.
\end{align}
Note that inside the sum, $m_s = \sigma_1 + \sigma_2$ and $m_\ell = m_j - m_s$.

This parameter has following properties
\begin{align}
  \theta_\lambda(-\bm{k},\sigma_2,\sigma_1)
  &= - \theta_\lambda(\bm{k}, \sigma_1, \sigma_2)
    \label{eq:theta-antisym}\,,\\
  \theta_\lambda^*(-\bm{k},-\sigma_1,-\sigma_2)
  &=(-1)^{1-(\sigma_1+\sigma_2)}\theta_\lambda(\bm{k},\sigma_1,\sigma_2)
    \label{eq:theta-tinv}\,.
\end{align}
The second equation follows from the time-reversal invariance of $\hat{S}$:
$\hat{T} i\hat{S} \hat{T}^{-1} = i\hat{S}$.
From Eqs.~\eqref{eq:theta-def} and \eqref{eq:theta-tinv}, we obtain
\begin{align}
  \label{eq:phi-t-inv}
  \phi_{\lambda m_j}^*(k) = (-1)^{j+m_j}\phi_{\lambda -m_j}(k)\,.
\end{align}

\subsection{Quasiparticle operators}
\label{sec:quasiparticle-op}
The creation/annihilation operators of quasiparticles on the true ground state
$\Ket{\Psi_0}$ is obtained by the Bogoliubov transformation of
$\hat{c}_{\bm{k}\sigma}$ and $\hat{c}_{\bm{k}\sigma}^\dagger$. To make the
notation simple, we write these operator as the vector
\begin{align}
  \label{eq:c-vec}
  \hat{\bm c}_{\bm{k}}
  =
  \begin{pmatrix}
    \hat{c}_{\bm{k} \uparrow} \\ \hat{c}_{\bm{k} \downarrow}
  \end{pmatrix},\quad
  \hat{\bm c}_{\bm{k}}^\dagger
  =
  \begin{pmatrix}
    \hat{c}_{\bm{k} \uparrow}^\dagger \\ \hat{c}_{\bm{k} \downarrow}^\dagger
  \end{pmatrix}
\end{align}
and the transformation coefficients $\theta_\lambda$ as the matrix
\begin{align}
  \label{eq:theta-mat}
  \Theta_\lambda(\bm k)
=
  \begin{pmatrix}
    \theta_\lambda(\bm k, \uparrow, \uparrow) & \theta_\lambda(\bm k, \uparrow, \downarrow) \\
    \theta_\lambda(\bm k, \downarrow, \uparrow) & \theta_\lambda(\bm k, \downarrow, \downarrow)
  \end{pmatrix}\,.
\end{align}
$\Theta_\lambda(-\bm k)^\intercal = -\Theta_\lambda(\bm k)$ follows from Eq.~\eqref{eq:theta-antisym}.

Then the Bogoliubov transformation on the creation/annihilation operators is
written as 
\begin{align}
  \label{eq:quasi-part-op}
  &\hat{\bm \alpha}_{\bm{k}} 
    = e^{i\hat S}\hat{\bm c}_{\bm{k}}e^{-i\hat S}
    = U_\lambda(\bm k)\hat{\bm c}_{\bm k} - V_\lambda(\bm k)\hat{\bm c}_{-\bm{k}}^\dagger
    \notag\\
  &\hat{\bm \alpha}_{\bm{k}}^\dagger
    = e^{i\hat S}\hat{\bm c}_{\bm{k}}^\dagger e^{-i\hat S}
    = U_\lambda^*(\bm k)\hat{\bm c}_{\bm k}^\dagger - V_\lambda^*(\bm k)\hat{\bm c}_{-\bm{k}}
\end{align}
where
\begin{align}
  \label{eq:u-v-vec}
  U_\lambda(\bm k) &= \cos\Theta_\lambda^\prime(\bm k)
                     \notag\\
  V_\lambda(\bm k) &=  \Theta_\lambda^{\prime -1}(\bm k)\sin\Theta_\lambda^{\prime}(\bm k)\Theta_\lambda(\bm k)
                     \notag\\
  \text{with } \Theta_\lambda^{\prime 2}(\bm k) &= \Theta_\lambda(\bm
                     k)\Theta_\lambda^\dagger(\bm k)\,.
\end{align}

\subsubsection{Singlet pairing}

Let us see these generalized Bogoliubov transformation on the $s$-wave spin-singlet pairing.
For the spin singlet state, $j$ and $\ell$ are even, $m_j=m_\ell$ and
$\phi_{\lambda m_\ell}^*(k) = (-1)^{m_\ell}\phi_{\lambda -m_\ell}(k)$. Thus
\begin{align}
  \label{eq:theta-singlet}
  \Theta_\lambda(\bm k)
  =
  \begin{pmatrix}
    0 & \theta_A(\bm k)\\
    -\theta_A(\bm k) & 0
  \end{pmatrix}
                       ,\quad
                       \text{where }
                       \theta_A(\bm k) = \sum_{m_\ell} Y_\ell^{m_\ell}(\hat k)\phi_{\lambda m_\ell}(k)\,.
\end{align}
Note that $\theta_A(\bm k)^* = \theta_A(\bm k)$ and $\theta_A(-\bm k) =
\theta_A(\bm k)$. The transformation matrix is
written as
\begin{align}
  \label{eq:u-v-singlet}
  U_\lambda(\bm k)
  &=
    \cos\theta_A(\bm k)\bm{1}
    \notag\,,\\
  V_\lambda(\bm k)
  &=
    \sin\theta_A(\bm k)
    \begin{pmatrix}
      0 & 1 \\ -1 & 0
    \end{pmatrix}
                    =i\sigma_2\,.
\end{align}
The Bogoliubov transformation is reduced to the well-known formula
\begin{align}
  \label{eq:bogoliubov-singlet}
  \hat\alpha_{\bm k \uparrow}
  &= \cos\theta_A(\bm k)\hat{c}_{\bm k \uparrow}  - \sin\theta_A(\bm k)\hat{c}^\dagger_{-\bm k \downarrow}
    \,,\notag\\
  \hat\alpha_{\bm k \downarrow}
  &= \cos\theta_A(\bm k)\hat{c}_{\bm k \downarrow}  + \sin\theta_A(\bm k)\hat{c}^\dagger_{-\bm k \uparrow}\,.
\end{align}

\subsubsection{Triplet pairing: \texorpdfstring{${}^3P_2$}{2P3}}
The triplet pairings has more complicated structures, so we focus on ${}^3P_2$
pairing which is important for nucleon pairing in a NS core.
This state has $s=\ell=1$ and $j=2$. The transformation matrix is written as
\begin{align}
  \label{eq:theta-3p2}
  &\Theta_\lambda(\bm k)
    \notag\\
  &=
  \begin{pmatrix}
    \sqrt{2}Y_1^1\phi_{\lambda 2} + Y_1^0\phi_{\lambda 1} + Y_1^{-1}\phi_{\lambda 0}/\sqrt{3}
    &
    Y_1^1\phi_{\lambda 1}/\sqrt{2} + \sqrt{2/3}Y_1^0\phi_{\lambda 0} + Y_1^{-1}\phi_{\lambda -1}/\sqrt{2}
    \\
    Y_1^1\phi_{\lambda 1}/\sqrt{2} + \sqrt{2/3}Y_1^0\phi_{\lambda 0} + Y_1^{-1}\phi_{\lambda -1}/\sqrt{2}
    &
    \sqrt{2}Y_1^{-1}\phi_{\lambda -2} + Y_1^0\phi_{\lambda -1} + Y_1^{1}\phi_{\lambda 0}/\sqrt{3}
  \end{pmatrix}\,.
\end{align}
This is symmetric and odd under the space inversion: $\Theta_\lambda(-\bm k) =
-\Theta_\lambda(\bm k)$.
Then we find $\Theta_\lambda^{\prime 2} = \theta_D^2\bm{1}$, where
\begin{align}
  \label{eq:theta-d}
  \theta_D^2(\bm k)
  =
  \left| \sqrt{2}Y_1^1\phi_{\lambda 2} + Y_1^0\phi_{\lambda 1} + Y_1^{-1}\phi_{\lambda 0}/\sqrt{3} \right|^2
  +
  \left| Y_1^1\phi_{\lambda 1}/\sqrt{2} + \sqrt{2/3}Y_1^0\phi_{\lambda 0} + Y_1^{-1}\phi_{\lambda -1}/\sqrt{2}\right|^2\,.
\end{align}
Note that the definition of $\theta_D$ has sign ambiguity, but that does not
affect the following discussion.
Therefore, the transformation to quasiparticles is performed by
\begin{align}
  \label{eq:u-v-triplet}
  U_\lambda(\bm k)
  &=
    \cos\theta_D(\bm k)\bm{1}\,,
    \notag\\
  V_\lambda(\bm k)
  &=
    \sin\theta_D(\bm k)\cdot\Gamma_\lambda(\bm k)\,,
\end{align}
where $\Gamma_\lambda(\bm k) \equiv \Theta_\lambda(\bm k)/\theta_D(\bm k)$ is a
symmetric unitary matrix depending on $\phi_{\lambda m_j}$ which is to be determined
later.

In the following,  we focus on general spin-singlet case and special
spin-triplet case ${}^3P_2$.
One can easily check the property $V^\intercal(-\bm k) = -V(\bm k)$ for both cases.
Thus the inverse Bogoliubov transformation is written as follows
\begin{align}
  \hat{\bm c}_{\bm k} &= U_\lambda(\bm k)\hat{\bm\alpha}_{\bm k} + V_\lambda(\bm k)\hat{\bm\alpha}^\dagger_{-\bm k}\,,
                    \notag\\
  \hat{\bm c}^\dagger_{\bm k} &= U_\lambda(\bm k)\hat{\bm\alpha}_{\bm k}^\dagger + V_\lambda^*(\bm k)\hat{\bm\alpha}_{-\bm k}\,.
                                \tag{\ref{eq:inverse-bogoliubov}}
\end{align}

\subsection{Energy}
\label{sec:energy}
We assume the system is isotropic, so there is no spin dependence in the
quasiparticle energy and its distribution function.  The
quasiparticle number operator is written as
\begin{align}
  \label{eq:num-op}
  \hat{n}_{\bm k\sigma\sigma^\prime} = \hat{\alpha}_{\bm{k}\sigma}^\dagger \hat{\alpha}_{\bm{k}\sigma^\prime}\,.
\end{align}
Using the following $2\times2$ matrix
$G_{\lambda m_j}(\hat k)$
\begin{align}
  \label{eq:g-matrix}
G_{\lambda m_j}(\hat k)_{\sigma_1\sigma_2} =
\Braket{\frac{1}{2}\frac{1}{2}\sigma_1\sigma_2|sm_s}\Braket{s\ell m_s m_j-m_s|jm_j}Y_\ell^{m_j-m_s}(\hat k)\,.
\end{align}
We rewrite the fermion pair operator as
\begin{align}
  \label{eq:fermion-pair-op-2}
  \hat{b}_{\lambda m_j}^\dagger(k)
  &=
  \frac{1}{\sqrt 2}\int d\hat{k}\,
    \left(  \hat{\bm c}_{\bm k}^\dagger\right)^\intercal G_{\lambda m_j}(\hat k) \hat{\bm c}_{-\bm k}^\dagger
    \notag\\
    &=
    \frac{1}{\sqrt 2}\int d\hat{k}\,\mathrm{tr}
    \left[ -U_\lambda(\bm k)G_{\lambda m_j}(\hat k) V_\lambda^\dagger(\bm k) \hat{n}_{\bm k}^\intercal
      + V_\lambda^\dagger(\bm k) G_{\lambda m_j}(\hat k) U_\lambda(\bm k)(1-\hat{n}_{\bm k})\right]
      \notag\\
  &+ (\hat\alpha_{\bm k\sigma}^\dagger\hat\alpha_{-\bm k\sigma^\prime}^\dagger\,,
    \hat\alpha_{\bm k\sigma}\hat\alpha_{-\bm k\sigma^\prime} \text{ terms})
    \,,
\end{align}
The thermodynamic energy $E = \braket{\hat H}$ is written as
\begin{align}
  E
  &= E_0 + \sum_{\lambda}E_\lambda\,,
    \label{eq:energy-decomp}\\
  E_0
  &=
  \sum_{\bm k}\eta_k\left[
    \mathrm{tr}\left( V^\dagger(\bm k) V(\bm k)\right)
    + \mathrm{tr}\left( U(\bm k)U(\bm k) - V^\dagger(\bm k) V(\bm k) \right)f_{\bm k}
    \right]\,,
    \label{eq:energy-0} \\
  E_\lambda
  &=
    \frac{(4\pi)^2}{2\Omega}\sum_{kk^\prime}\sum_{m_j}\Braket{k^\prime | V_\lambda | k}
    \notag\\
  &\times
    \left| \int d\hat{k}\,
    \left[
    -\mathrm{tr}\left(U_\lambda(\bm k)G_{\lambda m_j}(\hat k) V_\lambda^\dagger(\bm k) \right) f_{\bm k}
    +\mathrm{tr}\left(V_\lambda^\dagger(\bm k) G_{\lambda m_j}(\hat k) U_\lambda(\bm k)\right)(1-f_{\bm k})
    \right]
    \right|^2
    \label{eq:energy-lambda}\,,
\end{align}
where $f_{\bm k}$ is the given by Eq.\eqref{eq:dist-qpsh}, coming from the expectation value of
$\hat n_{\bm k}$, and we have dropped the terms such as $\vev{\hat\alpha_{\bm k\sigma}^\dagger\hat\alpha_{-\bm k\sigma^\prime}^\dagger\hat\alpha_{\bm k^\prime\sigma}\hat\alpha_{-\bm k^\prime\sigma^\prime}}$
because it is suppressed by the additional volume factor $1/\Omega$.

Thermodynamic equilibrium is determined by minimizing the energy for fixed entropy.
Since the entropy is given by the ordinary combinatorial expression Eq.~\eqref{eq:entropy}, this is the minimization of the energy for fixed $f_{\bm k}$ with respect to $U_\lambda$ and $V_\lambda$. 
For simplicity, we assume that the interaction is
dominated only one angular momentum state $\lambda$.
In order to treat singlet and triplet pairing collectively, we use the following notation:
\begin{align}
  \label{eq:notation}
  &{}^1S_0:\quad
  u_{\bm k} = \cos\theta_A(\bm k)\,,\quad
  v_{\bm k} = \sin\theta_A(\bm k)\,,\quad
  \Gamma_{\bm k} = i\sigma_2\,,
  \notag\\
  &{}^3P_2:\quad
  u_{\bm k} = \cos\theta_D(\bm k)\,,\quad
  v_{\bm k} = \sin\theta_D(\bm k)\,,\quad
  \Gamma_{\bm k} = \Gamma_\lambda(\bm k)\,.
\end{align}
Note the relation $u_{\bm k}^2 + v_{\bm k}^2  =1$. Then the energy for a
specific angular state is written as
\begin{align}
  \label{eq:energy-simple}
  E
  &= \sum_{\bm k}2\eta_k\left[ v_{\bm k}^2 + (u_{\bm k}^2 - v_{\bm k}^2)f_{\bm k} \right]
    \notag\\
  &+
  \frac{(4\pi)^2}{2\Omega}\sum_{kk^\prime}\Braket{k^\prime | V_\lambda | k}
\sum_{m_j}
    \left| \int d\hat{k}\,
    \left[
    u_{\bm k}v_{\bm k}\mathrm{tr}\left( G_{\lambda m_j}(\hat k)\Gamma_{\bm k}^\dagger \right)(1-2f_{\bm k})
    \right]
    \right|^2\,.
\end{align}

\subsection{Gap equation}
\label{sec:app-gap-eq}

Now we minimize Eq.~\eqref{eq:energy-simple}. It is convenient to introduce the gap function:
\begin{align}
  \label{eq:gap-def}
  \Delta_{\lambda m_j}(k)
  &=
  -\frac{(4\pi)^2(-1)^{1-s}}{\Omega}
  \sum_{\bm k^\prime}\Braket{k^\prime|V_\lambda|k}
    \mathrm{tr}\left[ U_\lambda(\bm k^\prime)G^*_{\lambda m_j}(\hat{k}^\prime)V_\lambda(\bm k^\prime) \right]
    (1-2f_{\bm k^\prime})
  \notag\\
  &=
  -\frac{(4\pi)^2(-1)^{1-s}}{\Omega}
  \sum_{\bm k^\prime}\Braket{k^\prime|V_\lambda|k}
    u_{\bm k^\prime}v_{\bm k^\prime}\mathrm{tr}\left[ G^*_{\lambda m_j}(\hat{k}^\prime)\Gamma_{\bm k^\prime}\right]
    (1-2f_{\bm k^\prime})
\end{align}
Note that $\Delta_{\lambda m_j}^* = (-1)^{m_j}\Delta_{\lambda -m_j}$ for ${}^1S_0$ and ${}^3P_2$ pairings.
Then the energy is written as
\begin{align}
  \label{eq:energy-w-gap}
  E
  &= \sum_{\bm k}2\eta_k\left[ v_{\bm k}^2 + (u_{\bm k}^2 - v_{\bm k}^2)f_{\bm k} \right]
    -\frac{1}{2}\sum_{\bm k}\sum_{m_j}\Delta_{\lambda m_j}(k)u_{\bm k}v_{\bm k}\mathrm{tr}\left[ G_{\lambda m_j}(\hat k)\Gamma_{\bm k}^\dagger \right] (1-2f_{\bm k})
\notag\\
&=
   \sum_{\bm k}2\eta_k\left[ v_{\bm k}^2 + (u_{\bm k}^2 - v_{\bm k}^2)f_{\bm k} \right]
    -\frac{1}{2}
           \sum_{\bm k}u_{\bm k}v_{\bm k}
           \mathrm{tr}\left[ \bm\Delta_\lambda (\bm k)\Gamma_{\bm k}^\dagger \right] (1-2f_{\bm k})\,,
\end{align}
where $\bm\Delta_\lambda(\bm k) \equiv \sum_{m_j}\Delta_{\lambda m_j}(k)G_{\lambda
  m_j}(\hat k)$.

\subsubsection{Singlet pairing}
The energy depends on the gap function only through $\Delta_\lambda(\bm k) \equiv \frac{1}{\sqrt 2}\sum_{m_\ell}\Delta_{\lambda m_\ell}(k)Y_\ell^{m_\ell}(\hat k)$ which is a real function.
Thus we write it as
\begin{align}
  \label{eq:energy-singlet}
  E
  &= \sum_{\bm k}2\eta_k\left[ v_{\bm k}^2 + (u_{\bm k}^2 - v_{\bm k}^2)f_{\bm k} \right]
    -\sum_{\bm k}u_{\bm k}v_{\bm k}\Delta_{\lambda}(\bm k)(1-2f_{\bm k})
\end{align}
Minimizing $E$ with respect to $u_{\bm k}$ leads to
\begin{align}
  \label{eq:minimize-singlet}
  (u_{\bm k}^2 - v_{\bm k}^2)\Delta_{\lambda}(\bm k) = 2\eta_ku_{\bm k}v_{\bm k}\,,
\end{align}
which gives
\begin{align}
  \label{eq:uv-sol-singlet}
  u_{\bm k}^2
  &= \frac{1}{2}\left( 1 + \frac{\eta_k}{\sqrt{\eta_k^2 + \Delta_\lambda(\bm k)^2}} \right)\,,\notag\\
  v_{\bm k}^2
  &= \frac{1}{2}\left( 1 - \frac{\eta_k}{\sqrt{\eta_k^2 + \Delta_\lambda(\bm k)^2}} \right)\,,\notag\\
  u_{\bm k}v_{\bm k}
  &= \frac{1}{2}\frac{\Delta_\lambda(\bm k)}{\sqrt{\eta_k^2 + \Delta_\lambda(\bm k)^2}}
\end{align}
Using Eqs.~\eqref{eq:gap-def} and \eqref{eq:uv-sol-singlet}, we obtain the gap
equation
\begin{align}
  \label{eq:gap-eq-singlet}
  \Delta_\lambda(\bm k)
  =
  -\frac{2}{\pi}
  \int d^3k^\prime\Braket{k^\prime | V_\lambda | k}
  \frac{\Delta_\lambda(\bm k^\prime)}{2\sqrt{\eta_{k^\prime}^2 + \Delta_\lambda(\bm k^\prime)^2}}
  \sum_{m_\ell}\left| Y_\ell^{m_\ell}(\hat k^\prime) \right|^2(1-2f_{\bm k^\prime})\,.
\end{align}

In particular, the gap equation for ${}^1S_0$ pairing ($\lambda = (s\ell j) =
(000)$) becomes
\begin{align}
  \label{eq:gap-eq-1s0}
  \Delta_{000}(\bm k)
  =
  -\frac{1}{4\pi^2}
  \int d^3k^\prime\Braket{k^\prime | V_{000} | k}
  \frac{\Delta_{000}(\bm k^\prime)}{\sqrt{\eta_{k^\prime}^2 + \Delta_{000}(\bm k^\prime)^2}}
  (1-2f_{\bm k^\prime})\,.
\end{align}
The integrand does not have angular dependence, thus the ${}^1S_0$ pairing is
isotropic: $\Delta_{000}(\bm k) = \Delta_{000}(k)$. 

\subsubsection{Tripelt pairing: \texorpdfstring{${}^3P_2$}{2p3}}
In the case of ${}^3P_2$ pairing, there is an additional degree of freedom in
$\Gamma_{\bm k} = \Gamma_\lambda(\bm k)$.
We need to minimize Eq.~\eqref{eq:energy-w-gap} with respect to both $\theta_D(\bm k)$ and $\Gamma_\lambda(\bm k)$.
This results in the same expressions as Eqs.~\eqref{eq:uv-sol-singlet} where $\Delta_\lambda(\bm k)$ in this case is given by%
\footnote{
  Varying Eq.~\eqref{eq:energy-w-gap} by $\Gamma_\lambda$ leads to $\bm\Delta_\lambda\Gamma_\lambda^\dagger = \Gamma_\lambda\bm\Delta_\lambda^\dagger$: $\bm\Delta_\lambda\Gamma_\lambda^\dagger$ is hermitian.
  The components of $\Gamma_\lambda$ satisfy $(\Gamma_\lambda^*)_{-\sigma,-\sigma^\prime} = (-1)^{\sigma+\sigma^\prime}(\Gamma_\lambda)_{\sigma,\sigma^\prime}$.
  The similar relation holds for $\bm\Delta_\lambda$.
  The hermiticity thus leads to $\bm\Delta_\lambda\Gamma_\lambda^\dagger = \Gamma_\lambda\bm\Delta_\lambda^\dagger = \Delta_\lambda \bm{1}$, where $\Delta_\lambda$ is a real number.}
\begin{align}
  \label{eq:gap-triplet}
  \Delta_\lambda^2(\bm k)
  \equiv \frac{1}{2}\mathrm{tr}\left[ \bm\Delta_\lambda(\bm k)\bm\Delta_\lambda^\dagger(\bm k) \right]
  = \sum_{\mu\nu}\Delta_{\lambda\mu}(k)\Delta_{\lambda\nu}^*(k)
  \mathrm{tr}\left[ G_{\lambda\mu}(\hat k)G_{\lambda\nu}^*(\hat k) \right]\,.
\end{align}
The gap equation \eqref{eq:gap-def} for $\lambda = (112)$ is
\begin{align}
  \label{eq:gap-eq-triplet}
  \Delta_{\lambda m_j}(k)
  =
  -\frac{1}{\pi}\int d^3k^\prime
  \Braket{k^\prime | V_\lambda |k}
  \frac{1}{\sqrt{\eta_{k^\prime}^2 + \Delta_\lambda(\bm k^\prime)^2}}
  (1-2f_{\bm k^\prime})
  \sum_\mu \Delta_{\lambda \mu}(k^\prime)
  \mathrm{tr}\left[ G_{\lambda\mu}(\hat k^\prime)G_{\lambda m_j}^*(\hat k^\prime) \right]\,.
\end{align}

The triplet pairing gap is generically complicated since different angular
momentum states are coupled. In the literature, it is
often assumed that the gap amplitude is determined by one angular momentum state
with definite $|m_j|$, which simplifies summation over $m_j$ appearing
Eqs.~\eqref{eq:gap-triplet} and \eqref{eq:gap-eq-triplet}.

Furthermore, the gap amplitude has angular dependence originating from $G_{\lambda m_j}$. In general the gap
depends on both polar and azimuthal angles with respect to the quantization
axis. But in the following two cases, the gap simplifies to the function which
only depends on the polar angle:
\begin{itemize}
\item
  For $m_j = 0$, the gap amplitude is calculated as
  \begin{align}
    \label{eq:gap-mj-0}
    \Delta_{\lambda}^2 =\frac{1}{16\pi}|\Delta_{\lambda 0}|^2(1+3\cos^2\theta)\,.
  \end{align}
  The gap $\Delta_{\lambda 0}$ is real and the gap equation is written as
  \begin{align}
    \label{eq:gap-eq-mj-0}
    \Delta_{\lambda 0}(k)
    =
    -\frac{1}{\pi}\int d^3k^\prime
    \Braket{k^\prime | V_\lambda |k}
    \frac{\Delta_{\lambda 0}(k^\prime)}{\sqrt{\eta_{k^\prime}^2 + \Delta_\lambda(\bm k^\prime)^2}}
    (1-2f_{\bm k^\prime})
    \frac{1}{8\pi}(1 + 3\cos^2\theta)\,.
  \end{align}

\item
  For $m_j=\pm2$, since $\Delta_{\lambda 2}^* = \Delta_{\lambda -2}$, we need to
  consider both $\Delta_{\lambda 2}$ and $\Delta_{\lambda -2}$. The gap function is
  \begin{align}
    \label{eq:gap-mj-2}
    \Delta_\lambda^2(\bm k) = \frac{3}{8\pi}|\Delta_{\lambda 2}(k)|^2\sin^2\theta\,,
  \end{align}
  and the gap equation becomes
  \begin{align}
  \label{eq:gap-eq-mj-2}
  \Delta_{\lambda m_j}(k)
  =
  -\frac{1}{\pi}\int d^3k^\prime
  \Braket{k^\prime | V_\lambda |k}
  \frac{\Delta_{\lambda m_j}(k^\prime)}{\sqrt{\eta_{k^\prime}^2 + \Delta_\lambda(\bm k^\prime)^2}}
    (1-2f_{\bm k^\prime})
    \frac{3}{8\pi}\sin^2\theta\,.
  \end{align}
  Thus the gap equation decouples for each $m_j=\pm2$, and we can take $\Delta_{\lambda 2} = \Delta_{\lambda -2}$ to be real.
\end{itemize}

\subsection{BCS limit}
\label{sec:app-bcs-limit}

The momentum integral in the gap equation is dominated near the
Fermi surface. In the BCS limit, we ignore the momentum dependence in the
potential and gap, keeping only the angular dependence of gap, and perform the
integration in the vicinity of the Fermi surface. Furthermore,
in order to collectively write the gap equations \eqref{eq:gap-eq-1s0}, \eqref{eq:gap-eq-mj-0} and
\eqref{eq:gap-eq-mj-2}, we define $\delta^2 = \Delta(T)^2F(\theta) = \Delta_{\lambda}^2(\bm k)$, and use
$\lambda$ and $F(\theta)$ defined in Tab.~\ref{tab:gap-angle}.
Consequently, we obtain the gap equation presented in Sec.\ref{sec:gap-eq}
\begin{align}
  \frac{g}{2}\int\frac{d^3p}{(2\pi\hbar)^3}
  \frac{1-2f(\bm p)}{\sqrt{\eta_p^2 + \delta^2}}\cdot\lambda F(\theta)
  = 1\,,
  \tag{\ref{eq:gap-eq}}
\end{align}
where we have defined the coupling constant
$g = -2(2\pi\hbar^3)\Braket{k_F^\prime |V_\lambda | k_F}/4\pi^2$.

\subsection{Quasiparticle energy spectrum}
\label{sec:app-quasi-energy-spectrum}

The gap $\delta$ corresponds to the actual energy gap of the
quasiparticle spectrum. By definition (see Eq.~\eqref{eq:energy}), the energy
spectrum is calculated by varying total energy $E$ by $f_{\bm k}$. Using
Eqs.~\eqref{eq:energy-singlet} and \eqref{eq:uv-sol-singlet}, we obtain
\begin{align}
  \label{eq:energy-spectrum-w-gap}
  \frac{\delta E}{\delta f(\bm p)}
  =
  2\sqrt{\eta_{p}^2 + \delta^2}\,.
\end{align}
The prefactor $2$ is due to our assumption of isotropy with which the
quasiparticle is created by a pair of up and down spin particles.
Thus $\sqrt{\eta_{p}^2 + \delta^2}$
corresponds to the energy of a single quasiparticle.
The hole part is calculated in the same way and provides the same gap equation and the energy spectrum.
This shows the existence of the finite energy gap between the ground state and excited state.

The energy spectrum for original quasiparticles, defined in Sec.~\ref{sec:fermi-liquid-theory}, is obtained as follows.
For $p > p_F$, the spectrum and distribution are the same, so that $\varepsilon = \mu + \sqrt{\eta_{p}^2 + \delta^2}$.
For $p < p_F$, excitation of a hole corresponds to the removal of a quasiparticle.
Thus the hole distribution $f_h$ and the quasiparticle distribution $f_p$ is related as $1-f_q = f_h$.
The energy spectrum for $p < p_F$ is then read as $\varepsilon = \mu - \sqrt{\eta_{p}^2 + \delta^2}$.
In summary, the energy spectrum is written as 
\begin{align}
  \tag{\ref{eq:energy-spectrum-gap}}
  \varepsilon(\bm p) \simeq \mu + \mathrm{sign}(p-p_F)\sqrt{v_F^2(p-p_F)^2 + \delta^2}\,.
\end{align}

\bibliographystyle{utphysmod}
\bibliography{thesis} 

\end{document}